\documentclass[twocolumn]{aastex631}
\usepackage{enumitem}
\usepackage{apjfonts,capt-of}
\usepackage{multirow}
\usepackage{enumitem}
\usepackage[section]{placeins}
\usepackage{hyperref}
\newcommand{\W}{{$\lambda$}}
\newcommand{\CH}[1]{\colhead{#1}}

\newcommand{\T}{\color{teal}$\checkmark$\color{black}}  
\newcommand{\X}{\color{black}$\times$}

\begin{document}
\shortauthors{Parker et al.}
\shorttitle{CLASSY~XI}

\author[0000-0002-8809-4608]{Kaelee S. Parker}
\affiliation{Department of Astronomy, The University of Texas at Austin, 2515 Speedway, Stop C1400, Austin, TX 78712, USA}

\author[0000-0002-4153-053X]{Danielle A. Berg}
\affiliation{Department of Astronomy, The University of Texas at Austin, 2515 Speedway, Stop C1400, Austin, TX 78712, USA}

\author[0000-0002-5659-4974]{Simon Gazagnes}
\affiliation{Department of Astronomy, The University of Texas at Austin, 2515 Speedway, Stop C1400, Austin, TX 78712, USA}

\author[0000-0002-0302-2577]{John Chisholm}
\affiliation{Department of Astronomy, The University of Texas at Austin, 2515 Speedway, Stop C1400, Austin, TX 78712, USA}

\author[0000-0003-4372-2006]{Bethan L. James}
\affiliation{AURA for ESA, Space Telescope Science Institute, 3700 San Martin Drive, Baltimore, MD 21218, USA}

\author[0000-0001-8587-218X]{Matthew Hayes}
\affiliation{Stockholm University, Department of Astronomy and Oskar Klein Centre for Cosmoparticle Physics, AlbaNova University Centre, SE-10691, Stockholm, Sweden}

\author[0000-0003-1127-7497]{Timothy Heckman}
\affiliation{Center for Astrophysical Sciences, Department of Physics \& Astronomy, Johns Hopkins University, Baltimore, MD 21218, USA}

\author[0000-0002-6586-4446]{Alaina Henry}
\affiliation{Space Telescope Science Institute, 3700 San Martin Drive, Baltimore, MD 21218, USA}

\author[0000-0002-8518-6638]{Michelle A. Berg}
\affiliation{Department of Astronomy, The University of Texas at Austin, 2515 Speedway, Stop C1400, Austin, TX 78712, USA}

\author[0000-0002-2644-3518]{Karla Z. Arellano-C\'{o}rdova}
\affiliation{Department of Astronomy, The University of Texas at Austin, 2515 Speedway, Stop C1400, Austin, TX 78712, USA}

\author[0000-0002-9217-7051]{Xinfeng Xu}
\affiliation{Center for Astrophysical Sciences, Department of Physics \& Astronomy, Johns Hopkins University, Baltimore, MD 21218, USA}

\author[0000-0001-9714-2758]{Dawn K. Erb}
\affiliation{Center for Gravitation, Cosmology and Astrophysics, Department of Physics, University of Wisconsin Milwaukee, 3135 N Maryland Ave., Milwaukee, WI 53211, USA}

\author[0000-0001-9189-7818]{Crystal L. Martin}
\affiliation{Department of Physics, University of California, Santa Barbara, Santa Barbara, CA 93106, USA}

\author[0000-0003-3424-3230]{Weida Hu}
\affiliation{Department of Physics and Astronomy, Texas A\&M University, College Station, TX 77843-4242, USA}
\affiliation{George P. and Cynthia Woods Mitchell Institute for Fundamental Physics and Astronomy, Texas A\&M University, College Station, TX 77843-4242, USA}

\author[0000-0003-0605-8732]{Evan D. Skillman}
\affiliation{Minnesota Institute for Astrophysics, University of Minnesota, 116 Church Street SE, Minneapolis, MN 55455, USA}

\author[0000-0001-5538-2614]{Kristen B. W. McQuinn}
\affiliation{Rutgers University, Department of Physics and Astronomy, 136 Frelinghuysen Road, Piscataway, NJ 08854, USA}
\affiliation{Space Telescope Science Institute, 3700 San Martin Drive, Baltimore, MD 21218, USA}

\author[0000-0002-2178-5471]{Zuyi Chen}
\affiliation{Steward Observatory, The University of Arizona, 933 N Cherry Ave, Tucson, AZ, 85721, USA}

\author[0000-0001-6106-5172]{Dan P. Stark}
\affiliation{Steward Observatory, The University of Arizona, 933 N Cherry Ave, Tucson, AZ, 85721, USA}

\correspondingauthor{Kaelee Parker} 
\email{kaelee.parker@utexas.edu}

\title{CLASSY XI: Tracing Neutral Gas Properties using UV Absorption Lines and 21-cm Observations\footnote{
Based on observations made with the NASA/ESA Hubble Space Telescope,
obtained from the Data Archive at the Space Telescope Science Institute, which
is operated by the Association of Universities for Research in Astronomy, Inc.,
under NASA contract NAS 5-26555.}}

\begin{abstract}
Rest-frame far-ultraviolet (FUV) observations from JWST are revolutionizing our understanding of the high-$z$ galaxies that drove reionization and the mechanisms by which they accomplished it. To fully interpret these observations, we must be able to diagnose how properties of the interstellar medium (ISM; e.g., column density, covering fraction, outflow velocity) directly relate to the absorption features produced. Using the high-S/N and high-resolution FUV spectra of 45 nearby star-forming galaxies from CLASSY, we present the largest uniform, simultaneous characterization of neutral and low-ionization state (LIS) interstellar UV absorption lines (\ion{O}{1}, \ion{Si}{2}, \ion{S}{2}, \ion{C}{2}, \ion{Al}{2}) across a wide range of galaxy properties. We also present 21-cm \ion{H}{1} observations for 35 galaxies, multiple of which are gas-poor or non-detected, possibly indicating the onset of a post-starburst phase. We find that our simultaneous 1-component Voigt profile fits are capable of accurately modeling the LIS absorption for $\sim$75\% of galaxies, mitigating challenges associated with saturation, infilling, and degeneracies. While the most massive galaxies require additional components, our 1-component fits return average properties of the absorbing gas and follow the scaling relations described by a single gas cloud. We explore connections between LIS absorption and direct tracers of the neutral ISM (\ion{O}{1}, Ly$\alpha$, \ion{H}{1} 21-cm), finding that \ion{C}{2} most closely traces the neutral gas trends while other ions exhibit weaker correlations. Given the challenges with directly observing \ion{H}{1} at higher-$z$, we demonstrate that LIS absorption can be a powerful means to study the neutral ISM and present empirical relationships for predicting neutral gas properties.
\end{abstract}


\section{Introduction} \label{intro}
The interstellar medium (ISM) is an important component of
star-forming galaxies and junction to the surrounding intergalactic medium (IGM). 
In particular, the neutral and low-energy ISM acts as a conduit for the exchange of matter and energy between these systems -- a role that is critical in governing how galaxies evolve.
Pristine neutral gas can fuel star formation, later mixing with the metals and dust that were produced by these stars. 
Furthermore, the \ion{H}{1} gas and dust reservoirs that host galaxies 
and active galactic nuclei are responsible for regulating the contribution of their ionizing photons into the surrounding IGM, thereby influencing characteristics of the universe's reionization, known as the Epoch of Reionization (EoR).

To approach answers for many of the open questions regarding EoR, it is crucial to 
measure properties of the neutral ISM (i.e., $\leq$ 13.6 eV). 
The amount of high-energy photons that can escape from star-forming galaxies and 
ionize the IGM depends largely on the distribution and density of the \ion{H}{1} gas 
within the galaxy and surrounding it (circumgalactic medium; CGM). 
Lyman continuum photons (LyC; $< 912$\AA) are responsible for ionizing hydrogen gas 
but they cannot be directly observed at high redshifts due to the opacity of the IGM 
\citep{Fan_2006,Bosman_2018,Bosman_2022,Yang_2020}. 
Many indirect diagnostics instead rely on features produced by the neutral gas, 
with Ly$\alpha$ being one of the most robust. 
The emergent spectral shape of Ly$\alpha$ emission is linked to the distribution 
and kinematics of neutral hydrogen gas in the galaxy's ISM and CGM; 
\citep[e.g.,][]{Verhamme_2015, Dijkstra_2016, Verhamme_2017, Izotov_2018, Marchi_2018,Hayes_2023}. 
While this encodes Ly$\alpha$ profiles with a wealth of information, 
it can also make the analysis of this line particularly complex and difficult to interpret.

As an alternative to Ly$\alpha$, a variety of rest-frame far-UV (FUV) absorption lines produced by low-ionization state (LIS) metals are often assumed to be closely associated with \ion{H}{1} gas.
Owing to the similar ionization energies of LIS ions and \ion{H}{1},
they are largely expected to exist in the same gas clouds when observed at similar kinematics.
LIS absorption profiles often share similar attributes that can be traced back 
to characteristics of the absorbing gas, such as its kinematics, column density, and geometry. This indicates that LIS absorption may act as a robust diagnostic for LyC escape. 
However, the formation of FUV metal lines is complex and the analysis of these lines can require high-S/N and high-spectral-resolution spectra. 
Furthermore, the investigation of nearby galaxies spanning diverse properties is necessary to disentangle how the properties of LIS absorption lines relate back to both the neutral gas in and around galaxies, as well as their larger-scale characteristics.

Interstellar absorption lines from \ion{Si}{2} and \ion{C}{2}, for example, are often assumed to originate in the same gas clouds that host \ion{H}{1}. However, this is not necessarily a realistic assumption because these ions can exist at energies higher than 13.6 eV, which means they can also co-exist with some \ion{H}{2} gas. In this work, we compare characteristics of multiple different LIS metals to evaluate whether-- or to what degree-- they trace the neutral gas in a sample of nearby star-forming galaxies. We characterize the neutral gas using multiple different approaches, specifically with \ion{O}{1} absorption, Ly$\alpha$ features, and 21-cm \ion{H}{1} emission. These features are all direct tracers of neutral gas but, until now, it has not been demonstrated whether they were produced by the same populations in the ISM.

In this paper, we use the diverse sample of 45 nearby star-forming galaxies from the COS Legacy Spectroscopic SurveY \citep[CLASSY;][]{Berg_2022} to investigate the use of LIS absorption features to trace properties of the neutral ISM. This involved using the high-resolution, high-S/N FUV spectra for CLASSY galaxies and performing new Green Bank Telescope observations to target 21-cm \ion{H}{1} emission. We present simultaneous 1-component fits to a total of 15 rest-frame UV LIS ISM absorption lines from multiple species (\ion{O}{1}, \ion{C}{2}, \ion{Si}{2}, \ion{S}{2}, and \ion{Al}{2}). Using a single component for this fit is a simplistic approach-- essentially an ``average"-- but it is often a necessary assumption in absorption-line studies due to limitations of the data or if a wide-aperture was used for the observations. Many studies have performed 1- and 2-component fits to the ISM absorption for galaxies at $z \sim 0-3$ \citep[e.g.,][]{Jenkins_1986, Martin_2005, Rupke_2005, Heckman_2015, Chisholm_2015, Reddy_2016, Steidel_2016, Gazagnes_2018, Xu_2022}, demonstrating that this approximation can be valid for some galaxies. Using simultaneous 1-component fits to the LIS absorption for CLASSY galaxies, we can explore the merits and limitations of this simplistic approach in the context of studying the neutral and low-ionization ISM in star-forming galaxies. These fits will also be used in a series of upcoming papers for analyzing the ionizing efficiency of the CLASSY galaxies based on their neutral gas properties.

The remainder of this paper is organized as follows: 
Section~\ref{sec:data} presents the CLASSY sample (\S~\ref{sub:classy}) and the data used in this work, with an overview of the FUV spectra (\S~\ref{sub:fuv}) and 21-cm \ion{H}{1} observations (\S~\ref{sub:GBT}). 
We present our simultaneous absorption-line fitting procedure in Section~\ref{UV}, including an overview of the absorption lines used in this work (\S~\ref{sub:lines}), the fitting methodology (\S~\ref{sub:strategy}), effects of saturated lines (\S~\ref{sub:saturation}), the overall fit results for CLASSY galaxies (\S~\ref{sub:overall_fits}), and a comparison to \cite{Xu_2022} (\S~\ref{sub:compXu}). 
In Section~\ref{results:LIS_OI} we present our \ion{O}{1} and LIS fit results, where we consider line opacities (\S~\ref{sub:tau}), compare equivalent widths (\S~\ref{sub:EW}), column densities (\S~\ref{sub:logN}), and covering fractions (\S~\ref{sub:cf}), and summarize our findings in \S~\ref{sub:LIS_as_OI_tracers} regarding \ion{O}{1} as a tracer of LIS absorption. 
In Section~\ref{sec:OI_HI}, we present multiple measurements of direct \ion{H}{1} properties, specifically the \ion{H}{1} gas mass and neutral gas fraction from 21-cm emission (\S~\ref{sub:HI_meas}) and the \ion{H}{1} column density from Ly$\alpha$ (\S~\ref{sub:Lya}), and investigate their relationship with \ion{O}{1} absorption (\S~\ref{sub:compOI_HI}).
Finally, we examine the scaling relations between our fit results and various global properties of the CLASSY galaxies in Section~\ref{results:scaling}, including stellar mass and star formation rate, and summarize our approach and results in Section~\ref{summary}.


\section{Data} \label{sec:data}
In this section, we discuss characteristics of the CLASSY galaxy sample (\S~\ref{sub:classy}) and present an overview of the data used in this analysis. Section~\ref{sub:GBT} describes the 21-cm \ion{H}{1} observations for CLASSY, both introducing the results of the HICLASS survey (\S~\ref{subsub:hiclass}), the available archival observations (\S~\ref{subsub:arch}), and our approach to the data reduction (\S~\ref{subsub:data_reduc}). Section \ref{sub:fuv} discusses the high-resolution FUV spectra we utilized for CLASSY.


\subsection{The CLASSY Sample} \label{sub:classy}
The COS Legacy Spectroscopic SurveY \citep[CLASSY;][]{Berg_2022} consists of high-S/N (S/N$_{1500} \gtrsim$ 5/resel), high-resolution ($R \sim 15,000$), FUV ($\sim1100-2000$ \AA) spectra from the Cosmic Origins Spectrograph (COS) on the Hubble Space Telescope (HST) for 45 low redshift ($z<0.18$) star-forming galaxies. These 45 galaxies span a diverse range of properties, with a relatively large range of stellar masses (6.2 $<$ log $M_\star(M_\odot) <$ 10.1), star formation rate (SFR; -2.0 $<$ log SFR ($M_\odot$ yr$^{-1}) < 1.6$), direct gas-phase metallicity (7.0 $<$ 12+log(O/H) $<$ 8.8), ionization (0.5 $<$ O$_{32} <$ 38.0), reddening ($0.02 < E_{B-V} < 0.67$), and electron density ($10 < n_e$ (cm$^{-3}$) $< 1120$). These galaxies were selected for the CLASSY Treasury, in part, for their high UV luminosities, resulting in a sample of galaxies that exhibit enhanced star formation, more comparable to galaxies at $z\sim2-3$ \citep[see Figure 8 in][]{Berg_2022}. These properties make CLASSY the ideal dataset to explore the connection between neutral and low ionization metal gas properties across a diverse range of objects that resemble higher redshift systems.

We present two datasets that are critical to interpreting the neutral and low-ionization gas dynamics present in the CLASSY galaxies: 1) radio 21-cm observations (\S~\ref{sub:GBT}) from the HICLASS survey and archival observations, and 2) FUV absorption-line spectra (\S~\ref{sub:fuv}). 
We use the 21-cm observations to characterize the neutral gas mass for a total of 35 of the CLASSY galaxies, while the UV absorption spectra provide information about the distribution of neutral and low-ionization gas in and around the galaxies, as well as properties of their feedback.


\subsection{FUV Spectra} \label{sub:fuv}
To map the neutral and low-ionization gas kinematics, we use the high level science product spectra from the CLASSY survey.\footnote{accessible via \url{https://archive.stsci.edu/hlsp/classy} and \url{https://mast.stsci.edu/search/ui/\#/classy}} In particular, we use the high-resolution (HR) co-added spectra, which consist of G130M+G160M grating observations. Although the resolution of these observations is nominally R $\sim 15000$, the actual resolution is lower than this because the targets are not point sources and from degradation of COS. Each of these gratings has a slightly different spectral resolution that must be accounted for when combining data. For technical details of the complex co-addition process, including optimal aperture extractions, wavelength and flux alignment, and co-addition weighting, see \citet[][Paper I]{Berg_2022} and \citet[][Paper II]{James_2022}. The HR co-added spectra cover wavelengths between 1200--1750\AA\ and have an average spectral resolution of 0.073 \AA/resel \citep{Berg_2022}, where 1 resel = 6 native pixels for COS \citep{Soderblom_2021}. 


\subsection{Radio 21-cm Observations}\label{sub:GBT}
Radio 21-cm observations with minimal radio-frequency interference (RFI) exist for 35/45 (78\%) CLASSY galaxies; their observing program IDs are listed in Table~\ref{tab:HI}. 
This includes 19 galaxies from the HICLASS Survey, presented here, and 16 galaxies with archival observations. 
Of the remaining 10 galaxies in CLASSY, nine lack archival 21-cm observations (J0021$+$0052, J1025$+$3622, J1112$+$5503, J1144$+$4012, J1416$+$1223, J1428$+$1653, J1429$+$0643, J1525$+$0757, J1612$+$0817) and one shows significant RFI that infers with measuring the \ion{H}{1} flux (J1448$-$0110; GBT/19A-301). 
With redshifts of $z > 0.1$, these galaxies were not included in HICLASS due to the limited sensitivity and increased RFI that is expected at these higher redshifts \citep[e.g.,][]{Springob_2005,Zwaan_2005,Martin_2010,Freudling_2011}.


\subsubsection{The HICLASS Survey} \label{subsub:hiclass}
The HICLASS Survey is a program using the Robert C. Byrd Green Bank Telescope (GBT) to observe \ion{H}{1} 21-cm emission from 19 CLASSY galaxies (AGBT/21B-323; PI: D. Berg). 
HICLASS was observed with the L-band receiver of the GBT, taking place between 2021 November and 2022 January. 
We used the VErsatile GBT Astronomical Spectrometer (VEGAS) in Mode 21 with a relatively narrow bandwidth of 23.44 MHz ($\sim 5000$ km s$^{-1}$) and a total of 8192 channels. 
These observations were centered on the redshifted \ion{H}{1} 21-cm frequency and performed using in-band \texttt{on/off} frequency switching. 
Each galaxy had a total time on source between 0.5-1.5 hours, depending on redshift and RFI conditions.


\begin{figure}
    \centering
    \includegraphics[width=\linewidth]{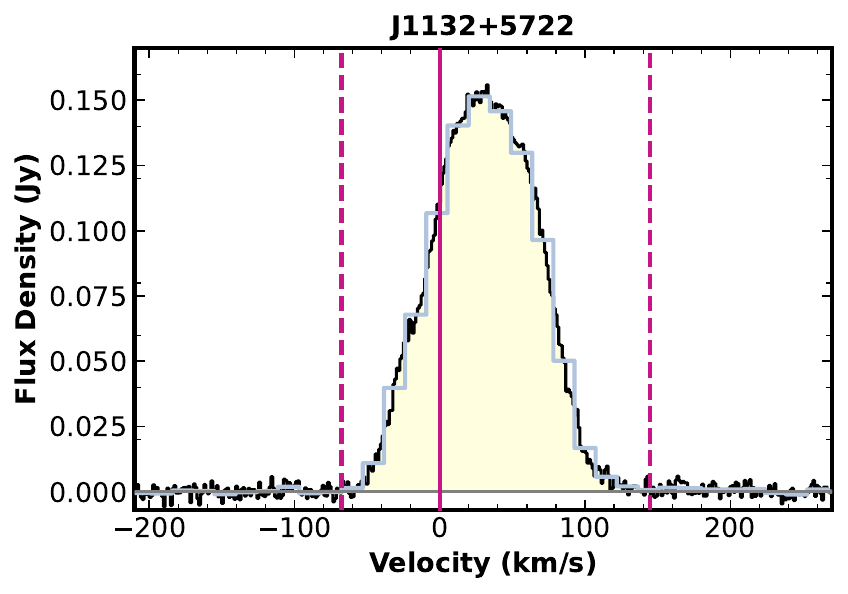} 
    \caption{An example of the stacked and reduced 21-cm \ion{H}{1} profile from the HICLASS observations of J1132+5722, shown in black. A smoothed version of this signal is over-plotted in grey. The bounds of integration used to calculate $S_{HI}$, chosen by eye to include the full emission profile, are shown by dashed pink lines while the area associated with $S_{HI}$ is shaded in light yellow. The solid pink line is noting the systemic velocity of the galaxy (v $= 0$ km/s). Similar plots for the other CLASSY galaxies with \ion{H}{1} observations are in 
    Appendix~\ref{appen21cm}. \label{fig-21cmexample}}
\end{figure}


\subsubsection{Archival Observations} \label{subsub:arch}
An additional 16 CLASSY galaxies have archival 21-cm observations: 11 galaxies were observed with the GBT 100-m telescope, while the five remaining galaxies were observed by the Nanc\'ay radio telescope \citep[J0127-0619, J0337-0502;][]{Thuan_1999}, the GBT 42-m \citep[J0934+5514;][]{Springob_2005}, the Parkes 64-m telescope \citep[J0944-0038;][]{Meyer_2004}, and the Hat Creek 85-ft telescope \citep[J1521+0759;][]{Heiles_1974}. 

The HICLASS observations used \texttt{on/off} frequency-switching to obtain the reference data needed to calibrate the 21-cm emission. This technique involves shifting the center frequency of the measurements to obtain the \texttt{off} observations. While this method is more efficient than many other calibration techniques, it requires a known frequency for the emission and a lack of other nearby features. With known redshifts for the CLASSY galaxies, observations of their 21-cm \ion{H}{1} emission meet these requirements. However, several archival observations of CLASSY galaxies used position-switching instead of frequency-switching for these calibrations, either moving the telescope or the subbeam between the source and an offset pointing. This technique requires careful pointing of the \texttt{on} and \texttt{off} observations and is most effective for compact sources, which is the case for many of the CLASSY sample. Since the galaxies with archival position-switching observations do not extend into the \texttt{off} positions, the method of calibration used is not expected to affect our final results.


\subsubsection{Data Reduction} \label{subsub:data_reduc}
For the 30 galaxies with HICLASS or archival 100-m GBT data, we performed a consistent reduction of the observations using \texttt{GBTIDL} \citep[NRAO;][]{Marganian_2013}.
For the 11 galaxies with archival GBT 21-cm observations, we accessed the raw data through the NRAO archive\footnote{https://data.nrao.edu/portal/\#/}. 
The observations for each galaxy were stacked and averaged into a single spectrum using standard \texttt{GBTIDL} data reduction algorithms, specifically \texttt{getps} or \texttt{getnod} depending on the \texttt{on/off} calibration method used. 
We shifted the stacked spectrum into the rest-frame and inspected them for signs of RFI contamination. 
The archival spectrum of J1448-0110 (AGBT19A-301) was excluded from our analysis due to extreme RFI contamination.

To remove the continuum, we used \texttt{polyfit} to fit the baseline on either side of the 21-cm emission with a 2$^{\rm nd}$ degree polynomial. 
The bounds for the fits were chosen manually in order to include large regions of the continuum while excluding artifacts from RFI and the bandpass edges. 
From the stacked spectrum, we subtracted the baseline continuum fit to obtain the reduced \ion{H}{1} 21-cm emission profile. Figure \ref{fig-21cmexample} shows this profile for J1132+5722, while all the others are included in Appendix~\ref{appen21cm}. By visual examination and considering the S/N, we classified eight of the 21-cm observations as non-detections of \ion{H}{1} emission (J0808+3948, J0926+4427, J0938+5428, J0942+3547, J1024+0524, J1323$-$0132, J1359+5726, and J1545+0858) and report their 21-cm measurements as upper limits.

The majority of the \ion{H}{1} emission profiles are centered near v $= 0$, the systemic velocity for each galaxy. 
The main exception was J1200+1343, which required a much larger offset ($\sim$1100 km/s). The SDSS imaging of this galaxy shows that it might be interacting with a nearby fainter companion, which could disturb the \ion{H}{1} gas and cause this large offset from the expected redshift.

After reducing and stacking the 21-cm observations, we measured $S_{HI}$ by integrating across the emission profiles in velocity-space. In Figure \ref{fig-21cmexample}, the bounds of this integration are denoted by pink dashed lines and the area under the profile is shaded in light yellow.
To quantify the uncertainties involved in our measurements of $S_{HI}$, we used the standard deviation of fluxes in our continuum windows to generate random, normally-distributed, variations to the observed emission in 1000 unique trials. 
We define the uncertainties of our GBT $S_{HI}$ measurements as the standard deviation of the \ion{H}{1} fluxes from these trials. 
We use these $S_{HI}$ measurements to calculate the \ion{H}{1} gas masses for the CLASSY galaxies, as discussed in Section \ref{sec:OI_HI}.


\section{Absorption Line Analysis} \label{UV}
In this section, we discuss our absorption line analysis strategy which involves fitting the observed absorption simultaneously using Voigt profiles.
The LIS absorption lines used in this work are introduced in \S~\ref{sub:lines}, our simultaneous fitting procedure methodology is discussed in \S~\ref{sub:strategy}, and an investigation of the saturated lines in our sample is in \S~\ref{sub:saturation}. Section~\ref{sub:overall_fits} presents an overall of our fit results for CLASSY, while \S~\ref{sub:compXu} compares these results with the two-component fits for the CLASSY galaxies by \citet{Xu_2022}.


\startlongtable
\begin{deluxetable}{c|C|rc|c} 
\setlength{\tabcolsep}{8pt}
\tabletypesize{\normalsize}
\tablecaption{Neutral and LIS Lines Used in Simultaneous Fits \label{tab:ions}}
\tablehead{
\CH{Ion} &\CH{$E_i$ - $E_k$} & \CH{\W}  &\CH{$f$-value} & \CH{\#} \\[-1.5ex] 
\CH{}    & \CH{(eV)}         & \CH{(\AA)} & \CH{} & \CH{Gal.}} 
\startdata
\ion{H}{1}$^*$  & 0.000-13.598  & 972.54    & 0.029   &  2  \\
            &               & 1025.72   & 0.079   & 10  \\
\hline
\ion{O}{1}  & 0.000-13.618  & 1039.23   & 0.009   & 8   \\
            &               & 1302.17   & 0.048   & 33  \\
\hline
\ion{C}{2}  & 11.260-24.383 & 1036.34   & 0.119   & 11  \\
            &               & 1334.53   & 0.129   & 36  \\
\hline
\ion{Si}{2} & 8.151-16.345  & 1190.42   & 0.256   & 24  \\
            &               & 1193.29   & 0.544   & 28  \\
            &               & 1260.42   & 1.200   & 32  \\
            &               & 1304.37   & 0.091   & 35  \\
            &               & 1526.71   & 0.144   & 34  \\
\hline
\ion{S}{2}  & 10.360-23.330 & 1250.58   & 0.006   & 31  \\
            &               & 1253.81   & 0.012   & 27  \\
            &               & 1259.52   & 0.018   & 19  \\
\hline
\ion{Al}{2} & 5.986-18.828  & 1670.79   & 1.740   & 30
\enddata    
\tablecomments{\footnotesize{
The suite of FUV neutral and LIS interstellar absorption lines used in the single-component simultaneous fit method to characterize the neutral gas in the CLASSY sample. 
For each observed absorption line, Columns 1--4 list the associated ion, ionization energy bounds ($E_i$, $E_k$), vacuum wavelength, and oscillator strength ($f-$value), respectively. 
Wavelengths and oscillator strengths for \ion{H}{1} were taken from \cite{Morton_2003}, while the metal line wavelengths and oscillator strengths are from \cite{Cashman_2017}. 
Column 5 specifies the number of CLASSY galaxies (out of the 45 objects) with a non-contaminated absorption line detection for each transition.\\
$^*$Although we include \ion{H}{1} lines in this table, they are not included in our fits because there are few with high S/N and that are uncontaminated.}}
\end{deluxetable}


\subsection{FUV absorption lines included in our analysis} \label{sub:lines}
The broad wavelength coverage of CLASSY ($\sim$1100--2000 \AA) provides a unique opportunity to simultaneously study the properties of multiple neutral atoms and low ionization ions that supposedly trace the same gas phases, across a sample of nearby galaxies that resemble those at higher redshifts ($z \sim 2$). In this work, we focus on FUV transitions produced by \ion{O}{1}, \ion{C}{2}, \ion{Si}{2}, \ion{S}{2}, and \ion{Al}{2}. We list the wavelengths, ionization potentials, and oscillator strengths for this series of lines in Table~\ref{tab:ions}. The final column contains the number of CLASSY galaxies for which a specific absorption line is detected without contamination by another feature such as Milky Way absorption, geocoronal emission, or blending between lines in the galaxy itself. This included lines that overlapped with MW features when accounting for the redshifts of each galaxy.

When using absorption lines to investigate the neutral gas properties, the ideal scenario is to look directly at \ion{H}{1} absorption via the Lyman series absorption features. However, this is complicated at high-$z$ ($z \gtrsim 5$) since the \ion{H}{1} absorption lines fall below 1025 \AA, and therefore are highly impacted by IGM absorption \citep{Fan_2006,Bosman_2018,Bosman_2022,Yang_2020}. Even in CLASSY, a $z < 0.2$ survey, this is not feasible for a large majority of galaxies due to the blue limit of the COS gratings used for the observations (only 10 galaxies have Ly-$\beta$ observed). Alternatively, with a similar ionization potential to \ion{H}{1}, \ion{O}{1} can be used as a proxy (13.598 eV). This allows us to compare properties of the low-ionization lines with a more direct indicator of the neutral gas, \ion{O}{1}. 


\subsection{Simultaneous Fitting Strategy} \label{sub:strategy}
To simultaneously fit our series of FUV absorption lines, we used an approach similar to those presented by \cite{Chisholm_2015} and \cite{Gazagnes_2018}. This first involves removing the stellar continuum from the observed spectra. To do so, we used the \texttt{Starburst99} \citep{Leitherer_1999} stellar continuum fits to the CLASSY spectra (a CLASSY high-level science product release; in prep.). We divided the observed spectra by the stellar continuum fits to remove the main stellar features that overlay the ISM lines in our analysis and to normalize the continuum. 

Similar to the methodology in \cite{Xu_2022}, we performed our fits using a partial covering fraction model. This involves approximating the absorption profiles using the following equation, where $C_f$(ion) is the overall covering fraction associated with the ion and $\tau$(line) is the opacity of the transition:
\begin{displaymath}
    F(\lambda) = 1 - C_f(\mathrm{ion}) + C_f(\mathrm{ion}) \times e^{-\tau_{line}}
\end{displaymath}
When multiple lines are fit for the same ion, the assumption that these lines share the same value for $C_f(\mathrm{ion})$ can break the degeneracy between the covering fraction and the opacity for each transition. 

In this work, we simultaneously fit a series of Voigt profiles to our suite of LIS absorption lines using the stellar continuum-normalized spectrum. We determined the opacity of the absorbing gas from its column density ($N_{ion}$) and Doppler broadening ($b$) using the following equation that approximates a Voigt profile:
\begin{displaymath}
    \tau_{line} \: = \: \frac{\sqrt{\pi} \: e^2 \: f \: \lambda_{0,obs}}{m_e \: c} \times \frac{N_{ion}}{b} \hspace{5cm}
\end{displaymath}
\begin{displaymath}
    \hspace{0.7cm} \times \left[H - \frac{\lambda_{0,obs} \: \gamma}{4\pi^{3/2} \: b \: P} \times (H^2 \times (4P^2 + 7P + 4 + Q) - Q - 1)\right]
\end{displaymath}
where
\begin{displaymath}
    H = e^{-x^2}, \hspace{1cm} P = x^2, \hspace{1cm} Q = \frac{1.5}{x^2}
\end{displaymath}
and
\begin{displaymath}
    x = \frac{c}{b} \times (1 - \frac{\lambda_{0,obs}}{\lambda})
\end{displaymath}
In these equations, $\lambda_{0,obs}$ is the observed wavelength of the absorption line, i.e., $\lambda_{0,obs} = \lambda_0 \times (\frac{v_{min}}{c} + 1)$ where $\lambda_0$ is its rest wavelength and $v_{min}$ is the velocity offset of the line from systemic. Additionally, in these equations, $f$ is the oscillator strength ($f$-value) and $\gamma$ is the Einstein coefficient for each line.

Overall, this corresponds to four free parameters for each Voigt profile: the ion-specific column density and covering fraction, and the overall Doppler parameter and velocity offset of the full series of LIS absorption lines. We detail each of these parameters in the following subsections.


\subsubsection{Column Density}
The column density represents the amount of gas between the observer and the dominant sources of emission from the host galaxy. For our fits, this quantity can vary between ions but is fixed for all transitions of the same ion. Our fits assume that \ion{O}{1} and the LIS lines have the same $b$-parameter, which enables us to estimate $\log(N_{OI})$ even when it is optically-thick. We use relatively large priors for the ion column densities, with $\log(N)$ bounds from 13.0-18.5 for \ion{O}{1} and 12.0-18.0 for the other elements. 

Accurately constraining $N_{ion}$ is particularly complex when absorption lines are saturated, though it can be possible if the spectra are at a very-high-S/N and -resolution \citep{Gazagnes_2018}. Saturation is an important aspect of absorption line analyses and, in \S~\ref{sub:saturation}, we discuss how it seems to impact our final measurements and results. 


\subsubsection{Covering Fraction}
The covering fraction serves as a theoretical indicator of the spatial distribution of optically dense gas along a given line-of-sight. It can range in value from 0 (no interactions between the gas and incident light) to 1 (full coverage). 
For saturated absorption lines, $C_f$ is consistent with 1 - $R_f$, where $R_f$ is the flux at the line's minimum depth and is known as the residual flux.

In practice, the $C_f$ estimated from absorption lines can underestimate the actual geometric covering fraction when the gas has complex kinematics. For instance, if two gas clumps each cover 50\% of the line-of-sight but have distinct velocities, the inferred covering fraction would also be $C_f = 0.5$ even though the geometric covering fraction is actually 1.0. Despite such challenges, the trends observed between $C_f$-- for both \ion{H}{1} and low-ionization species-- and the escape of ionizing photons suggest that $C_f$ can still robustly trace the geometric gas covering, notwithstanding strong kinematic biases \citep{Gazagnes_2020}.

In our fits, we adopt a distinct $C_f$ for each ion, assuming a uniform prior ranging between 0 and 1 for all fitted elements. This choice was motivated by previous absorption-line studies that found $C_f$ to vary between different elements \citep{Henry_2015, Reddy_2016, Gazagnes_2018, Mauerhofer_2021}. Although this could be inconsistent with the assumption that the neutral and LIS species coexist, this discrepancy could also be an effect of in-filling, wherein the absorbed photons are re-emitted along the line-of-sight. In-filling offers a plausible explanation for minor discrepancies observed in the covering fractions among saturated transitions of the same ion, which should exhibit identical residual fluxes. Further discussion on this topic, specifically related to our fit results, is provided in \S~\ref{sub:cf}.


\subsubsection{Offset Velocity} \label{sub:line_kinematics}
\begin{figure*}[!htbp]
    \centering
    \includegraphics[width=0.9\linewidth]{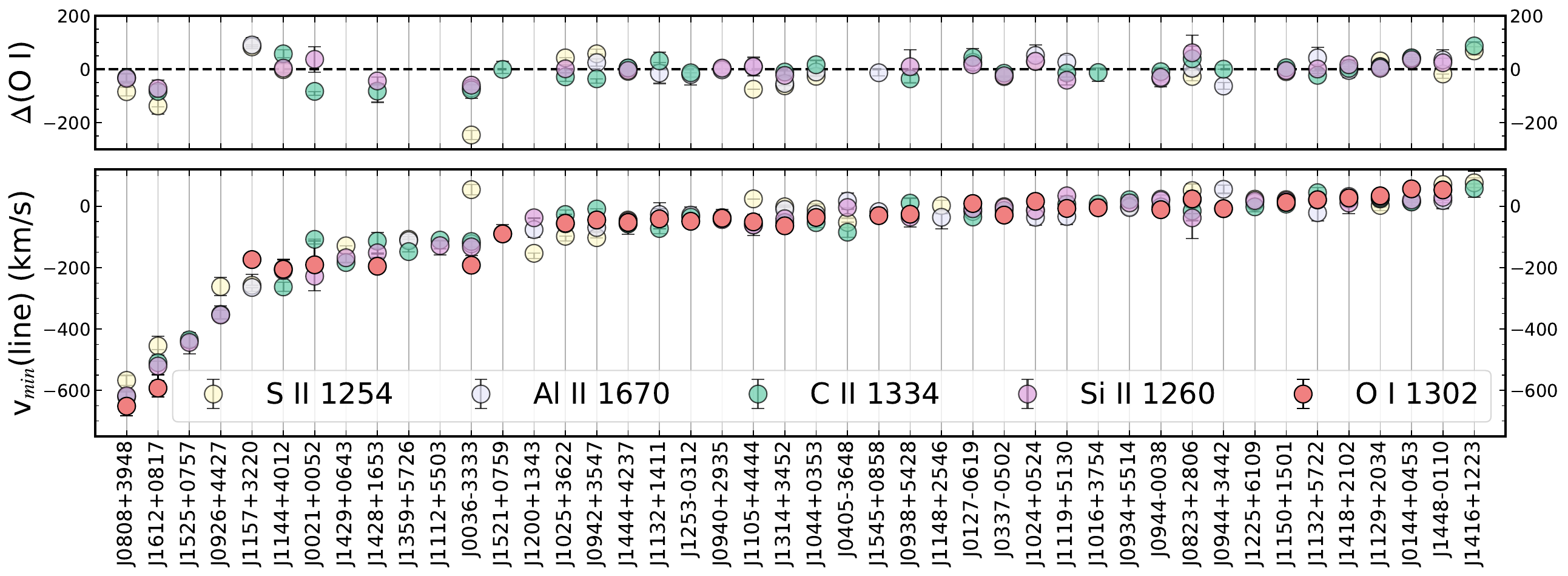}
    \figcaption{
    The velocity offsets of the strongest line for each ion (\ion{O}{1} \W1302: red points, \ion{C}{2} \W1334: green points, \ion{S}{2} \W1254: yellow points, \ion{Si}{2} \W1260: pink points, \ion{Al}{2} \W1670: purple points). The $x$-axis lists the galaxies in our sample, ordered by the average value of these points. \textit{Top:} The difference in velocity compared to \ion{O}{1}, for those where it is measured. Overall, this shows that the velocity offsets are generally consistent between these ions, which is expected if they were produced by the same clouds of gas. This further motivates the use of a single component of gas to simultaneously fit this series of absorption lines.
    \label{fig3}}
\end{figure*}

The offset velocity represents the typical shift of the absorption lines from the systemic of the galaxy, as determined using the redshifts reported by \cite{Berg_2022}. To apply a 1-component fit to our series of neutral and LIS absorption lines, it assumes that the associated gas is moving together, meaning all lines would have a similar $v_{min}$. We assess the validity of this assumption by comparing the kinematics of the strongest individual lines from each species used here, for the full sample of CLASSY galaxies. To measure the velocity offset $v_{min}$ for individual lines, we determined the median offset in wavelength that corresponded to a line's minimum flux across 1000 Monte Carlo (MC) variations of the spectral window. 
In Figure \ref{fig3} we compare the average velocities of the \ion{O}{1}, \ion{C}{2}, \ion{Si}{2}, \ion{S}{2}, \ion{Al}{2} profiles on a galaxy-by-galaxy basis. The top panel shows the relative velocity difference of the lines with respect to the \ion{O}{1} \W1302 line, taken as a reference. 

For the majority of the CLASSY galaxies, the velocity offsets of individual LIS lines are consistent with one another, with the galaxies with the weakest outflows having the lowest dispersion of velocities ($\Delta v \lesssim 20$ km/s). The velocity offset of \ion{S}{2} \W1254 is sometimes significantly different than the other lines, yet this is a relatively weak line as compared to the other lines.
Overall, the scaling between $v_{min}$ for the different lines tested supports our assumption that all Voigt profiles can be modeled with a single $v_{min}$ parameter.  It also supports the idea that the ions may coexist within the same populations of gas in the ISM.


\subsubsection{Doppler Parameter} \label{subsub:b}
The Doppler parameter $b$ describes the line broadening. It is generally modeled at the combination of two terms, a turbulence velocity ($b_{turb}$), also sometimes referred to as the random motion velocity, and the thermal energy ($b_{therm}$) of the gas. The full expression for $b$ is: 
\begin{equation} \label{eq:b}
    b = \sqrt{b_{turb}^2 + b_{therm}^2} = \sqrt{b_{turb}^2 +  \frac{2k_B \times T}{m_{ion}}}
\end{equation}
where $k_B$ is the Boltzmann constant ($1.3807 \times 10^{-16}$ erg/K), $T$ is the temperature of the gas, and $m_{ion}$ is the atomic mass of the ion involved. 

In this study, we focus on fitting species with comparable masses, ranging from carbon with the lowest atomic mass ($\sim$12) to sulfur with the largest mass among the considered ion ($\sim$32) (the hydrogen lines, when present, are fit separately). For this set of species, assuming a T = 10$^4$ K, $b_{therm}$ ranges from 3 to 3.5 km/s, rendering the differences in ion masses negligible on the final $b$ values. Further, we assume that $b_{turb}$ remains constant across these elements, given their similar susceptibility to turbulence. Although each ion theoretically should have its own $b$, in our context, we consider $b$ to be constant for all species due to their similar masses. It is important to exercise caution when applying this assumption to other scenarios, as it only holds here because of the similar masses of the considered species. In Appendix~\ref{appen:doppler}, we demonstrate that when considering ions with significantly different masses, such as \ion{Fe}{2} (with an atomic mass of approximately 55), assuming a constant $b$ parameter leads to inconsistent line reproductions.

Initially, we treated both forms of $b_{therm}$ and $b_{turb}$ as free parameters in our estimate for $b$. However, despite employing high-resolution spectra and simultaneous fits to a broad suite of absorption lines, we could not constrain these parameters effectively, resulting in significant degeneracies between both terms. Therefore, we assumed a single dominant component of $b$, considering only the collective impact of broadening without distinguishing between turbulence and thermal broadening.

Overall, concurrently examining a set of species with similar mass and ionization potential enables us to standardize some of the fit parameters across all elements. This approach effectively simplifies the fitting process by reducing overall complexity without affecting the accuracy of the fit within the uncertainties. 
 
In addition to the four free parameters ($N_{ion}$, $C_f$(ion), $b$, $v_{min}$), the Voigt profiles require a set of fixed parameters that are specific to the line (the oscillator strength, rest wavelength, and damping parameter) and to the data (spectral resolution). We adopt oscillator strengths, rest wavelengths, and damping parameters from \cite{Morton_2003} for \ion{H}{1} and from \cite{Cashman_2017} for the metal lines, accessed via the NIST Atomic Spectra Database \citep{Kramida_2023}. For the spectral resolution, \cite{Berg_2022} used the width of the cleanest non-saturated Milky Way absorption feature present to measure the spectral resolution for the CLASSY observations. Depending on the galaxy, the line used for this measurement was \ion{Si}{2} \W1260, \ion{C}{2} \W1334, or \ion{C}{4} \W1548 (see their Table 3). We repeated this analysis, using multiple of our LIS lines when possible, and generally found results that were consistent with \cite{Berg_2022}. 

We note that, for the resolution of the CLASSY spectra, the shape of the Voigt profiles in these fits should agree with the Gaussian models used to describe the absorption in some other studies.


\subsubsection{Fitting the Spectra}
The code\footnote{Available on Github: \url{https://github.com/sgazagnes/AbsorpFit}} used to perform the fitting is built around the package \texttt{lmfit}, using least squares minimization \citep{Newville_2014}. We performed 300 iterations of these fits for each galaxy \citep[as recommended by][]{Chisholm_2019}, using the uncertainties of the observed fluxes to vary the spectrum between runs and also making small perturbations to the initial parameters. The final best-fit parameters correspond to the median values returned by these 300 runs and have uncertainties based on their standard deviation.
 
For each spectrum, we fit all lines that are not contaminated by Milky Way absorption or geocoronal emission, or that were too close to the edge of the COS grating segment. From these initial fits, we found some cases where a single line would skew the quality of fit for all the other included lines. When this happens, it often seemed to be with lines that show some evidence of in-filling, which can happen when resonant photons are re-emitted along the line-of-sight, changing characteristics of the observed absorption profile \citep[e.g.,][]{Prochaska_2011,Erb_2012,Scarlata_2015}. \ion{Si}{2} \W1304 in J1119+5130 (Figure E21, Appendix) is a clear example of this. One benefit of our simultaneous fitting approach, particularly when multiple lines for an ion are included, is that we can easily detect when a single line does not agree with the characteristics set by the others. In cases where this line skews the fit to do poorly regarding the other lines in the fit, we exclude that single line and assume it is likely contaminated by in-filling.


\subsection{Dealing with Saturation} \label{sub:saturation}
As discussed above, retrieving accurate column density estimates when fitting saturated lines is complex because the opacity of the absorbing material reaches a point where additional absorption does not lead to a proportional increase in line strength. Further, when all lines considered are saturated, there also exists a $\log(N)$/$b$ degeneracy such that multiple combinations of $b$ and $N$ may result in similar line profiles that cannot be distinguished when considering the typical noise and resolution of the instrument. In the following subsections, \S~\ref{subsub:test_mock_lines} and \S~\ref{subsub:test_varyN}, we test the accuracy of our $\log(N)$ fit results for saturated transitions and demonstrate that our fitting procedure is capable of constraining these measurements.

One notable advantage of our approach to simultaneous absorption line fitting is its ability to resolve the $N$/$b$ degeneracy by incorporating a single unsaturated line. Since it is relatively straightforward to recover $N$ and $b$ from unsaturated lines, a single unsaturated line is adequate to establish a consistent $b$ value across our full set of lines. We used the Curve of Growth (CoG) to evaluate whether individual lines in our fits are saturated. Appendix \ref{appen:cog} contains the CoG diagrams for CLASSY. 

We find that 17 of the 44 galaxies ($\sim$40\%) in our analysis have at least one unsaturated line included in the fit. Table \ref{tab:saturated2} in the Appendix \ref{appen:cog} lists the individual lines that are saturated vs. unsaturated for each galaxy. Most often, the lines that are unsaturated are \ion{S}{2} (\W\W1251,4) for which \ion{S}{2} \W1251 is unsaturated in 43\% of those where it is fit and \ion{S}{2} \W1254 is unsaturated in 21\%. When present and uncontaminated, \ion{O}{1} \W1039 and \ion{Si}{2} \W1304 also seem to have a better chance of being unsaturated compared to many of the other lines in our sample (i.e., 29\% and 20\% unsaturated, respectively). 

Since we expect $b$ to be set by the unsaturated line(s) in the fit, we can test our fitting procedure by constraining $b$ from a fit solely to unsaturated \ion{S}{2} \W1251 or \W1254 absorption. These \ion{S}{2} lines were the most likely lines to be unsaturated, which make them powerful in breaking the degeneracy between b and N in many of our fits. We then performed the full fit, but with $b$ constrained to the value and uncertainty set by \ion{S}{2}. Overall, we found the results of these test fits to be in agreement with those returned by our normal fits, which supports the idea that our fitting approach is using the unsaturated lines to determine $b$ anyways.

For elements with all lines fully saturated, significant uncertainty can persist in $\log(N)$, irrespective of whether $b$ is determined by an unsaturated line from a different ion. \citet{Gazagnes_2018} quantified the typical expected error on $N$ as a function of resolution and S/N. For our CLASSY spectra with observed S/N $\sim$ 5 and resolution R $\sim$ 5000, the typical uncertainty is around 1 order of magnitude. This uncertainty decreases notably when multiple lines from the same ion are included, as each saturates at different $\log(N)$ values. 


\begin{figure}[!htbp]
    \centering
    \includegraphics[width=\linewidth]{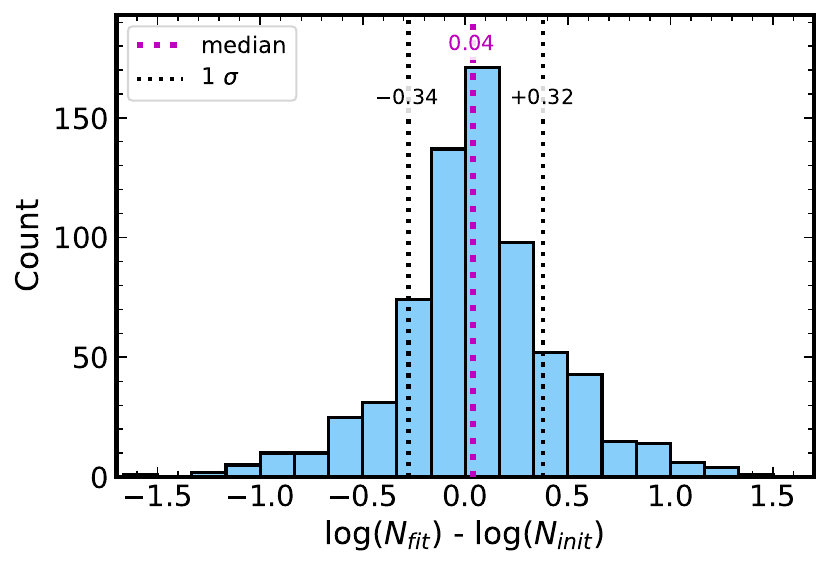}
    \figcaption{Saturation Test 1. We performed our MC least-squares fitting procedure on a series of 1000 models of saturated \ion{Si}{2} \W1260 absorption across a grid of input parameters ($b$, $\log(N)$, $C_f$), adding noise and assuming a resolution similar to those typical of the CLASSY observations. This figure shows a histogram of the differences between the results of these fits, $\log(N_{fit})$, and the actual values, $\log(N_{init})$. Since this distribution is centered on zero and is relatively symmetric, it demonstrates that our MC approach is capable of constraining $\log(N)$ for saturated lines. \label{fig:Nerr_from_model}}
\end{figure}


\subsubsection{Saturation Test 1: Fitting Simulated Lines}\label{subsub:test_mock_lines}
Although fits to saturated lines largely require the resulting column densities to be treated as lower limits, we posit that our Monte Carlo approach is still capable of returning reasonable uncertainties associated with $\log(N)$, since it explores the entire range of possible $\log(N)$ values and estimates the final uncertainties from these varied fits. To evaluate whether this truly allows us to constrain $\log(N)$ for saturated transitions, we created models of \ion{Si}{2} \W1260 absorption using a grid of input parameters and adopting the same resolution and estimated noise as one of the CLASSY observations. These input parameters range in $b_{init}$ (50 km/s to 200 km/s; 10 evenly-spaced values), in $\log(N_{init})$ (14.8 to 16.2; 10 evenly-spaced values), and $C_f^{init}$ (0.2 to 1.0; 10 evenly-spaced values). For the 1000 combinations of these initial parameters, we performed 300 iterations of our MC least-squares fitting procedure on each resulting model. Based on their location on the CoG, all models from this parameter-space correspond to saturated \ion{Si}{2} \W1260 absorption.

Figure \ref{fig:Nerr_from_model} shows the distribution of log$(N_{fit})-$log$(N_{init})$, the differences between column densities returned by these test fits and the initial values used to create the models. This distribution is centered near zero difference and is relatively symmetrical, which indicates that our MC approach is capable of consistently covering the range of values expected by the saturated part of the CoG. Overall, this demonstrates that our fitting procedure is capable of constraining the column density of the absorbing gas, even for saturated transitions, rather than merely returning a lower limit.

We note that this test was performed on the models for an individual line while our fitting code is most accurate when it is fitting many lines simultaneously. This indicates that its performance in this test is likely to improve when more lines are considered, placing these test results as a lower limit on its accuracy.


\subsubsection{Saturation Test 2: Varying Column Densities}\label{subsub:test_varyN}
\begin{figure*}[!htbp]
    \centering
    \includegraphics[width=\linewidth]{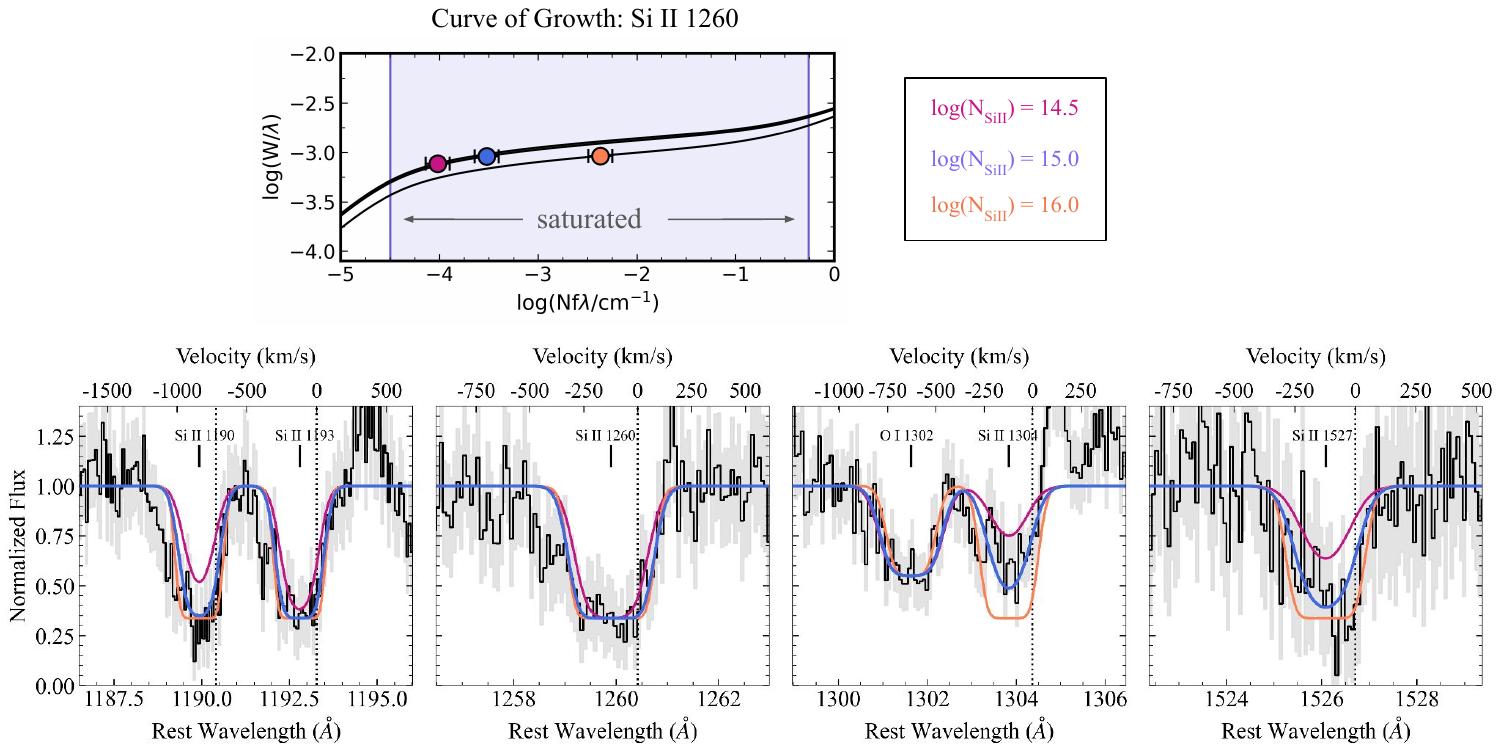}
    \figcaption{Saturation Test 2. J1025+3622 fits with $\log(N_{SiII})$ fixed to different values in the saturated regime (14.5, 15.0, and 16.0). For these test fits, the covering fractions, column densities associated with other elements, and $v_{min}$ are fixed at the best fit values for J1025+3622 while $b$ is allowed to vary.
    \textit{Top}: The curve of growth for \ion{Si}{2} \W1260 for these test fits (i.e., the black curves), pinpointing three locations in the saturated zone: the best fit in purple, a lower column density in pink, and a higher column density in orange. The light purple shading corresponds to the saturated regime. 
    \textit{Bottom:} The models corresponding to these fixed \ion{Si}{2} column densities compared with the observed spectrum for J1025+3622. As shown here, we are able to distinguish between the accuracy of these fits to the data even though they all correspond to saturated absorption. This demonstrates that our fitting procedure is capable of constraining the column density of saturated lines. \label{fig:test_sat_SiII}}
\end{figure*}

\begin{figure*}[!htbp]
    \centering
    \includegraphics[width=1.0\linewidth]{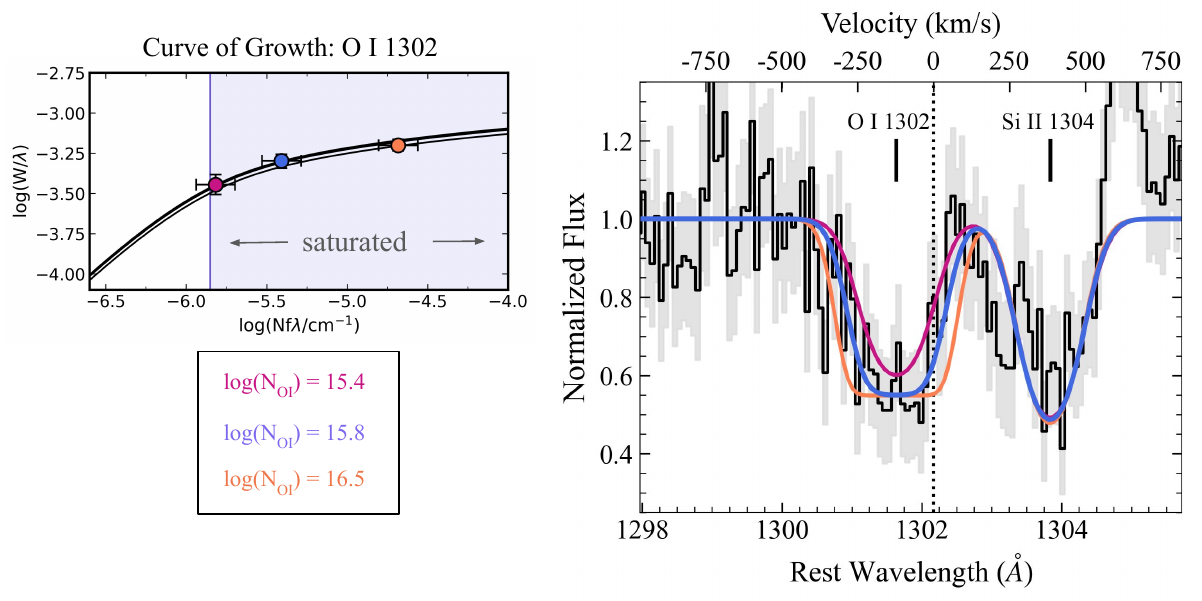}
    \figcaption{Saturation Test 2 (cont.). Similar to Figure \ref{fig:test_sat_SiII} except $\log(N_{OI})$ is fixed to different values in the saturated regime (15.4, 15.8, and 16.5) while $\log(N_{SiII})$ is fixed to 15.0 (the best-fit value for this galaxy). Once again, $b$ is allowed to vary. This shows that even for a single saturated line, we can still distinguish between these fits when compared with the data. This further supports our ability to constrain the column densities for saturated lines. \label{fig:test_sat_OI}}
\end{figure*}

To further investigate the accuracy of our fits with regards to saturated lines, we performed alternate fits with column densities fixed to other values in the saturated regime. Figures \ref{fig:test_sat_SiII} and \ref{fig:test_sat_OI} show an example of this test for J1025+3622, where we vary $\log(N_{SiII})$ in the former and $\log(N_{OI})$ in the later. By comparing these models, we can evaluate whether they are distinguishable when all the transitions are saturated.

In the top rows of these figures, we show three different locations in the saturated regime of the Curve of Growth (specifically for \ion{Si}{2} \W1260 and \ion{O}{1} \W1302) which correspond to different column densities. Our best fit for J1025+3622 resulted in $\log(N_{SiII})$ = 15.0 and $\log(N_{OI})$ = 15.8, which are shown with purple in these figures. We reran these fits, now fixing the majority of parameters to those from the best-fit, with the exception of 1) $\log(N_{ion})$, set to an alternate column densities, and 2) $b$, which was allowed to vary. For \ion{Si}{2}, we ran this fit with $\log(N_{SiII})$ fixed at 14.5 and at 16.0, which are shown with pink and orange in Figure \ref{fig:test_sat_SiII}, respectively. Similarly, in Figure \ref{fig:test_sat_OI}, the fit is fixed with $\log(N_{OI})$ = 15.4 and 16.5. As shown by the CoG diagrams in these figures, these parameters all correspond to the saturated regime (light purple shading).

We chose to perform these fits with $b$ as a free parameter, letting it settle towards the value that is ideal for modeling the data, when assuming the column density adopted for each test case. If our fitting technique is unable to break the $\log(N)$/$b$ degeneracy for saturated lines, we expect not to be able to distinguish between these different fits, i.e., that they could all model the data with similar success. The bottom panels of Figures \ref{fig:test_sat_SiII} and \ref{fig:test_sat_OI} show the performance of these tests for \ion{Si}{2} and \ion{O}{1}, respectively. Although a lower/higher value of $b$ can compensate for higher/lower $\log(N)$ values due to their degeneracy, these panels demonstrate that we are still able to distinguish between the resulting fits. This is particularly true when multiple transitions for an ion are included in the fit, which can be seen by comparing the \ion{Si}{2} and \ion{O}{1} models from these panels. Although our best-fit column densities (in purple) appear to outperform the alternate fits (orange+pink) in both cases, we can say this more confidently for \ion{Si}{2} with its five lines than for the single line considered here for \ion{O}{1}.

This further demonstrates that our MC approach is capable of placing actual constraints on the column densities when the lines are saturated. Overall, our fitting strategy enables us to mitigate, although not suppress, the potential biases from saturated line fitting.


\begin{figure*}[!htbp]
    \centering
    \includegraphics[width=0.9\linewidth]{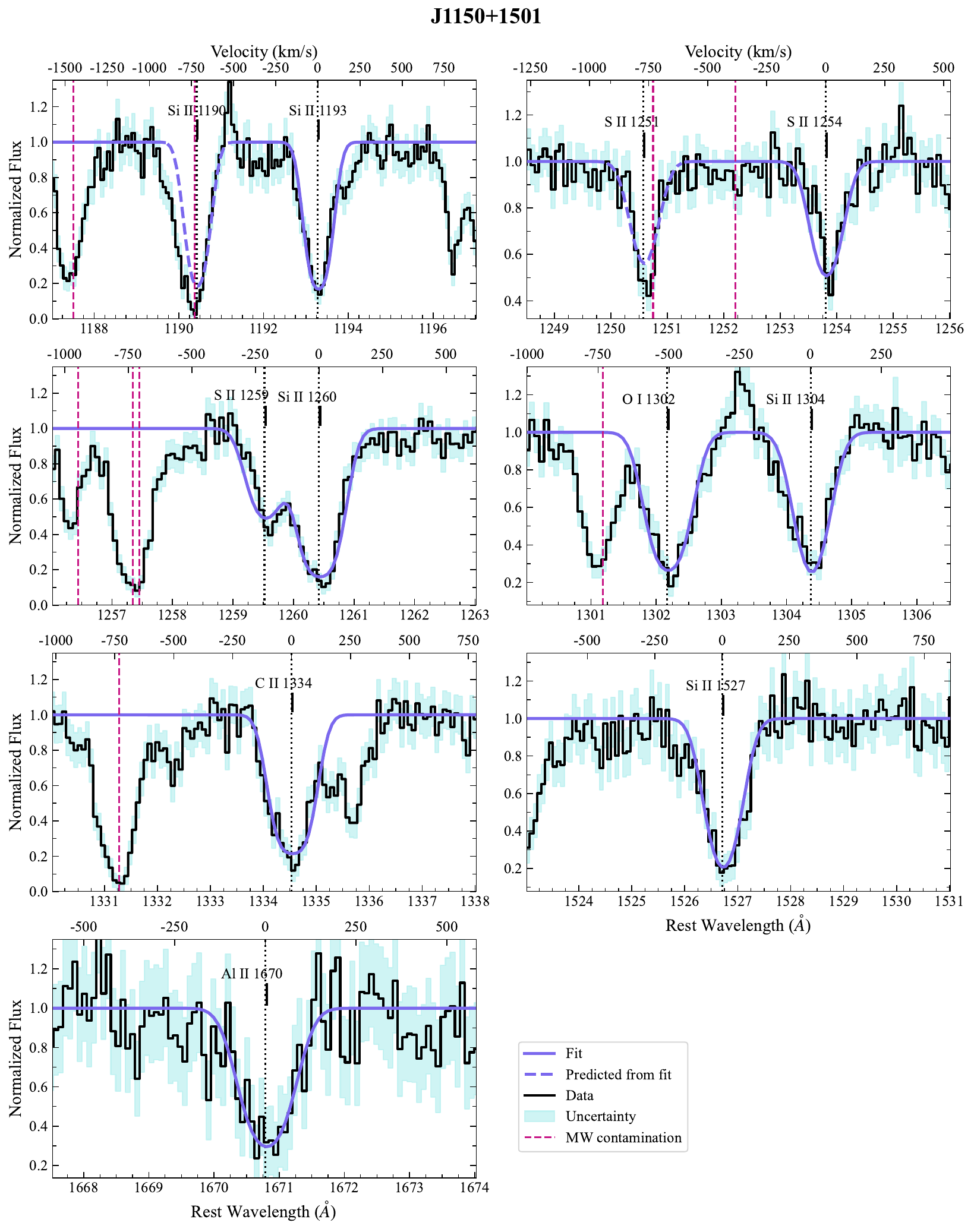}
    \figcaption{
    Simultaneous one-component fits (solid purple line) to the LIS features in the high-S/N, high-spectral-resolution CLASSY spectrum for J1150+1501 (solid black line). 
    Lines not included in the fit but predicted using the fit parameters are shown as dashed-purple line profiles. 
    The systemic wavelengths of the LIS absorption lines are marked by  vertical dotted lines, while the line's trough ($v_{min}$) is denoted by the solid black ticks. 
    Milky Way features, denoted by vertical dashed-magenta lines, were not included in the fit.
    Figures presenting the fits for the other CLASSY galaxies are included in Section \ref{appenD} of the Appendix. 
    \label{fig2}}
\end{figure*}

\begin{figure*}[!htbp]
    \centering
    \includegraphics[width=0.95\linewidth]{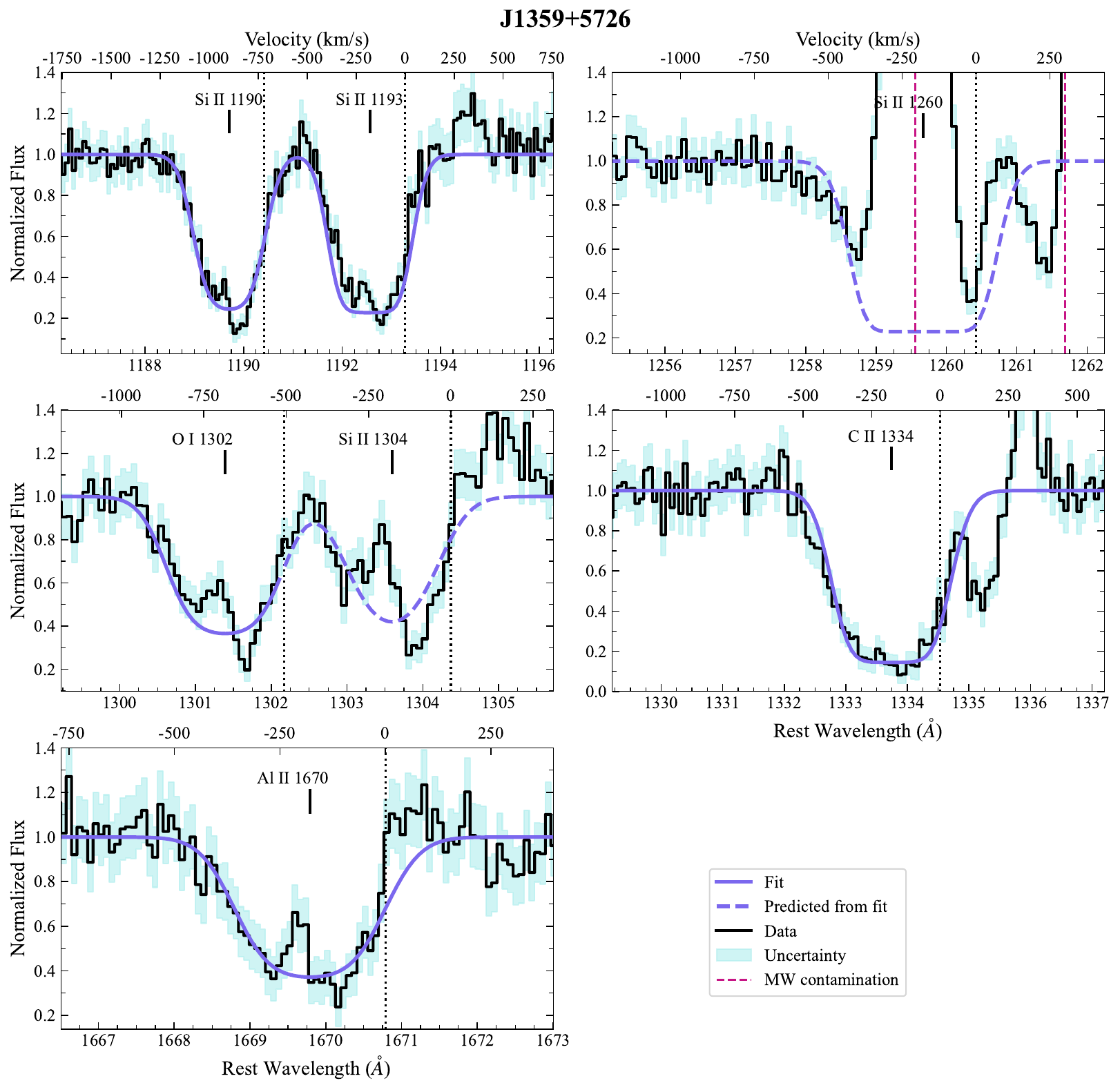}
    \figcaption{The simultaneous one-component fit for J1359+5726, a member of the multi-component sample, which requires more than one component to adequately fit the observed LIS absorption. See caption of Figure \ref{fig2} for details.
    \label{fig2b}}
\end{figure*}


\subsection{Overall Fit Results} \label{sub:overall_fits}
We applied our fitting approach on the 44 CLASSY galaxies with at least one observed, non-contaminated, absorption line. One galaxy-- J1323$-$0132-- does not display any LIS absorption lines and is not included in this analysis. Figure ~\ref{fig2} shows an example of the simultaneous fit for CLASSY galaxy J1150+1501. This fit included lines from four different LIS ions: \ion{O}{1} \W1302, \ion{Si}{2} \W1193, \W1260, \W1304, \W1527, \ion{S}{2} \W1254, \W1259, and \ion{Al}{2} \W1671. We note that \ion{S}{2} \W1259 and \ion{Si}{2} \W1260 are blended in some galaxies and that \ion{O}{1} \W1302 can sometimes have fluorescent emission (\ion{O}{1}$^*$ \W1302) that contaminates its absorption. We consider these complications when performing and interpreting our fits. For multiple galaxies, we also found that the \ion{S}{2} lines were poorly fit when included in the simultaneous fit, such that $\log(N_{SII})$ was driven towards larger values. For these cases, we chose to exclude the \ion{S}{2} lines from the fit. 

To assess whether the LIS absorption for the CLASSY galaxies can be adequately modeled using a single dominant cloud of gas, we performed both a 1-component and a 2-component fit to an individual LIS line for each galaxy. We performed these fits on the highest-S/N uncontaminated LIS line available, which was often \ion{C}{2} $\lambda$1334. We consider the difference in the Bayesian Information Criterion (\texttt{bic}), as returned by \texttt{lmfit}, to evaluate the degree by which a 2nd component can improve the fit to the line. The \texttt{bic} is similar to $\chi^2$ but it can be more useful when comparing the quality of fit between models where the number of variables are different \citep{Schwarz_1978,Raftery_1995}. Lower values of the \texttt{bic} correspond to a better fit to the data, even when the fits have varying degrees of freedom. Following the guidelines outlined by \cite{Raftery_1995}, when the 2-component \texttt{bic} is less than the one-component \texttt{bic} by a difference of 6 or more, it is strong evidence that a single component is not ideal for modeling the absorption.

Based on the difference in \texttt{bic} values, we find that our suite of LIS absorption lines can be well-modeled by a single dominant cloud of gas for 75\% of the CLASSY galaxies. We refer to this subset as the {\it single-component sample}. Two (or more) components appear necessary to describe the observed profiles for the remaining 11/44 CLASSY galaxies (J0823+2806, J0938+5428, J1025+3622, J1112+5503, J1157+3220, J1359+5726, J1416+1223, J1448$-$0110, J1525+0757, J1612+0817), hereafter referred to as the {\it multi-component sample}. The need to use multiple components to fit these galaxies was expected from previous work \citep{Xu_2022} showing that some galaxies are better fit by the combination of a static and an outflow component. We discuss our results in the context of this previous study in the next section.

Although using one component might be an oversimplification of the absorption lines shape of the multi-component sample, we still consider these objects in our analysis. This is because we are interested in the overall average ISM properties of the galaxy when assumed to be composed of a single population of gas. Figure~\ref{fig2b} shows an example of our 1-component fit for J1359+5726, a multi-component galaxy, and although the details of the absorption profiles are missed by the fit, the overall shape to the profiles is still captured by the single-component model. We surmise that, although these absorption profiles may not be accurately reproduced, our approach may still enable us to infer average properties that are consistent with the overall scaling relations that relate the properties of different neutral and LIS metal species, as well as their connection with galaxy properties. This aspect is further explored in Sections~\ref{results:LIS_OI} and \ref{results:scaling}.

Using the best-fit Voigt profile models, we measured the equivalent widths, $W_\lambda$, and residual fluxes, $R_f$, of the individual lines. Equivalent widths were determined by integrating across the modeled line profile, while residual fluxes were estimated as the median value of the continuum-normalized model flux near the line minimum (thereby limited to between 0.0 and 1.0). We made these measurements from our fits rather than the observed spectrum in order to maintain consistency with the other measurements in our analysis. Additionally, measuring $W_\lambda$ from the fits allows for a larger number of measurements because we can include lines that are blended in the observations and those that are partially contaminated. In Appendix \ref{appen:EW_comp}, we show that the equivalent widths for \ion{C}{2} and \ion{Si}{2} lines, as measured from our fits, are with 10\% of those measured from the data even when the multi-component galaxies are included. The lines associated with \ion{O}{1} and \ion{Al}{2} generally have percent errors $<$ 20\% while those for \ion{S}{2} lines are generally $<$ 35\%. This indicates that the measurements of $W_\lambda$ from our fits are largely in agreement with those measured directly from the data.

The complete figure set showing the simultaneous single-component fits for all 44 CLASSY galaxies is available online (in Appendix~\ref{appenD}). The best-fit parameters are also listed in Table~\ref{tab:fit_results} in Appendix~\ref{appenD}.


\begin{figure*}[!ht] 
    \centering
    \includegraphics[width=0.85\linewidth]{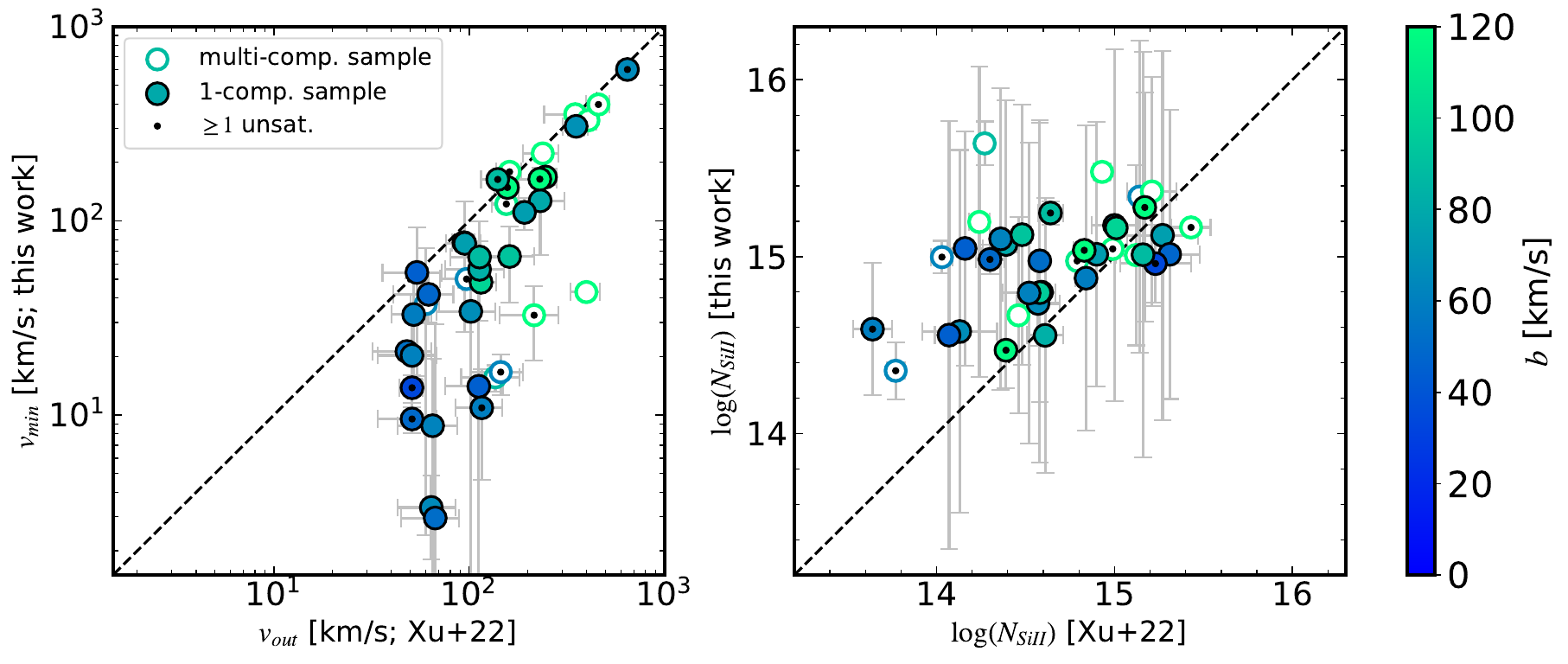}
    \figcaption{Comparing our 1-component fit results ($y$-axis) with the previous 2-component fits from \citet{Xu_2022} ($x$-axis) for CLASSY. 
    \textit{Left:} $v_{min}$ versus the outflow velocities from \citet{Xu_2022}. The dashed line corresponds to 1-to-1. The filled points correspond to the 1-component sample while the empty points are the multi-component sample. The points with a central dot correspond to the galaxies with at least one unsaturated line in their fit. The points are color-coded by the Doppler broadening.
    \textit{Right:} $\log(N_{SiII}$ versus the column densities of the \ion{Si}{2} outflows as reported by \citet{Xu_2022}. Overall, we see good agreement for galaxies that have wind-dominated absorption profiles (large $v_{out}$) while our 1-component fits extend to smaller velocities for galaxies with ISM-dominated profiles. \label{figB3}}
\end{figure*}


\subsection{Comparison to Xu et al. 2022} \label{sub:compXu}
\citet{Xu_2022} presented two-component absorption-line fits for the CLASSY galaxies for the purpose of studying their warm ionized outflows.
They used two Gaussian profiles convolved with the line PSF to simultaneously fit the static (fixed at the systemic velocity) and outflowing ISM components.
As a result, \citet{Xu_2022} find that five galaxies exhibit absorption lines consistent with the absence of outflows, with four of them belonging to our single-component sample and one being J1323-0132, where no absorption line is detected. 

Among the 23 galaxies in our single-component sample, \citet{Xu_2022} find outflow parameters between 50 and 100 km/s. 
However, we note that the presence of an outflow component does not necessarily preclude a single-component fit. In many CLASSY observations, the spectra presents profiles that can still be adequately described by a single Voigt, such as J1150+1501, seen in Figure~\ref{fig2}. \citet{Xu_2022} report an outflow velocity of 67 km/s for this galaxy while our Voigt profile fit returns $v_{min} \sim 0$ km/s. This suggests that the 2-component approach is likely combining a significant static component and a weaker outflowing component even though the spectrum of J1150+1501 does not exhibit clear evidence of such distinct components. Our findings demonstrate that a one-component Voigt profile, centered at the systemic velocity, offers a satisfactory representation of this profile.

Our multi-component sample corresponds to galaxies in \citet{Xu_2022} with $v_{outflow} > 100$ km/s and FWHM$_{outflow} > 350$ km/s. This means that the galaxies with the largest and strongest outflows are notably harder to reproduce using a single component, for which cases, a 2-component model is more appropriate. We note that the galaxies in the multi-component sample tend to have higher masses, suggesting that their environment might be more complex leading to more intricate absorption profiles \citep[see also][]{Gazagnes_2023}.

Finally, in Figure \ref{figB3}, we compare our values of $v_{LIS}$ and $\log(N_{SiII})$ with the outflow-component of the velocity ($v_{out}$) and \ion{Si}{2} column densities from the 2-component fit from \cite{Xu_2022}. It is important to note that the values reported by \cite{Xu_2022} focus exclusively on the outflow characteristics and do not account for any static component contribution.

In the left panel, we note a consistent alignment of velocity measurements from both methods with the 1-to-1 line at higher velocities. This alignment stems from the dominant influence of the outflow component on the overall absorption profile. Consequently, our single-component model yields a $v_{min}$ value consistent with the $v_{out}$ of the outflow component reported in \cite{Xu_2022}. However, at lower velocities ($\lesssim$ 100 km/s), the $v_{out}$ derived by \cite{Xu_2022} tends to converge towards similar values for all galaxies, whereas our approach finds a broader range of $v_{min}$. In addition, we find that for most of these galaxies, a single Voigt profile adequately represents the spectra. 

In the right panel, we observe an overall agreement between the derived $\log(N_{SiII})$ values, but we also note that our approach typically yields values of $\log(N_{SiII})$ that are higher than those reported in \cite{Xu_2022}. This discrepancy is expected, as our analysis encompasses the overall $\log(N_{SiII})$, whereas \cite{Xu_2022} specifically focused on the outflow component. Instances where our measurements notably exceed those of \cite{Xu_2022} likely correspond to cases where a substantial static component was identified in their study. Although \cite{Xu_2022} does not provide details on the parameters of the static component fits, it is notable that the most discrepant data points are part of our multi-component sample. This supports that the two-component fit of these galaxies combines a prominent static component in addition to the outflow.

Overall, it is important to emphasize that one purpose of this study is to evaluate the general effectiveness of the single-component fitting approach. While it is evident that a single component alone cannot precisely replicate all the CLASSY observations, we aim to assess whether such a fit still offers reasonable constraints on the average properties of the gas in these galaxies. This evaluation is particularly valuable for, e.g., estimating the expected escape fraction of ionizing photons. In such cases, kinematic effects are relatively insignificant, and it has been found that the overall gas column density and covering fraction primarily regulates the amount of escaping photons \citep{Gazagnes_2018, Chisholm_2018, Gazagnes_2020}. Hence the absence of strict consistency between \citet{Xu_2022} and our approach is not preventive for the aim of the current paper. 


\begin{figure}
    \centering
    \includegraphics[width=\linewidth]{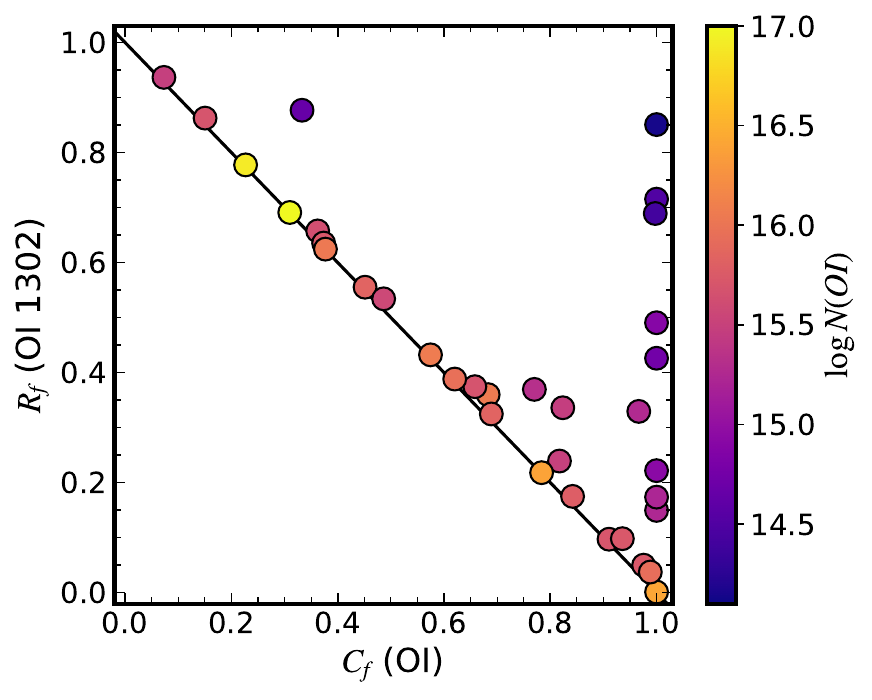}
    \figcaption{The \ion{O}{1} covering fractions versus \ion{O}{1} \W1302 residual fluxes, color-coded by $\log(N_{OI})$. This plot shows two main subsets: those that roughly follow $R_f \sim 1 - C_f$ and those with $C_f \approx 1$ and relatively large $R_f$ values. These seem to correspond to lines in the optically-thick and optically-thin regimes, respectively. This is evidenced by the correlation that column density seems to have with these measurements: the galaxies with optically thin \ion{O}{1} tend to be those with low \ion{O}{1} column densities, often with $\log(N_{OI}) < 15.5$.} 
    \label{fig:OI_RfCf}
\end{figure}


\section{LIS metal and OI gas properties} \label{results:LIS_OI}
In this section, we explore the connection between different ion properties ($\log(N)$, $C_f$, $W_\lambda$), particularly between low-ionization gas and neutral gas as traced by LIS and \ion{O}{1} absorption, respectively. Numerous studies have used interstellar LIS metal absorption lines to trace the neutral ISM and CGM \citep[e.g.,][]{Grimes_2009,James_2014,Heckman_2015,Chisholm_2016,James_2018,Gazagnes_2020,Hernandez_2020,Xu_2022} since these ions have similar ionization potentials to \ion{H}{1}. When using absorption-line studies to investigate the neutral ISM, the ideal scenario would be to look directly at \ion{H}{1} absorption via the Lyman series absorption features. However, this is not feasible for the majority of CLASSY galaxies due to their low redshifts and the blue limit of the COS detector. 

Alternatively, with a similar ionization potential to \ion{H}{1} (13.618 eV), \ion{O}{1} can be used as a proxy (13.598 eV).
However, \ion{O}{1} $\lambda$1302 is the only strong neutral ISM absorption line that is present in the majority of the CLASSY spectra and we found it to be saturated for $\sim$86\% of our sample. This motivates our inclusion of singly-ionized absorption lines here, ones chosen for their low ionization energies that largely overlap with \ion{H}{1}. The range of energies associated with these lines, however, extends beyond the upper limit of \ion{H}{1} gas and means they are also able to coexist with ionized H gas. We investigate the opacity of these lines in \S~\ref{sub:tau} and the extent by which their equivalent widths (\S~\ref{sub:EW}), column densities (\S~\ref{sub:logN}), and covering fractions (\S~\ref{sub:cf}) agree with properties of the \ion{O}{1} gas. 
Finally, Section~\ref{sub:LIS_as_OI_tracers} presents our recommendations as to the use of various LIS absorption lines to trace properties of the \ion{O}{1} gas and considers the physical interpretation behind their varying levels of agreement.


\begin{figure*}[t]
    \centering
    \includegraphics[width=\linewidth]{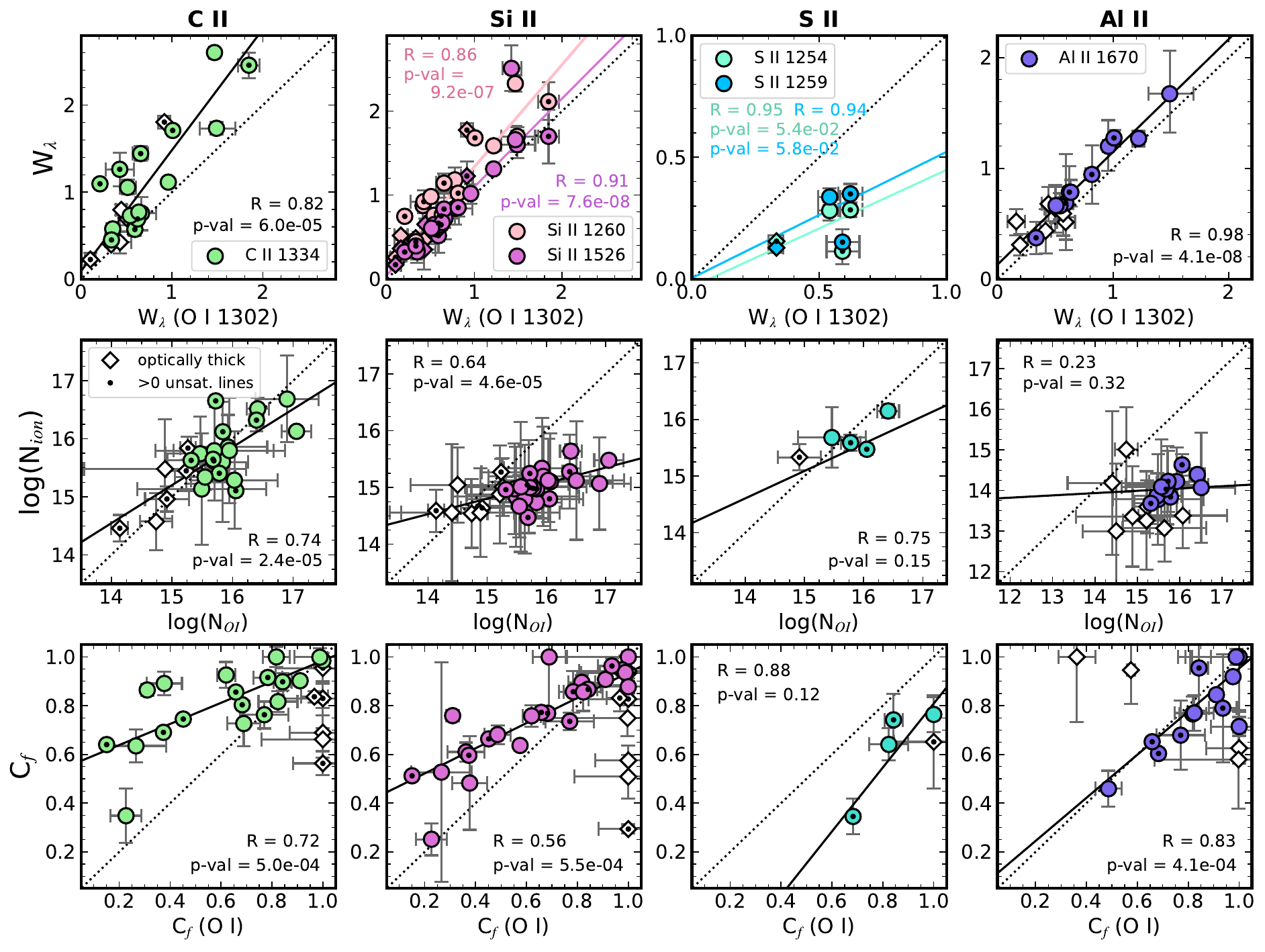}   
    \figcaption{Comparing \ion{O}{1} and LIS ion fit results: $W_\lambda$ (\textit{top row}), $\log(N)$ (\textit{middle row}), and $C_f$ (\textit{bottom row}). The $x$-axis of all panels correspond to \ion{O}{1} while the $y$-axis corresponds to \ion{C}{2}, \ion{Si}{2}, \ion{S}{2}, and \ion{Al}{2}. The filled points represent galaxies with $R_f \sim 1 - C_f$ while the empty diamond points correspond to those with $C_f \sim 1$ and a non-zero $R_f$. We expect that the difference between these subsets is related to their opacity, with the optically-thick lines shown by filled points while the empty diamonds are optically-thin, at least based on classifications described in Figure \ref{fig:OI_RfCf} for \ion{O}{1}. The galaxies with $\geq$1 unsaturated line in the fit are distinguished by a point in their center. The $W_\lambda$ values are measured for individual lines (\ion{C}{2} \W1334, \ion{S}{2} \W1260 and \W1526, \ion{S}{2} \W1254 and \W1259, and \ion{Al}{2} \W1671), while $\log(N)$ and $C_f$ are determined by all lines for that ion that were included in the fit. The linear fits to the solid points are shown in each panel by the solid black line (pink/blue to differentiate for \ion{S}{2} and \ion{Si}{2} $W_\lambda$ panels) while 1-to-1 is the dotted line. The measurements shown in this figure and used in these fits are limited to those with S/N $>$ 1. The Pearson correlation coefficients ($R$) and their associated $p$-values are listed in the panels.
    \label{fig5}}
\end{figure*}

\begin{figure*}
    \centering
    \includegraphics[width=0.9\linewidth]{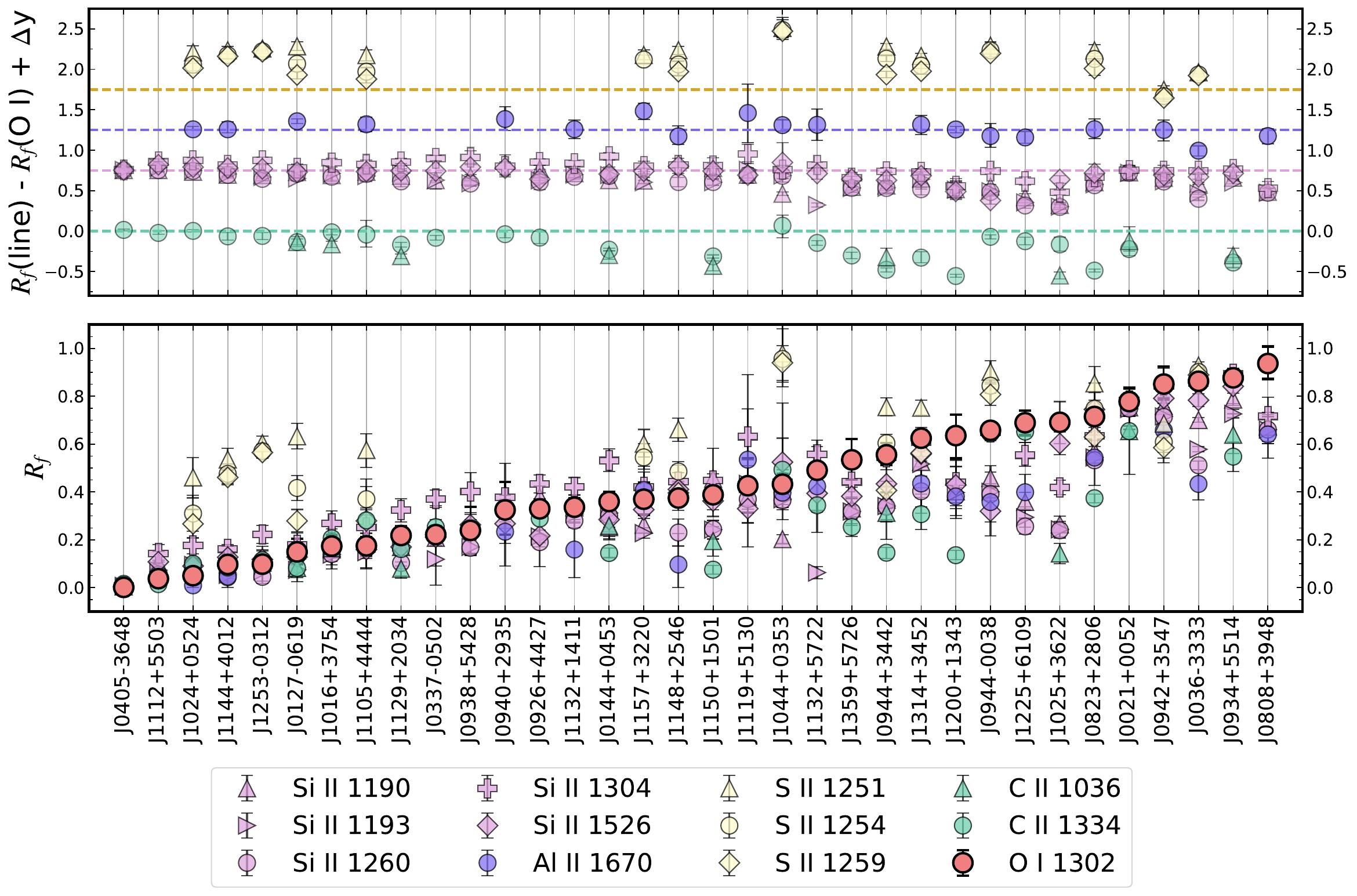}
    \figcaption{The residual fluxes $R_f$ of the lines for each ion (\ion{O}{1} \W 1302: red points, \ion{C}{2}: green points, \ion{S}{2}: yellow points, \ion{Si}{2}: pink points, \ion{Al}{2}: purple points). The $x$-axis lists the galaxies in our sample with at least 1 \ion{O}{1} line in the fit, ordered by $R_f$(\ion{O}{1}). \textit{Top:} The difference between $R_f$(line) and $R_f$(\ion{O}{1}). Overall, this shows how the residual fluxes of different ions relate to that of \ion{O}{1}. \label{fig6b}}
\end{figure*}


\subsection{Opacity} \label{sub:tau}
In the following subsections, we discuss the effects of optically-thin versus -thick absorption on the trends between LIS and \ion{O}{1} measurements. We investigate these two subsets and then explore how they relate to other line properties in both this subsection and the following ones.

Figure \ref{fig:OI_RfCf} shows how our fit results for $C_f$(\ion{O}{1}) relate to the residual fluxes measured from the \ion{O}{1} \W1302 models. Two distinct populations can be seen on this plot: 1) those that approximately follow $C_f \sim 1 - R_f$ (shown by the solid black line) and 2) those with a large covering fraction ($C_f \sim 1$) and non-zero measurements of $R_f$. This figure is color-coded by the \ion{O}{1} column density, which indicates that the former subset has larger column densities than the later. We therefore conclude that the galaxies with high opacities (optically-thick) follow the trend $C_f \sim 1 - R_f$ while those with low opacities (optically-thin) tend to have a variety of residual flux values despite covering fractions near unity. 


\subsection{Equivalent Widths} \label{sub:EW}
We measured the equivalent widths, $W_\lambda$, from the best-fit Voigt profile models for our series of absorption lines (see \S~\ref{sub:overall_fits}). These measurements are shown in the top row of Figure \ref{fig5}, which compares $W_\lambda$(\ion{O}{1} \W1302) with the $W_\lambda$ for the strongest lines of each ion in our sample (\ion{C}{2} \W1334, \ion{Si}{2} \W1260, \ion{Si}{2} \W1526, \ion{S}{2} \W1254, \ion{S}{2} \W1259, \ion{Al}{2} \W1671). Since \ion{S}{2} \W1259 and \ion{Si}{2} \W1260 tend to be blended together for many CLASSY galaxies, with weak \ion{Fe}{2} \W1260 likely also contributing in some cases, we chose to also include \ion{S}{2} \W1254 and \ion{Si}{2} \W1526 in this figure. Since these measurements for $W_\lambda$ are based on the individual models (i.e., unblended) for each of these lines, we would not expect contamination between $W_\lambda$ for \ion{S}{2} \W1259 and \ion{Si}{2} \W1260, but \ion{S}{2} \W1254 and \ion{Si}{2} \W1526 are included to help evaluate this assumption.

In Figure \ref{fig5}, the filled points corresponding to measurements from our 1-component sample while the empty ones are for the multi-component sample. The fits that include at least one unsaturated line are distinguished with a dot at their center. We fit each trend with a linear regression using the \texttt{scipy.polyfit} package and weights based on the $x$ and $y$ uncertainties, added in quadrature. 
The linear models are shown by solid lines in the figure and the dotted lines correspond to the 1-to-1 relationship. We note, however, that we do not expect our $W_\lambda$ measurements to lie directly along the 1-to-1 line due to differences in the chemical abundances and oscillator strengths of these elements. These panels include the Pearson correlation coefficients ($R$) and their associated $p$-values, determined using the package \texttt{scipy.stats.pearsonr}, with values ranging from -1 (monotonically decreasing) to 1 (monotonically increasing), where 0 corresponds to no correlation. The $p$-value for these trends is indicative of the probability that an uncorrelated system could return a value for $R$ that is at least as extreme as the one it computes. For all $W_\lambda$ trends in this figure, the $p$-values indicate that the estimates for $R$ are robust with a confidence level greater than 3$\sigma$.

For all four LIS ions (\ion{C}{2}, \ion{Si}{2}, \ion{S}{2}, and \ion{Al}{2}), we found significant correlations between the equivalent width of their strongest line(s) and that of \ion{O}{1} \W1302, both for our 1-component and multi-component samples. The large values of $R$ (with low $p$-values) is evidence for the strength of these trends. While the \ion{Si}{2} \W1526 and \ion{Al}{2} \W1671 measurements tend to lie along the 1-to-1 line with \ion{O}{2} \W1302, the trends with \ion{Si}{2} \W1260, \ion{C}{2} \W1334, and \ion{S}{2} show some deviations. The slope differs for \ion{C}{2} \W1334 and \ion{Si}{2}, steeper than the 1-to-1 line for the former and flatter for the later lines. Unlike \ion{Si}{2} \W1526, \ion{Si}{2} \W1260 seems to be offset from, but relatively parallel to, the 1-to-1 line. This could suggest that blending with \ion{S}{2} \W1259 and/or \ion{Fe}{2} \W1260 could be responsible for the deviation. 
The trends for \ion{S}{2} \W1254 and \W1259 are similar to one another, but have substantially lower slopes than the other ions ($\sim 0.65$). 
This suggests that the amount of \ion{S}{2} absorption is not as sensitive to the \ion{O}{1} absorption.

In general, we see a tight correlation between interstellar \ion{O}{1} $W_\lambda$(\W1302) and the equivalent widths measured from \ion{Si}{2} \W1260, \ion{Si}{2} \W1526, \ion{C}{2} \W1334, \ion{S}{2} \W1254, \ion{S}{2} \W1259, and \ion{Al}{2} \W1671, for the vast majority of CLASSY galaxies. The relationship between $W_\lambda$ and $b$ for these lines seems to be one of the main drivers for the trends we see between $W_\lambda$ measurements.   


\subsection{Column Densities} \label{sub:logN}
The middle row of Figure~\ref{fig5} shows a comparison between the \ion{O}{1} column densities with those for the ions in our sample (\ion{C}{2}, \ion{Si}{2}, \ion{S}{2}, \ion{Al}{2}). These panels show a general agreement between column density measurements, particularly between \ion{C}{2}, \ion{S}{2}, and \ion{O}{1}. We find the \ion{Al}{2} column densities to be lower than \ion{O}{1} column densities, though roughly parallel to the 1-to-1 line. Meanwhile, the \ion{Si}{2} column densities are approximately equal to the \ion{O}{1} column densities at $\log(N) \lesssim 15$, while the slope in this trend is significantly flatter than the 1-to-1 line with increasing column densities. Based on the correlation coefficients and associated $p$-values for these trends, however, \ion{C}{2} and \ion{Si}{2} seem to have the most robust correlations with \ion{O}{1}. The \ion{S}{2} and \ion{Al}{2} column densities may also correlate with \ion{O}{1} but since we have fewer measurements from these ions for CLASSY, it corresponds to less statistically significant trends. Overall, $\log(N_{CII})$ seems particularly sensitive to $\log(N_{OI})$, even though we expect these measurements to be less reliable than $\log(N_{SiII})$ since fewer \ion{C}{2} transitions were included in the fits and it is therefore more sensitive to error due to saturation.


\subsection{Covering Fraction + Residual Flux} \label{sub:cf}
The bottom row of Figure~\ref{fig5} compares the \ion{O}{1} and ion covering fractions as derived from our fits. The optically thick measurements (empty points) show large scatter from the trends established by the optically thin lines (filled points). For this row, we only included the filled points in our linear fits. When the lines are not optically thick, we see that the \ion{O}{1} and LIS covering fractions scale strongly with one another.

In addition to comparing the covering fractions associated with different ions, we also consider the relationship between the residual fluxes of these lines. Figure \ref{fig6b} shows this comparison, particularly with regards to comparing the LIS residual fluxes with that for \ion{O}{1} \W1302. The lower panel of this figure shows the $R_f$ measurements for the galaxies in our sample with at least one \ion{O}{1} line included in the fit (\ion{O}{1} \W 1302: red points, \ion{C}{2}: green points, \ion{S}{2}: yellow points, \ion{Si}{2}: pink points, \ion{Al}{2}: purple points). The $x$-axis is ordered by increasing $R_f$(\ion{O}{1}). The top panel has the same $x$-axis as the bottom panel, but it is showing the difference between $R_f$(line)-$R_f$(\ion{O}{1}), with an offset in $y$ to organize the points by their associated ion. The dashed lines correspond to zero difference for the ion of the same color points. Overall, this figure demonstrates how the residual fluxes of different ions might relate to that of \ion{O}{1}. We see that \ion{S}{2} tends to have larger residual fluxes than the other ions in our sample, while \ion{Si}{2} and \ion{C}{2} can present smaller residual fluxes compared to \ion{O}{1} \W1302 (i.e. are deeper absorption profiles). 

We found the scatter between $R_f$ for \ion{O}{1} \W1302 and \ion{C}{2} \W1334 to be related to the Doppler broadening of the lines, with galaxies with larger $b$ values presenting lower \ion{C}{2} \W1334 residual fluxes for the same values of $R_f$(\ion{O}{1}). This relationship between the residual fluxes and their scatter is not as prominent or even present for the other LIS lines in our sample. We infer that this trend between scatter and $b$ for \ion{C}{2} \W1334 could be due to flourescent \ion{C*}{2} \W1335 that would preferentially contaminate the \ion{C}{2} \W1334 absorption in the galaxies with broader absorption lines. This contamination would cause $R_f$ for \ion{C}{2} \W1334 to be lower (i.e. deeper absorption lines) than $R_f$ for \ion{O}{1} \W1302.


\begin{figure}
    \centering
    \includegraphics[width=\linewidth]{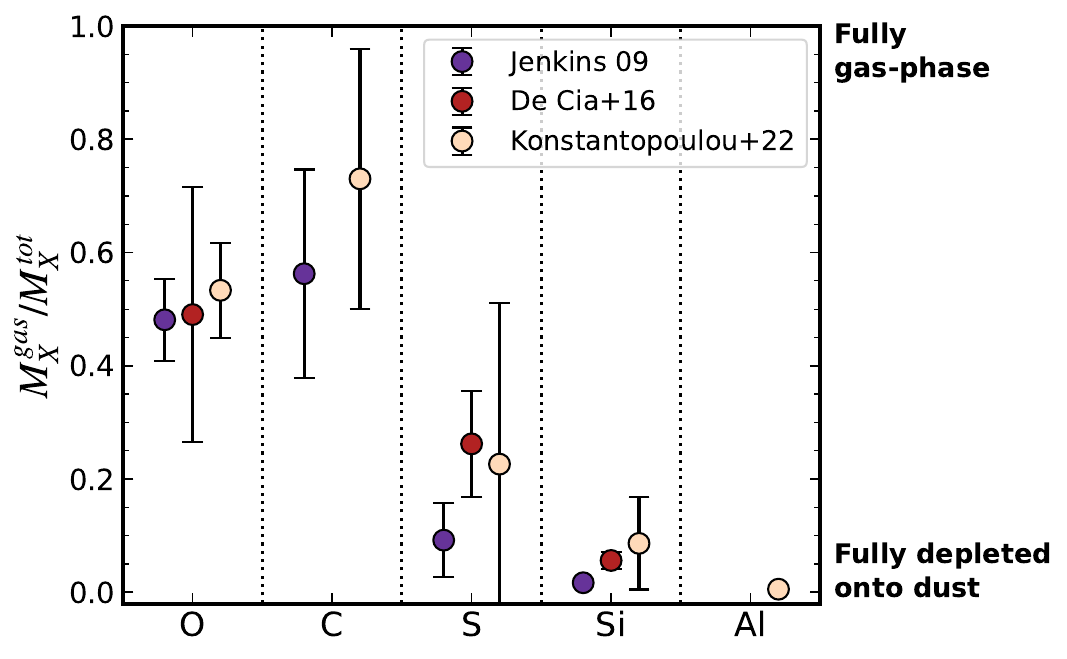}
    \figcaption{The median gas-phase fractions $(M_X^{gas}/M_X^{tot})$ of the metals in our analysis for the CLASSY galaxies, as estimated from their depletion factor $F_\star$ and the empirical trends from \cite{Jenkins_2009}, \cite{deCia_2016}, and \cite{Konstantopoulou_2022}. This shows that the total fraction of carbon in the gas (versus in the dust) is most similar to that of oxygen, which may explain why \ion{C}{2} properties have a stronger correlation with \ion{O}{1} compared to the other ions.} \label{fig:depletion}
\end{figure}


\subsection{LIS Lines as Tracers of \ion{O}{1}} \label{sub:LIS_as_OI_tracers}
In general, we find a strong correlation between the observed absorption-line properties of the neutral \ion{O}{1} gas and some characteristics of the low-ionization gas, particularly with \ion{C}{2} \W1334. Although the trends with \ion{Si}{2} also seem significant, the relationship with the \ion{O}{1} column density is quite flat compared to the \ion{C}{2} trend. This suggests that the density of \ion{C}{2} may be more closely linked to the density of the neutral ISM than is the case with \ion{Si}{2}. The trends we find for \ion{S}{2} and \ion{Al}{2} may also be robust but we are limited by the fact there are fewer measurements for these ions.

We theorize that differences in the strengths of these trends-- and specifically why \ion{C}{2} seems to trace \ion{O}{1} better than the other ions do-- could be the result of different dust depletion rates. To investigate this idea, we estimated the dust depletion factor ($F_\star$) of CLASSY galaxies using the empirical relationship presented by \cite{Jenkins_2009} and shown below (Equation \ref{eq:Fstar}).
\begin{equation}\label{eq:Fstar}
    F_\star = 0.772 \: + \: 0.461 \times \log(N_{HI})
\end{equation}
\cite{Jenkins_2009} parameterized $F_\star$ as the overall strength of dust depletion across 243 MW sightlines, ranging from $F_\star = 0$ (the least depleted sightline) and $F_\star = 1$ (the most depleted sightline). To estimate $F_\star$ for the CLASSY galaxies, we used the values of $\log(N_{HI})$ presented by \cite{Hu_2023}, as measured from fits to the Ly$\alpha$ emission (discussed more in Section~\ref{sec:OI_HI}).\looseness=-2

In Figure \ref{fig:depletion} we examine the fraction of an ion that remains in the gas phase in our sample, $M_X^{gas}/M_X^{tot}$, using the empirical trends presented by \cite{Jenkins_2009}, \cite{deCia_2016}, and \cite{Konstantopoulou_2022}. 
This shows that the total fraction of carbon in the gas (compared to that in the dust) is most similar to that of oxygen $(M_O^{gas}/M_O^{tot} \sim 0.5)$, which may explain why \ion{C}{2} properties have a stronger correlation with \ion{O}{1} compared to the other ions. Regardless of the empirical trend used, the gas-phase fractions for sulfur, silicon, and aluminum, are significantly lower than those for carbon and oxygen. This means that higher fractions of these elements are bound to dust in the ISM, rather than existing more fully in the gas-phase. Given the complex distribution and behavior of dust in the ISM, similar fractions of depletion onto dust grains may be a reason for the close agreement between the properties we derived from \ion{O}{1} and \ion{C}{2} absorption.\looseness=-2


\section{OI as a tracer of HI} \label{sec:OI_HI}
Having established correlations between the fit results for a majority of the LIS ions in this work and \ion{O}{1}, 
here we introduce a few direct measurements from \ion{H}{1} based on 21-cm emission (\S~\ref{sub:HI_meas}) and Ly$\alpha$ \citep[\S~\ref{sub:Lya}; analyzed for CLASSY galaxies by][]{Hu_2023}. Section \ref{sub:compOI_HI} presents a comparison between our \ion{O}{1} fit results and these \ion{H}{1} properties. 
Since \ion{O}{1} does not present the same concerns regarding ionization energy as the LIS ions in this analysis, it is a unique avenue to investigate how the complex geometry and dynamics of the ISM of these galaxies can still produce relatively simple absorption profiles that can be well-explained by a single population of neutral gas in the majority of cases. In this section, we investigate how certain characteristics of the neutral ISM can influence which population of gas will dominate the results returned by a simultaneous 1-component fit. 

 
\subsection{21-cm emission: FWHM$_{21cm}$,} $M_{HI}$, $\mu$  \label{sub:HI_meas} 
In this section, we present measurements of the full-width-half-maximums of the 21-cm profiles (FWHM$_{21cm}$), the \ion{H}{1} gas masses ($M_{HI}$), and the neutral gas fractions ($\mu$) for the 35 CLASSY galaxies with 21-cm \ion{H}{1} observations. We measured FWHM$_{21cm}$ by fitting the emission profiles with Gaussian models, for which the FWHM = 2.355 $\times \sigma$. A few of the CLASSY galaxies have double-horned \ion{H}{1} profiles, a sign of more systemic rotation, and for these cases, the Gaussian model still provides a reasonable estimate of FWHM$_{21cm}$ even if it does not accurate reproduce the profile's shape. All emission profiles used for these 21-cm measurements are included in Appendix~\ref{appen21cm}, along with notes regarding the \ion{H}{1} characteristics of individual galaxies.

We calculated $M_{HI}$ from the 21-cm line fluxes ($S_{HI}$; introduced in $\S$~\ref{subsub:data_reduc}), using the following equation:
\begin{equation} \label{eq:HI}
    M_{HI} = 2.36 \times 10^5 \: \left(\frac{D}{\mathrm{Mpc}}\right)^2 \: \left(\frac{S_{HI}}{\mathrm{Jy \: km/s}}\right) \: \mathrm{M_\odot}
\end{equation}
where $D$ is the distance to the galaxy. Uncertainties in $M_{HI}$ were determined from the $S_{HI}$ uncertainties. This equation assumes the optically-thin emission limit of the radiative transfer equation, requiring the source to be entirely captured within the telescope's beam. This assumption holds for CLASSY galaxies since their angular sizes are much smaller than the GBT's primary beam at 1.4 GHz \citep[FWHM $\sim 9\arcmin$; ][]{Boothroyd_2011}. We are not concerned about contamination from nearby objects in the beam because the CLASSY survey specifically chose targets in isolated systems. 

The neutral gas fraction, which is defined as $\mu = M_{HI}/(M_{HI}+M_\star)$, characterizes the quantity of ``untapped" gas available as reservoirs for star formation, in contrast to the amount of stars that have already formed. It is therefore an essential parameter to comprehend the complex interplay between gas dynamics, star formation, and galaxy evolution. 


\begin{deluxetable}{l|r|cr|rr|cr}
\tablewidth{\textwidth}
\setlength{\tabcolsep}{1pt}
\renewcommand{\arraystretch}{0.9}
\tablecaption{HI 21-cm measurements \label{tab:HI}}
\tabletypesize{\footnotesize}
\tablehead{
\CH{Galaxy} & \CH{$z$} & \CH{FWHM}   & \CH{$S_{HI}$}  & \CH{$\log(M_{HI})$} & \CH{$\mu$}            & \multicolumn{2}{c}{Program} \\ [-2ex]
\CH{}       & \CH{}    & \CH{(km/s)} & \CH{(Jy km/s)} & \CH{($M_\odot$)}    & \CH{($\times 100\%$)} & \CH{} & \CH{} }
\startdata
J0036$-$3333                    & 0.021 & 91 & 0.31$\pm$0.02 & 8.77$\pm$0.02  & 29.7$^{+13.9}_{-10.9}$ & {14B-306} & $\star$\\ 
\hline
\multirow{2}{*}{J0127$-$0619}   & \multirow{2}{*}{0.005}    &95 & 1.13$\pm$0.12 & 8.16$\pm$0.05  & 20.8$^{+7.7}_{-6.3}$   & {HICLASS} & $\star$\\
                                & \multicolumn{1}{r|}{}     & - & 1.26$\pm$0.20 & 8.21$\pm$0.07  & 22.8$^{+8.5}_{-7.0}$   &  (a) & \\
\hline
\multirow{2}{*}{J0144$+$0453}   & \multirow{2}{*}{0.005}    & 75 & 3.85$\pm$0.02 & 8.66$\pm$0.01  & 91.1$^{+3.6}_{-5.6}$   & {HICLASS} & $\star$\\
                                & \multicolumn{1}{r|}{}     & - & 3.65$\pm$0.16 & 8.63$\pm$0.02  & 90.6$^{+3.7}_{-6.0}$   & (b) & \\
\hline
\multirow{2}{*}{J0337$-$0502}   & \multirow{2}{*}{0.014} & - & 1.45$\pm$0.25 & 9.07$\pm$0.08  & 99.0$_{-0.8}^{+0.4}$   & (a) & \\
                                & \multicolumn{1}{r|}{}   & - & 0.61$\pm$0.25 & 8.69$\pm$0.18  & 97.7$^{+1.1}_{-2.2}$   & (c) & \\ 
\hline
J0405$-$3648                    & 0.003 & 34 & 1.96$\pm$0.02 & 7.83$\pm$0.01  & 94.3$^{+2.6}_{-4.7}$   & {HICLASS} & $\star$\\
\hline
J0808$+$3948                    & 0.091 & - & $<$ 0.003      & $<$ 8.10       & $<$ 8.7              & {06B-047} & $\star$\\
\hline
\multirow{2}{*}{J0823$+$2806}   & \multirow{2}{*}{0.047} & 332 & 1.51$\pm$0.30 & 10.19$\pm$0.08 & 86.7$^{+6.7}_{-11.8}$  & {11A-057} & $\star$\\ 
                                & \multicolumn{1}{r|}{}   & - & 1.20$\pm$0.10 & 10.09$\pm$0.04 & 83.9$^{+7.9}_{-12.8}$  & (d) & \\
\hline
\multirow{2}{*}{J0926$+$4427}   & \multirow{2}{*}{0.181} & - &  $<$ 0.16      & $<$ 10.45      & $<$ 98.0               & {11A-057} & $\star$\\  
                                & \multicolumn{1}{r|}{}  & - & $<$ 1.70      & $<$ 11.49      & $<$ 99.8               & (d) & \\
\hline
J0934$+$5514                    & 0.003 & 40 &  2.79$\pm$0.12 & 7.88$\pm$0.02  & 97.6$^{+0.7}_{-0.9}$   & (e) & $\star$\\ 
\hline
J0938$+$5428                    & 0.102 & - & $<$ 0.002      & $<$ 8.07      &  $<$ 7.8           & {06B-047} & $\star$\\ 
\hline
\multirow{2}{*}{J0940$+$2935}   & \multirow{2}{*}{0.002} & 69 & 2.06$\pm$0.03 & 7.40$\pm$0.01  & 82.9$^{+6.3}_{-8.7}$   & {HICLASS} & $\star$\\
                                & \multicolumn{1}{r|}{} & - & 2.05$\pm$0.25 & 7.40$\pm$0.05  & 83.0$^{+6.3}_{-9.0}$   & (f) & \\
\hline
J0942$+$3547                    & 0.015 & - & $<$ 0.004      & $<$ 6.56       & $<$ 9.1             & {HICLASS} & $\star$ \\
\hline
J0944$-$0038                    & 0.005 & 118 & 12.24$\pm$0.65 & 9.09$\pm$0.02  & 99.4$^{+0.4}_{-0.9}$   & (g) & $\star$\\
\hline
J0944$+$3442                    & 0.020 & 169 & 2.72$\pm$0.04 & 9.69$\pm$0.01  & 96.9$^{+1.8}_{-4.4}$   & {HICLASS} & $\star$\\
\hline
\multirow{2}{*}{J1016$+$3754}  & \multirow{2}{*}{0.004} & 69 & 1.10$\pm$0.02 & 7.86$\pm$0.01  & 93.2$^{+3.0}_{-5.2}$   & {HICLASS} & $\star$\\ 
                               & \multicolumn{1}{r|}{} & - & 1.51$\pm$0.39 & 8.00$\pm$0.11  & 94.9$^{+2.4}_{-4.4}$   & (f) & \\
\hline
J1024$+$0524                    & 0.033 & - & $<$ 0.04      & $<$ 8.28       & $<$ 71.0               & {19A-301} & $\star$\\
\hline
J1044$+$0353                    & 0.013 & 63 & 0.31$\pm$0.03 & 8.36$\pm$0.04  & 97.3$^{+1.6}_{-3.9}$   & {HICLASS} & $\star$\\
\hline
J1105$+$4444                    & 0.022 & 112 & 3.34$\pm$0.05 & 9.84$\pm$0.01  & 87.9$^{+5.5}_{-8.8}$   & {HICLASS} & $\star$\\
\hline
J1119$+$5130                    & 0.005 & 97 & 1.48$\pm$0.05 & 8.11$\pm$0.01  & 95.6$^{+1.2}_{-1.7}$   & {05A-034} & $\star$\\
\hline
J1129$+$2034                    & 0.005 & 82 & 6.27$\pm$0.02 & 8.78$\pm$0.01  & 83.2$^{+8.8}_{-15.0}$  & {HICLASS} & $\star$ \\
\hline
J1132$+$1411                    & 0.018 & 94 & 2.02$\pm$0.02 & 9.45$\pm$0.01  & 85.4$^{+5.4}_{-7.8}$   & {HICLASS} & $\star$\\
\hline
\multirow{3}{*}{J1132$+$5722}   & \multirow{3}{*}{0.005} & 86 & 14.09$\pm$0.04 & 9.19$\pm$0.01  & 98.7$^{+0.6}_{-1.1}$   & {HICLASS} & $\star$\\
                                & \multicolumn{1}{r|}{} & - & 13.85         & 9.19           & 98.7                   & (h) & \\
                                & \multicolumn{1}{r|}{} & - & 3.9           & 8.80           & 96.9                   & (i)  & \\
\hline
J1148$+$2546                    & 0.045 & 171 & 2.21$\pm$0.12 & 10.32$\pm$0.02 & 99.3$^{+0.4}_{-0.8}$   & {19A-301} & $\star$\\
\hline
J1150$+$1501                    & 0.002 & 47 & 1.20$\pm$0.02 & 7.50$\pm$0.01  & 82.0$^{+7.7}_{-11.6}$  & {HICLASS} & $\star$\\
\hline
J1157$+$3220                    & 0.011 & 189 & 28.14$\pm$0.07 & 10.17$\pm$0.01 & 93.1$^{+3.4}_{-6.6}$   & {HICLASS} & $\star$\\
\hline
J1200$+$1343                    & 0.067 & 153 & 4.30$\pm$0.04 & 10.96$\pm$0.01 & 99.9$^{+0.1}_{-0.3}$   & {HICLASS} & $\star$\\
\hline
J1225$+$6109                    & 0.002 & 63 & 2.24$\pm$0.03 & 7.73$\pm$0.01  & 80.4$^{+9.5}_{-15.5}$  & {HICLASS} & $\star$\\
\hline
J1253$-$0312                    & 0.023 & 44 & 0.17$\pm$0.02 & 8.61$\pm$0.05  & 89.9$^{+6.7}_{-16.1}$  & {19A-301} & $\star$\\
\hline
J1314$+$3452                    & 0.003 & 62 & 3.19$\pm$0.02 & 8.06$\pm$0.01  & 76.2$^{+10.2}_{-14.0}$ & {HICLASS} & $\star$\\
\hline
J1323$-$0132                    & 0.023 & - & $<$ 0.005      & $<$ 7.05       & $<$ 84.5              & {HICLASS} & $\star$\\
\hline
\multirow{2}{*}{J1359$+$5726}   & \multirow{2}{*}{0.034} & - & $<$ 0.07      & $<$ 8.68       & $<$ 70.1               & {19A-301} & $\star$ \\
                                & \multicolumn{1}{r|}{} & - & $<$ 0.017      & $<$ 7.94       & $<$ 25.3               & (d) & \\
\hline
J1418$+$2102                    & 0.009 & 51 & 0.47$\pm$0.02 & 8.18$\pm$0.01  & 98.9$^{+0.7}_{-2.1}$   & {HICLASS} & $\star$\\
\hline
\multirow{2}{*}{J1444$+$4237}   & \multirow{2}{*}{0.002} & 77 & 6.17$\pm$0.03 & 8.15$\pm$0.01  & 97.9$^{+0.7}_{-0.9}$   & {HICLASS} & $\star$\\
                                & \multicolumn{1}{r|}{}   & -& 7.05$\pm$0.16 & 8.21$\pm$0.01  &  98.2$^{+0.6}_{-0.8}$  & (f)  &  \\
\hline
J1521$+$0759                    & 0.094 & 13 & 0.53$\pm$0.01 & 10.37$\pm$0.01 & 95.9$_{-3.5}^{+2.0}$   & (j) & $\star$\\
\hline
J1545$+$0858                    & 0.038 & - & $<$ 0.02      & $<$ 8.13       & $<$ 80.2                   & {19A-301} & $\star$
\enddata 
\tablecomments{
Columns 1 and 2 list the CLASSY galaxy names and redshifts.
Columns 3--4 list the \ion{H}{1} FWHMs and fluxes measured from observations associated
with the programs in Column 7. 
Columns 5--6 contains the gas masses and the gas fractions $\mu$ = $M_{HI}$/($M_{HI}$+$M_\star$), using the total $M_\star$ estimates from \cite{Berg_2022}. All HICLASS observations are from the program GBT/21B-323. 
Rows with $\star$ correspond to our measurements from the HICLASS observations and archival data. These rows correspond to the values used in our analysis. 
\newline
$^{(a)}$\cite{Thuan_1999}, $^{(b)}$\cite{Ekta_2010}, $^{(c)}$\cite{Ekta_2009}, $^{(d)}$\cite{Pardy_2014}, $^{(e)}$\cite{Springob_2005}, $^{(f)}$\cite{Pustilnik_2007}, $^{(g)}$\cite{Meyer_2004}, $^{(h)}$\cite{Huchtmeier_2005}, $^{(i)}$\cite{Ekta_2006}, $^{(j)}$\cite{Heiles_1974}.}
\end{deluxetable}


\begin{figure}
    \centering
    \includegraphics[width=0.9\linewidth]{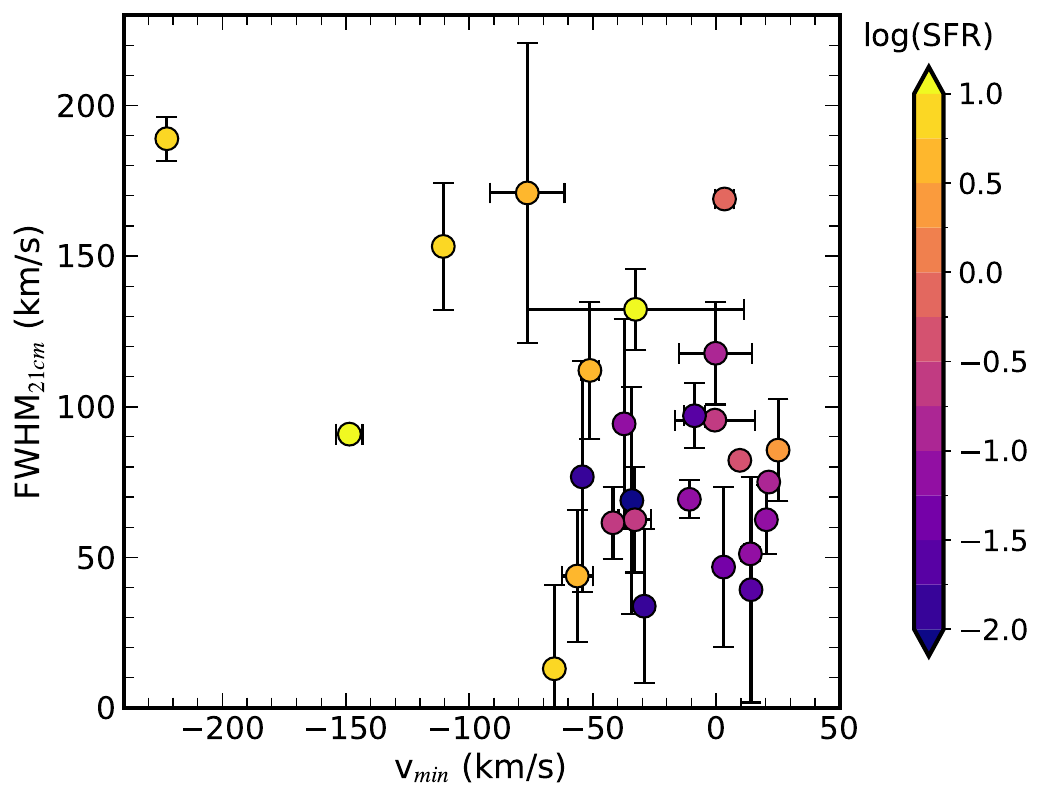}
    \figcaption{The FWHM of the 21-cm profile as a function of the velocity offset of the LIS absorption lines, color-coded by the total SFRs. This plot relates kinematic information as measured from the 21-cm emission and the LIS absorption and shows a loose anti-correlation between the FWHM$_{21cm}$ and $v_{min}$. This indicates that the galaxies with stronger outflows of neutral and LIS gas (i.e., more negative values for $v_{min}$) also tend to have broader 21-cm \ion{H}{1} profiles-- indicating more organized gas flows-- and higher star formation rates.
    \label{fig:FWHM_v}}
\end{figure}

\begin{figure*}
    \centering
    \includegraphics[width=0.7\linewidth]{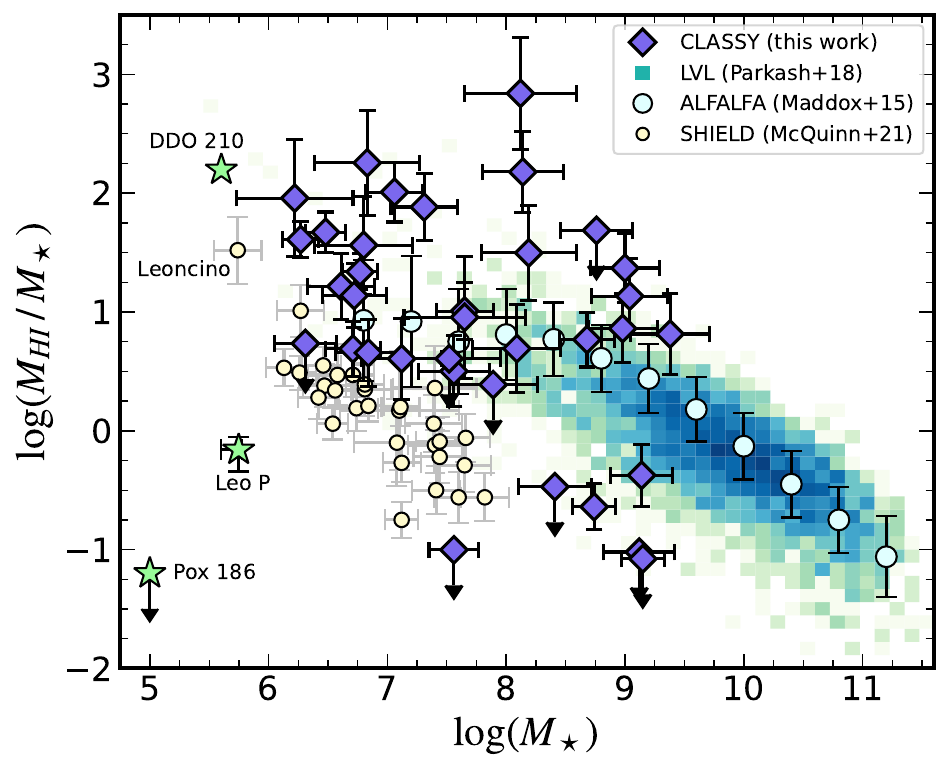}
    \figcaption{The $\log(M_{HI}$/$M_\star)$ measurements for the 27 CLASSY galaxies with \ion{H}{1} 21-cm detections and the 8 galaxies with non-detections (shown as upper limits) as a function of the stellar mass (purple diamonds). Comparing our measurements with those from the LVL \citep{Parkash_2018}, ALFALFA \citep{Saintonge_2011,Maddox_2015}, and SHIELD surveys \citep{McQuinn_2021}, we find that the majority of CLASSY galaxies are gas-rich. The green stars correspond to the low-mass star-forming galaxies Pox 186 \citep{Corbin_2002,Begum_2005,Eggen_2021}, Leo P \citep{McQuinn_2015}, and DDO 210 \citep{Hunter_2012}. Six CLASSY galaxies are particularly gas-poor compared to the rest of the sample:  J0036-3333 ($\mu = 0.297$), J0127-0619 ($\mu = 0.208$), J0808+3948 ($\mu < 0.087$), J0938+5428 ($\mu < 0.078$), J0942+3547 ($\mu < 0.091$), and J1359+5726  ($\mu < 0.253$).\label{fig1}}
\end{figure*}


Our \ion{H}{1} flux measurements, FWHMs, \ion{H}{1} gas masses, and gas fractions (assuming total $M_\star$ values from \citealt{Berg_2022}) are reported in Table~\ref{tab:HI}. This table also includes literature values for galaxies with non-GBT \ion{H}{1} 21-cm observations. 
For the seven galaxies with 21-cm non-detections, we list the $3\sigma$ upper limits on $S_{HI}$ in Table \ref{tab:HI}, where the \ion{H}{1} flux and its uncertainty were estimated from a window of the spectra where the 21-cm emission was expected to be located. 
The rows with $\star$ in their final column correspond to our measurements, made directly from the observations, and are those used in our analysis. The rows without a star refer to previous published measurements. An archival spectrum was not available for J0337-0502, making it the only galaxy in Table \ref{tab:HI} for which we used a previous literature value-- specifically the measurements from \citet{Thuan_1999}.

In Figure \ref{fig:FWHM_v}, we compare the FWHMs of the 21-cm profiles to the velocity offsets of the LIS absorption lines, color-coded based on the total SFR. This relates the kinematics derived from the 21-cm emission to that of the LIS absorption lines. Although the trend is not strong, we notice an anti-correlation between the FWHM$_{21cm}$ and $v_{min}$, which indicates that the galaxies with stronger outflows of neutral and LIS gas will also tend to have broader 21-cm \ion{H}{1} profiles (i.e., more organized gas flows) and a higher star formation rate. We also compared the FWHM$_{21cm}$ to $b$ but found less of a correlation between these kinematic parameters.

Overall, we find that the majority of CLASSY galaxies are gas-rich, with gas fractions $\mu\gtrsim75\%$ for 25 of the 27 galaxies with \ion{H}{1} detections. Figure \ref{fig1} compares our measurements of $\log(M_{HI}/M_\star)$ for CLASSY with those for other nearby galaxies \citep{Hunter_2012,McQuinn_2015,Maddox_2015,Parkash_2018,Eggen_2021,McQuinn_2021}. This shows that many of the CLASSY galaxies have large fractions of neutral gas, as they tend to lie above many of the over-plotted relations from other studies. The CLASSY sample was expected to have particularly high neutral gas fractions, given their offset towards higher SFRs on the star-forming main sequence \citep{Berg_2022}. This indicates that the majority of CLASSY galaxies are undergoing an intense star formation burst that is fueled by their large reservoirs of neutral gas. 

There are six galaxies, however, that stand out with significantly lower gas fractions than the rest of the sample (J0036-3333, J0127-0619, J0808+3948, J0938+5428, J0942+3547, and J1359+5726) and an additional four that are non-detected but may also be gas-poor, albeit with higher-upper limits (J0926+4427, J1024+0524, J1323-0132, and J1545+0858). This is interesting because (1) we would not expect starburst galaxies to have low measurements of $\log(M_{HI}/M_\star)$ and (2) there are very few nearby star-forming galaxies with 21-cm \ion{H}{1} non-detections-- prior to this, the only well-studied example was the very low-mass ($\log M_\star \sim 10^5$) galaxy, Pox 186 \citep{Corbin_2002,Begum_2005,Eggen_2021}. We note, however, that a few of the archival non-detections have relatively short integrated exposure times (particularly  J1545+0858 and J1024+0524) and therefore may not be as gas-poor as these observations suggest. We plan to perform follow-up observations to determine whether they are truly non-detected in \ion{H}{1} 21-cm. Of the eight non-detections, we are most confident in this classification for J0942+3547 and J1323-0132, which were observed through HICLASS for $>$30 minutes each.

For a significant number of CLASSY galaxies to be gas-poor or have non-detected \ion{H}{1}, it would indicate that there may be something truly unique about the sample. Considering that these gas-poor galaxies tend to have lower specific star formation rates (sSFR = SFR / $M_\star$) than their gas-rich counterparts, we theorize that they may be are nearing a post-starburst phase where much of the \ion{H}{1} reservoirs that began the starburst have now been driven out and/or ionized. Compared with Pox 186 (one of the green stars in Figure \ref{fig1}), these CLASSY galaxies are much more massive and would require far higher feedback energies to drive $\log(M_{HI}/M_\star)$ towards such low values.


\subsection{Ly$\alpha$ emission: $\log(N_{HI})$} \label{sub:Lya}
From the Ly$\alpha$ profiles for CLASSY, \cite{Hu_2023} measured the \ion{H}{1} column densities through two different methods: 1) using simple shell radiative transfer models of the Ly$\alpha$ emission profiles, and 2) Voigt profiles matched to the damped Ly$\alpha$ absorption (DLA) wings. The former -- $\log(N_{HI}^{Ly\alpha})$ -- corresponds to lower densities of \ion{H}{1} gas (18 $< \log N_{HI} <$ 20), while the later -- $\log(N_{HI}^{DLA})$ -- seems to probe \ion{H}{1} gas at higher densities ($\log N_{HI} >$ 20). Another difference between these measurements relates to the physical scale that these features probe. Since Ly$\alpha$ photons are easily scattered by neutral gas in the ISM and CGM, the observed Ly$\alpha$ emission profiles will correspond to a larger extent of the galaxy compared with the DLA wings (which are produced solely along the line-of-sight). Unlike BPASS, STARBURST99 is not yet capable of robustly estimating the stellar continuum across regions with strong Ly$\alpha$ damping. Therefore, in this analysis, we chose to focus on the Ly$\alpha$ emission measurements since we could not confidently constrain the stellar populations spanning the DLA wings.


\begin{figure}
    \centering
    \includegraphics[width=0.9\linewidth]{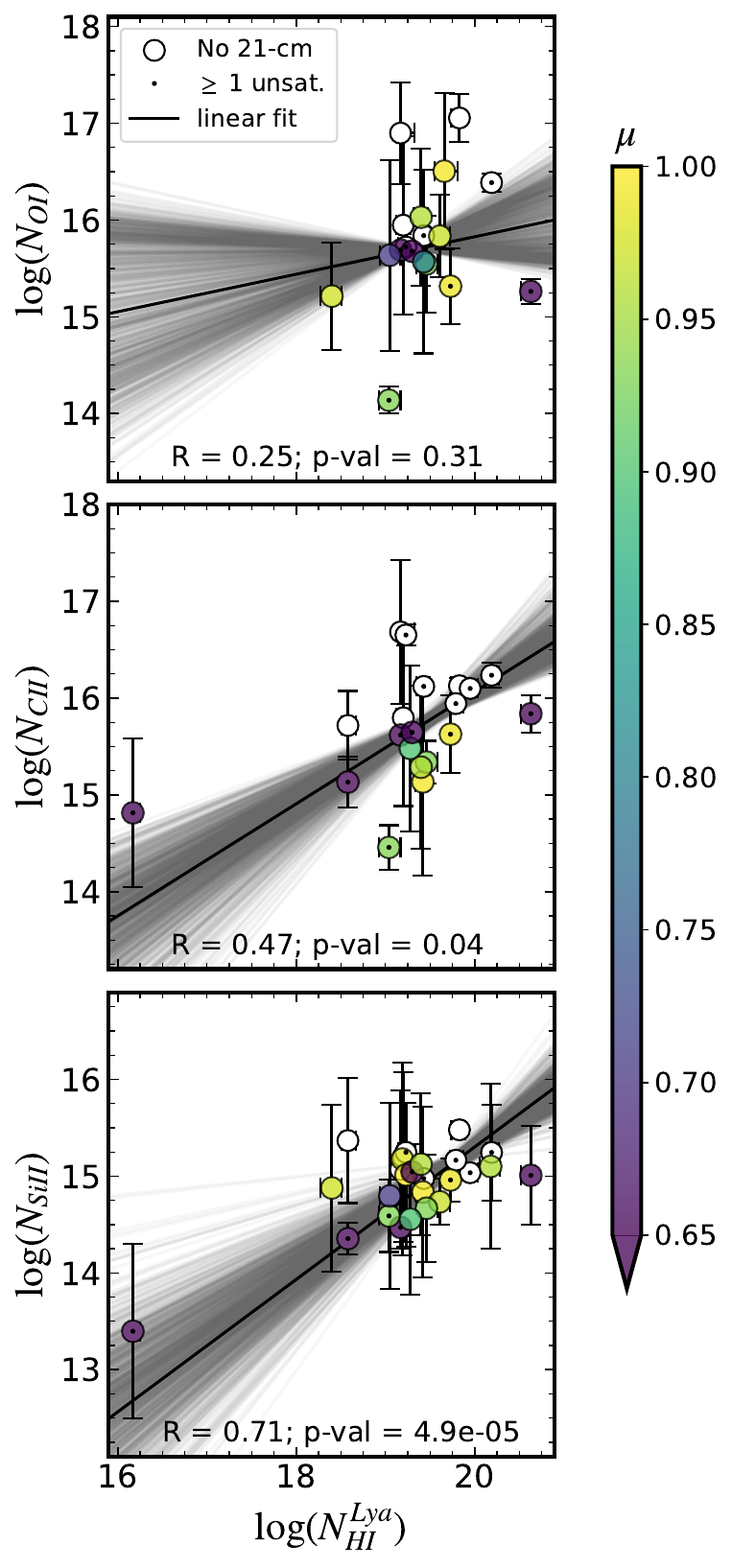}
     \figcaption{
     Column densities of neutral and low-ionization metals versus \ion{H}{1} column densities measured from Ly$\alpha$ emission \citep{Hu_2023}.
     The points are color-coded by gas fraction, $\mu$, or are white for galaxies lacking 21-cm detections. The central black dots correspond to the galaxies with at least 1 unsaturated line in their fit. The solid black lines are linear fits to each panel while the light grey lines correspond to fits from 1000 MC variations to the measurements based on their uncertainties. 
     This figure shows a lack of a significant trend between the \ion{O}{1} and Ly$\alpha$-derived \ion{H}{1} column densities, at least for this relatively small range in $\log(N_{HI}^{Ly\alpha})$. The trends in the middle and lower panels, however, have statistically significant correlations at $\gtrsim 95\%$ confidence interval based on their Pearson correlation $p$-values (listed at the bottom of each panel). Of these three panels, \ion{Si}{2} exhibits the strongest correlations, both with $N_{HI}^{Ly\alpha}$ and $\mu$. Since \ion{O}{1} and \ion{Si}{2} trace similar populations of gas (see Section~\ref{results:LIS_OI}), a trend may still exist between the \ion{O}{1} column densities and $N_{HI}^{Ly\alpha}$. \label{fig8}}
\end{figure}


\subsection{Comparing \ion{O}{1} and \ion{H}{1} properties} \label{sub:compOI_HI}
Using $M_{HI}$, $\mu$, and $\log(N_{HI}^{Ly\alpha})$, we now investigate the relationship between the \ion{O}{1} properties derived from our simultaneous fits and direct measurements from the \ion{H}{1} gas. This means we can evaluate the use of \ion{O}{1} absorption as a tracer of the neutral ISM, and furthermore, consider whether \ion{O}{1} absorption, 21-cm emission, and Ly$\alpha$ emission are produced within the same population of interstellar gas.

In the top panel of Figure \ref{fig8}, we compare our \ion{O}{1} column densities with the \ion{H}{1} column densities measured from Ly$\alpha$ emission. The points are color-coded by $\mu$ or are white for galaxies that lack a 21-cm \ion{H}{1} detection. 
Once again, we distinguish the measurements for galaxies with at least unsaturated line in their fit by a black dot at the center of these points. A linear fit to all points in this figure is shown by the solid black line, while 1000 MC variations based on the uncertainties of the points are shown with light grey. 

Somewhat surprisingly, we did not find a strong correlation between the \ion{O}{1} column density and the Ly$\alpha$-derived \ion{H}{1} column density, which we would have initially expected to see if \ion{O}{1} and \ion{H}{1} are co-spatial in the ISM. That said, the galaxies with $\log(N_{OI})$ measurements span a relatively small range in $N_{HI}^{Ly\alpha}$ which makes it difficult to conclude whether or not a trend actually exists.

In Section \ref{results:LIS_OI}, we showed both \ion{C}{2} and \ion{Si}{2} trace properties of the \ion{O}{1} gas. To investigate this further, we compare the \ion{C}{2} and \ion{Si}{2} column densities with $N_{HI}^{Ly\alpha}$ in the middle and lower panels of Figure \ref{fig8}. These two panels show stronger correlations with $N_{HI}^{Ly\alpha}$ (at $\gtrsim 95\%$ confidence) than \ion{O}{1} does, particularly in the case of \ion{Si}{2}. This indicates that there may exist a trend between \ion{O}{1} column density and $N_{HI}^{Ly\alpha}$, albeit with significant scatter. We also color-code by gas fraction in Figure~\ref{fig8}, revealing a trend with $\log(N_{SiII})$ where galaxies with lower neutral and LIS column densities also have lower neutral gas fractions.

Comparing with a different direct tracer of \ion{H}{1}-- the 21-cm emission-- we see a much more significant trend with \ion{O}{1}. This can be seen in the upper right panel of Figure \ref{fig7} which compares our measurements of $\log(N_{OI})$ and $\log(M_{HI})$. We found this trend to be significant at $>3\sigma$ confidence level ($p$-value = 0.0048) which indicates that galaxies with higher \ion{O}{1} column densities also have larger \ion{H}{1} gas masses. 
This trend between $\log(N_{OI})$ and $\log(M_{HI})$ is stronger than those between $\log(M_{HI})$ and any of the ions in our sample (\ion{Si}{2}, \ion{C}{2}, \ion{S}{2}, and \ion{Al}{2}). 
Overall, this indicates that properties of \ion{O}{1} gas can be reliable tracers of \ion{H}{1}, at least with regards to 21-cm emission.

Since the strength of the 21-cm based trends with \ion{O}{1} seem to be stronger than those with Ly$\alpha$, we need to consider the differences in \ion{H}{1} populations that they trace. 
The 21-cm observations contain emission from the \ion{H}{1} gas spanning a wide field-of-view (FWHM $\sim 9\arcmin$ at 1.4 GHz). Although this same gas is responsible for scattering the Ly$\alpha$ emission (both into and out of the line-of-sight), the smaller aperture of HST/COS (2.5$\arcsec$ diameter) will only encapsulate signal from a portion of this gas. Lastly, even though it was also observed by COS, the \ion{O}{1} absorption does not resonantly scatter like Ly$\alpha$ does and is therefore linked most directly to the line-of-sight, probing the smallest scale of the three. Likely, the reason we find \ion{O}{1} to scale more strongly with 21-cm emission compared to Ly$\alpha$ despite their large differences in scale is due to the density of gas traced by each signal. Both interstellar \ion{O}{1} absorption and 21-cm emission are predominantly produced by dense clumps of gas in the ISM whereas Ly$\alpha$ emission is more closely tied to channels of diffuse gas.


\section{Scaling relations with Galaxy Properties} \label{results:scaling}
\begin{figure*}
    \centering
    \includegraphics[width=\linewidth]{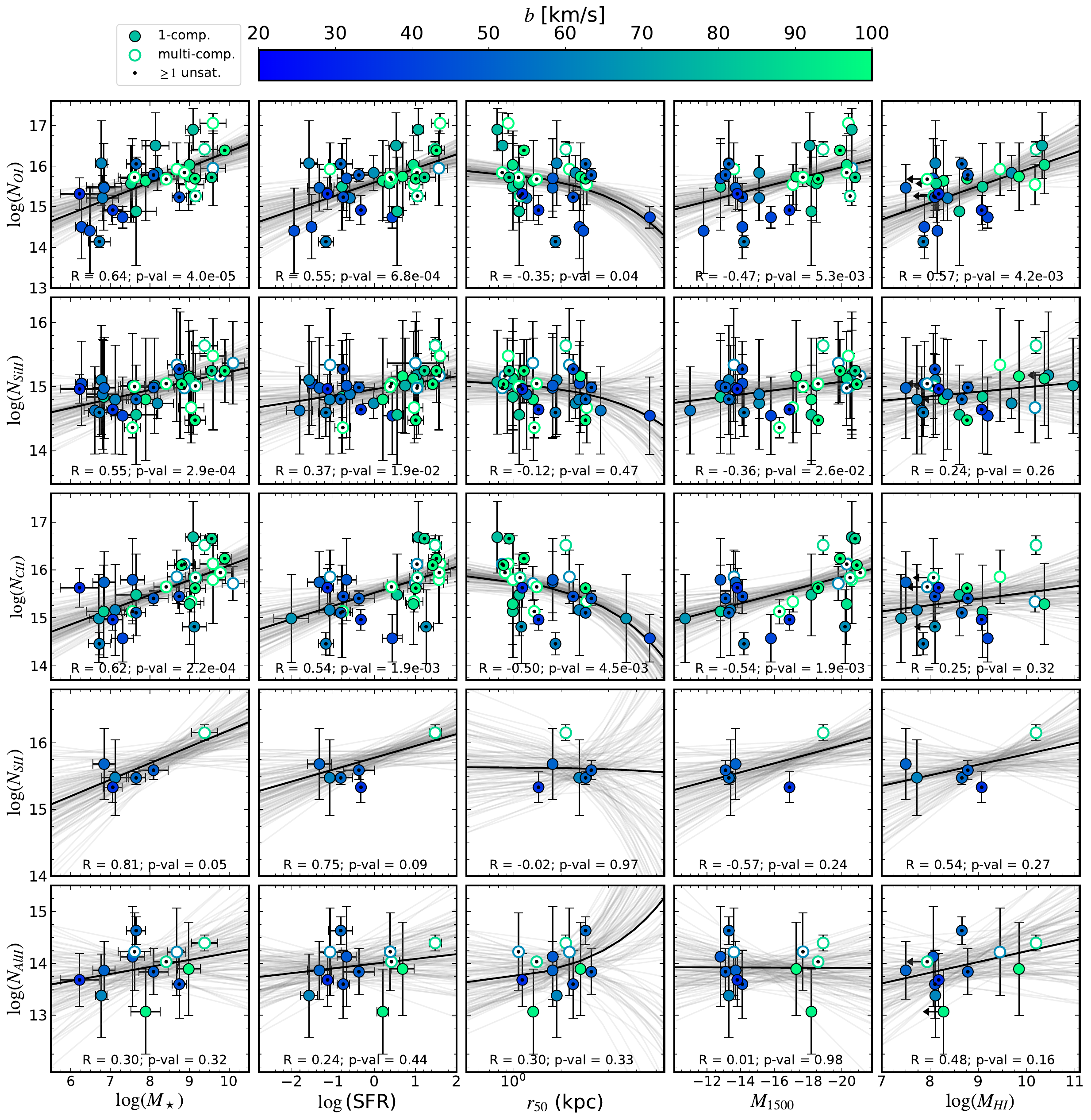}
    \figcaption{Global properties of the CLASSY galaxies ($\log(M_\star)$, SFR, $r_{50}$, $M_{1500}$, and $M_{HI}$) versus the fitted column densities for \ion{O}{1}, \ion{Si}{2}, \ion{C}{2}, \ion{S}{2}, and \ion{Al}{2}, color-coded by $b$ from our fits. The filled points correspond to measurements for the galaxies in the 1-component sample while the empty points are for those in the multi-component sample. Galaxies with at least one unsaturated line are distinguished by a black dot at their center. The solid black line is the linear fit to all points (filled + empty) with uncertainties represented by the light grey curves. From these panels, we find the strongest correlations to be with \ion{O}{1} and \ion{C}{2} column densities and these galactic properties. In general, we show that the galaxies with higher column densities of absorbing gas tend to be those that are more massive, have higher SFRs, are more compact, and are intrinsically brighter. \label{fig7}}
\end{figure*}

In this section, we explore correlations between the fitted densities of different ions and various global properties of the CLASSY galaxies, specifically $M_\star$, SFR, half-light radius ($r_{50}$), magnitude at 1500\AA\ ($M_{1500}$), and $M_{HI}$. These trends are shown in Figure \ref{fig7} with points that are color-coded by the Doppler parameter. The measurements for the galaxies that seem to be well-fit by a single component are shown with filled points while those that require more components are represented by empty points. A linear fit to all points (filled + empty) in each panel is shown by a solid black line. 
We used 100 MC runs for these fits (represented by light grey lines) to estimate the uncertainties in the linear fit coefficients, perturbing the data points according to their uncertainties for each run. 
Each panel also includes the Pearson correlation coefficient $R$ and $p$-value for each trend.

We found that multiple of these global properties scale strongly with $b$. In general, we found that the galaxies with broader absorption profiles tend to be those that are more massive ($\log M_\star > 8$), have higher star formation rates ($\log$SFR $>$ 0), are intrinsically brighter ($M_{1500}$ $<$ -16), and have larger \ion{H}{1} masses ($\log M_{HI} > 9$). 

For all five neutral and LIS elements (\ion{O}{1}, \ion{C}{2}, \ion{Si}{2}, \ion{S}{2}, and \ion{Al}{2}), we see correlations between $\log(N_{ion})$ and both $\log(M_\star)$ and SFR. This is particularly true for \ion{O}{1}, \ion{Si}{2}, and \ion{C}{2}, which are the species for which we have the most measurements. These trends indicate that the galaxies with stronger ISM absorption for these lines tend to be the more massive galaxies and those with higher SFRs. 

For \ion{O}{1}, \ion{Si}{2}, and \ion{C}{2}, there is also evidence of a trend with r$_{50}$ and M$_{1500}$, though the former is relatively weak and shows substantial scatter. These trends indicate that the more extended galaxies (those with larger r$_{50}$ values) have lower column densities and that those that are intrinsically brighter (lower values of M$_{1500}$) have higher column densities of absorbing gas. We also see a particularly strong correlation between $\log(N_{OI})$ and $\log(M_{HI})$, as was discussed in \S~\ref{sec:OI_HI}.

We also see that the multi-component galaxies are often the more massive galaxies with higher SFRs. 
These galaxies tend to have much more gas than their less massive counterparts, which agrees with the extra complexity required to accurately model them. 
In contrast, we find low-mass galaxies ($\log M_\star < 8$) to have relatively simple ISM absorption profiles that can be modeled using properties for a single population of gas.

Overall, the strongest correlations in Figure \ref{fig7} seem to be with $\log(N_{OI})$ and $\log(N_{CII})$. 
Despite the large quantity of \ion{Si}{2} measurements, the relationships with $\log(N_{SiII})$ seem to exhibit more scatter with the global properties of our sample, compared to the trends with \ion{O}{1} and \ion{C}{2}. 
As discussed in \S~\ref{sub:LIS_as_OI_tracers}, this may be a result of the high fraction of Si that is depleted onto dust, compared with C and O which have lower, and more similar, fractions. 
Both \ion{S}{2} and \ion{Al}{2} show the potential of strong correlations with many of the global properties explored here, but the smaller quantity of measurements for these lines and their strong depletion fractions make it difficult to interpret their strength relative to the trends with \ion{C}{2} and \ion{Si}{2}. 
If dust depletion is the reason for Si's poorer performance as a tracer of \ion{O}{1}, we expect this to also affect the performance of \ion{S}{2} and, especially, \ion{Al}{2} as tracers.


Below, we list the equations for our fits that have a $p$-value less than 0.05 (i.e., $\geq$95\% confidence level). Since the 1-component sample and the multi-component sample appear to follow the same trends, these equations are based on the measurements from both samples. 
The agreement between the global trends from these two subsets is further evidence that 1-component fits are capable of returning average properties of the absorbing gas, even when the absorption profiles appear to have $>$1 component of absorption.
\begin{displaymath}
    \log(N_{OI}) \: = \: (0.38 \pm 0.09) \times \log M_\star \: + \: (12.54 \pm 0.80)
\end{displaymath}
\begin{displaymath}
    \hspace{1.55cm} = \: (0.35 \pm 0.10) \times \log\mathrm{SFR} \: + \: (15.60 \pm 0.11)
\end{displaymath}
\begin{displaymath}
    \hspace{1.05cm} = \: (-0.19 \pm 0.06) \times r_{50} \: + \: (15.97 \pm 0.17)
\end{displaymath}
\begin{displaymath}
    \hspace{1.45cm} = \: (-0.10 \pm 0.04) \times M_{1500} \: + \: (13.88 \pm 0.72)
\end{displaymath} 
\begin{displaymath}
    \hspace{1.55cm} = \: (0.41 \pm 0.16) \times \log M_{HI} \: + \: (11.83 \pm 1.47)
\end{displaymath}

\begin{displaymath}
    \log(N_{CII}) \: = \: (0.31 \pm 0.09) \times \log M_\star \: + \: (13.01 \pm 0.81)
\end{displaymath}
\begin{displaymath}
    \hspace{1.6cm} = \: (0.27 \pm 0.10) \times \log\mathrm{SFR} \: + \: (15.52 \pm 0.09)
\end{displaymath}
\begin{displaymath}
    \hspace{1.1cm} = \: (-0.20 \pm 0.08) \times r_{50} \: + \: (15.97 \pm 0.17)
\end{displaymath}
\begin{displaymath}
    \hspace{1.5cm} = \: (-0.09 \pm 0.03) \times M_{1500} \: + \: (14.01 \pm 0.60)
\end{displaymath}

\begin{displaymath}
    \log(N_{SiII}) \: = \: (0.14 \pm 0.08) \times \log M_\star \: + \: (13.82 \pm 0.73)
\end{displaymath}
\begin{displaymath}
    \hspace{1.65cm} = \: (0.10 \pm 0.10) \times \log\mathrm{SFR} \: + \: (14.96 \pm 0.11)
\end{displaymath}
\begin{displaymath}
    \hspace{1.55cm} = \: (-0.03 \pm 0.04) \times M_{1500} \: + \: (14.41 \pm 0.63)
\end{displaymath}

\begin{displaymath}
    \log(N_{SII}) \: = \: (0.25 \pm 0.14) \times \log M_\star \: + \: (13.72 \pm 1.21)
\end{displaymath} 
These equations can be used to predict the neutral ISM column densities for star-forming galaxies based on their global properties, and vice versa. This may be particularly useful in estimating neutral gas characteristics or global properties of galaxies at high-redshifts. Since the ionizing radiation from galaxies at $z > 4$ can often not be observed directly due to the IGM's opacity, measuring properties of its neutral gas can present an alternative method to study the ionizing photon escape in galaxies during the EoR.

Although LIS absorption-lines may provide a way to trace neutral gas without interference from the IGM, detecting absorption at high redshifts introduces other difficulties, such as requiring much longer exposure times than is needed for emission features. When absorption is detected, the lower S/N and resolution of high-$z$ spectra also limit how well measurements from these lines can be constrained. In these cases, it can sometimes still be possible to estimate the column density of a galaxy's ISM (+CGM) from the observed LIS lines, making this a powerful tool by which to study low-ionization gas-- and perhaps the escape of ionizing radiation-- in galaxies from the distant universe.


\section{Summary} \label{summary}
We present an analysis of the neutral ISM properties for the 45 nearby star-forming galaxies from the CLASSY treasury, as determined via LIS metal absorption lines and 21-cm \ion{H}{1} emission. Despite their low redshift, the CLASSY galaxies are forming stars at rates comparable with those at $z\sim2$. Galaxies at $z\sim2$ are often prime targets in the study of galaxy evolution due to their extreme diversity of characteristics, attributed to the peak of star formation that occurred at this redshift in the universe's past. The CLASSY galaxies present this diversity but at a low redshift, which makes it possible for far more detailed study. 

We simultaneously fit a series of Voigt profiles to a subset of 15 interstellar LIS absorption lines, corresponding to \ion{C}{2}, \ion{Si}{2}, \ion{S}{2}, and \ion{Al}{2}, as well as neutral absorption lines produced by \ion{O}{1}. These fits were conducted using \texttt{lmfit} minimization, following the procedure from \cite{Gazagnes_2018} and \cite{Chisholm_2018}. Our simultaneous fits returned parameters related to the overall kinematics of the LIS gas, $v_{min}$(LIS) and $b$, and ion-specific characteristics based on the included lines, $C_f$(ion) and $N_{ion}$.

In this paper, we test the capabilities of our simultaneous 1-component absorption-line fitting procedure and find it to be a powerful method for bypassing some of the typical challenges that are faced when fitting absorption lines individually. These tests demonstrated that our fitting methodology is capable of:
\begin{enumerate}
    \item \textit{Reducing the influence of saturated lines on the returned fit results.} This is because a single unsaturated line in the fit will be able to help constrain $b$, assuming the masses of the ions are similar. This reduces the degeneracy between $b$ and $\log(N_{ion})$ for the fitted ions. 
    \item \textit{Constraining $\log{N}$ even when all the lines in the fit are saturated.} This is possible because our fits involve an MC approach to estimating uncertainties.
    \item \textit{Well-fitting the LIS absorption in $\sim$75\% of the CLASSY galaxies using a single component of neutral gas}, despite the complex structure of the ISM and CGM that is predicted by simulations. 
    The most massive galaxies -- particularly those with high SFRs-- are often the ones that require additional components, likely a consequence of their more complex ISM structure and dynamics. For galaxies with $\log(M_\star) \lesssim 8.5$, we demonstrate that a single component fit can often suffice to accurately model their LIS absorption profiles. 
    \item \textit{Determining average properties of the low-ionization and neutral ISM} for galaxies where a single component of absorbing gas is not ideal to explain the observed absorption profiles (i.e., for high-mass galaxies).
\end{enumerate}
Overall, these tests illustrate that there are multiple benefits to using simultaneous 1-component fits to study the low-ionization and neutral ISM.

Using the results of our absorption-line fits, we aim to evaluate (1) whether LIS absorption line properties accurately trace characteristics of the interstellar \ion{O}{1} gas, (2) whether \ion{O}{1} absorption can be used as a tracer of the \ion{H}{1} gas, and (3) how properties of LIS absorption relate to large-scale properties of star-forming galaxies. Our main findings for each of these points are summarized below.
\begin{enumerate}[label=(\arabic*)]
    \item \textit{LIS ions as tracers of \ion{O}{1}:} Although LIS ions are often assumed to co-exist with the neutral ISM, the range of ionization energies associated with these ions indicates that they can also coexist with ionized H gas. In this work, we test the validity of using interstellar LIS metal absorption lines to trace properties of the neutral ISM using the high-resolution, high S/N FUV spectra for the 45 galaxies in CLASSY. We compare the fitted properties of the LIS ions with those for \ion{O}{1} and find that $W_\lambda$, $\log(N_{ion})$, and $C_f$ for interstellar \ion{O}{1} absorption scales with the majority of these properties. 
    We found that, of the ions in this work, \ion{C}{2} is a particularly good indicator of \ion{O}{1}, perhaps due to the similar fraction of mass in their gas-phases.
    
    \item \textit{\ion{O}{1} as a tracer of \ion{H}{1}:} After demonstrating that the majority of ions in our analysis are closely tied to \ion{O}{1} in the ISM, we compare our \ion{O}{1} fit results with two direct measurements of \ion{H}{1}: the 21-cm \ion{H}{1} emission and Ly$\alpha$ emission. Of note, we present uniformly measured \ion{H}{1} gas masses for 35 of the CLASSY galaxies, using a combination of new 21-cm GBT observations from the HICLASS survey and archival GBT data. We find that eight of these galaxies are non-detections, indicating they may be approaching a post-starburst phase.
    
    We find a significant trend between our \ion{O}{1} column densities and the \ion{H}{1} mass measured from 21-cm emission. We find the correlation with Ly$\alpha$ emission, however, to be far weaker. We interpret this as being due to differences in the gas populations probed by these \ion{H}{1} indicators, since Ly$\alpha$ emission is more closely tied to channels of diffuse gas whereas \ion{O}{1} and 21-cm emission are predominantly produced by dense clumps of neutral gas. Overall, this helps to establish the validity of using LIS absorption lines to trace the neutral ISM but also demonstrates that direct tracers of \ion{H}{1} gas relate to different density regimes of the ISM.
    
    \item \textit{Scaling relations between LIS absorption properties and global galaxy characteristics:} 
    We present the main correlations we find between properties of the neutral and low-ionization ISM from our 1-component fits with global properties of the galaxies in the CLASSY sample ($M_\star$, SFR, r$_{50}$, $M_{1500}$, and $M_{HI}$). These equations can be used to predict the neutral ISM column densities for galaxies based on their global properties, and vice versa. 
    This may be especially useful in predicting the neutral gas properties of galaxies at high-redshifts.
\end{enumerate}

The use of LIS absorption lines to constrain properties of the neutral ISM is particularly powerful for observations of galaxies at higher redshifts, where we often cannot directly observe the \ion{H}{1} gas or the ionizing radiation these galaxies produce. Using simultaneous 1-component MC fitting to model low-ionization and neutral interstellar absorption lines, we demonstrate the validity and limitations of this technique in tracing properties of the \ion{H}{1} gas for the 45 nearby star-forming galaxies in the CLASSY galaxy survey.

\begin{acknowledgements}
We would like to thank the referee for their thoughtful review and suggestions which have greatly improved the quality of this work. Additionally, K.S.P. would like to acknowledge and thank Nissim Kanekar for his guidance with reducing the GBT observations. K.S.P. is also grateful to the GBT Observer Training Workshop and those that organized and led the program.

We acknowledge the GBT Legacy Archive and the teams responsible for the following archival programs: AGBT/05A-034, AGBT/06B-047 (PI: David Schiminovich), AGBT/11A-057 (PI: Goran Ostlin), AGBT/14B-306 (PI: Stephen Pardy), and AGBT/19A-301 (PI: Sangeeta Malhotra). The Green Bank Observatory is a facility of the U.S. National Science Foundation operated under cooperative agreement by Associated Universities, Inc.
\end{acknowledgements}

\software{polyfit, scipy.stats.pearsonr, GBTIDL \citep{Marganian_2013}, asymmetric\_uncertainty \citep{Gobat_2022}, spectres.spectres \citep{Carnall_2017}}
\facilities{HST (COS), GBT}

\bibliographystyle{aasjournal}
\bibliography{thbib.bib}


\begin{appendix}
\vspace{-0.5cm}
\noindent Section \ref{appen21cm} contains the 21-cm \ion{H}{1} observations: HICLASS in \S~\ref{appen:hiclass} and archival ones in \S~\ref{appen:archival}. These sections include the stacked and smoothed 21-cm profiles that were used to measure the values for $S_{HI}$ that were report in Table \ref{tab:HI}, as well as notes regarding the interpretation for individual galaxies in our sample. \\

\noindent Section \ref{appen:doppler} presents additional information regarding our assumption of a constant Doppler parameter in our absorption line fits.\\

\noindent Section \ref{appen:cog} includes information about saturation of the LIS lines for CLASSY galaxies. Table \ref{tab:saturated2} lists which lines we characterize as saturated for each fit, while the figures show the curve of growth diagnostics we used to make these determinations. \\

\noindent Section \ref{appen:EW_comp} shows a comparison between the equivalent widths measured from the best-fit models from this work compared to those measured directly from the data for our series of neutral and LIS absorption lines. \\

\noindent Section \ref{appenD} contains the results from our simultaneous fitting procedure, with Table \ref{tab:fit_results} containing our best-fit parameters. This section also contains the figures with the best-fit models for each of the CLASSY galaxies.\\

\vspace{1.5cm}
\section{HI 21-cm Observations}\label{appen21cm}
\restartappendixnumbering

\subsection{The HICLASS Survey} \label{appen:hiclass}
\noindent Here we present the uniformly observed, reduced, and stacked 21-cm \ion{H}{1} emission profiles from the GBT HICLASS observing program (AGBT/21B-323; PI: Berg). The resulting profiles are shown in the following pages and were used to calculate $S_{HI}$. For each panel, the flux (black line) is continuum-subtracted and plotted in velocity-space, based on the nominal 21-cm frequency at the galaxy redshifts. The bounds used to integrate across the emission profiles are shown by pink dashed lines, the systemic velocity of the galaxy is the solid pink line, and the area integrated for each profile is shaded in light yellow. A smoothed version of each spectrum (using the \texttt{spectres} package) is overplotted in light grey. For transparency and to help interpret the \ion{H}{1} profiles, we mention here any unique notes for individual CLASSY galaxies beside each of the profiles below.

\begin{minipage}{.40\linewidth}
    \includegraphics[width=\linewidth]{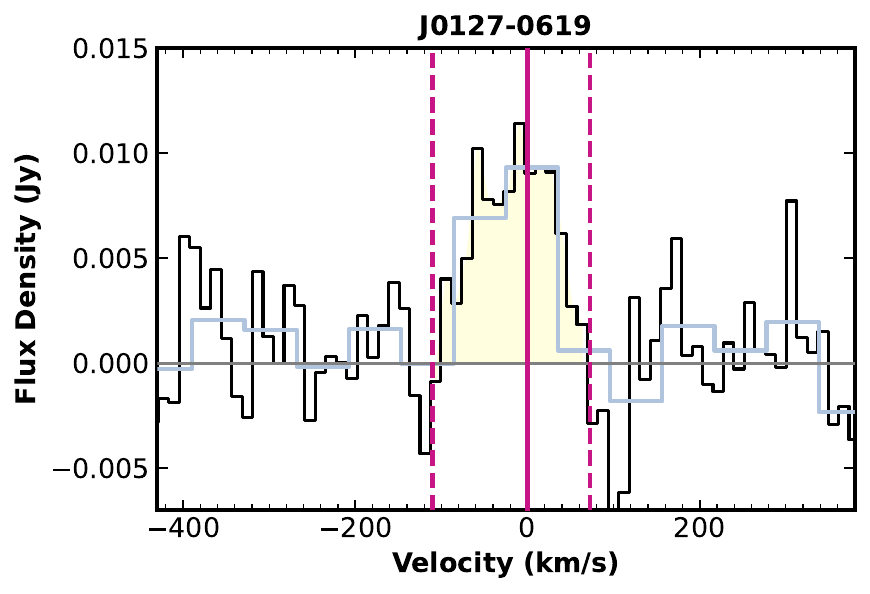}
\end{minipage}\hfill
\begin{minipage}{.55\linewidth}
\captionof{figure}{{\it J0127--0619}:
The common name of J0127--0619 is Mrk 996. This galaxy is known to have complex emission line kinematics, a large Wolf Rayet (WR) population, and a high electron density within its inner core region \citep[][]{Thuan_1996,James_2009,Telles_2014}. Galaxies with WR spectral features often have more highly ionized ISM than other star-forming galaxies \citep[e.g.,][]{Brinchmann_2008}. Additionally, its broad FWHM ($\sim$ 95 km s$^{-1}$) could be further evidence of WR stars and/or an AGN. Using a large sample of AGN from the xGASS and ALFALFA surveys, \citet{Ellison_2018} found that low-mass ($\log M_\star < 9.6$) galaxies with AGN are typically \ion{H}{1}-poor. Together, these factors indicate that J0127--0619's low gas fraction ($\mu \sim 0.2$) is the result of its extremely ionized conditions and high-energy sources.}
\label{fig:layout1}
\end{minipage} \hfill

\begin{minipage}{.40\linewidth}
    \includegraphics[width=\linewidth]{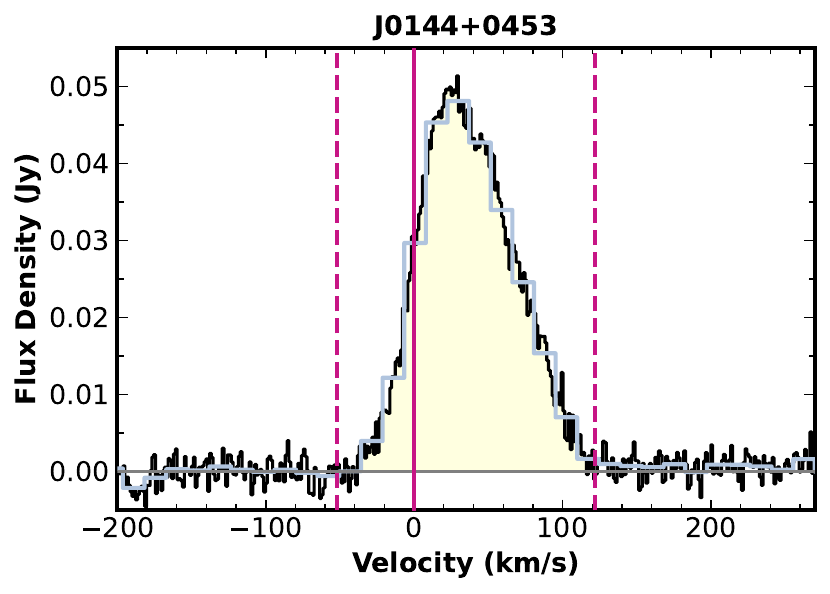}
\end{minipage}\hfill
\begin{minipage}{.55\linewidth}
\captionof{figure}{{\it J0144+0453}:
J0144+0453 has a Gaussian \ion{H}{1} profile characteristic of non-rotation or random gas motions of low-mass galaxies, but with a slight blue-skew, possibly due to its extended tail to the North.}
\label{fig:layout2}
\end{minipage} \hfill

\begin{minipage}{.40\linewidth}
    \includegraphics[width=\linewidth]{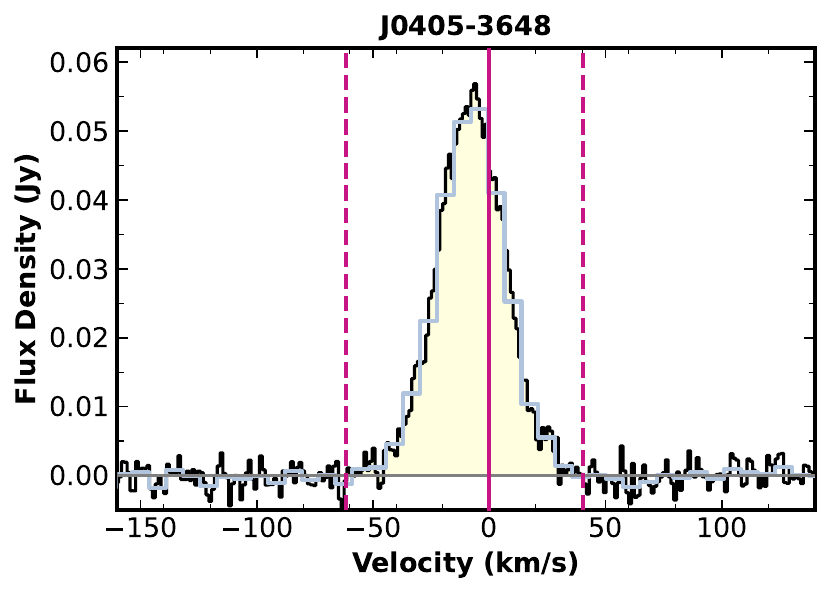}
\end{minipage}\hfill
\begin{minipage}{.55\linewidth}
\captionof{figure}{{\it J0405--3648}: 
J0405--3648 is a low-mass ($M_\star < 10^7 M_\odot$) galaxy with a Gaussian \ion{H}{1} profile characteristic of non-rotation or random gas motions of low-mass galaxies. Despite being fairly extended, with many small ionized clumps seen in the optical imaging, the \ion{H}{1} profile width of J0405--3648 is narrow (FWHM $\sim 35$ km s$^{-1}$), suggesting little to no organized gas flows.}
\label{fig:layout3}
\end{minipage} \hfill

\begin{minipage}{.40\linewidth}
    \includegraphics[width=\linewidth]{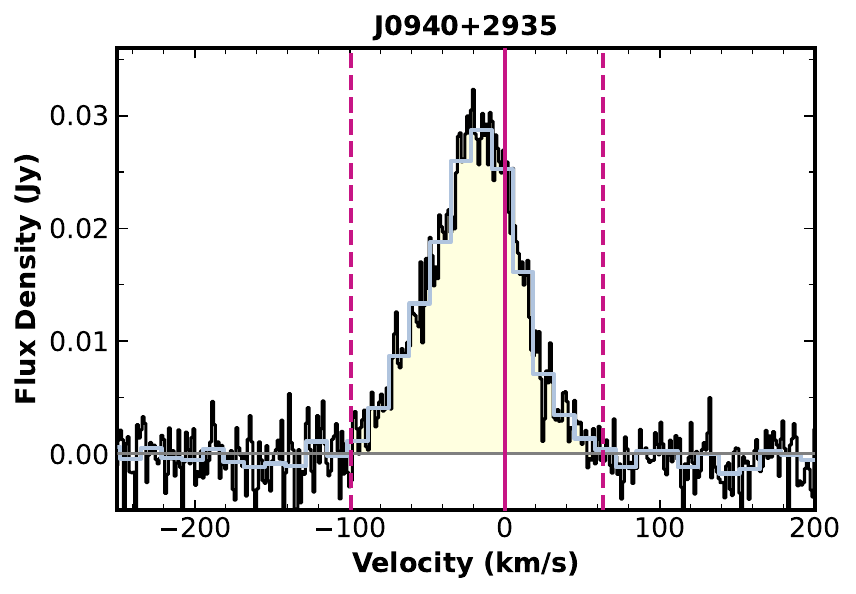}
\end{minipage}\hfill
\begin{minipage}{.55\linewidth}
\captionof{figure}{{\it J0940+2935}: 
J0940+2935 is a low-mass ($M_\star < 10^7 M_\odot$) galaxy with a Gaussian \ion{H}{1} profile characteristic of non-rotation or random gas motions of low-mass galaxies. The profile width is a bit broader than other low-mass CLASSY galaxies (FWHM $\sim 70$ km s$^{-1}$), suggesting some organized gas flows.}
\label{fig:layout4}
\end{minipage} \hfill

\begin{minipage}{.40\linewidth}
    \includegraphics[width=\linewidth]{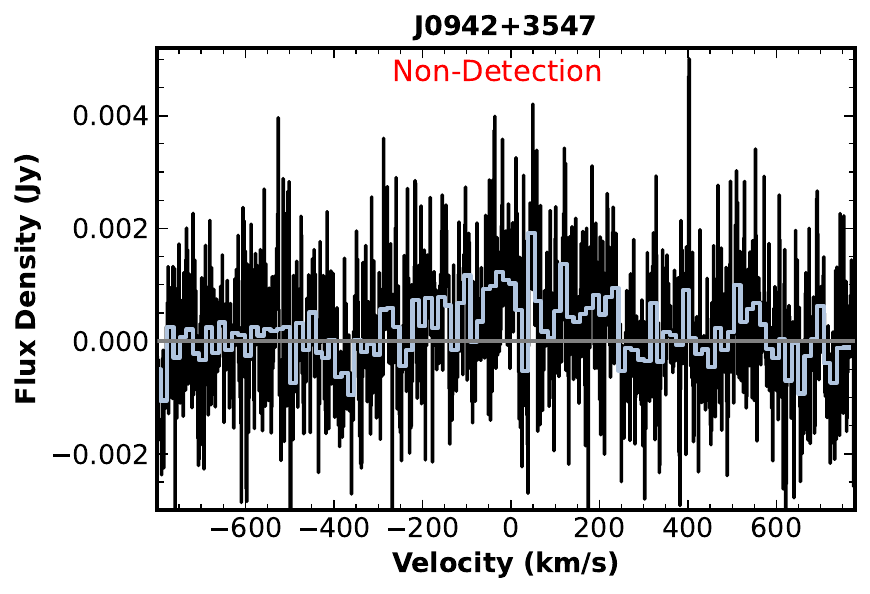}
\end{minipage}\hfill
\begin{minipage}{.55\linewidth}
\captionof{figure}{{\it J0942+3547}: 
The HICLASS 21-cm observations of J0942+3547 have insufficient SNR to characterize the \ion{H}{1} properties. The upper limit on $S_{HI}$ corresponds to a very low gas fraction ($\mu < 0.09$).}
\label{fig:layout5}
\end{minipage} \hfill

\begin{minipage}{.40\linewidth}
    \includegraphics[width=\linewidth]{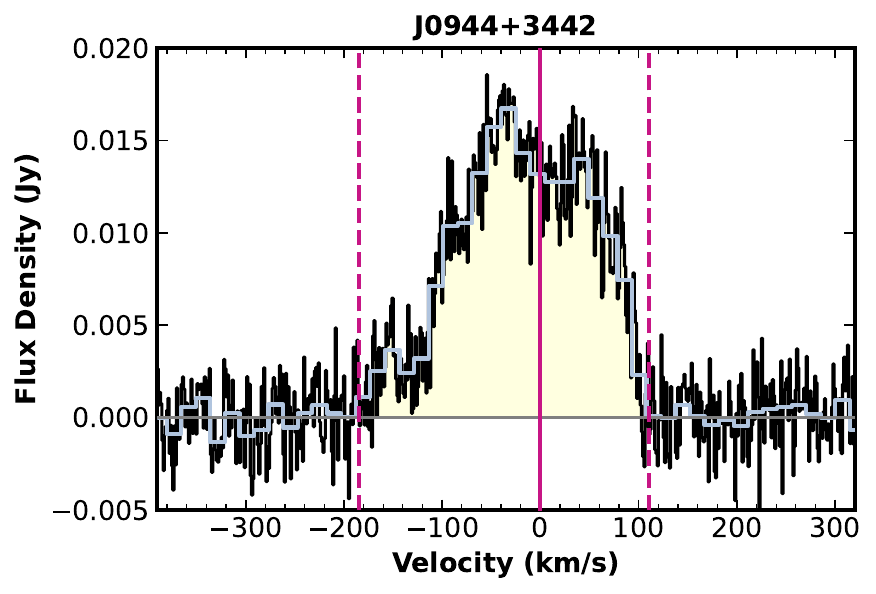}
\end{minipage}\hfill
\begin{minipage}{.55\linewidth}
\captionof{figure}{{\it J0944+3442}: J0944+3424 has extended disk-like optical emission. As a result, the \ion{H}{1} profile appears like a broadened double-horn profile, which is typical of disk rotation.}
\label{fig:layout6}
\end{minipage} \hfill

\begin{minipage}{.40\linewidth}
    \includegraphics[width=\linewidth]{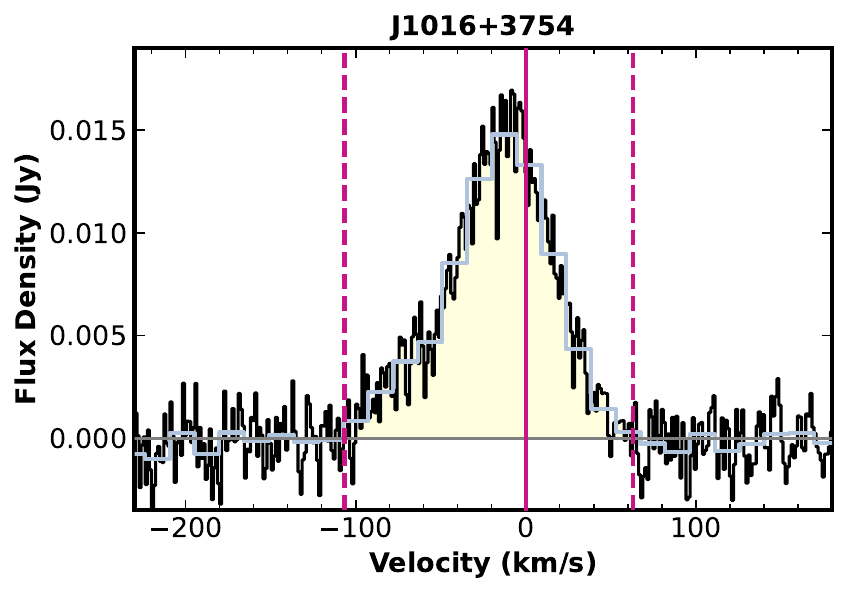}
\end{minipage}\hfill
\begin{minipage}{.55\linewidth}
\captionof{figure}{{\it J1016+3754}: 
J1016+3754 is a low-mass ($M_\star < 10^7 M_\odot$) galaxy with a Gaussian \ion{H}{1} profile characteristic of non-rotation or random gas motions of low-mass galaxies. The profile width is a bit broader than other low-mass CLASSY galaxies (FWHM $\sim 70$ km s$^{-1}$), suggesting some organized gas flows.}
\label{fig:layout7}
\end{minipage} \hfill

\begin{minipage}{.40\linewidth}
    \includegraphics[width=\linewidth]{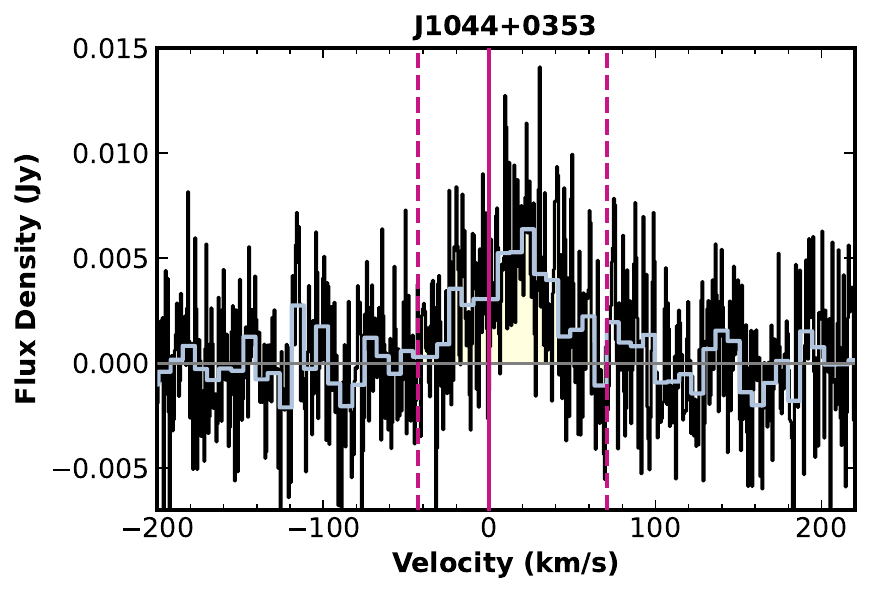}
\end{minipage}\hfill
\begin{minipage}{.55\linewidth}
\captionof{figure}{{\it J1044+0353}: J1044+0353 is a low-mass ($M_\star < 10^7 M_\odot$) galaxy with a Gaussian \ion{H}{1} profile characteristic of non-rotation or random gas motions of low-mass galaxies. The profile width is a bit broader than other low-mass CLASSY galaxies (FWHM $\sim 65$ km s$^{-1}$), suggesting some organized gas flows, possibly related to a second star-forming clump \citep[e.g.,][]{Martin_2024}.}
\label{fig:layout8}
\end{minipage} \hfill

\begin{minipage}{.40\linewidth}
    \includegraphics[width=\linewidth]{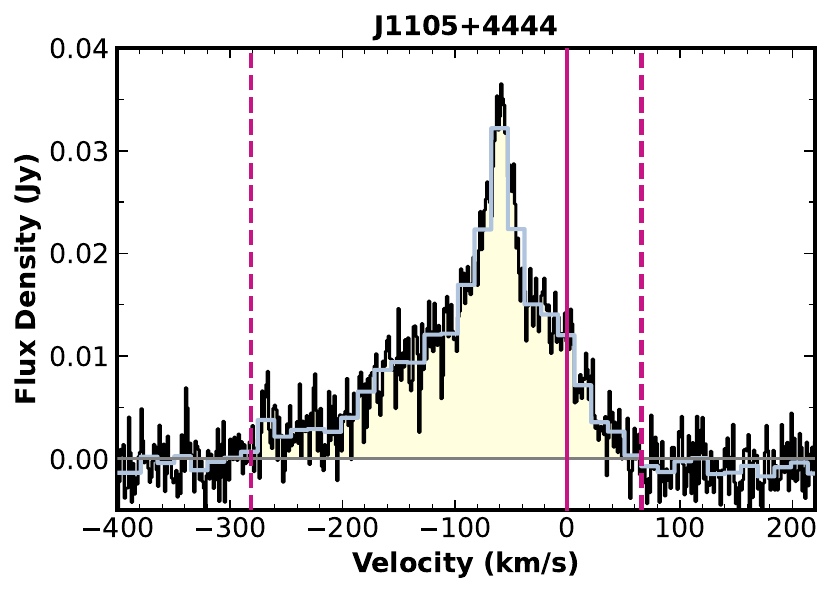}
\end{minipage}\hfill
\begin{minipage}{.55\linewidth}
\captionof{figure}{{\it J1105+4444}: J1105+445 is dominated by a UV-bright knot at its center, but has extended optical emission, possibly due to interacting clumps. The \ion{H}{1} profile shows a very narrow Gaussian component atop a broader profile with a redshifted skew.}
\label{fig:layout9}
\end{minipage} \hfill

\begin{minipage}{.40\linewidth}
    \includegraphics[width=\linewidth]{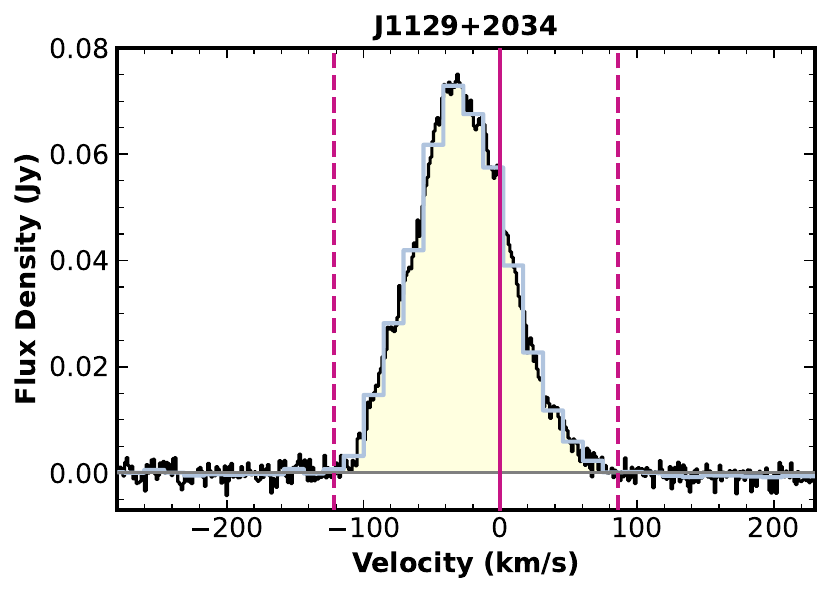}
\end{minipage}\hfill
\begin{minipage}{.55\linewidth}
\captionof{figure}{{\it J1129+2034}: 
Has a Gaussian \ion{H}{1} profile characteristic of non-rotation or random gas motions of diskless galaxies. J1129+2034 also has a faint extended optical tail that could be the source of the slight asymmetry in its \ion{H}{1} profile.}
\label{fig:layout10}
\end{minipage} \hfill

\begin{minipage}{.40\linewidth}
    \includegraphics[width=\linewidth]{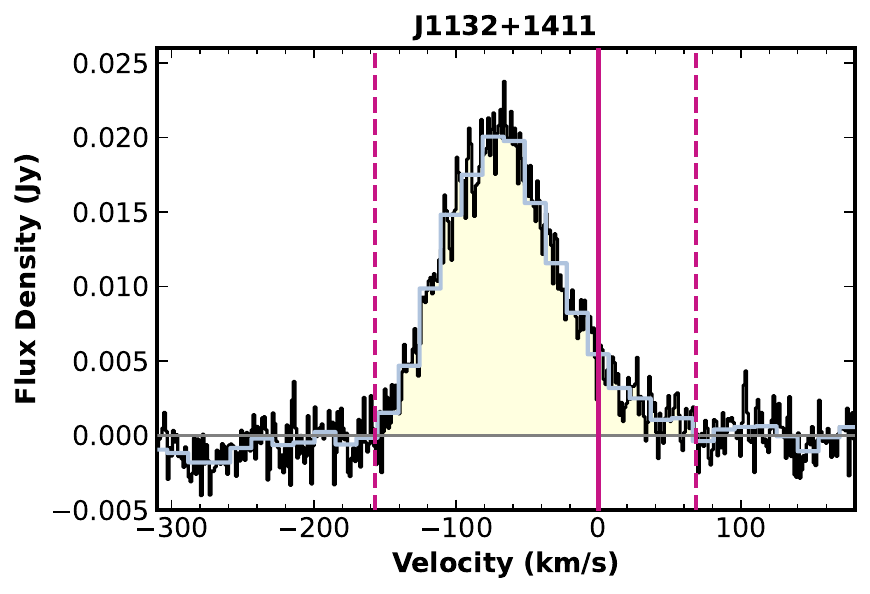}
\end{minipage}\hfill
\begin{minipage}{.55\linewidth}
\captionof{figure}{{\it J1132+1411}: J1132+1411 is dominated by a compact UV-bright star-forming region, but appears to lie within a larger disk. The observed \ion{H}{1} profile is fairly broad (FWHM $\sim 95$ km s$^{-1}$) and skewed to the blue, but the gas dynamics are not clear.}
\label{fig:layout11}
\end{minipage} \hfill

\begin{minipage}{.40\linewidth}
    \includegraphics[width=\linewidth]{J1132+5722_zoom.pdf}
\end{minipage}\hfill
\begin{minipage}{.55\linewidth}
\captionof{figure}{{\it J1132+5722}: J1132+5722 has extended optical emission, but a symmetric \ion{H}{1} profile. With a FWHM $\sim 85$ km s$^{-1}$, some organized gas flows may be present.}
\label{fig:layout12}
\end{minipage} \hfill

\begin{minipage}{.40\linewidth}
    \includegraphics[width=\linewidth]{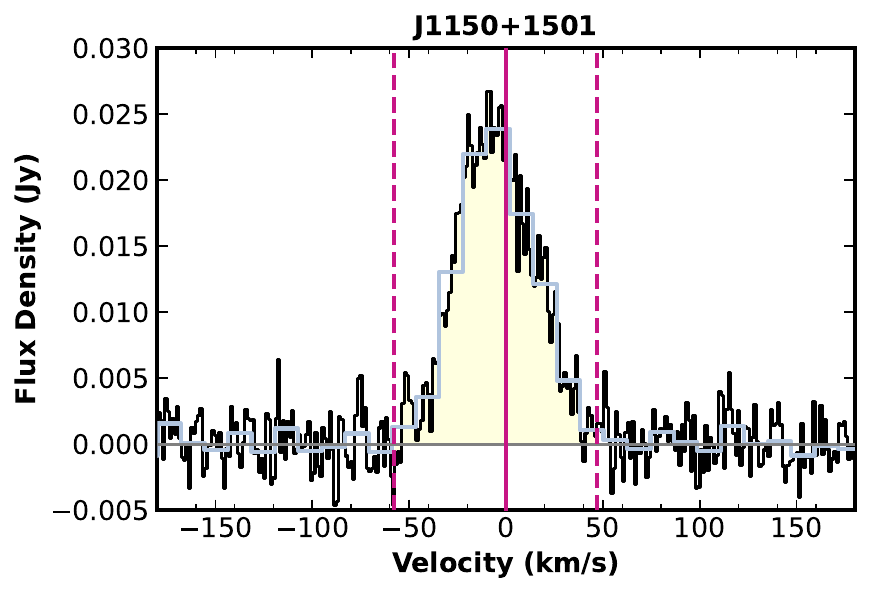}
\end{minipage}\hfill
\begin{minipage}{.55\linewidth}
\captionof{figure}{{\it J1150+1501}: J1150+1501 is a low-mass ($M_\star < 10^7 M_\odot$) galaxy with a narrow Gaussian (FWHM $\sim 45$ km s$^{-1}$) \ion{H}{1} profile characteristic of non-rotation or random gas motions of low-mass galaxies.}
\label{fig:layout13}
\end{minipage} \hfill

\begin{minipage}{.40\linewidth}
    \includegraphics[width=\linewidth]{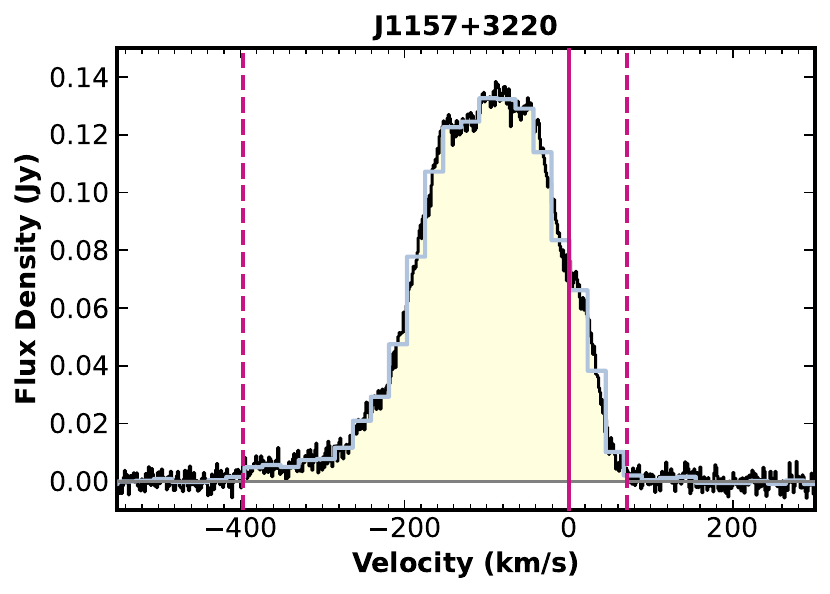}
\end{minipage}\hfill
\begin{minipage}{.55\linewidth}
\captionof{figure}{{\it J1157+3220}: J1157+3220 is one of the more massive CLASSY galaxies ($M_\star \sim 10^9 M_\odot$), with disk-like emission. As a result, the \ion{H}{1} profile is approaching a double-horn shape, with a width of FWHM $\sim 190$ km s$^{-1}$.}
\label{fig:layout14}
\end{minipage} \hfill

\begin{minipage}{.40\linewidth}
    \includegraphics[width=\linewidth]{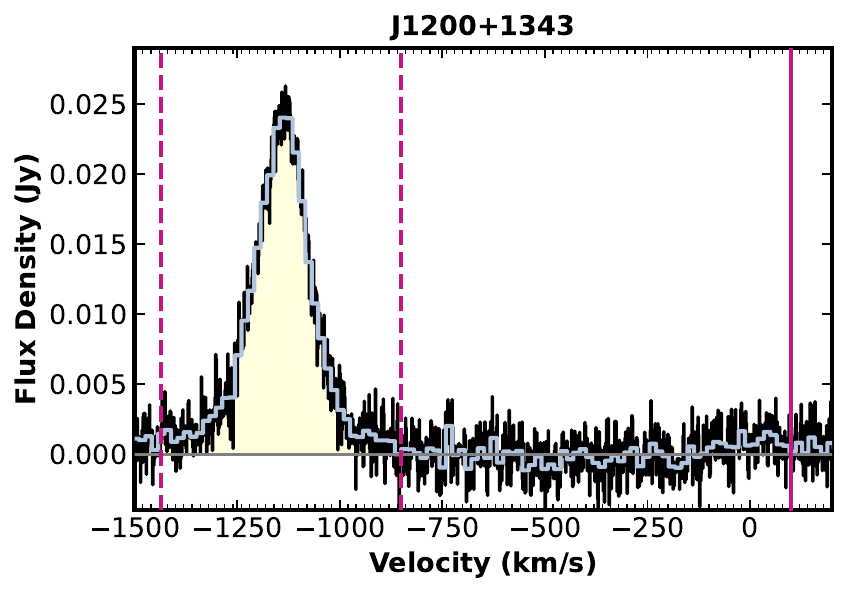}
\end{minipage}\hfill
\begin{minipage}{.55\linewidth}
\captionof{figure}{{\it J1200+1343}: J1200+1343 has a Gaussian \ion{H}{1} profile characteristic of non-rotation or random gas motions. However, the profile has extended wings and is broad (FWHM $\sim 155$ km s$^{-1}$), suggesting some organized gas flows. This is the only \ion{H}{1} profile in our sample that shows a significant deviation from the systemic velocity. This may be a result of its nearby companion.}
\label{fig:layout15}
\end{minipage} \hfill

\begin{minipage}{.40\linewidth}
    \includegraphics[width=\linewidth]{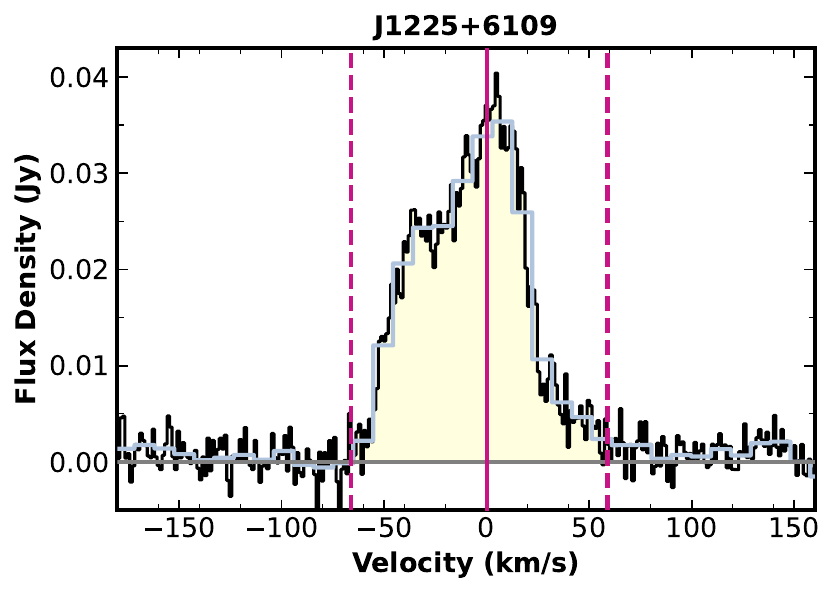}
\end{minipage}\hfill
\begin{minipage}{.55\linewidth}
\captionof{figure}{{\it J1225+6109}: J1225+6109 is a low-mass galaxy ($M_\star \sim 10^7 M_\odot$) with minimal extended optical emission. However, while the \ion{H}{1} profile is fairly narrow (FWHM $\sim 60$ km s$^{-1}$), the profile looks double-horned and asymmetric, suggesting distorted organized gas flows.}
\label{fig:layout16}
\end{minipage} \hfill

\begin{minipage}{.40\linewidth}
    \includegraphics[width=\linewidth]{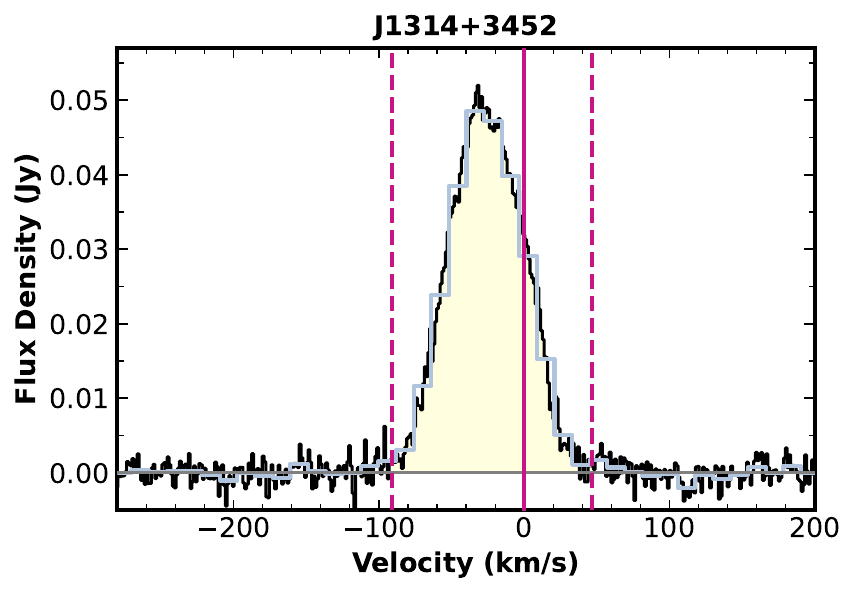}
\end{minipage}\hfill
\begin{minipage}{.55\linewidth}
\captionof{figure}{{\it J1314+3452}: J1314+3452 is a relatively low-mass ($M_\star < 10^{7.5} M_\odot$) galaxy with a Gaussian \ion{H}{1} profile characteristic of non-rotation or random gas motions of low-mass galaxies. Despite having extended faint optical emission and one other UV-bright know, the \ion{H}{1} profile width is moderate (FWHM $\sim 60$ km s$^{-1}$), suggesting minimal organized gas flows.}
\label{fig:layout17}
\end{minipage} \hfill

\begin{minipage}{.40\linewidth}
    \includegraphics[width=\linewidth]{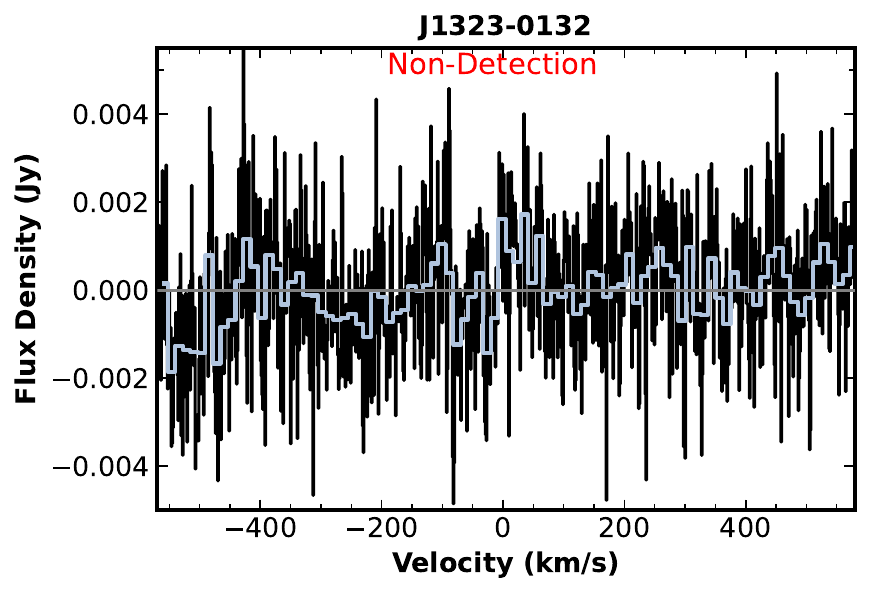}
\end{minipage}\hfill
\begin{minipage}{.55\linewidth}
\captionof{figure}{{\it J1323-0132}: The HICLASS 21-cm observations of J1323--0132 have insufficient SNR to characterize the \ion{H}{1} properties.}
\label{fig:layout18}
\end{minipage} \hfill

\begin{minipage}{.40\linewidth}
    \includegraphics[width=\linewidth]{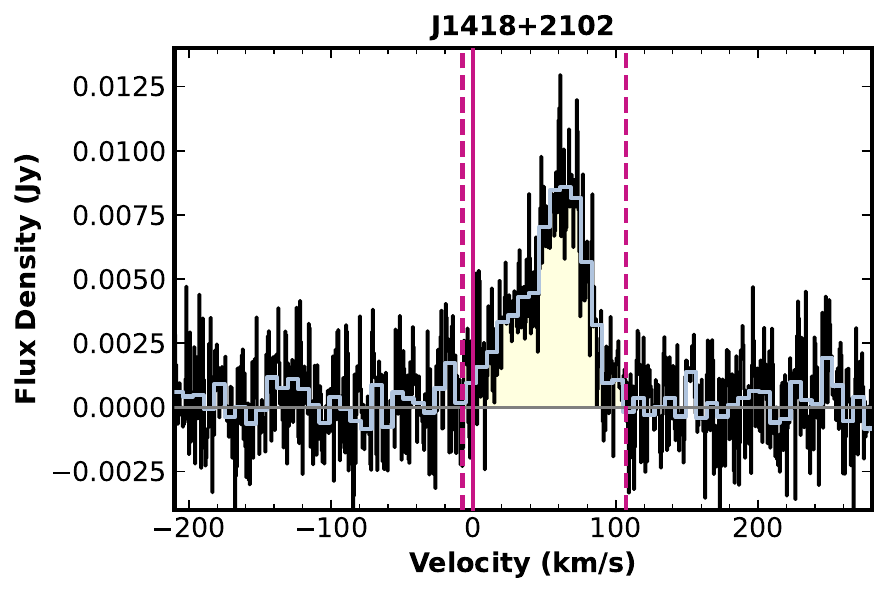}
\end{minipage}\hfill
\begin{minipage}{.55\linewidth}
\captionof{figure}{{\it J1418+2102}: J1418+2102 is a low-mass ($M_\star < 10^7 M_\odot$) galaxy with an \ion{H}{1} profile dominated by a narrow Gaussian (FWHM $\sim 50$ km s$^{-1}$) that is typical of non-rotation or random gas motions of low-mass galaxies. J1418+2102 has a second knot of star-formation that may be responsible for redshifted emission in the 21-cm profile.}
\label{fig:layout19}
\end{minipage} \hfill

\begin{minipage}{.40\linewidth}
    \includegraphics[width=\linewidth]{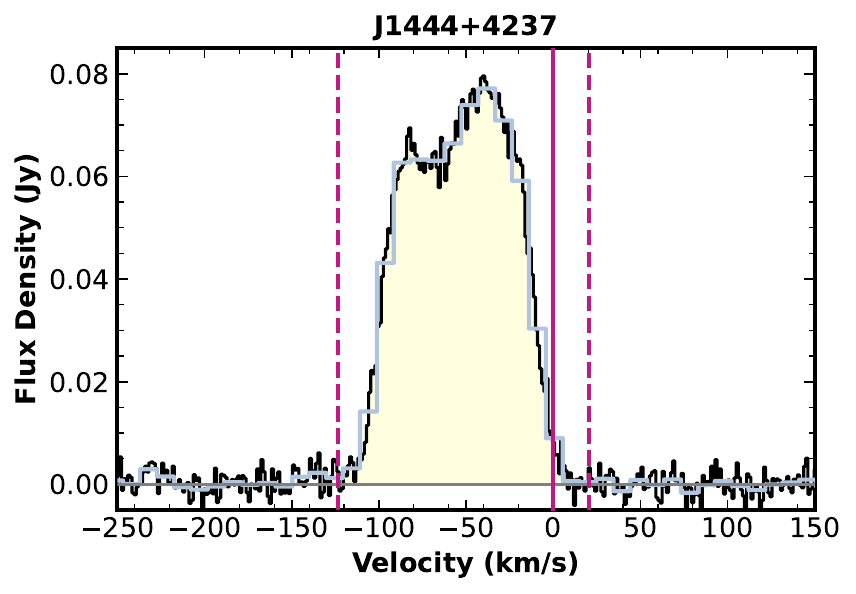}
\end{minipage}\hfill
\begin{minipage}{.55\linewidth}
\captionof{figure}{{\it J1444+4237}: J1444+4237 is a low-mass ($M_\star < 10^7 M_\odot$) galaxy, but with an extended disk-like nature that seems to correspond to organized disk rotation in its double-horned \ion{H}{1} profile.}
\label{fig:layout20}
\end{minipage} \hfill


\clearpage

\subsection{Archival Observations} \label{appen:archival}
\noindent We divide the archival observations for CLASSY galaxies into three different categories in order to provide as much consistency between measurements as possible:
\setlist{nolistsep}
\begin{enumerate}[noitemsep]
    \item \textit{Archival GBT observations that lack strong RFI:} The 11 CLASSY galaxies with archival GPT observations lacking strong  RFI and their program IDs are listed in Table~\ref{tab:HI}. For these galaxies, we rereduced the archival observations consistently with the HICLASS observations.
    \item \textit{Other reduced 21-cm observations:}
    In addition to the GBT observations, 21-cm observations exist for five additional galaxies from other radio telescopes. For these galaxies, we acquired the spectrum, removed the baseline (if it has not already been done), and performed the measurement for $S_{HI}$ ourselves. Although this allows for some similar procedures to be performed, there is less consistency due to potential differences in observing strategy, data reduction, and the stacking of the spectra.
    \item \textit{Published measurements for $S_{HI}$:} 
    For J0337-0502, the archival \ion{H}{1} measurement is available but the observed spectrum is not.
    For this galaxy, we adopted the archival $S_{HI}$ value to determine $M_{HI}$ and $\mu$. This option provides the least consistency with the other measurements and was only used when the previous two options weren't available.
\end{enumerate}

\noindent Using these archival observations, we uniformly reduce and stack the available spectra. The resulting profiles are shown over the following pages and were used to calculate $S_{HI}$, when possible. 

\begin{minipage}{.40\linewidth}
    \includegraphics[width=\linewidth]{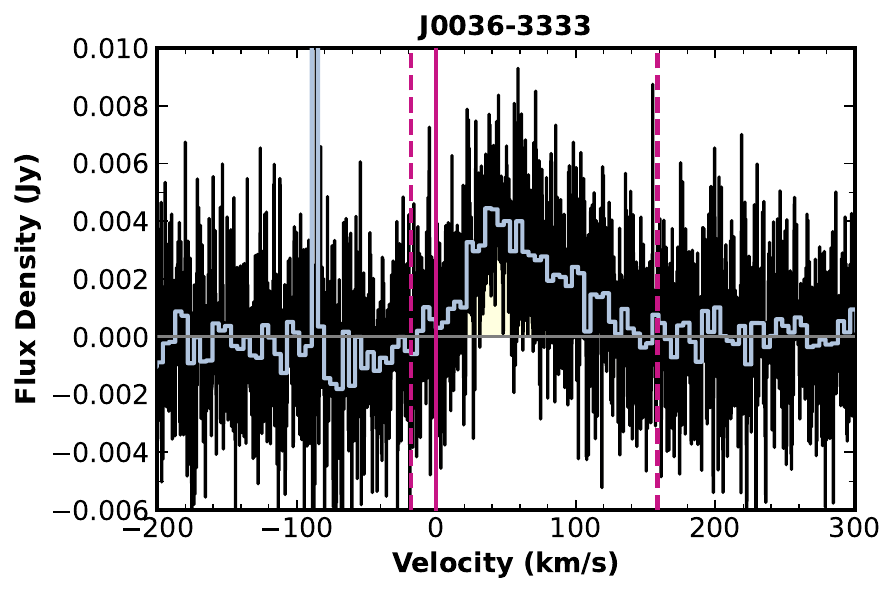}
\end{minipage}\hfill
\begin{minipage}{.55\linewidth}
\captionof{figure}{{\it J0036--3333} (AGBT/14B-306): Also commonly known as the galaxy Haro 11, J0036--3333 contains multiple merging stellar populations, where the CLASSY coadded spectrum only focuses on the most UV luminous source, knot C \citep[][]{Hayes_2007}. The \ion{H}{1} profile represents the integrated \ion{H}{1} gas flows for J0036--3333 and shows a substantial blue wing. This is consistent with the recent report by \citet{LeReste_2024} who find that merger-driven interactions have caused a bulk offset of the neutral gas from the galaxy center and that the extended component is consistent with outflowing gas in the tidal tail. \cite{Pardy_2016} found J0036-3333 to be gas-deficient compared with other nearby galaxies with similar optical properties, which \cite{LeReste_2024} determined was the result of a merger. Much of the gas that remained, post-merger, was photoionized by the interaction \citep{Menacho_2019}, further driving $\mu$ towards a lower value ($\mu \sim 0.3$).}
\label{fig:layout21}
\end{minipage} \hfill

\begin{minipage}{.40\linewidth}
    \includegraphics[width=\linewidth]{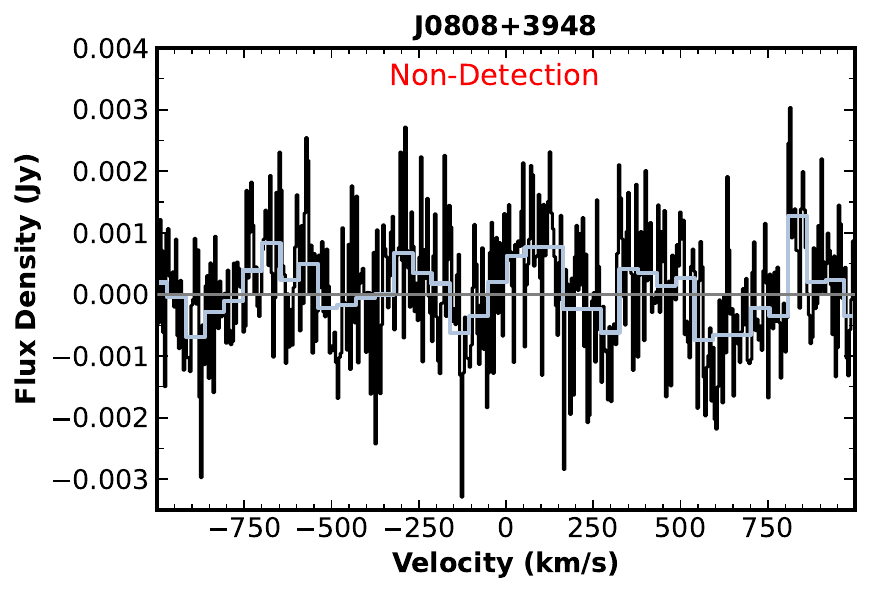}
\end{minipage}\hfill
\begin{minipage}{.55\linewidth}
\captionof{figure}{{\it J0808+3948} (AGBT/06B-047): J0808+3948 tends to be a general outlier from the CLASSY sample. It lies at the upper end of the CLASSY mass-metallicity relationship, with a metallicity of 12+log(O/H) = 8.77 and a stellar mass of  $M_\star = 10^{9.12} M_\odot$, it appears to have a substantial older stellar population in its imaging and spectral energy distribution fitting, and has the most significant outflows amongst the CLASSY sample ($v_{min} \sim 600$ km s$^{-1}$). For these reasons, J0808+3948 seems to have consumed most of its neutral gas reservoir, resulting in only an upper limit detection and a small gas fraction of $\mu \sim 0.09$.}
\label{fig:layout22}
\end{minipage} \hfill

\begin{minipage}{.40\linewidth}
    \includegraphics[width=\linewidth]{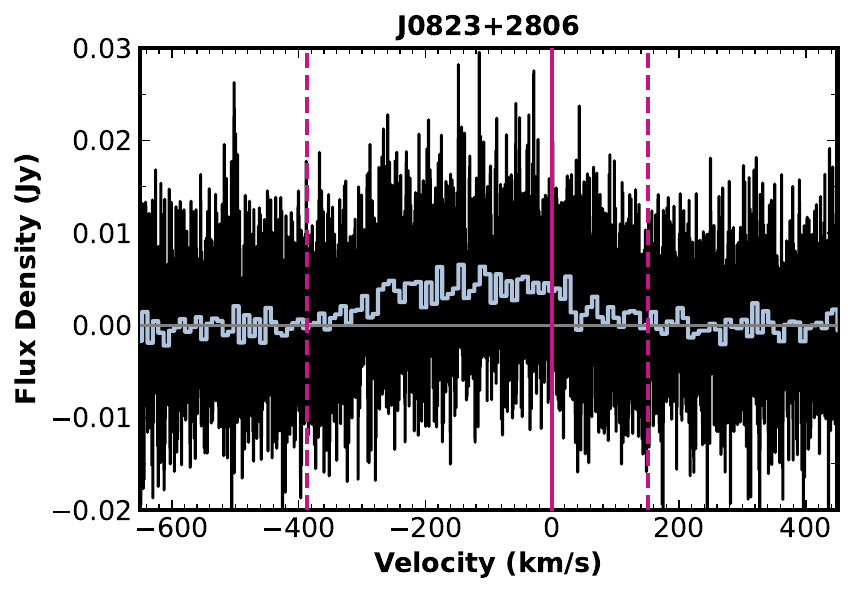}
\end{minipage}\hfill
\begin{minipage}{.55\linewidth}
\captionof{figure}{{\it J0823+2806} (AGBT/11A-057): The CLASSY coadded spectrum of J0823+2806 is focused on UV-bright emission from a single massive star cluster, however, the total galaxy shows extended tails to both the N and S. This structure, likely due to interactions, produces the broad \ion{H}{1} profile that's observed.}
\label{fig:layout23}
\end{minipage} \hfill

\begin{minipage}{.40\linewidth}
    \includegraphics[width=\linewidth]{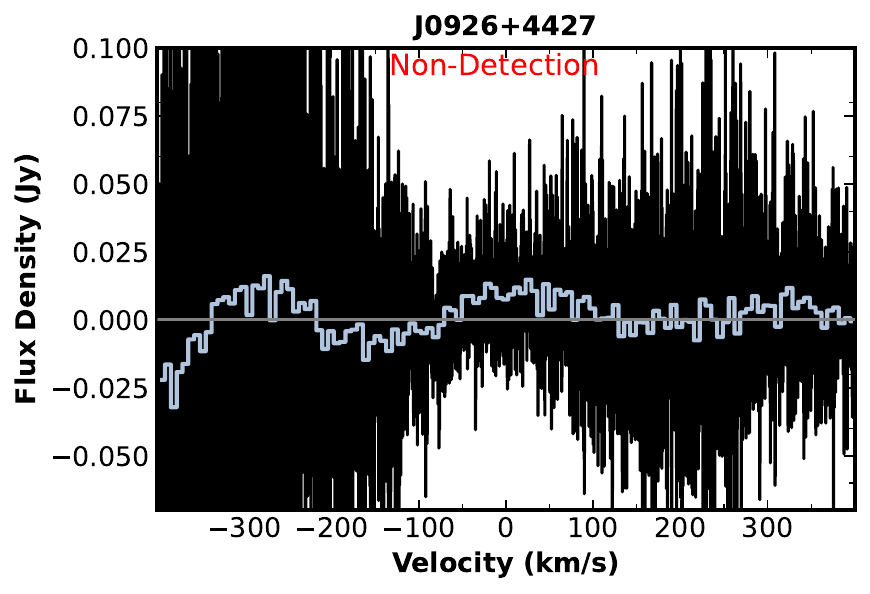}
\end{minipage}\hfill
\begin{minipage}{.55\linewidth}
\captionof{figure}{{\it J0926+4427} (AGBT/11A-057): Despite having a high SFR and a large upper limit on the \ion{H}{1} gas mass and gas fraction, 21-cm emission is not significantly detected in the archival GBT data for J0926+4427. This may simply be due to its higher redshift ($z = 0.18$) and the rather strong RFI that's present in its continuum.}
\label{fig:layout24}
\end{minipage} \hfill

\begin{minipage}{.40\linewidth}
    \includegraphics[width=\linewidth]{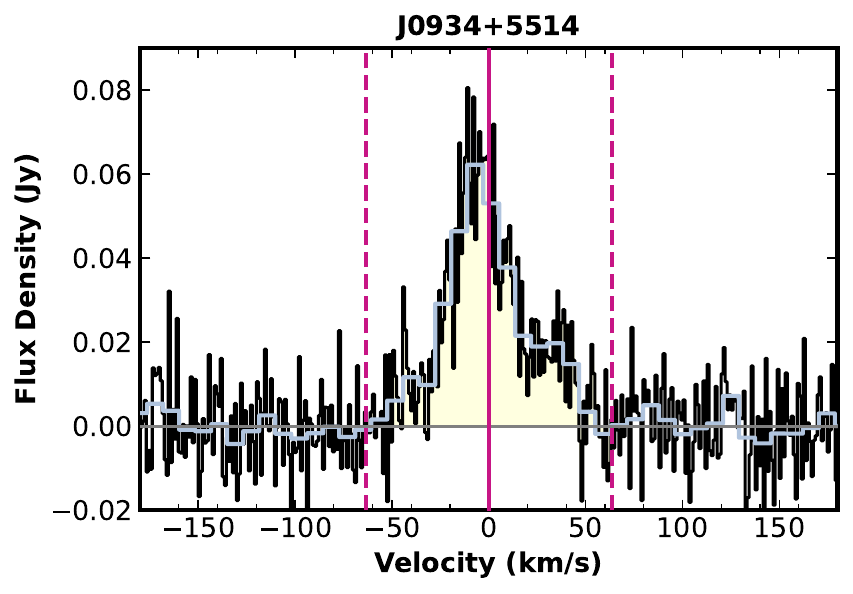}
\end{minipage}\hfill
\begin{minipage}{.55\linewidth}
\captionof{figure}{{\it J0934+5514} \citep{Springob_2005}: This is the famous low-metallicity dwarf galaxy, I~Zw~18. I~Zw~18 has two main components, NW and SE, where the GBT observations encompass both. The main body of the \ion{H}{1} profile appears Gaussian, characteristic of the random gas motions of diskless low-mass galaxies. However, a weaker blue tail is present that may be associated with gas flows between the NW and SE components.}
\label{fig:layout25}
\end{minipage} \hfill

\begin{minipage}{.40\linewidth}
    \includegraphics[width=\linewidth]{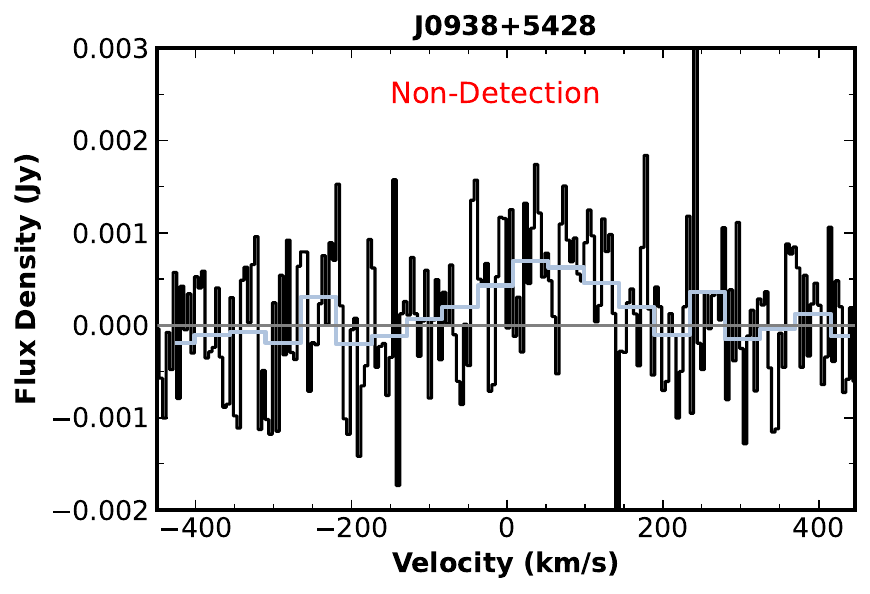}
\end{minipage}\hfill
\begin{minipage}{.55\linewidth}
\captionof{figure}{{\it J0938+5428} (AGBT/06B-047): The archival 21-cm observations of J0938+5428 have insufficient SNR to characterize the \ion{H}{1} properties. The upper limit on $S_{HI}$ corresponds to a very low gas fraction ($\mu < 0.08$).}
\label{fig:layout26}
\end{minipage} \hfill

\begin{minipage}{.40\linewidth}
    \includegraphics[width=\linewidth]{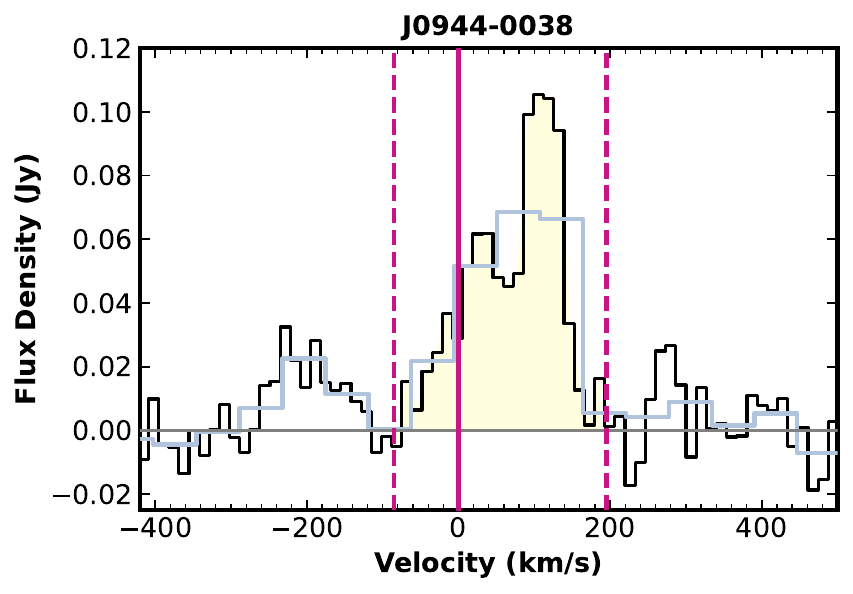}
\end{minipage}\hfill
\begin{minipage}{.55\linewidth}
\captionof{figure}{{\it J0944-0038} \citep{Meyer_2004}: Commonly known as the metal-poor galaxy CGCG 007-025, J0944--0038 contains multiple star-forming knots that may be interacting, causing kinematic disturbances to the \ion{H}{1} gas. As a result, the \ion{H}{1} profile appears asymmetric and double-peaked, with a broad width (FWHM $\sim 120$ km s$^{-1}$).}
\label{fig:layout27}
\end{minipage} \hfill

\begin{minipage}{.40\linewidth}
    \includegraphics[width=\linewidth]{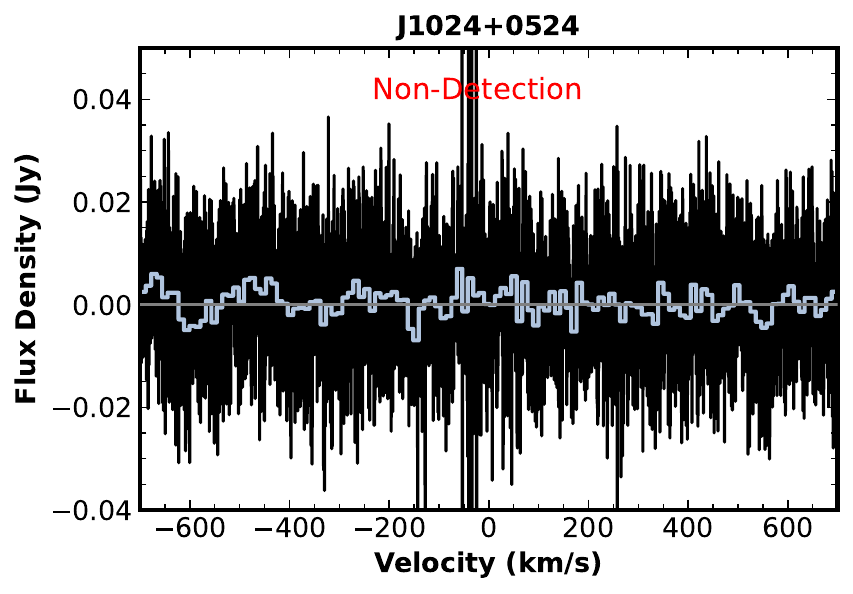}
\end{minipage}\hfill
\begin{minipage}{.55\linewidth}
\captionof{figure}{{\it J1024+0524} (AGBT/19A-301): The archival 21-cm observations of J1024+0524 have insufficient SNR to characterize the \ion{H}{1} properties. With a total exposure time of about 3 minutes on target, we are not confident that this galaxy would still be non-detected if it was observed for longer.}
\label{fig:layout28}
\end{minipage} \hfill

\begin{minipage}{.40\linewidth}
    \includegraphics[width=\linewidth]{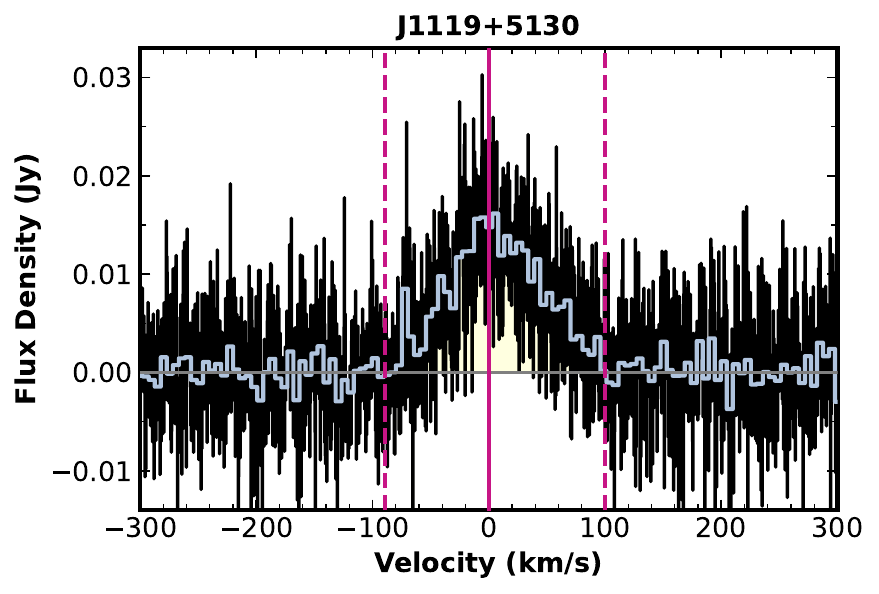}
\end{minipage}\hfill
\begin{minipage}{.55\linewidth}
\captionof{figure}{{\it J1119+5130} (AGBT/05A-034): J1119+5130 is a low-mass ($M_\star < 10^7 M_\odot$) galaxy with a Gaussian \ion{H}{1} profile characteristic of non-rotation or random gas motions of low-mass galaxies. The profile width is a bit broader than other low-mass CLASSY galaxies (FWHM $\sim 100$ km s$^{-1}$), suggesting some organized gas flows possibly related to its extended optical emission.}
\label{fig:layout29}
\end{minipage} \hfill

\begin{minipage}{.40\linewidth}
    \includegraphics[width=\linewidth]{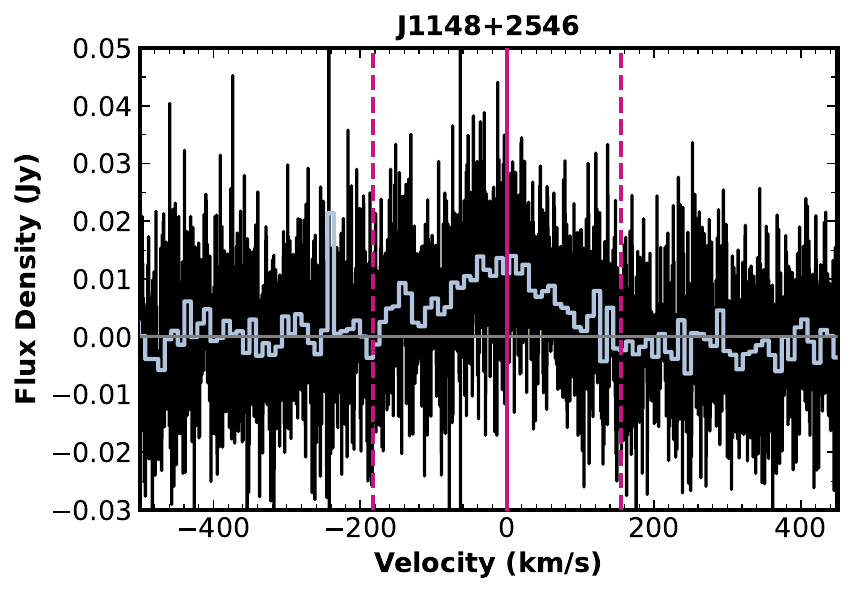}
\end{minipage}\hfill
\begin{minipage}{.55\linewidth}
\captionof{figure}{{\it J1148+2546} (AGBT/19A-301): J1148+2546 has a broad 21-cm profile (FWHM $\sim$ 170 km s$^{-1}$), particularly compared with other CLASSY galaxies of a similar mass ($M_\star \sim 10^8 M_\odot$).}
\label{fig:layout30}
\end{minipage} \hfill

\begin{minipage}{.40\linewidth}
    \includegraphics[width=\linewidth]{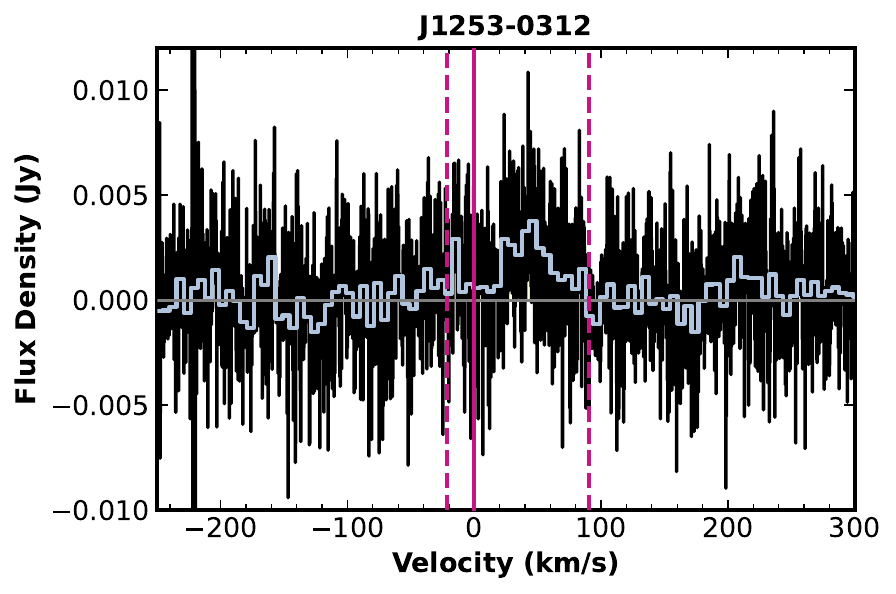}
\end{minipage}\hfill
\begin{minipage}{.55\linewidth}
\captionof{figure}{{\it J1253-0312} (AGBT/19A-301): J1253-0312 is a relatively low-mass ($M_\star \sim 10^{7.5} M_\odot$) that has a relatively weak detection of 21-cm emission. It exhibits one of the lower FWHM among the CLASSY sample (FWHM $\sim$ 45 km s$^{-1}$).}
\label{fig:layout31}
\end{minipage} \hfill

\begin{minipage}{.40\linewidth}
    \includegraphics[width=\linewidth]{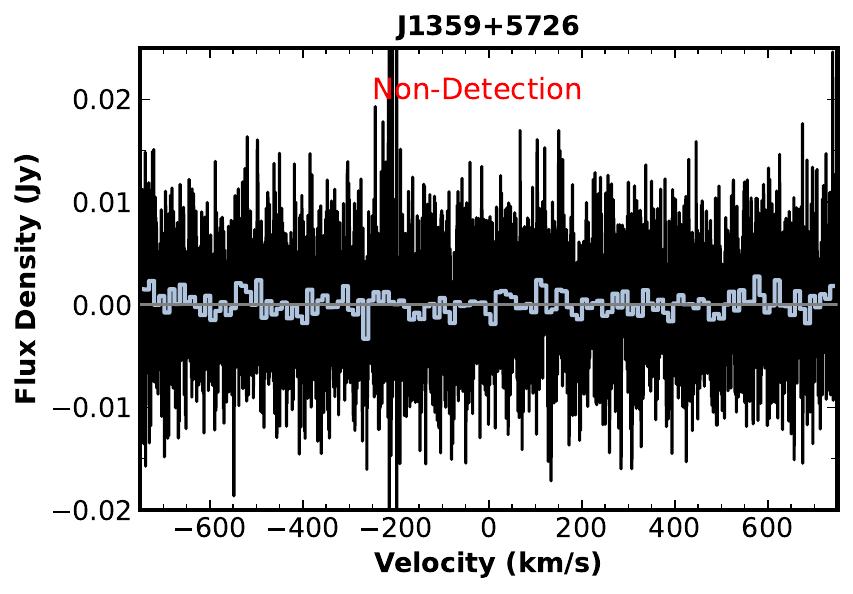}
\end{minipage}\hfill
\begin{minipage}{.55\linewidth}
\captionof{figure}{{\it J1359+5726} (AGBT/19A-301): The archival 21-cm observations of J1359+5726 have insufficient SNR to characterize the \ion{H}{1} properties. The upper limit on $S_{HI}$ corresponds to a low gas fraction ($\mu < 0.25$).}
\label{fig:layout32}
\end{minipage} \hfill

\begin{minipage}{.40\linewidth}
    \includegraphics[width=\linewidth]{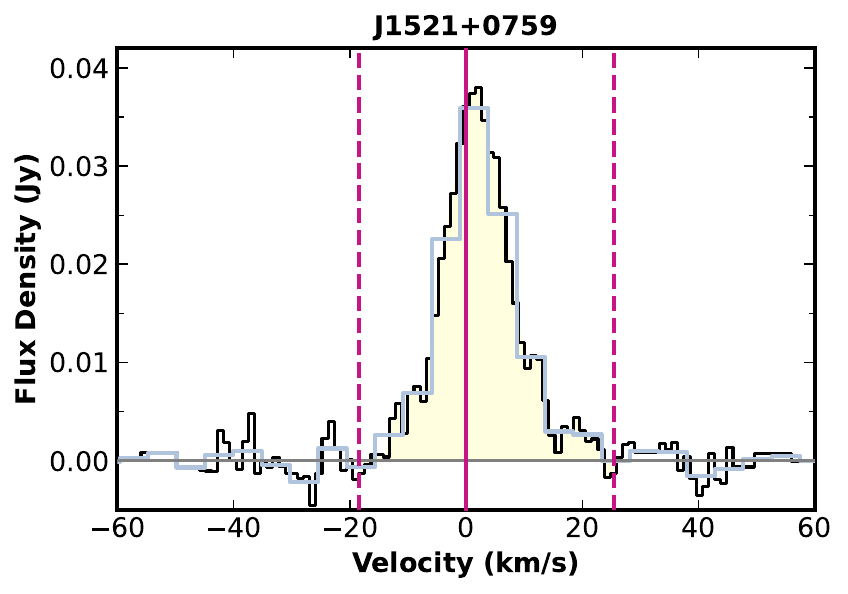}
\end{minipage}\hfill
\begin{minipage}{.55\linewidth}
\captionof{figure}{{\it J1521+0759} \citep{Heiles_1974}: J1521+0759 is a more massive CLASSY galaxy ($M_\star \sim 10^9 M_\odot$), but has an \ion{H}{1} profile dominated by a narrow Gaussian (FWHM $\sim 15$ km s$^{-1}$) suggesting little to no organized gas motions.}
\label{fig:layout33}
\end{minipage} \hfill

\begin{minipage}{.40\linewidth}
    \includegraphics[width=\linewidth]{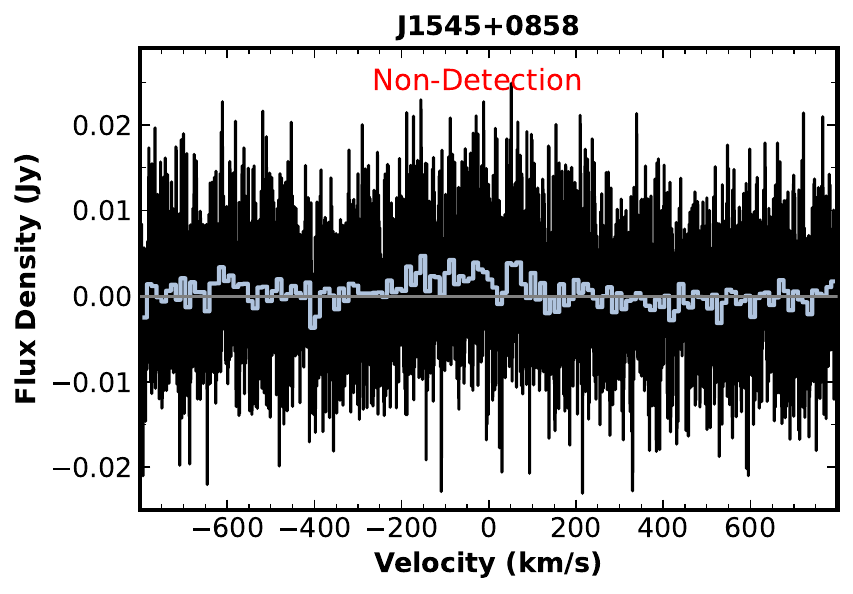}
\end{minipage}\hfill
\begin{minipage}{.55\linewidth}
\captionof{figure}{{\it J1545+0858} (AGBT/19A-301): The archival 21-cm observations of J1545+0858 have insufficient SNR to characterize the \ion{H}{1} properties. With a total exposure time of about 3 minutes on target, we are not confident that this galaxy would still be non-detected if it was observed for longer.}
\label{fig:layout34}
\end{minipage} \hfill


\clearpage

\section{On assuming a constant Doppler parameter} \label{appen:doppler}

\begin{figure}[!h]
    \centering
    \includegraphics[width=0.4\linewidth]{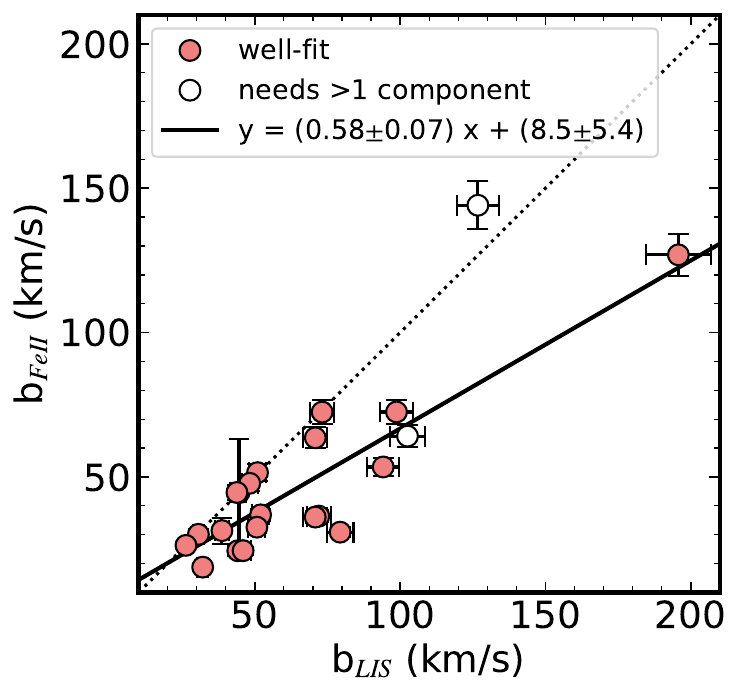}
    \figcaption{The Doppler broadening from a simultaneous fit to solely the \ion{Fe}{2} lines ($y$-axis) compared to that for the other LIS lines ($x$-axis). The solid orange points correspond to the galaxies that are well-fit by a single component while the empty points require at least 2 components. The dashed line corresponds to 1-to-1 while the solid line is the linear best-fit to the solid orange points. We found that the Doppler broadening of the \ion{Fe}{2} lines were nearly always smaller than those for the other lines. This indicates a non-negligible thermal component to $b$. \label{fig4}}
\end{figure}

In Section~\ref{subsub:b}, we discuss our assumption of using a fixed $b$ for all species considered in this work. Here we show that the validity of this assumption breaks when considering higher-mass ions such as Fe$^+$.
In Figure \ref{fig4} we compare the fit kinematics returned by fitting the \ion{Fe}{2} lines alone to those determined by a simultaneous fit to all other LIS absorption lines in our sample (i.e., excluding \ion{Fe}{2}). 
The solid points correspond to the galaxies that are well-fit by a single component (the single-component sample), while the open points correspond to the galaxies that require more than 1-component of gas (the multi-component sample). 
We used the \texttt{polyfit} package to fit a linear trend to the solid points,
finding a significant offset from the 1-to-1 trend (dotted line) that is characterized by $b_{FeII}=(0.58\pm0.07)\times b_{LIS} + (8.5\pm5.4)$.
In general, the \ion{Fe}{2} absorption lines are characterized by smaller $b$ values than the other LIS lines in our spectra. 
This indicates that the Doppler parameter is {\it not} consistent between {\it all} of our LIS absorption lines, especially when the masses of the ions vary greatly. 
When \ion{Fe}{2} was included in the simultaneous LIS fits, 
the broadening was driven towards lower values, which caused the column densities to be overestimated. 
Due to these complications, we did not include \ion{Fe}{2} in our reported fits, instead focusing on the ions that could be well-described by similar values of $b$.

\clearpage


\section{Curve of Growth} \label{appen:cog}
\restartappendixnumbering

\begin{figure}[!ht]
    \centering
    \textbf{\ion{O}{1} \W1039}\par\medskip
    \includegraphics[width=0.9\linewidth]{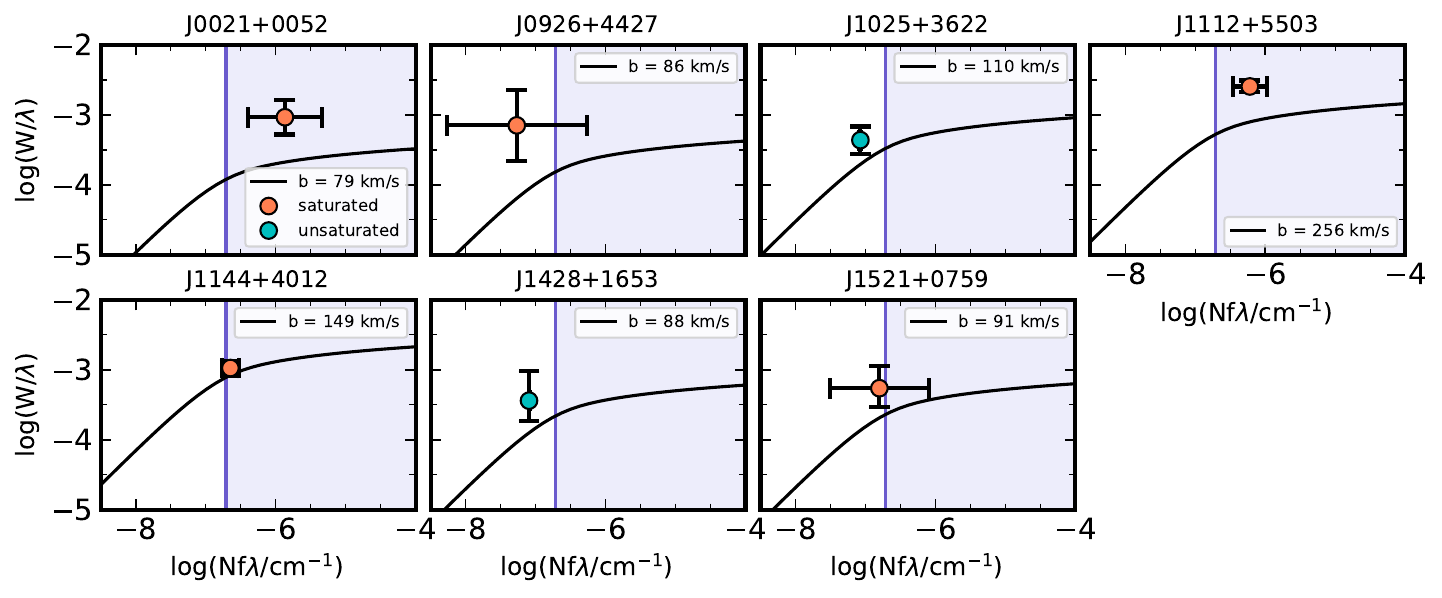}
    \figcaption{\ion{O}{1} \W1039 curve of growth diagrams. The purple shaded regions correspond to the saturated regime of the CoG, assuming $b$ from our fits. Galaxies with $W_\lambda$ and $\log(N_{ion})$ values (as measured from our fits) that lie in this saturated regime or are within 1$\sigma$ of it are classified as `saturated' and shown here as an orange point. Those that are at least 1$\sigma$ from the saturated regime are classified as `unsaturated' and shown by cyan points. \label{fig:firstCoG}}
\end{figure}

Table \ref{tab:saturated2} below lists which lines are characterized as saturated for each galaxy, as well as the total number of unsaturated lines include in each fit. Figures \ref{fig:firstCoG}-\ref{fig:lastCoG} contain the CoG diagrams for the absorption lines in our fits. These figures are in a Figure Set in the journal version of this document.

\startlongtable
\begin{deluxetable*}{lr|cc|cc|ccc|ccccc|c}
	 \tablewidth{0pt}
	 \setlength{\tabcolsep}{2pt}
	 \tabletypesize{\footnotesize}
	 \tablecaption{\footnotesize Saturated Lines}
	 \tablehead{
  \CH{}         & \CH{} \vline &  
        \multicolumn{2}{c|}{\ion{O}{1}} & 
        \multicolumn{2}{c|}{\ion{C}{2}} & 
        \multicolumn{3}{c|}{\ion{S}{2}} & 
        \multicolumn{5}{c|}{\ion{Si}{2}} & \CH{\ion{Al}{2}} \\ 
   \CH{Galaxy}  & \CH{\# unsat.} \vline & 
        \CH{1039} & \CH{1302} \vline & 
        \CH{1036} & \CH{1334} \vline & 
        \CH{1251} & \CH{1254} & 
        \CH{1259} \vline & \CH{1190} & 
        \CH{1193} & \CH{1260} & 
        \CH{1304} & \CH{1526} \vline & \CH{1671}}
	 \startdata
	 J0021+0052 & 0 & \X & \X & \X & \X & - & - & - & - & - & \X & \X & \X & - \\ 
	 J0036-3333 & 2 & - & \X & - & \X & - & - & - & - & - & \X & \T & \T & - \\ 
	 J0127-0619 & 1 & - & \X & - & \X & \T & - & \X & - & - & \X & \X & \X & \X \\ 
	 J0144+0453 & 1 & - & \X & - & \X & \T & - & \X & \X & - & \X & \X & \X & \X \\ 
	 J0337-0502 & 1 & - & \X & - & \X & \T & \X & \X & \X & - & \X & \X & \X & - \\ 
	 J0405-3648 & 0 & - & - & - & - & - & \X & \X & - & \X & \X & - & \X & - \\ 
	 J0808+3948 & 2 & - & - & - & \X & - & - & - & \T & - & \X & \T & - & - \\ 
	 J0823+2806 & 0 & - & \X & - & \X & \X & \X & \X & \X & \X & \X & \X & \X & \X \\ 
	 J0926+4427 & 0 & \X & - & \X & \X & - & - & - & \X & \X & \X & - & - & - \\ 
	 J0934+5514 & 3 & - & - & - & - & \T & \T & \T & - & \X & \X & \X & \X & \X \\ 
	 J0938+5428 & 0 & - & \X & \X & \X & - & - & - & \X & \X & \X & \X & \X & - \\ 
	 J0940+2935 & 1 & - & - & - & - & - & \T & \X & - & \X & \X & \X & \X & - \\ 
	 J0942+3547 & 1 & - & - & - & \X & - & - & - & \X & \X & - & \T & - & - \\ 
	 J0944-0038 & 0 & - & \X & - & - & - & - & - & \X & - & \X & \X & \X & - \\ 
	 J0944+3442 & 0 & - & \X & - & \X & - & - & - & - & - & - & \X & \X & \X \\ 
	 J1016+3754 & 1 & - & \T & - & \X & - & - & - & \X & \X & - & \X & \X & - \\ 
	 J1024+0524 & 0 & - & \X & - & - & - & - & - & \X & \X & \X & - & - & \X \\ 
	 J1025+3622 & 2 & \T & \X & \X & \X & - & - & - & \X & \X & \X & \T & \X & - \\ 
	 J1044+0353 & 0 & - & \X & - & - & \X & \X & - & \X & \X & - & \X & \X & \X \\ 
	 J1105+4444 & 0 & - & \X & - & - & \X & \X & \X & - & - & \X & \X & \X & \X \\ 
	 J1112+5503 & 0 & \X & - & \X & \X & - & - & - & - & - & \X & \X & \X & - \\ 
	 J1119+5130 & 0 & - & \X & - & - & - & - & - & \X & - & \X & \X & - & \X \\ 
	 J1129+2034 & 2 & - & \X & - & \X & \T & \T & \X & \X & - & \X & \X & \X & \X \\ 
	 J1132+1411 & 0 & - & \X & - & \X & - & - & - & - & - & - & \X & \X & \X \\ 
	 J1132+5722 & 0 & - & \X & - & \X & - & - & - & \X & - & \X & \X & \X & - \\ 
	 J1144+4012 & 0 & \X & \X & \X & \X & \X & \X & \X & \X & \X & \X & \X & \X & - \\ 
	 J1148+2546 & 0 & - & - & - & - & - & - & - & \X & \X & - & - & \X & \X \\ 
	 J1150+1501 & 0 & - & \X & - & \X & \X & \X & \X & - & \X & \X & \X & \X & \X \\ 
	 J1157+3220 & 0 & - & - & - & - & - & - & - & \X & - & - & \X & \X & \X \\ 
	 J1200+1343 & 0 & - & - & - & - & - & - & - & \X & \X & \X & - & \X & \X \\ 
	 J1225+6109 & 0 & - & - & - & \X & \X & \X & \X & - & \X & \X & - & \X & \X \\ 
	 J1253-0312 & 0 & - & \X & - & - & - & - & - & - & - & - & - & \X & \X \\ 
	 J1314+3452 & 0 & - & \X & - & \X & - & \X & \X & - & \X & \X & \X & \X & \X \\ 
	 J1359+5726 & 1 & - & - & - & \X & - & - & - & \X & \X & - & \T & - & \X \\ 
	 J1416+1223 & 0 & - & \X & \X & \X & - & - & - & - & \X & - & \X & \X & - \\ 
	 J1418+2102 & 0 & - & \X & - & \X & \X & - & \X & - & \X & \X & \X & \X & \X \\ 
	 J1428+1653 & 1 & \T & \X & \X & \X & - & - & - & \X & \X & \X & \X & - & - \\ 
	 J1429+0643 & 1 & - & - & - & \X & - & - & - & \X & \X & \X & \T & - & - \\ 
	 J1444+4237 & 0 & - & \X & - & - & \X & \X & \X & - & \X & \X & \X & \X & \X \\ 
	 J1448-0110 & 1 & - & \X & - & - & \T & \X & \X & - & \X & \X & \X & \X & \X \\ 
	 J1521+0759 & 1 & \X & \X & \X & \X & - & - & - & - & - & - & \T & - & - \\ 
	 J1525+0757 & 0 & - & - & \X & \X & - & - & - & \X & \X & \X & - & \X & - \\ 
	 J1545+0858 & 0 & - & \X & - & - & - & - & - & \X & \X & - & \X & - & \X \\ 
	 J1612+0817 & 1 & - & - & - & \X & - & - & - & \X & \X & \X & - & \T & - \\ 
	 \hline
	 \hline
	 {}         & unsaturated & 29\% & 3\% & 0\% & 0\% & 43\% & 21\% & 6\% & 4\% & 0\% & 0\% & 20\% & 6\% & 0\%  
	 \enddata
	 \tablecomments{\T Not saturated. \X Saturated. - Not included in fit.}
	 \label{tab:saturated2}
\end{deluxetable*}

\begin{figure}
    \centering
    \textbf{\ion{O}{1} \W1302}\par\medskip
    \includegraphics[width=0.9\linewidth]{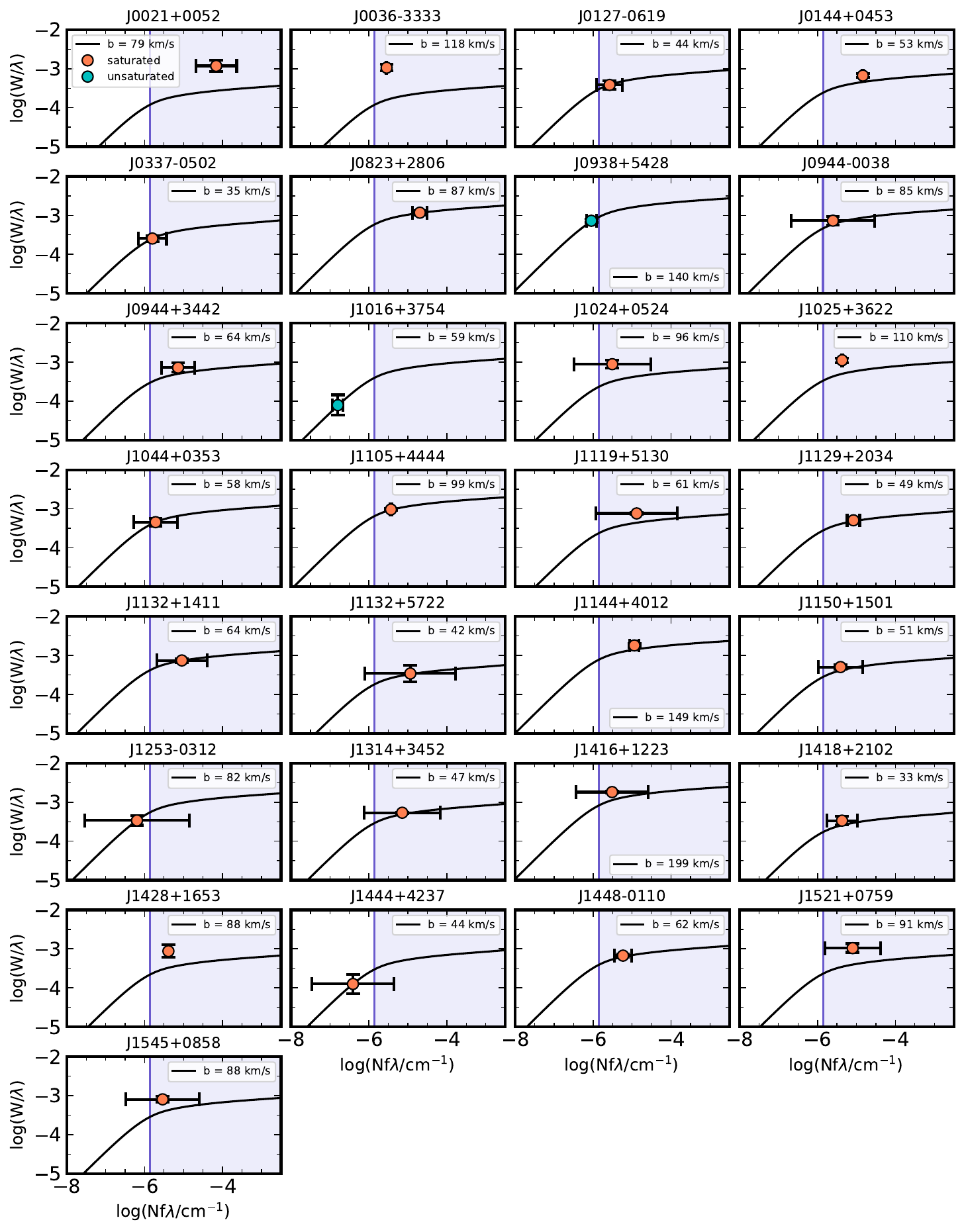}
    \figcaption{\ion{O}{1} \W1302 curve of growth diagrams. See description from Figure \ref{fig:firstCoG} for details.}
\end{figure}

\begin{figure}
    \centering
    \textbf{\ion{C}{2} \W1036}\par\medskip
    \includegraphics[width=0.9\linewidth]{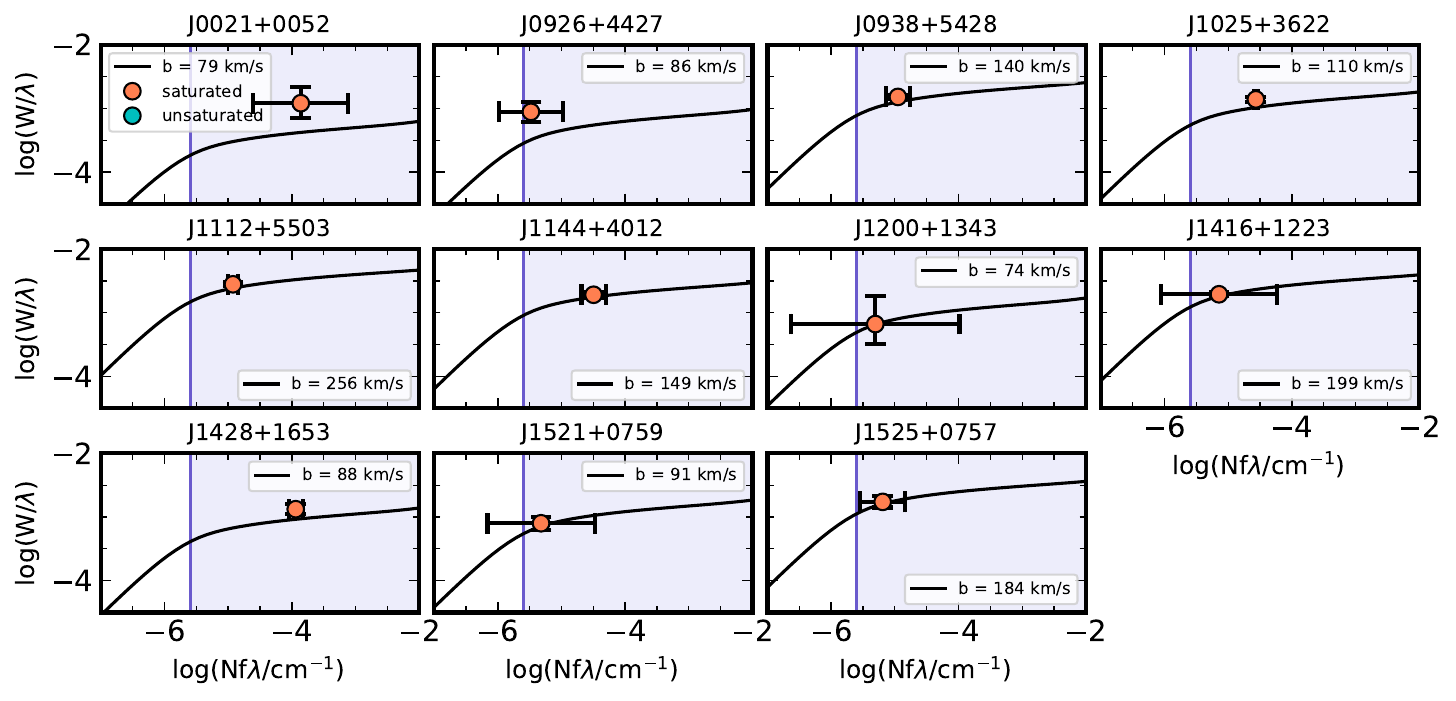}
    \figcaption{\ion{C}{2} \W1036 curve of growth diagrams. See description from Figure \ref{fig:firstCoG} for details.}
\end{figure}

\begin{figure}
    \centering
    \textbf{\ion{C}{2} \W1334}\par\medskip
    \includegraphics[width=0.9\linewidth]{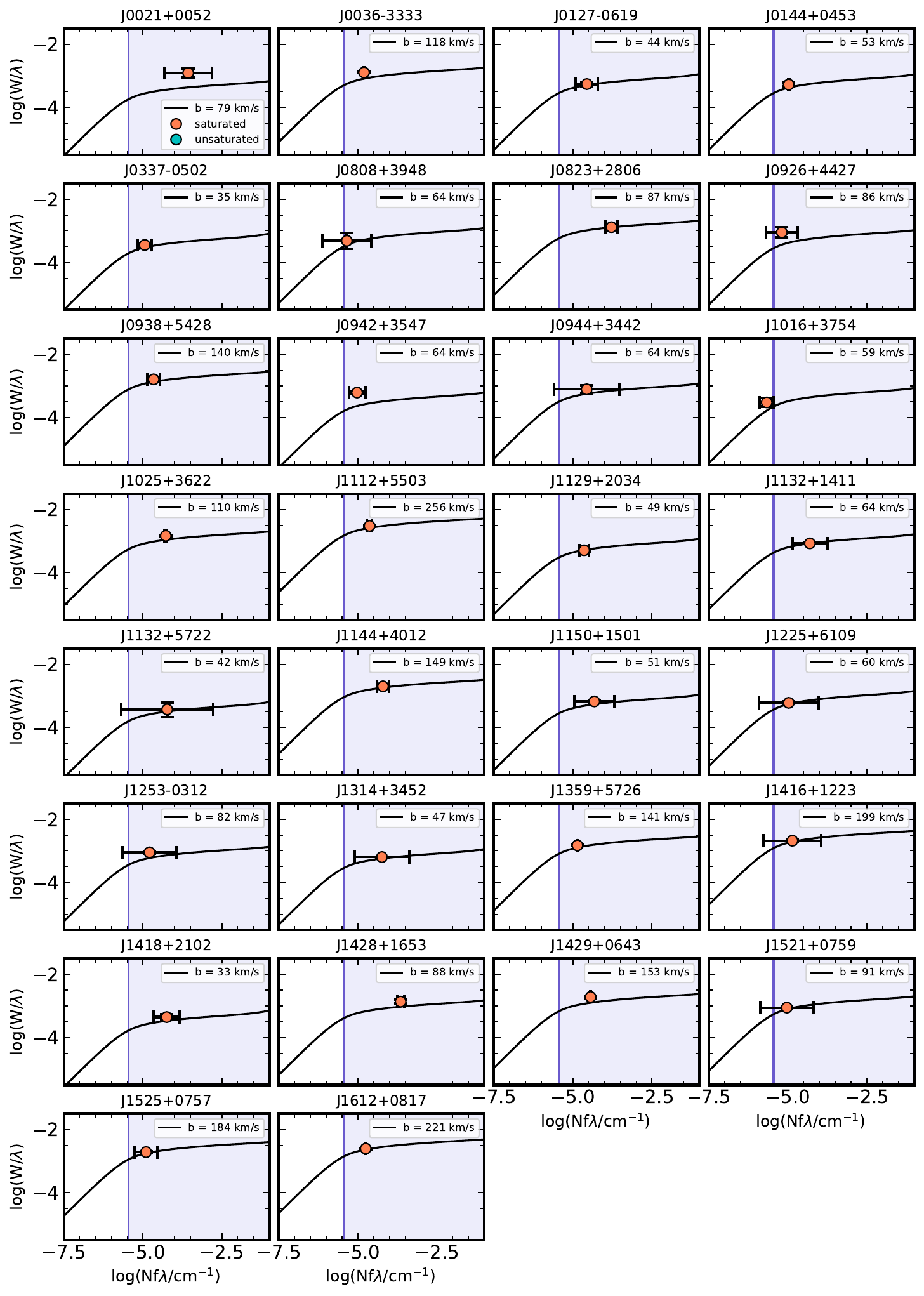}
    \figcaption{\ion{C}{2} \W1334 curve of growth diagrams. See description from Figure \ref{fig:firstCoG} for details.}
\end{figure}

\begin{figure}
    \centering
    \textbf{\ion{Si}{2} \W1190}\par\medskip
    \includegraphics[width=0.9\linewidth]{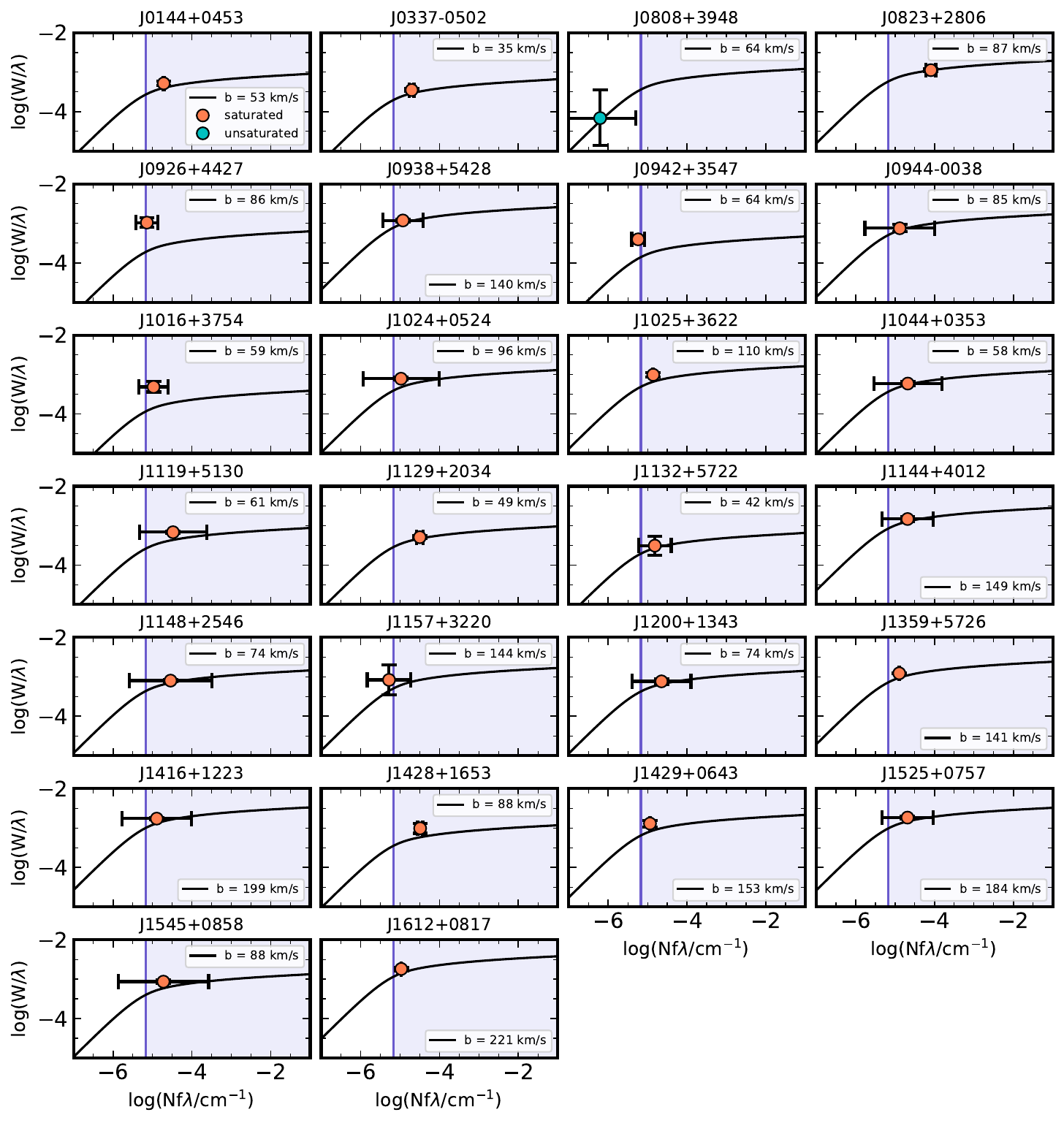}
    \figcaption{\ion{Si}{2} \W1190 curve of growth diagrams. See description from Figure \ref{fig:firstCoG} for details.}
\end{figure}

\begin{figure}
    \centering
    \textbf{\ion{Si}{2} \W1193}\par\medskip
    \includegraphics[width=0.9\linewidth]{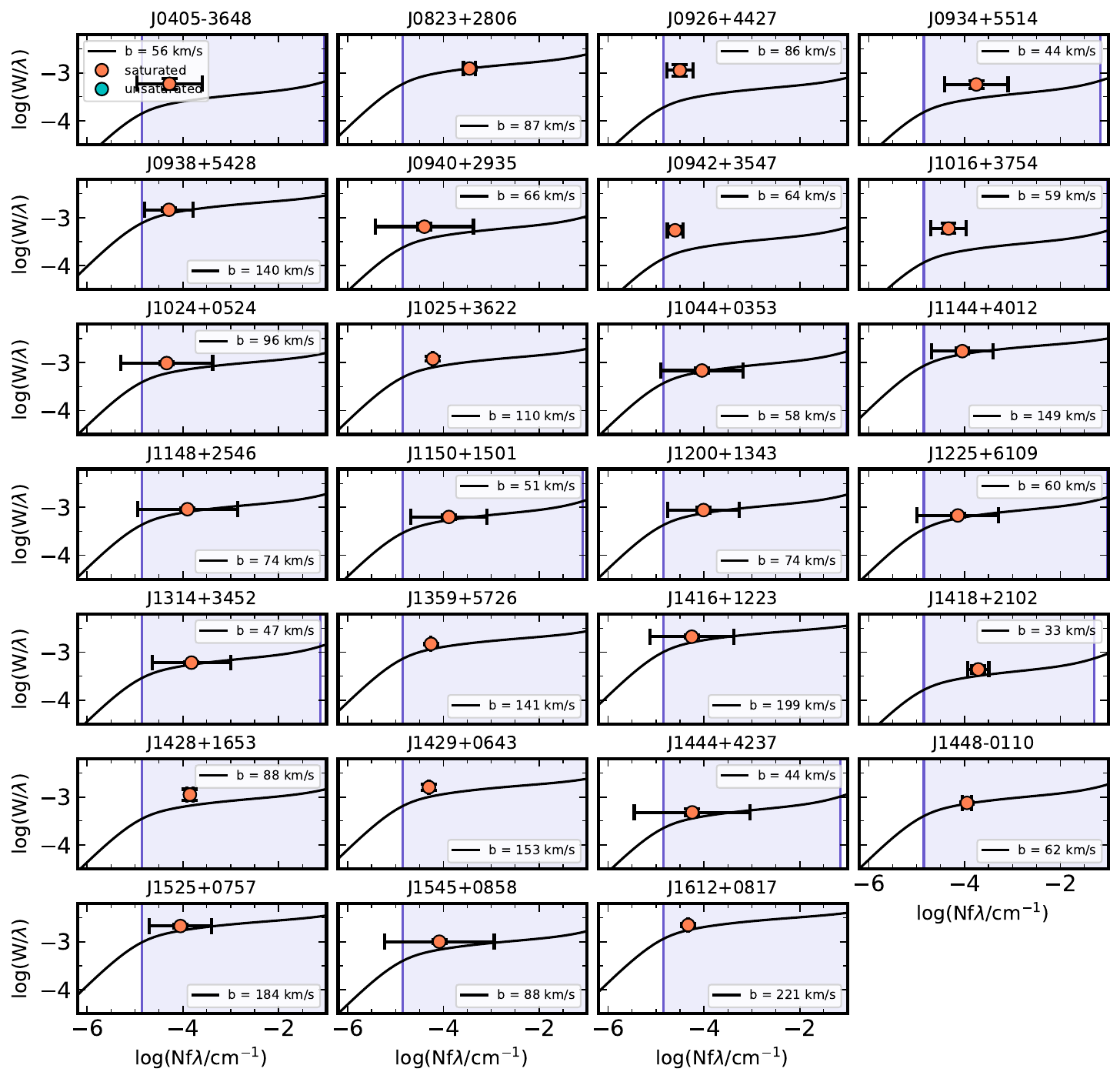}
    \figcaption{\ion{Si}{2} \W1193 curve of growth diagrams. See description from Figure \ref{fig:firstCoG} for details.}
\end{figure}

\begin{figure}
    \centering
    \textbf{\ion{Si}{2} \W1260}\par\medskip
    \includegraphics[width=0.9\linewidth]{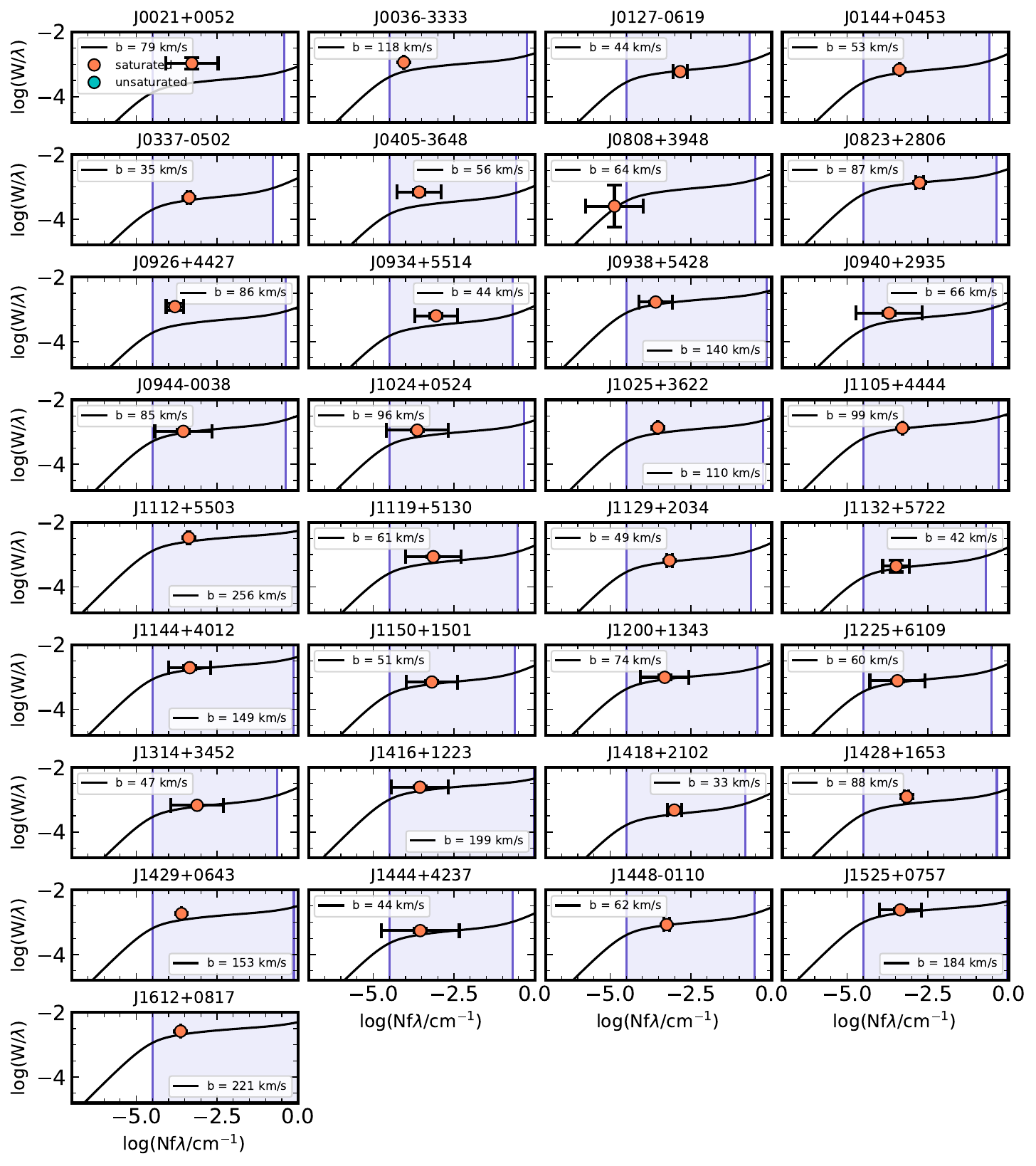}
    \figcaption{\ion{Si}{2} \W1260 curve of growth diagrams. See description from Figure \ref{fig:firstCoG} for details.}
\end{figure}

\begin{figure}
    \centering
    \textbf{\ion{Si}{2} \W1304}\par\medskip
    \includegraphics[width=0.9\linewidth]{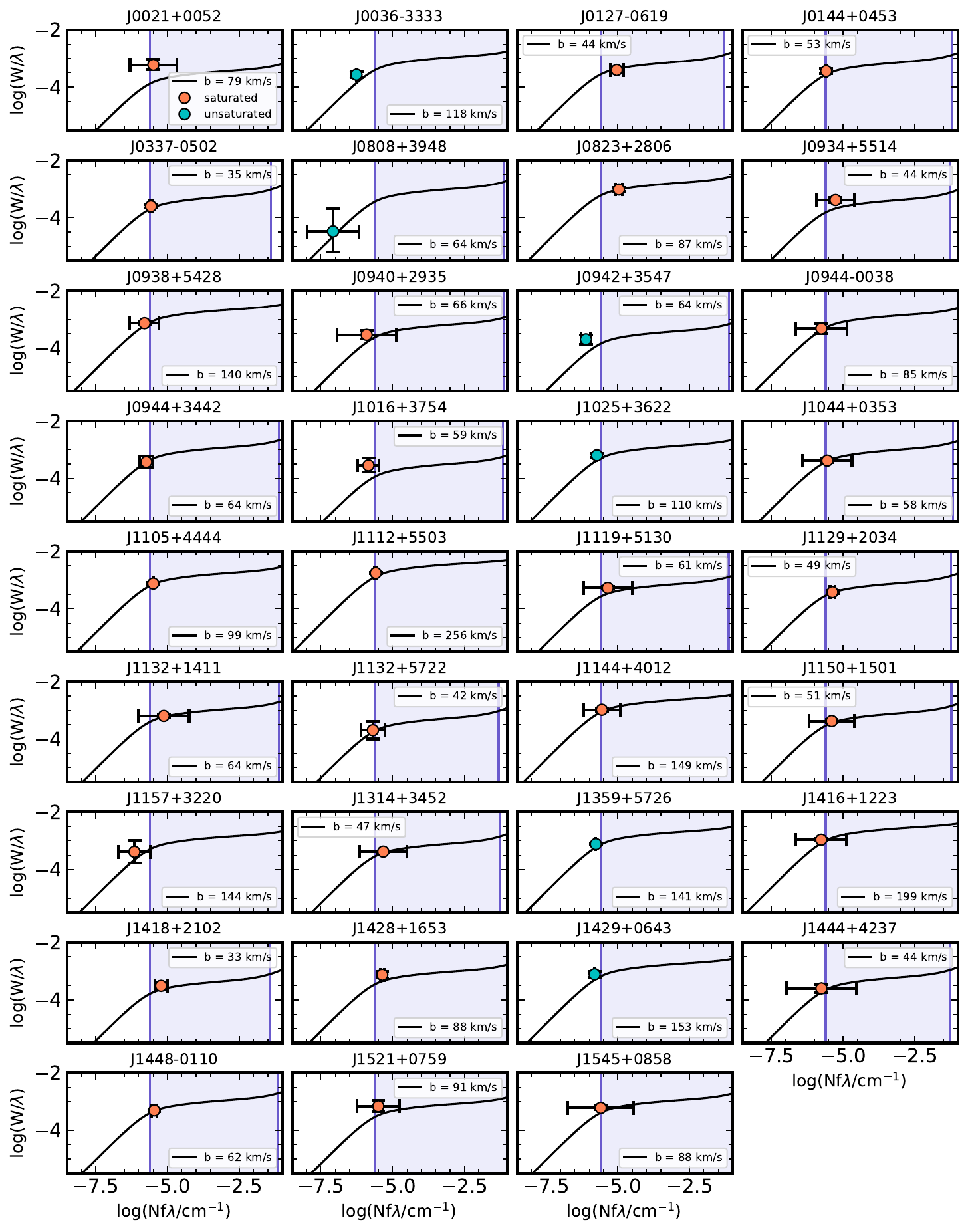}
    \figcaption{\ion{Si}{2} \W1304 curve of growth diagrams. See description from Figure \ref{fig:firstCoG} for details.}
\end{figure}

\begin{figure}
    \centering
    \textbf{\ion{Si}{2} \W1526}\par\medskip
    \includegraphics[width=0.9\linewidth]{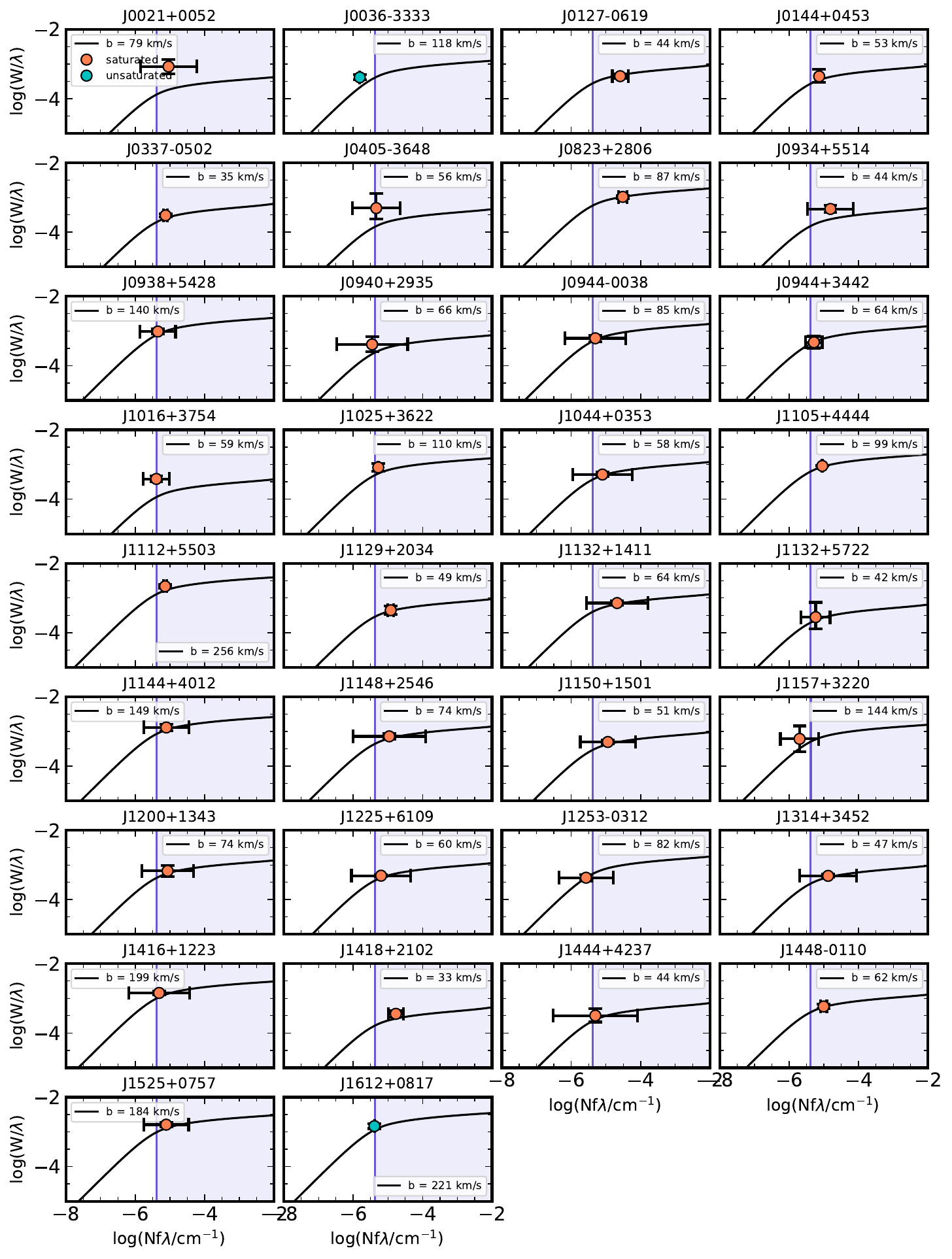}
    \figcaption{\ion{Si}{2} \W1526 curve of growth diagrams. See description from Figure \ref{fig:firstCoG} for details.}
\end{figure}

\begin{figure}
    \centering
    \textbf{\ion{S}{2} \W1251}\par\medskip
    \includegraphics[width=0.9\linewidth]{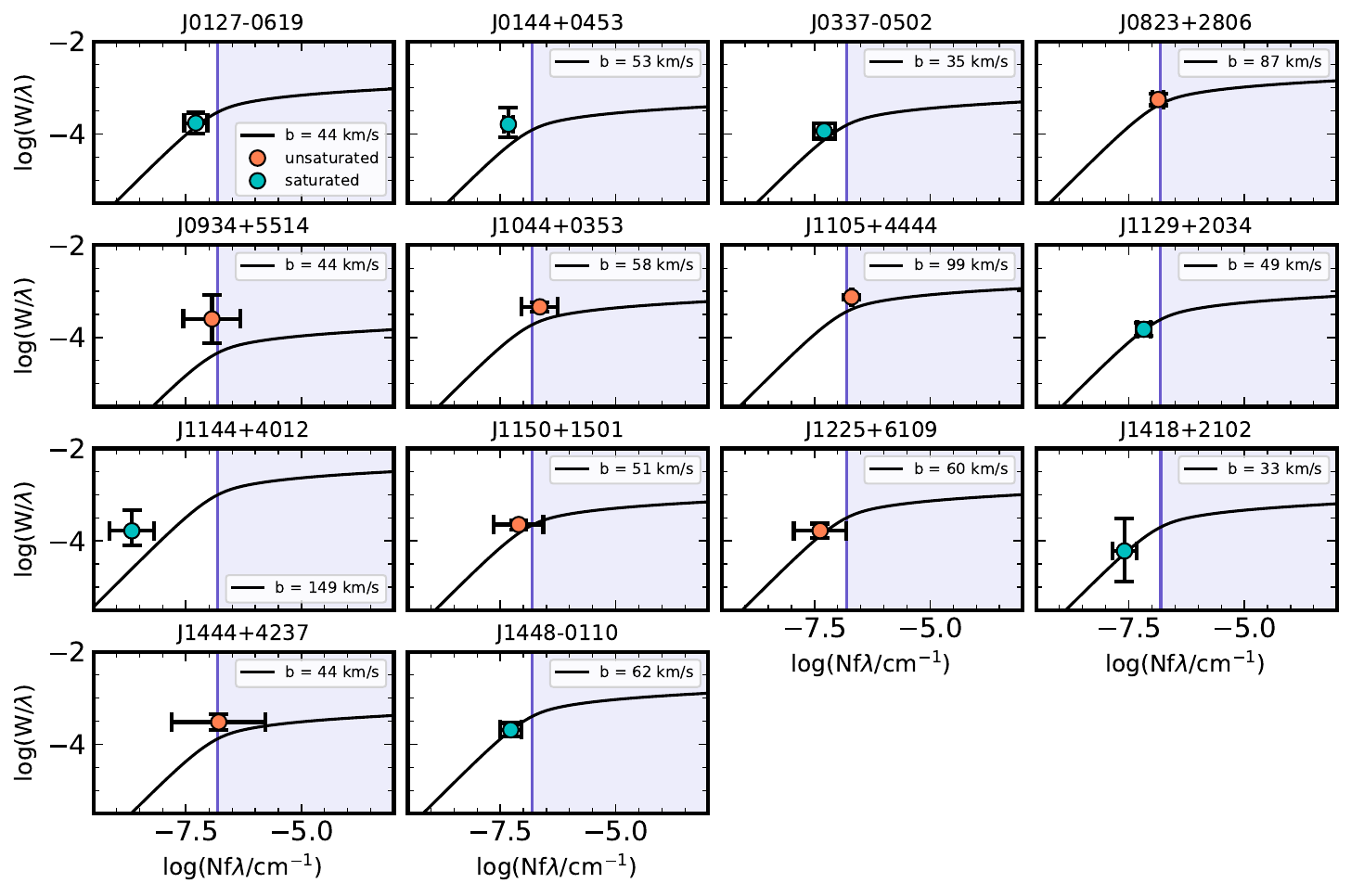}
    \figcaption{\ion{S}{2} \W1251 curve of growth diagrams. See description from Figure \ref{fig:firstCoG} for details.}
\end{figure}

\begin{figure}
    \centering
    \textbf{\ion{S}{2} \W1254}\par\medskip
    \includegraphics[width=0.9\linewidth]{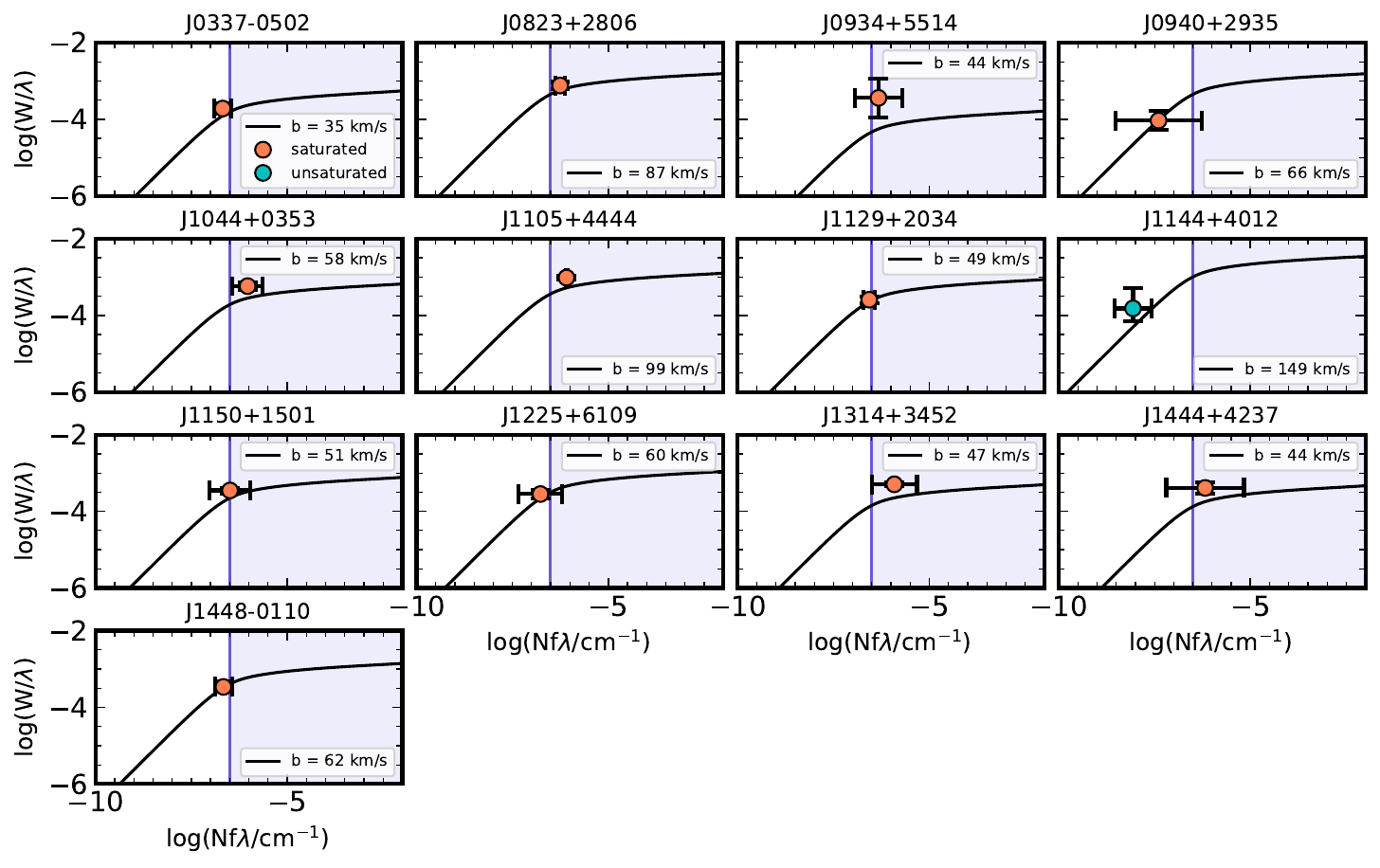}
    \figcaption{\ion{S}{2} \W1254 curve of growth diagrams. See description from Figure \ref{fig:firstCoG} for details.}
\end{figure}

\begin{figure}
    \centering
    \textbf{\ion{S}{2} \W1259}\par\medskip
    \includegraphics[width=0.9\linewidth]{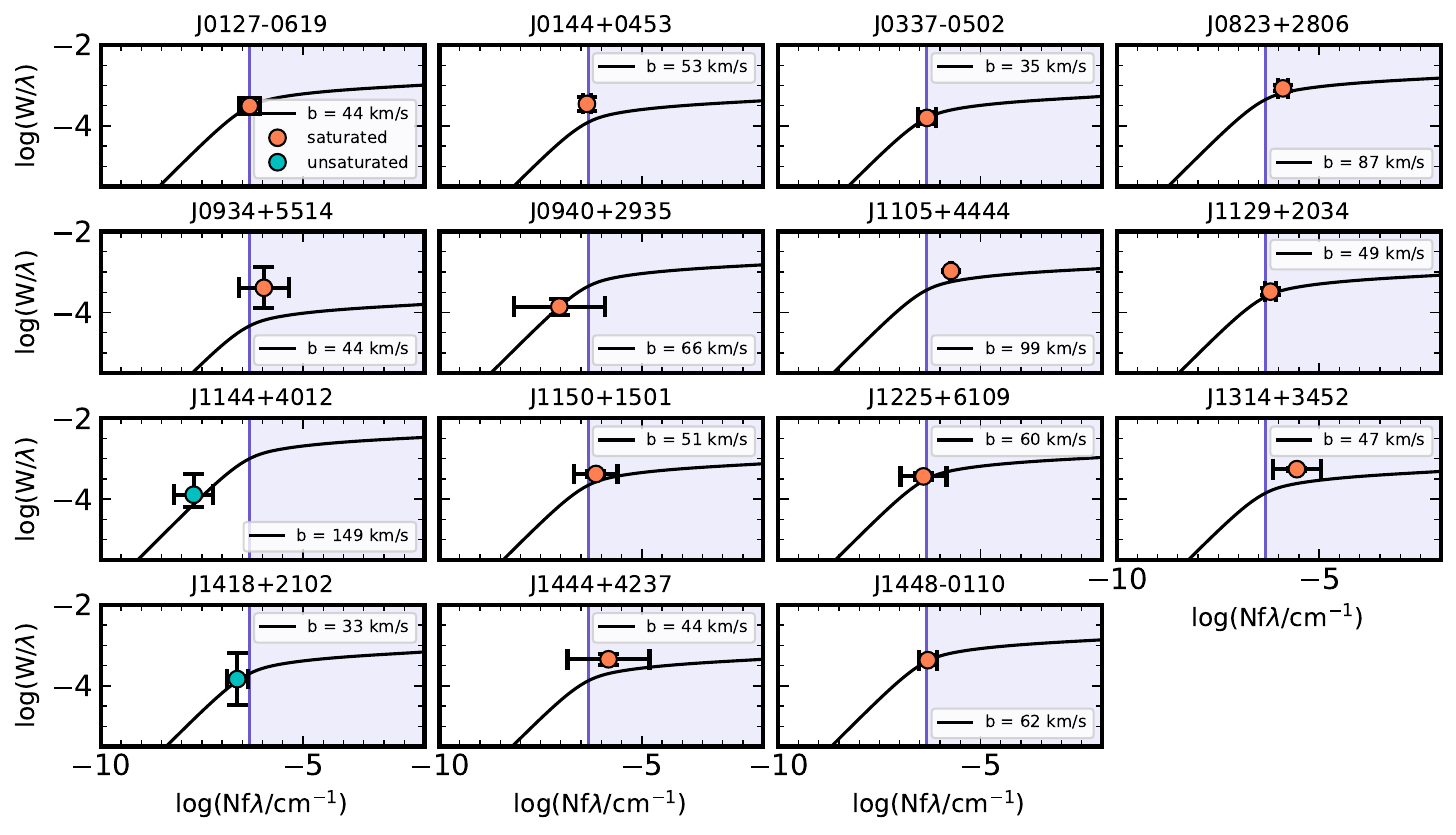}
    \figcaption{\ion{S}{2} \W1259 curve of growth diagrams. See description from Figure \ref{fig:firstCoG} for details.}
\end{figure}

\begin{figure}
    \centering
    \textbf{\ion{Al}{2} \W1670}\par\medskip
    \includegraphics[width=0.9\linewidth]{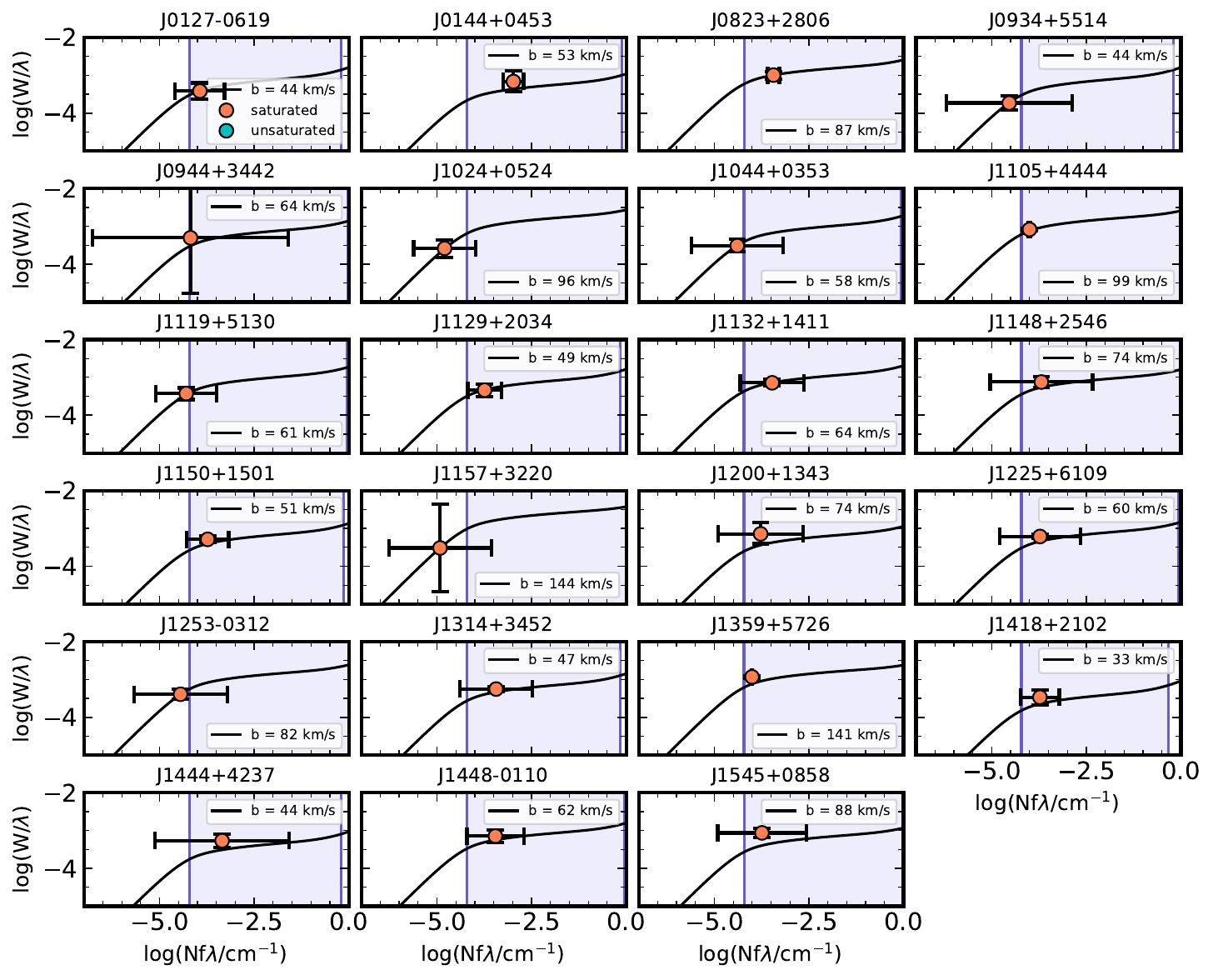}
    \figcaption{\ion{Al}{2} \W1670 curve of growth diagrams. See description from Figure \ref{fig:firstCoG} for details. \label{fig:lastCoG}}
\end{figure}


\section{Equivalent Width Comparison} \label{appen:EW_comp}
\restartappendixnumbering

\begin{figure*}[!h]
    \centering
    \includegraphics[width=\linewidth]{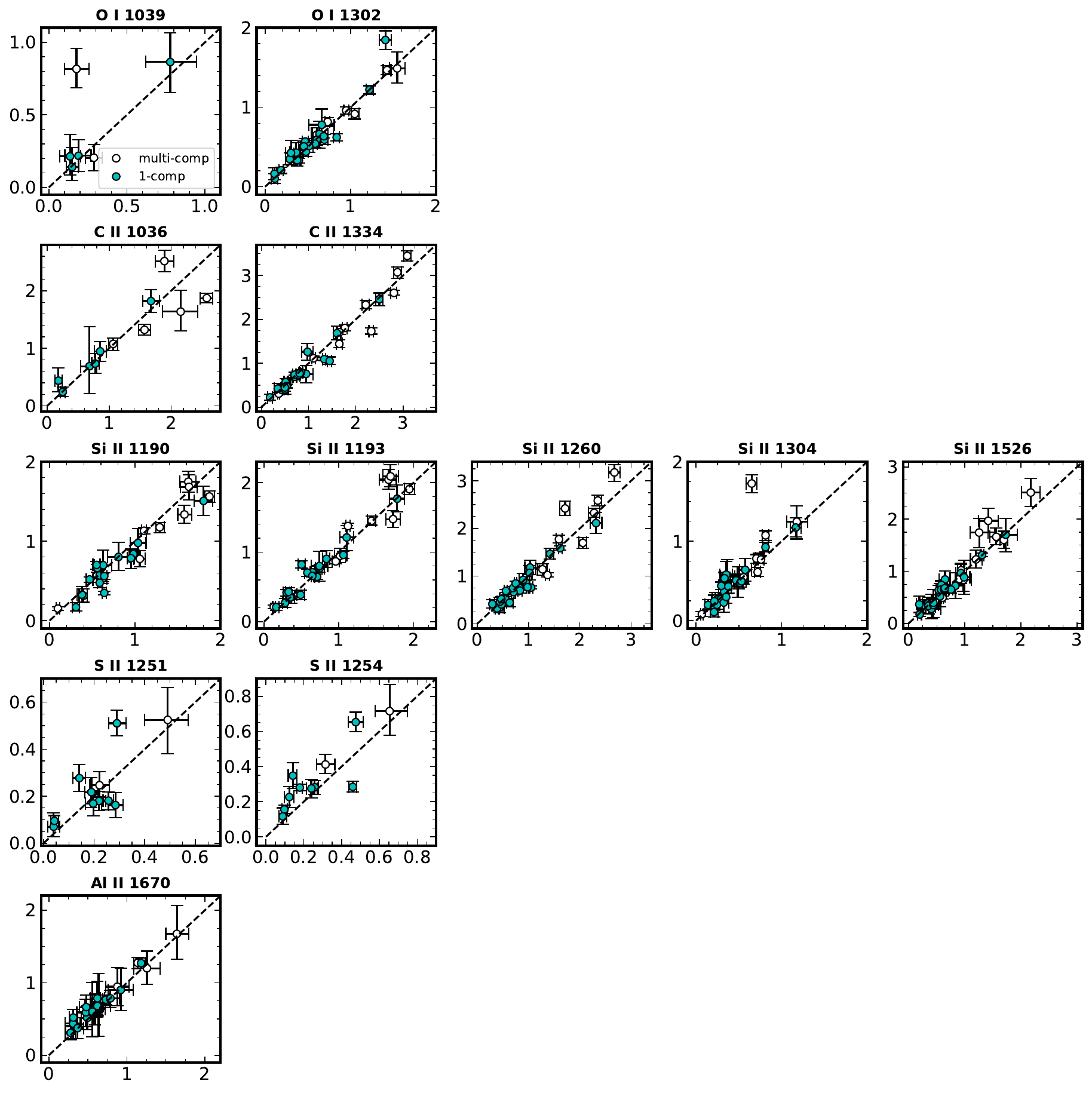}
    \figcaption{Comparing the equivalent widths measured from the observed data with those measured from the models. \ion{C}{2} and \ion{Si}{2} lines, as measured from our fits, are with 10\% of those measured from the data even when the multi-component galaxies are included. The lines associated with \ion{O}{1} and \ion{Al}{2} generally have percent errors $<$ 20\% while those for \ion{S}{2} lines are generally $<$ 35\%. This indicates that the measurements of $W_\lambda$ from our fits are largely in agreement with those measured directly from the data. \label{fig:compEWmeas}}
\end{figure*}

Figure \ref{fig:compEWmeas} compares the equivalent widths measured from the best-fit models from this work with those measured directly from the data for our series of neutral and LIS absorption lines. This plot shows general agreement between these measurements of $W_\lambda$, for both the 1- and multi-component samples. \\

\clearpage
\section{Fit Results} \label{appenD}
\restartappendixnumbering

\begin{figure*}[h!]
    \centering
    \includegraphics[width=\linewidth]{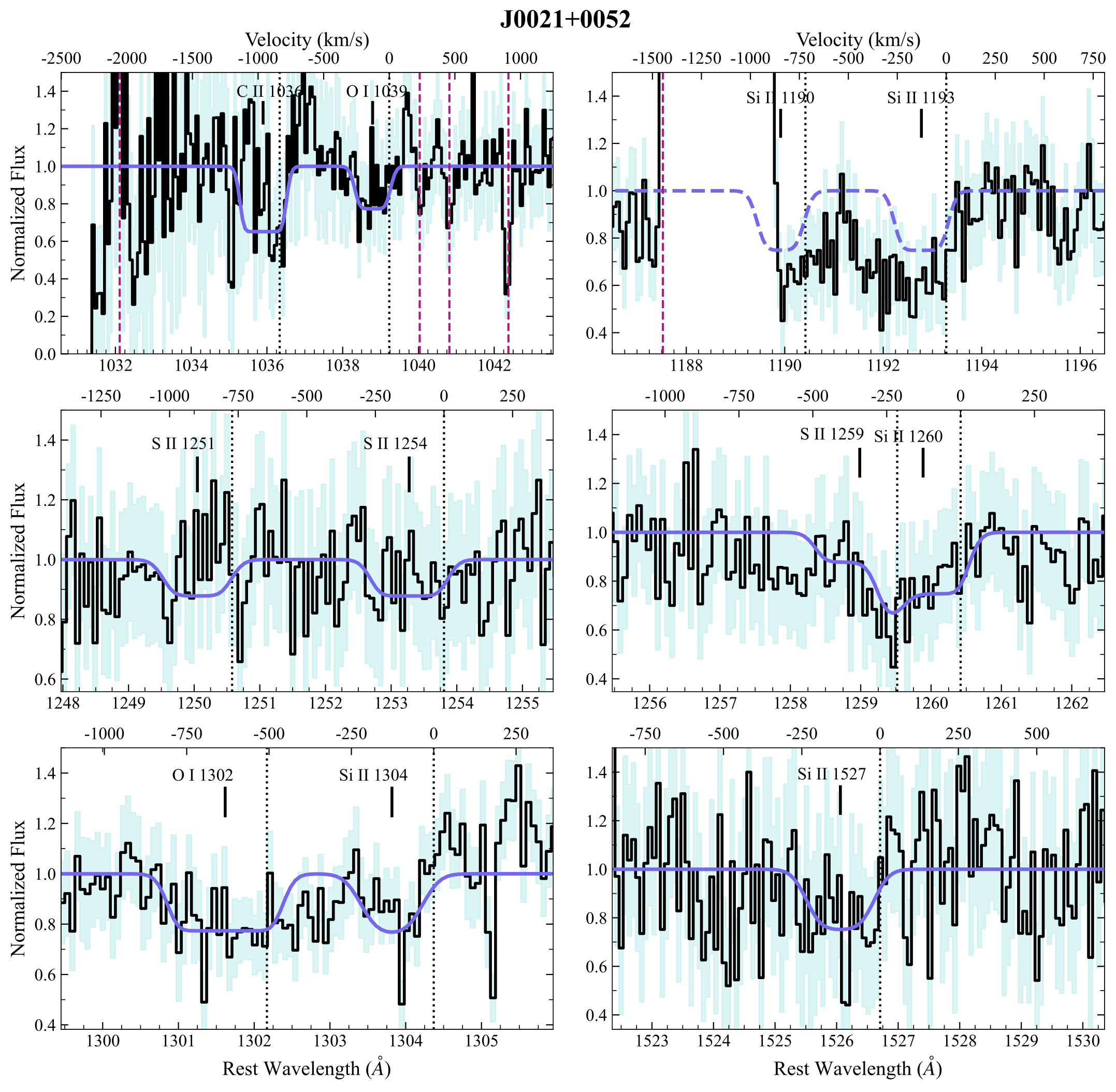}
    \figcaption{Simultaneous one-component fits (solid purple line) to the LIS features in the high-S/N, high-spectral-resolution CLASSY spectrum for J0021+0052 (solid black line). 
    Lines not included in the fit but predicted using the fit parameters are shown as dashed-purple line profiles. 
    The systemic wavelengths of the LIS absorption lines are marked by  vertical dotted lines, while the outflow trough ($v_{min}$(LIS)) is denoted by the solid black ticks. 
    Milky Way features, denoted by vertical dashed-magenta lines, were not included in the fit. \label{fig:firstfit}}
\end{figure*}

Table \ref{tab:fit_results} contains the results from our simultaneous 1-component LIS absorption line fits. Figures \ref{fig:firstfit}-\ref{fig:lastfit} contain the best fits to the LIS absorption lines for the CLASSY galaxies. The fits for J1150+1501 and J1359+5726 were shown in the body of this paper as Figures \ref{fig2} and \ref{fig2b}, respectively. These figures are in a Figure Set in the journal version of this document.

\startlongtable
\begin{longrotatetable}
\setlength{\tabcolsep}{3pt}
\begin{deluxetable*}{lc|CC|CCCCC|CCCCC}
	 \tablehead{
	 \CH{} & \CH{} & \CH{$v_{min}$(LIS)} & \CH{$b$} \vline & \multicolumn{5}{c}{\underline{log $N_{ion}$ }} \vline & \multicolumn{5}{c}{\underline{$C_f$(ion)}} \\ [-2ex]
	 \CH{Galaxy} & \CH{Notes} & \CH{(km/s)} & \CH{(km/s)} \vline & \CH{\ion{O}{1}} & \CH{\ion{C}{2}} & \CH{\ion{Si}{2}} & \CH{\ion{S}{2}} & \CH{\ion{Al}{2}} \vline & \CH{\ion{O}{1}} & \CH{\ion{C}{2}} & \CH{\ion{Si}{2}} & \CH{\ion{S}{2}} & \CH{\ion{Al}{2}}
	 }
	 \tablecaption{Best-Fit Parameters}
	 \label{tab:fit_results}
	 \startdata
	 J0021+0052 & (1) & $-127 \pm 60$ & $79 \pm 29$ & $16.90 \pm 0.53$ & $16.68 \pm 0.75$ & $15.07 \pm 0.81$ & $16.79 \pm 1.58$ & \nodata & $0.23 \pm 0.06$ & $0.35 \pm 0.11$ & $0.25 \pm 0.07$ & $0.12 \pm 0.18$ & \nodata \\ 
	 J0036-3333 & (2) & $-149 \pm 1$ & $118 \pm 1$ & $15.69 \pm 0.01$ & $15.62 \pm 0.01$ & $14.47 \pm 0.01$ & \nodata & \nodata & $0.15 \pm 0.01$ & $0.64 \pm 0.01$ & $0.51 \pm 0.01$ & \nodata & \nodata \\ 
	 J0127-0619 & (2) & $+0 \pm 1$ & $31 \pm 1$ & $15.79 \pm 0.01$ & $16.91 \pm 0.01$ & $16.88 \pm 0.01$ & $15.58 \pm 0.01$ & $13.91 \pm 0.01$ & $0.97 \pm 0.01$ & $0.94 \pm 0.01$ & $0.95 \pm 0.01$ & $0.98 \pm 0.01$ & $1.00 \pm 0.01$ \\ 
	 J0144+0453 & (2) & $+21 \pm 1$ & $53 \pm 1$ & $16.05 \pm 0.02$ & $15.11 \pm 0.02$ & $14.80 \pm 0.01$ & $15.47 \pm 0.10$ & $14.63 \pm 0.27$ & $0.68 \pm 0.01$ & $0.80 \pm 0.01$ & $0.77 \pm 0.01$ & $0.35 \pm 0.07$ & $0.60 \pm 0.01$ \\ 
	 J0337-0502 & (2) & $-17 \pm 1$ & $36 \pm 2$ & $14.92 \pm 0.36$ & $14.96 \pm 0.22$ & $14.64 \pm 0.06$ & $15.33 \pm 0.23$ & \nodata & $1.00 \pm 0.16$ & $0.83 \pm 0.07$ & $0.83 \pm 0.03$ & $0.65 \pm 0.19$ & \nodata \\ 
	 J0405-3648 & & $-27 \pm 31$ & $28 \pm 24$ & \nodata & \nodata & $16.92 \pm 0.19$ & $17.50 \pm 0.62$ & \nodata & \nodata & \nodata & $0.38 \pm 0.04$ & $0.09 \pm 0.08$ & \nodata \\ 
	 J0808+3948 & (2) & $-602 \pm 1$ & $65 \pm 32$ & \nodata & $14.81 \pm 0.77$ & $13.40 \pm 0.90$ & \nodata & \nodata & \nodata & $0.75 \pm 0.42$ & $0.84 \pm 1.25$ & \nodata & \nodata \\ 
	 J0823+2806 & (1) & $+16 \pm 2$ & $88 \pm 5$ & $16.41 \pm 0.18$ & $16.52 \pm 0.19$ & $15.64 \pm 0.12$ & $16.15 \pm 0.12$ & $14.39 \pm 0.16$ & $1.00 \pm 0.02$ & $0.98 \pm 0.01$ & $1.00 \pm 0.01$ & $0.76 \pm 0.07$ & $1.00 \pm 0.01$ \\ 
	 J0926+4427 & (1) & $-306 \pm 50$ & $86 \pm 6$ & $15.53 \pm 1.00$ & $15.10 \pm 0.50$ & $14.58 \pm 0.27$ & \nodata & \nodata & $0.27 \pm 0.30$ & $0.50 \pm 0.16$ & $0.33 \pm 0.06$ & \nodata & \nodata \\ 
	 J0934+5514 & (2) & $+15 \pm 1$ & $45 \pm 4$ & $14.50 \pm 1.10$ & \nodata & $15.38 \pm 0.34$ & $15.02 \pm 0.16$ & $15.76 \pm 0.57$ & $0.94 \pm 0.34$ & \nodata & $0.43 \pm 0.05$ & $0.51 \pm 0.01$ & $0.16 \pm 0.12$ \\ 
	 J0938+5428 & (1) & $-30 \pm 11$ & $135 \pm 21$ & $15.33 \pm 0.22$ & $15.87 \pm 0.15$ & $15.03 \pm 0.55$ & \nodata & \nodata & $0.88 \pm 0.06$ & $0.84 \pm 0.02$ & $0.82 \pm 0.05$ & \nodata & \nodata \\ 
	 J0940+2935 & (2) & $-38 \pm 2$ & $66 \pm 5$ & \nodata & \nodata & $14.85 \pm 0.45$ & $15.29 \pm 0.42$ & \nodata & \nodata & \nodata & $1.00 \pm 0.20$ & $0.41 \pm 0.06$ & \nodata \\ 
	 J0942+3547 & (2) & $-50 \pm 5$ & $64 \pm 10$ & $14.65 \pm 2.05$ & $15.13 \pm 0.26$ & $14.35 \pm 0.16$ & \nodata & \nodata & $0.33 \pm 1.19$ & $0.38 \pm 0.04$ & $0.33 \pm 0.05$ & \nodata & \nodata \\ 
	 J0944-0038 & & $+6 \pm 2$ & $60 \pm 2$ & $15.43 \pm 1.59$ & \nodata & $15.18 \pm 0.05$ & \nodata & \nodata & $0.87 \pm 0.11$ & \nodata & $0.87 \pm 0.01$ & \nodata & \nodata \\ 
	 J0944+3442 & & $+13 \pm 2$ & $77 \pm 6$ & $15.65 \pm 0.15$ & $16.23 \pm 0.34$ & $14.77 \pm 0.25$ & \nodata & $13.20 \pm 1.33$ & $0.71 \pm 0.04$ & $0.65 \pm 0.03$ & $1.00 \pm 0.07$ & \nodata & $1.00 \pm 0.22$ \\ 
	 J1016+3754 & (2) & $-11 \pm 6$ & $59 \pm 19$ & $14.14 \pm 0.14$ & $14.46 \pm 0.23$ & $14.59 \pm 0.37$ & \nodata & \nodata & $1.00 \pm 0.12$ & $0.56 \pm 0.05$ & $0.29 \pm 0.02$ & \nodata & \nodata \\ 
	 J1024+0524 & & $-77 \pm 2$ & $96 \pm 24$ & $15.64 \pm 0.99$ & \nodata & $14.80 \pm 0.96$ & \nodata & $13.07 \pm 0.82$ & $0.36 \pm 0.07$ & \nodata & $0.61 \pm 0.04$ & \nodata & $1.00 \pm 0.27$ \\ 
	 J1025+3622 & (1,2) & $-122 \pm 1$ & $111 \pm 1$ & $15.84 \pm 0.01$ & $16.12 \pm 0.03$ & $14.98 \pm 0.01$ & \nodata & \nodata & $0.45 \pm 0.01$ & $0.75 \pm 0.01$ & $0.66 \pm 0.01$ & \nodata & \nodata \\ 
	 J1044+0353 & & $-33 \pm 18$ & $59 \pm 47$ & $15.21 \pm 0.56$ & \nodata & $14.88 \pm 0.86$ & $16.20 \pm 0.39$ & $13.26 \pm 1.21$ & $1.00 \pm 0.19$ & \nodata & $0.94 \pm 0.08$ & $0.48 \pm 0.05$ & $1.00 \pm 0.23$ \\ 
	 J1105+4444 & & $-48 \pm 1$ & $100 \pm 1$ & $15.73 \pm 0.02$ & \nodata & $15.16 \pm 0.01$ & $16.37 \pm 0.02$ & $13.88 \pm 0.03$ & $0.98 \pm 0.01$ & \nodata & $0.95 \pm 0.01$ & $0.54 \pm 0.01$ & $0.92 \pm 0.02$ \\ 
	 J1112+5503 & (1) & $-354 \pm 4$ & $256 \pm 10$ & $17.06 \pm 0.25$ & $16.13 \pm 0.08$ & $15.48 \pm 0.05$ & \nodata & \nodata & $0.31 \pm 0.02$ & $0.86 \pm 0.02$ & $0.76 \pm 0.02$ & \nodata & \nodata \\ 
	 J1119+5130 & & $-6 \pm 18$ & $60 \pm 10$ & $16.05 \pm 0.92$ & \nodata & $15.08 \pm 0.75$ & \nodata & $13.30 \pm 1.43$ & $0.58 \pm 0.01$ & \nodata & $0.64 \pm 0.02$ & \nodata & $1.00 \pm 0.17$ \\ 
	 J1129+2034 & (2) & $+8 \pm 1$ & $49 \pm 1$ & $15.59 \pm 0.01$ & $15.19 \pm 0.01$ & $14.90 \pm 0.01$ & $15.46 \pm 0.04$ & $13.78 \pm 0.09$ & $0.82 \pm 0.01$ & $0.87 \pm 0.01$ & $0.84 \pm 0.01$ & $0.75 \pm 0.03$ & $0.92 \pm 0.02$ \\ 
	 J1132+1411 & & $-39 \pm 2$ & $74 \pm 4$ & $15.75 \pm 0.09$ & $15.55 \pm 0.11$ & $15.16 \pm 0.08$ & \nodata & $14.06 \pm 0.22$ & $0.99 \pm 0.01$ & $1.00 \pm 0.01$ & $0.97 \pm 0.02$ & \nodata & $1.00 \pm 0.04$ \\ 
	 J1132+5722 & & $+13 \pm 8$ & $43 \pm 16$ & $15.86 \pm 1.17$ & $15.75 \pm 1.46$ & $14.60 \pm 0.42$ & \nodata & $15.12 \pm 2.67$ & $0.64 \pm 0.14$ & $0.56 \pm 0.14$ & $0.68 \pm 0.24$ & \nodata & $0.69 \pm 0.35$ \\ 
	 J1144+4012 & & $-169 \pm 1$ & $149 \pm 2$ & $16.38 \pm 0.02$ & $16.33 \pm 0.04$ & $15.28 \pm 0.01$ & $14.67 \pm 2.92$ & \nodata & $0.78 \pm 0.01$ & $0.92 \pm 0.01$ & $0.86 \pm 0.01$ & $0.99 \pm 0.49$ & \nodata \\ 
	 J1148+2546 & & $-73 \pm 1$ & $77 \pm 1$ & $16.51 \pm 0.83$ & \nodata & $15.18 \pm 0.02$ & \nodata & $14.06 \pm 0.49$ & $1.00 \pm 0.12$ & \nodata & $0.89 \pm 0.01$ & \nodata & $0.72 \pm 0.03$ \\ 
	 J1150+1501 & & $+4 \pm 1$ & $50 \pm 1$ & $15.92 \pm 0.01$ & $16.14 \pm 0.01$ & $15.07 \pm 0.01$ & $16.16 \pm 0.01$ & $14.09 \pm 0.13$ & $0.75 \pm 0.01$ & $0.79 \pm 0.01$ & $0.84 \pm 0.01$ & $0.53 \pm 0.01$ & $0.73 \pm 0.01$ \\ 
	 J1157+3220 & (1) & $-223 \pm 7$ & $144 \pm 20$ & $15.55 \pm 0.50$ & $15.34 \pm 0.22$ & $14.67 \pm 0.55$ & \nodata & $13.12 \pm 1.35$ & $0.27 \pm 0.12$ & $0.63 \pm 0.07$ & $0.53 \pm 0.45$ & \nodata & $1.00 \pm 2.67$ \\ 
	 J1200+1343 & (1) & $-124 \pm 1$ & $59 \pm 1$ & \nodata & \nodata & $16.00 \pm 0.01$ & \nodata & $14.87 \pm 0.17$ & \nodata & \nodata & $0.81 \pm 0.01$ & \nodata & $0.45 \pm 0.03$ \\ 
	 J1225+6109 & & $+20 \pm 11$ & $61 \pm 18$ & \nodata & $15.16 \pm 0.95$ & $14.79 \pm 0.85$ & $15.48 \pm 0.57$ & $13.94 \pm 1.06$ & \nodata & $0.92 \pm 0.03$ & $0.87 \pm 0.04$ & $0.78 \pm 0.11$ & $0.75 \pm 0.06$ \\ 
	 J1253-0312 & & $-50 \pm 6$ & $82 \pm 12$ & $14.94 \pm 0.46$ & \nodata & $14.56 \pm 0.46$ & \nodata & $13.50 \pm 0.81$ & $0.91 \pm 0.20$ & \nodata & $0.99 \pm 0.13$ & \nodata & $0.83 \pm 0.18$ \\ 
	 J1314+3452 & & $-42 \pm 12$ & $48 \pm 9$ & $15.70 \pm 0.97$ & $15.79 \pm 0.86$ & $15.01 \pm 0.82$ & $16.23 \pm 0.59$ & $14.13 \pm 0.96$ & $0.91 \pm 0.01$ & $0.90 \pm 0.02$ & $0.91 \pm 0.02$ & $0.44 \pm 0.03$ & $0.84 \pm 0.02$ \\ 
	 J1359+5726 & (1,2) & $-179 \pm 1$ & $141 \pm 1$ & $15.68 \pm 0.01$ & $15.65 \pm 0.01$ & $15.04 \pm 0.01$ & \nodata & $14.03 \pm 0.01$ & $0.66 \pm 0.01$ & $0.86 \pm 0.01$ & $0.77 \pm 0.01$ & \nodata & $0.65 \pm 0.01$ \\ 
	 J1416+1223 & (1) & $-43 \pm 2$ & $199 \pm 11$ & $15.95 \pm 0.92$ & $15.80 \pm 0.91$ & $15.20 \pm 0.88$ & \nodata & \nodata & $0.62 \pm 0.04$ & $0.93 \pm 0.05$ & $0.76 \pm 0.04$ & \nodata & \nodata \\ 
	 J1418+2102 & & $+9 \pm 1$ & $31 \pm 7$ & $15.92 \pm 0.52$ & $16.08 \pm 0.65$ & $16.05 \pm 0.79$ & $15.87 \pm 0.57$ & $13.68 \pm 0.37$ & $0.76 \pm 0.05$ & $0.76 \pm 0.02$ & $0.75 \pm 0.02$ & $0.43 \pm 0.08$ & $0.75 \pm 0.13$ \\ 
	 J1428+1653 & (1,2) & $-163 \pm 1$ & $89 \pm 2$ & $15.72 \pm 0.01$ & $16.65 \pm 0.11$ & $15.25 \pm 0.06$ & \nodata & \nodata & $0.37 \pm 0.01$ & $0.69 \pm 0.01$ & $0.60 \pm 0.01$ & \nodata & \nodata \\ 
	 J1429+0643 & (1,2) & $-164 \pm 1$ & $154 \pm 1$ & \nodata & $16.10 \pm 0.01$ & $15.04 \pm 0.01$ & \nodata & \nodata & \nodata & $0.66 \pm 0.01$ & $0.63 \pm 0.01$ & \nodata & \nodata \\ 
	 J1444+4237 & & $-54 \pm 38$ & $45 \pm 32$ & $14.40 \pm 1.05$ & \nodata & $14.56 \pm 1.21$ & $15.93 \pm 1.01$ & $14.19 \pm 1.77$ & $1.00 \pm 0.33$ & \nodata & $0.75 \pm 0.08$ & $0.44 \pm 0.10$ & $0.58 \pm 0.20$ \\ 
	 J1448-0110 & (1,2) & $+17 \pm 4$ & $62 \pm 3$ & $15.72 \pm 0.23$ & \nodata & $15.00 \pm 0.09$ & $15.60 \pm 0.23$ & $14.22 \pm 0.75$ & $0.94 \pm 0.06$ & \nodata & $0.96 \pm 0.03$ & $0.96 \pm 0.27$ & $0.79 \pm 0.22$ \\ 
	 J1521+0759 & (1,2) & $-69 \pm 33$ & $131 \pm 15$ & $15.91 \pm 0.43$ & $15.05 \pm 0.47$ & $14.62 \pm 0.42$ & \nodata & \nodata & $0.36 \pm 0.03$ & $1.00 \pm 0.07$ & $1.00 \pm 0.16$ & \nodata & \nodata \\ 
	 J1525+0757 & (1) & $-333 \pm 34$ & $193 \pm 33$ & \nodata & $15.69 \pm 0.58$ & $15.31 \pm 0.86$ & \nodata & \nodata & \nodata & $0.92 \pm 0.02$ & $0.81 \pm 0.06$ & \nodata & \nodata \\ 
	 J1545+0858 & (1) & $-65 \pm 35$ & $88 \pm 31$ & $15.57 \pm 0.95$ & \nodata & $15.01 \pm 1.15$ & \nodata & $14.09 \pm 1.17$ & $0.49 \pm 0.05$ & \nodata & $0.68 \pm 0.04$ & \nodata & $0.46 \pm 0.07$ \\ 
	 J1612+0817 & (1,2) & $-403 \pm 1$ & $222 \pm 1$ & \nodata & $15.84 \pm 0.01$ & $15.15 \pm 0.01$ & \nodata & \nodata & \nodata & $0.95 \pm 0.01$ & $0.78 \pm 0.01$ & \nodata & \nodata \\ 
	 \enddata
	 \tablecomments{\textit{Column 1:} Galaxy name (J1323-0132 is not included due to its lack of LIS absorption). \textit{Column 2}: (1) for galaxies in our multi-component sample, (2) for galaxies with $\geq$1 unsaturated line in the fit. \textit{Column 3}: the neutral gas outflow velocity, as traced by LIS absorption lines. \textit{Column 4}: the Doppler broadening parameter. \textit{Columns 5-9}: the fit results for $\log(N_{ion})$. \textit{Column 10-14}: the fit results for $C_f$(ion). When all lines are saturated, the uncertainties associated with $b$ and $N_{ion}$ are likely to be underestimated. See more information on our treatment of saturation in $\S$\ref{sub:saturation}.}
\end{deluxetable*}
\end{longrotatetable}


\begin{figure*}
    \centering
    \includegraphics[width=\linewidth]{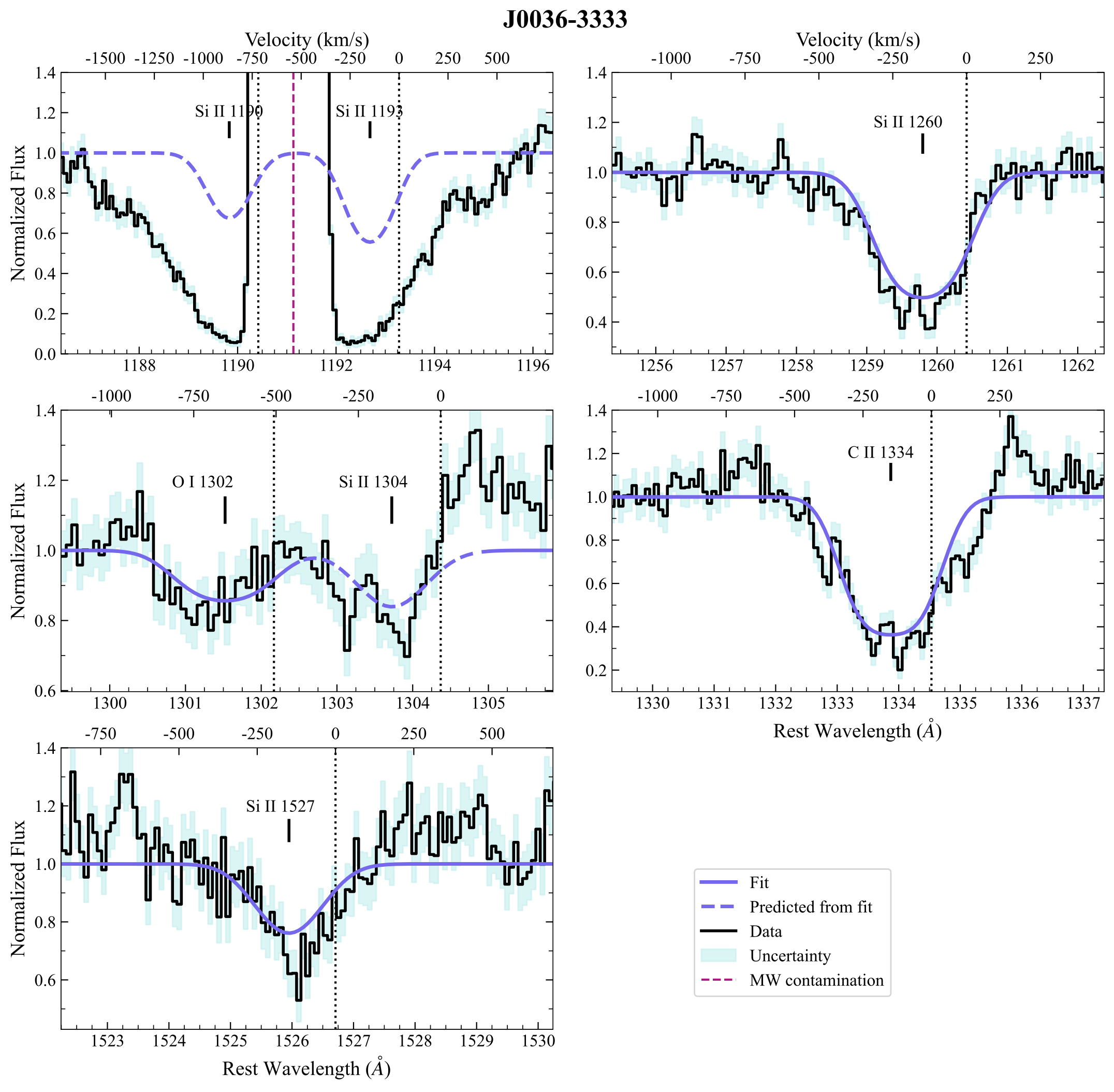}
    \figcaption{Simultaneous one-component fit to the LIS features in the CLASSY spectrum for J0036-3333. See description from Figure \ref{fig:firstfit} for details.}
\end{figure*}

\begin{figure*}
    \centering
    \includegraphics[width=\linewidth]{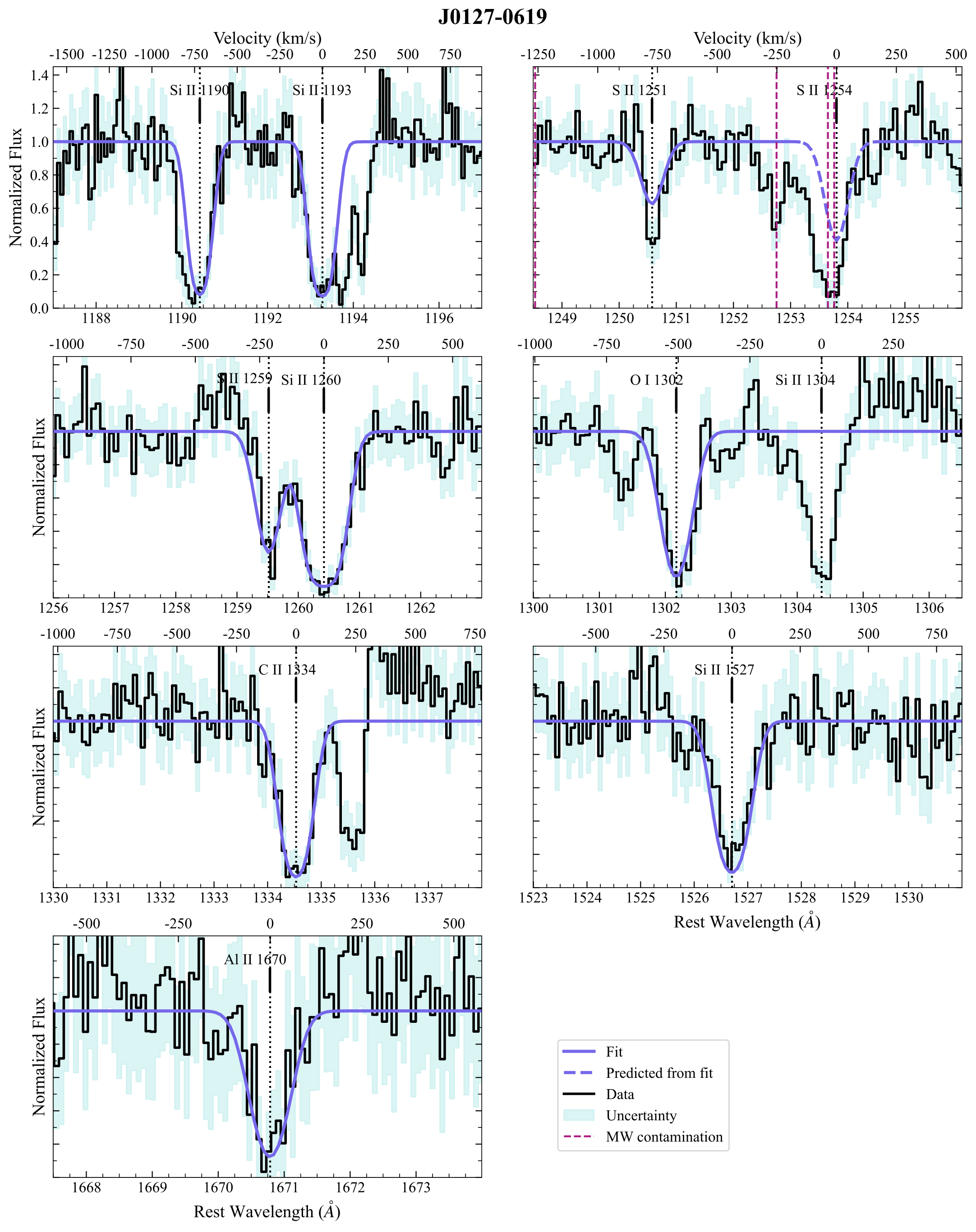}
    \figcaption{Simultaneous one-component fit to the LIS features in the CLASSY spectrum for J0127-0619. See description from Figure \ref{fig:firstfit} for details.}
\end{figure*}

\begin{figure*}
    \centering
    \includegraphics[width=\linewidth]{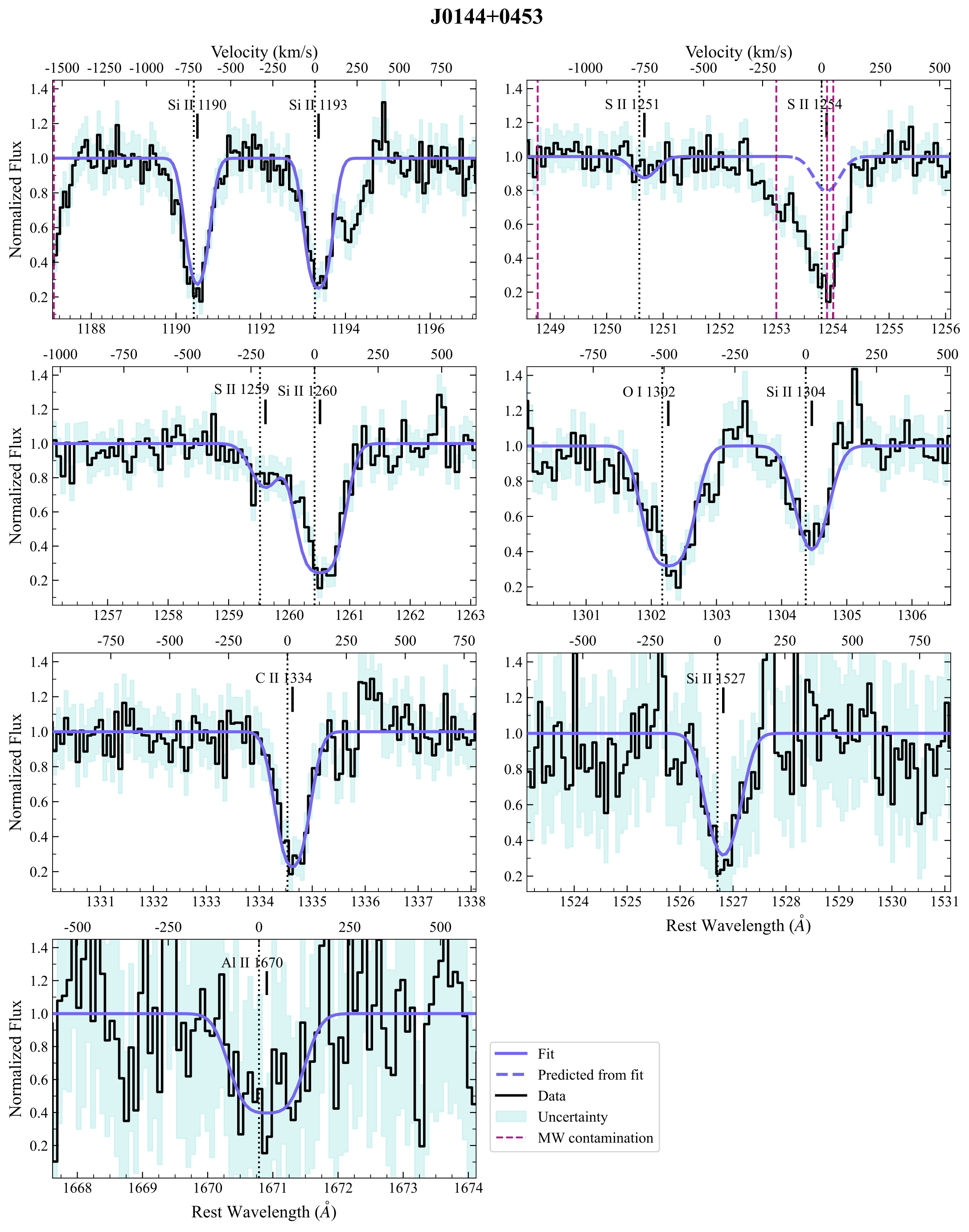}
    \figcaption{Simultaneous one-component fit to the LIS features in the CLASSY spectrum for J0144+0453. See description from Figure \ref{fig:firstfit} for details.}
\end{figure*}

\begin{figure*}
    \centering
    \includegraphics[width=\linewidth]{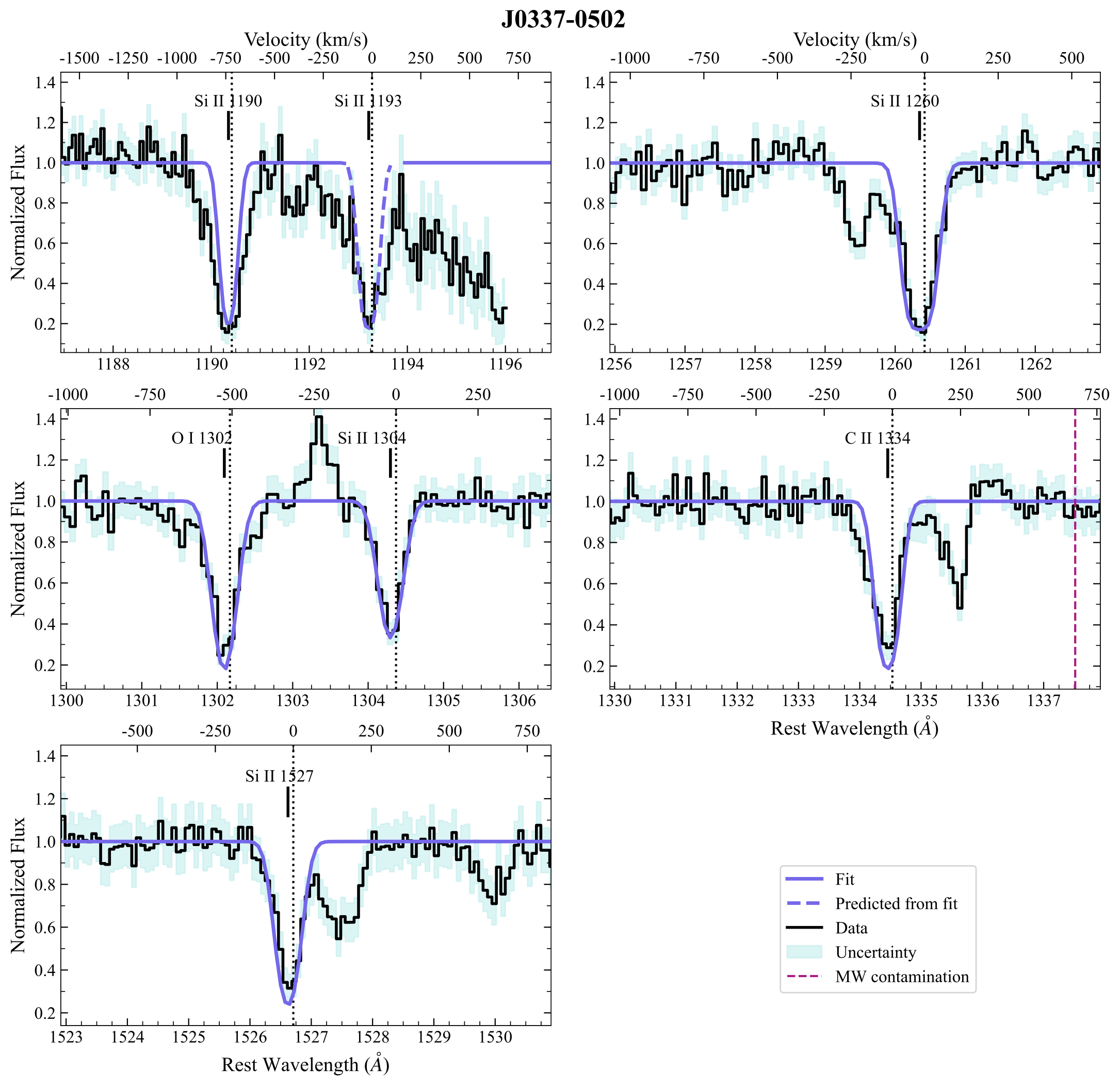}
    \figcaption{Simultaneous one-component fit to the LIS features in the CLASSY spectrum for J0337-0502. See description from Figure \ref{fig:firstfit} for details.}
\end{figure*}

\begin{figure*}
    \centering
    \includegraphics[width=\linewidth]{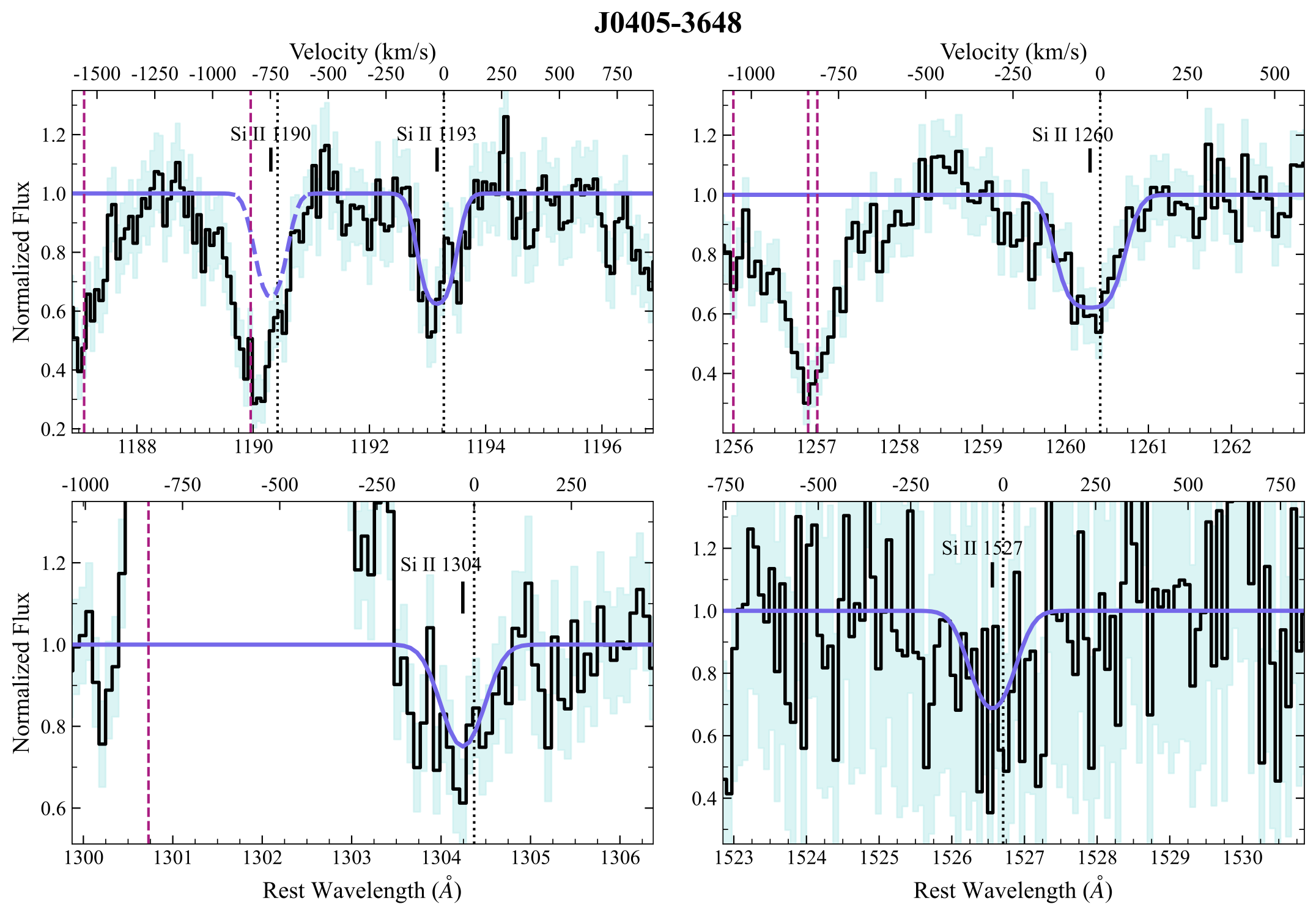}
    \figcaption{Simultaneous one-component fit to the LIS features in the CLASSY spectrum for J0405-3648. See description from Figure \ref{fig:firstfit} for details.}
\end{figure*}

\begin{figure*}
    \centering
    \includegraphics[width=\linewidth]{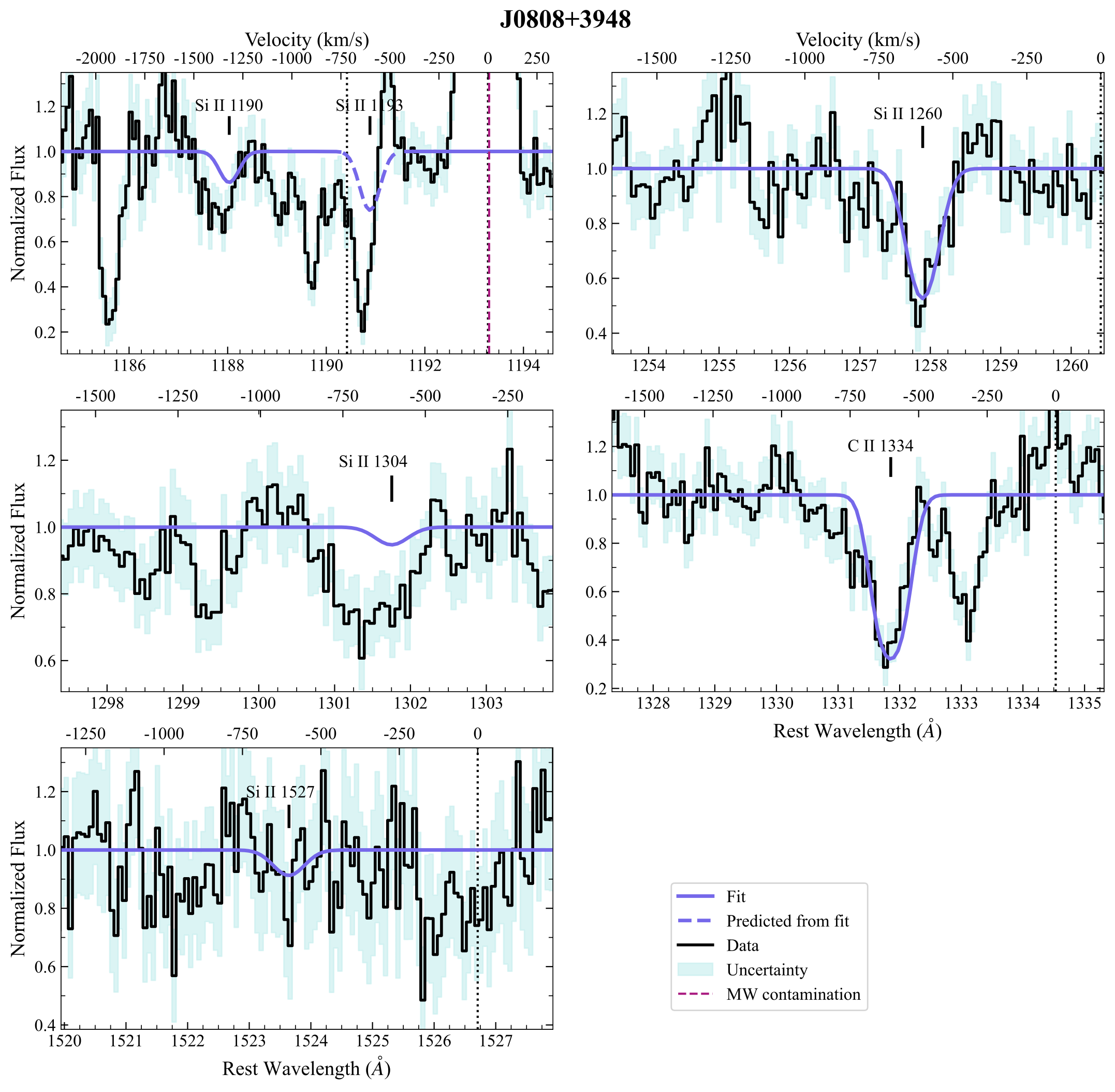}
    \figcaption{Simultaneous one-component fit to the LIS features in the CLASSY spectrum for J0808+3948. See description from Figure \ref{fig:firstfit} for details.}
\end{figure*}

\begin{figure*}
    \centering
    \includegraphics[width=\linewidth]{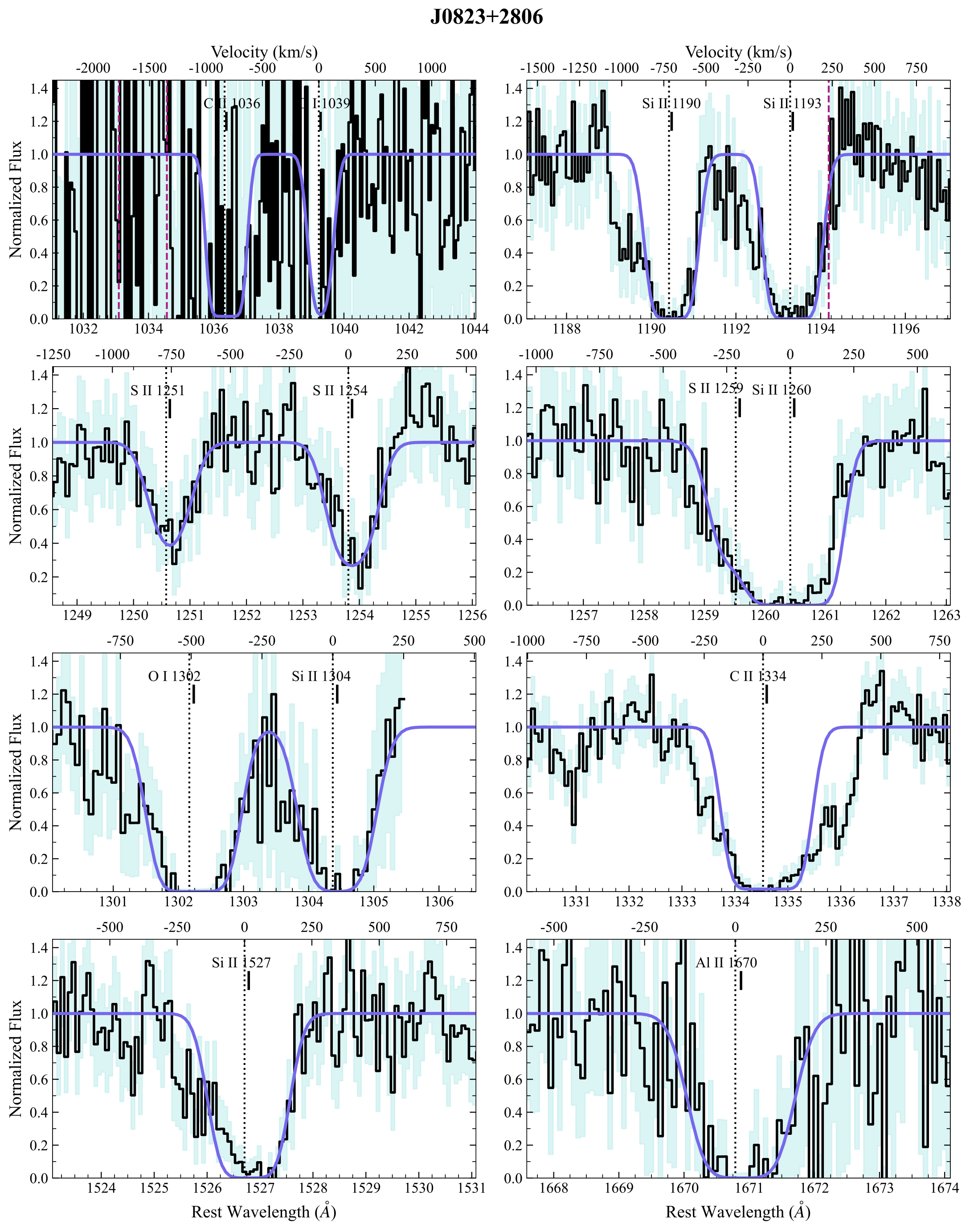}
    \figcaption{Simultaneous one-component fit to the LIS features in the CLASSY spectrum for J0823+2806. See description from Figure \ref{fig:firstfit} for details.}
\end{figure*}

\begin{figure*}
    \centering
    \includegraphics[width=\linewidth]{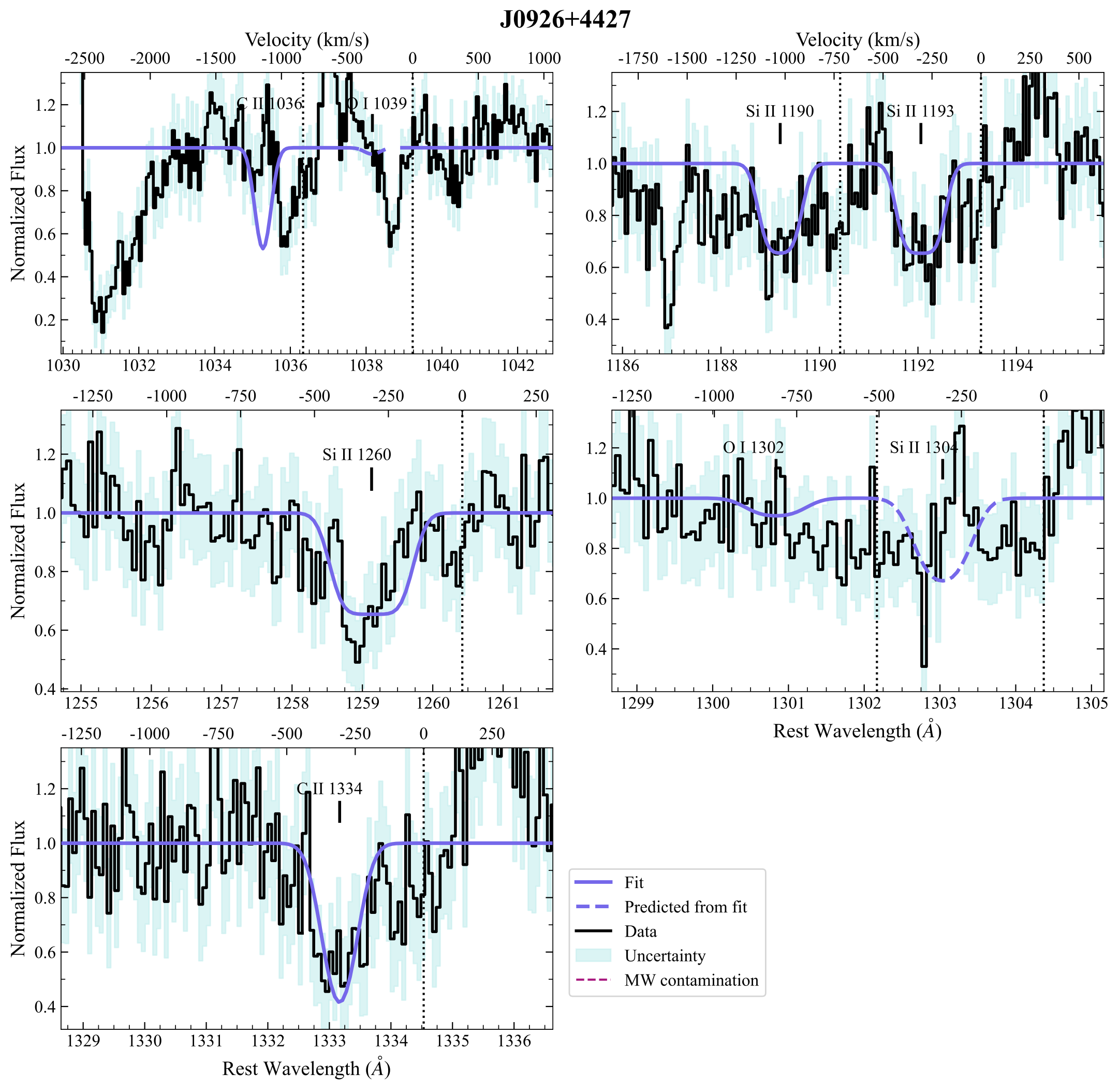}
    \figcaption{Simultaneous one-component fit to the LIS features in the CLASSY spectrum for J0926+4427. See description from Figure \ref{fig:firstfit} for details.}
\end{figure*}

\begin{figure*}
    \centering
    \includegraphics[width=\linewidth]{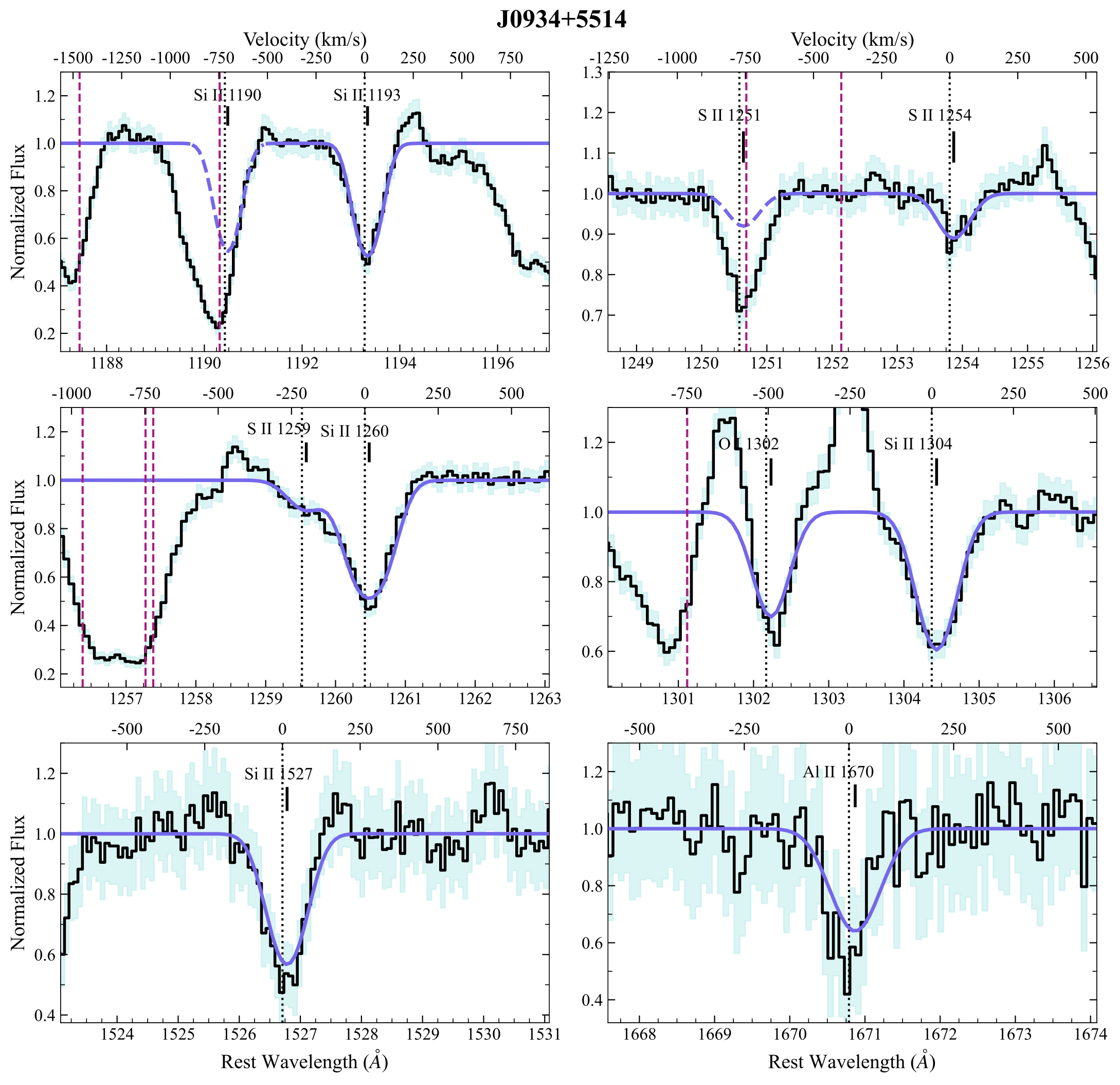}
    \figcaption{Simultaneous one-component fit to the LIS features in the CLASSY spectrum for J0934+5514. See description from Figure \ref{fig:firstfit} for details.}
\end{figure*}

\begin{figure*}
    \centering
    \includegraphics[width=\linewidth]{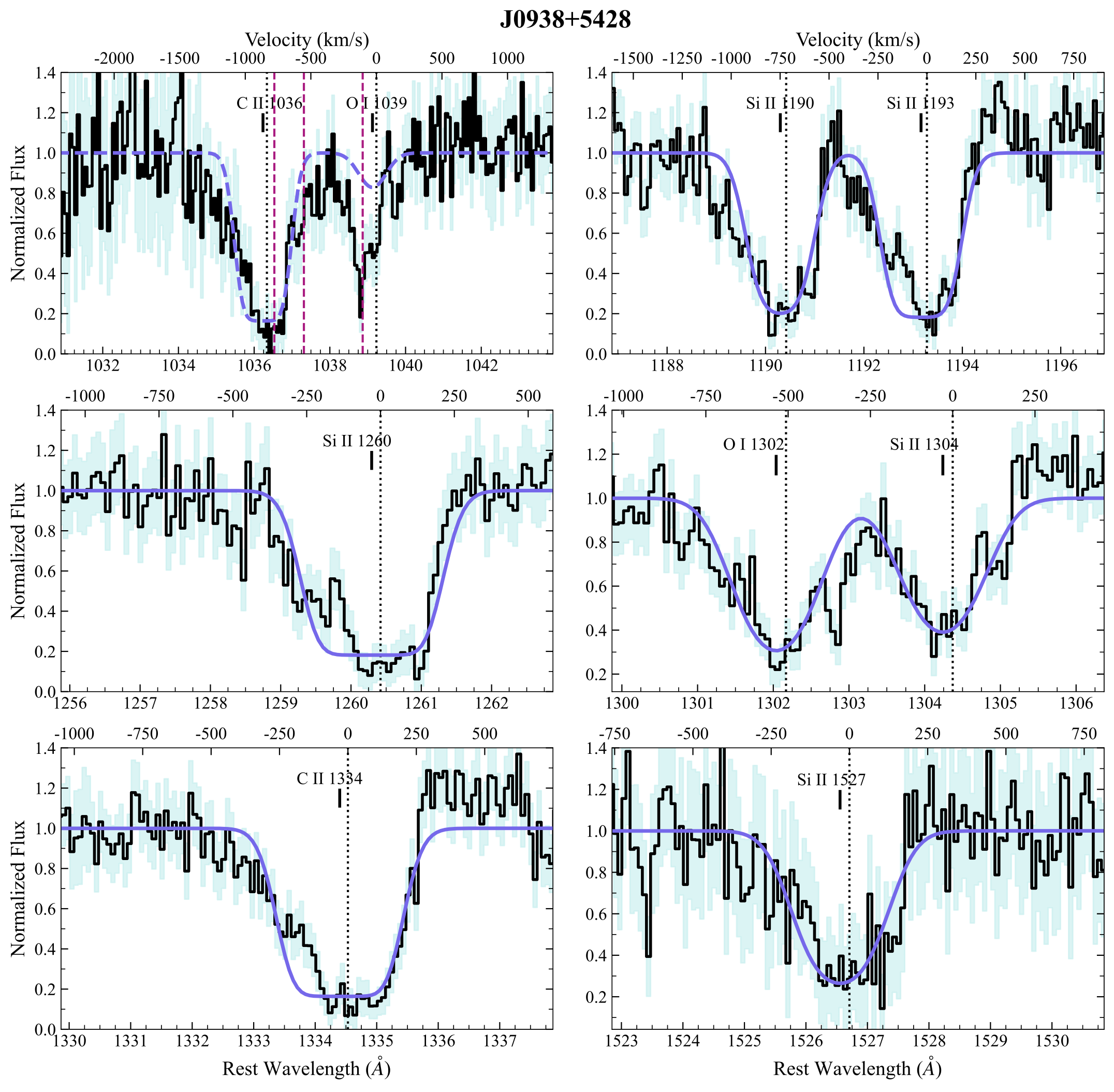}
    \figcaption{Simultaneous one-component fit to the LIS features in the CLASSY spectrum for J0938+5428. See description from Figure \ref{fig:firstfit} for details.}
\end{figure*}

\begin{figure*}
    \centering
    \includegraphics[width=\linewidth]{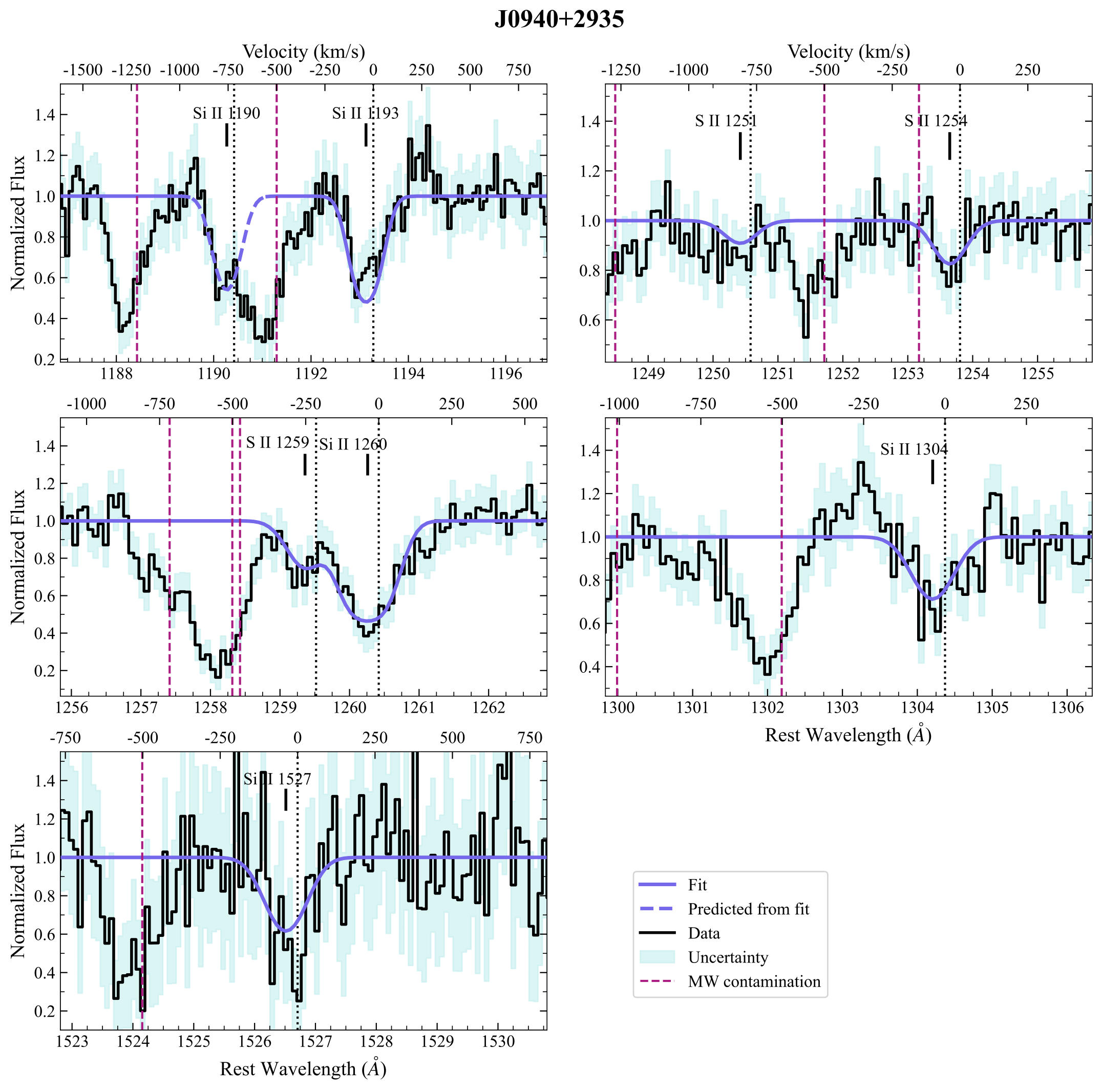}
    \figcaption{Simultaneous one-component fit to the LIS features in the CLASSY spectrum for J0940+2935. See description from Figure \ref{fig:firstfit} for details.}
\end{figure*}

\begin{figure*}
    \centering
    \includegraphics[width=\linewidth]{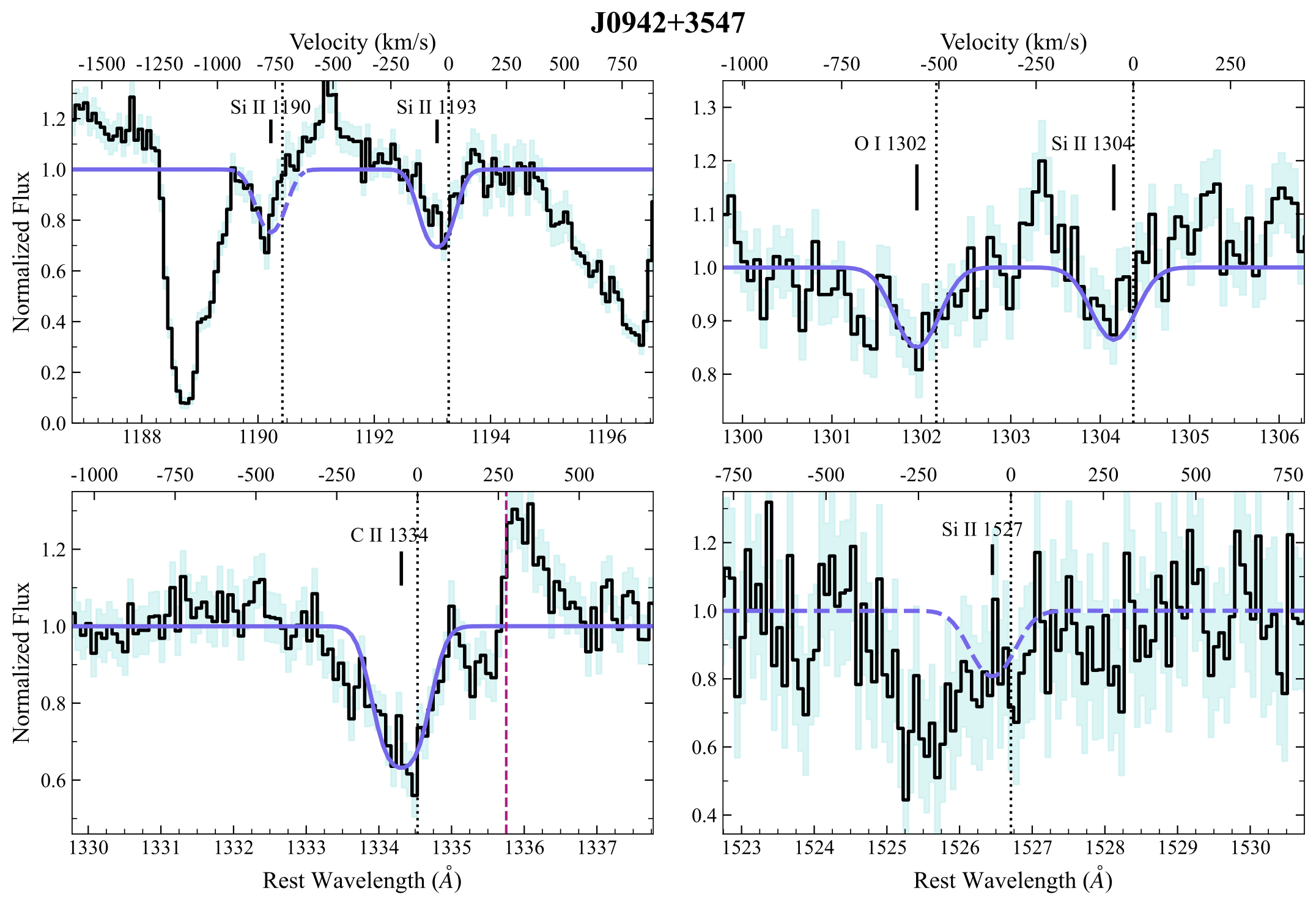}
    \figcaption{Simultaneous one-component fit to the LIS features in the CLASSY spectrum for J0942+3547. See description from Figure \ref{fig:firstfit} for details.}
\end{figure*}

\begin{figure*}
    \centering
    \includegraphics[width=\linewidth]{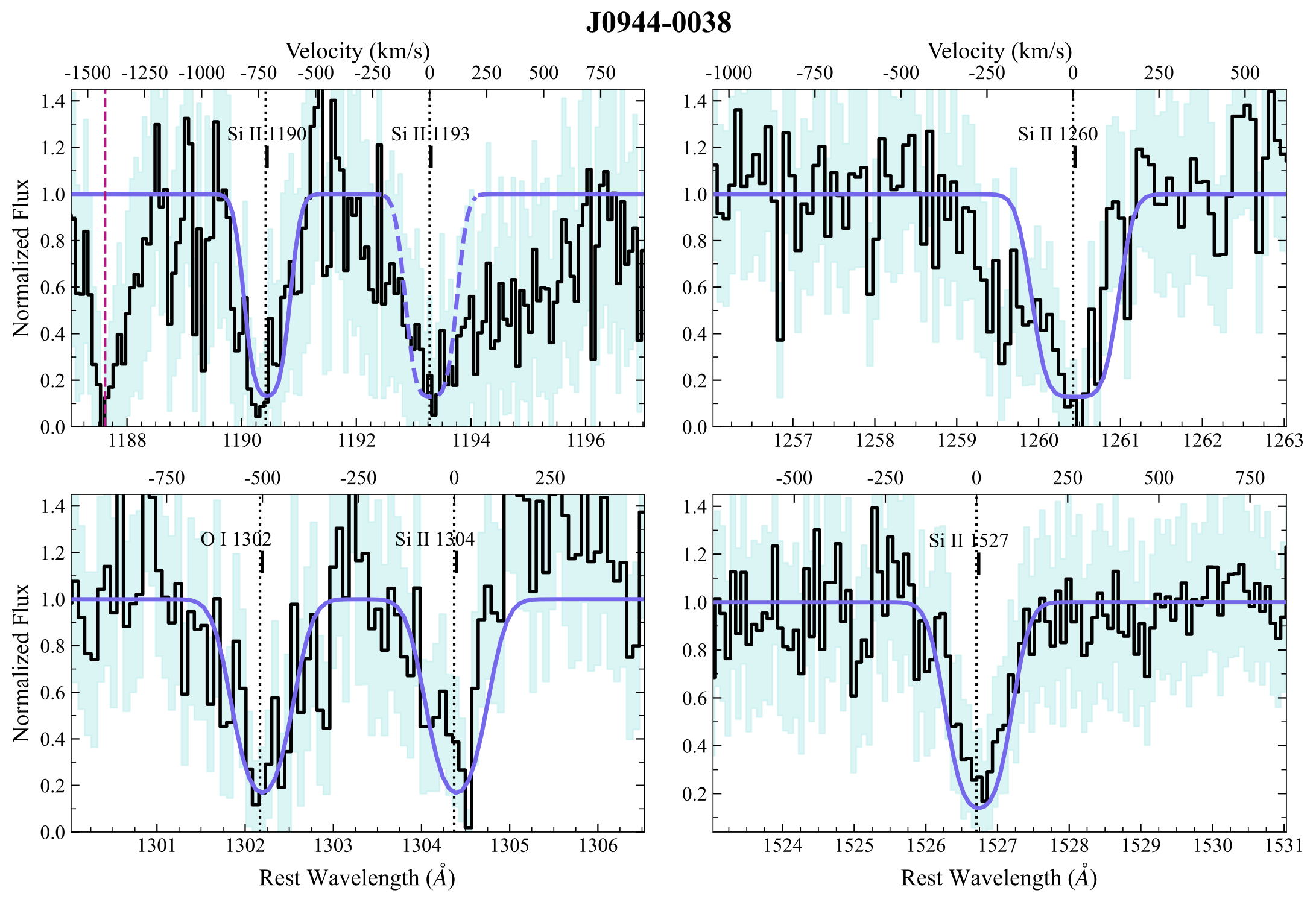}
    \figcaption{Simultaneous one-component fit to the LIS features in the CLASSY spectrum for J0944-0038. See description from Figure \ref{fig:firstfit} for details.}
\end{figure*}

\begin{figure*}
    \centering
    \includegraphics[width=\linewidth]{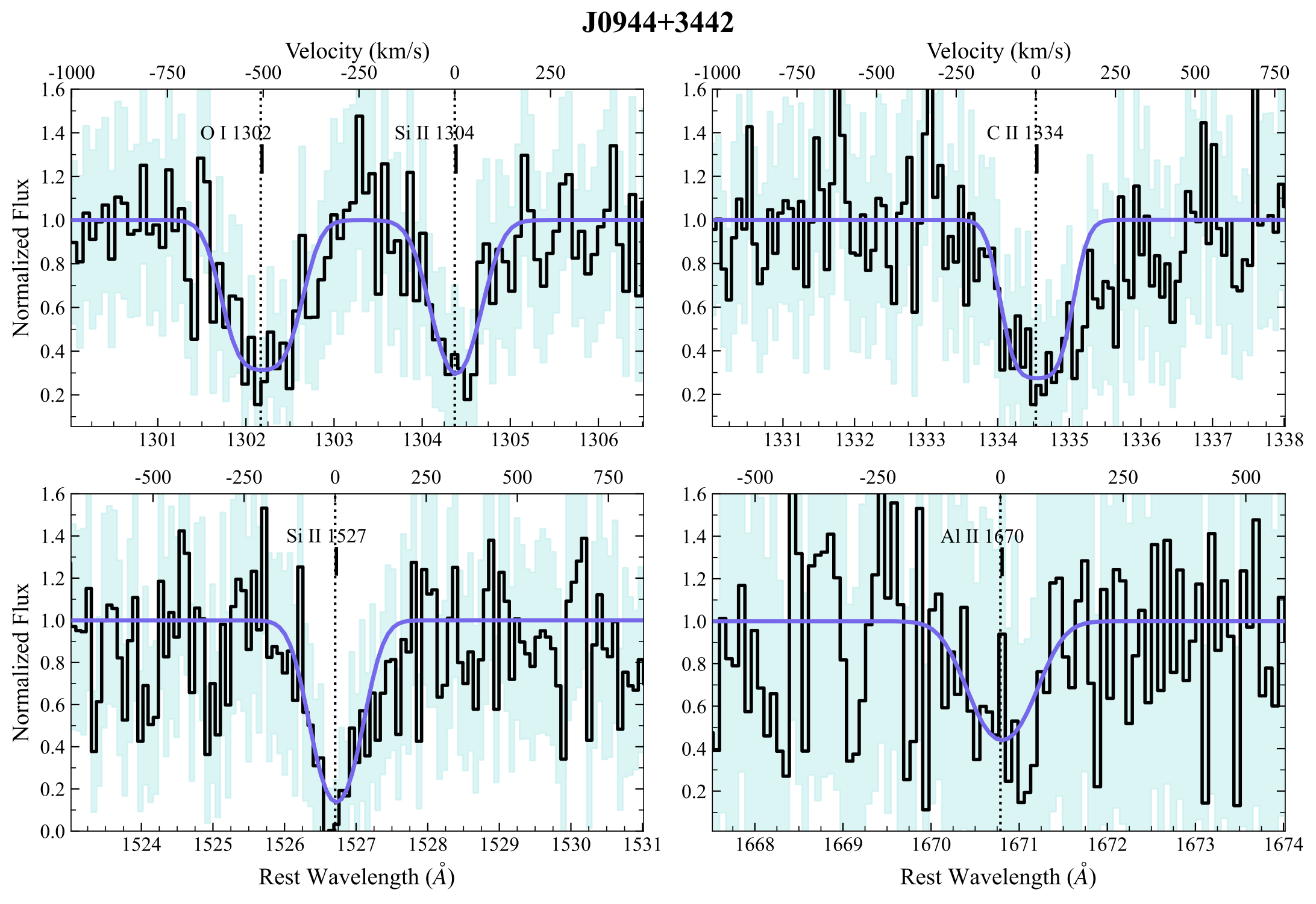}
    \figcaption{Simultaneous one-component fit to the LIS features in the CLASSY spectrum for J0944+3442. See description from Figure \ref{fig:firstfit} for details.}
\end{figure*}

\begin{figure*}
    \centering
    \includegraphics[width=\linewidth]{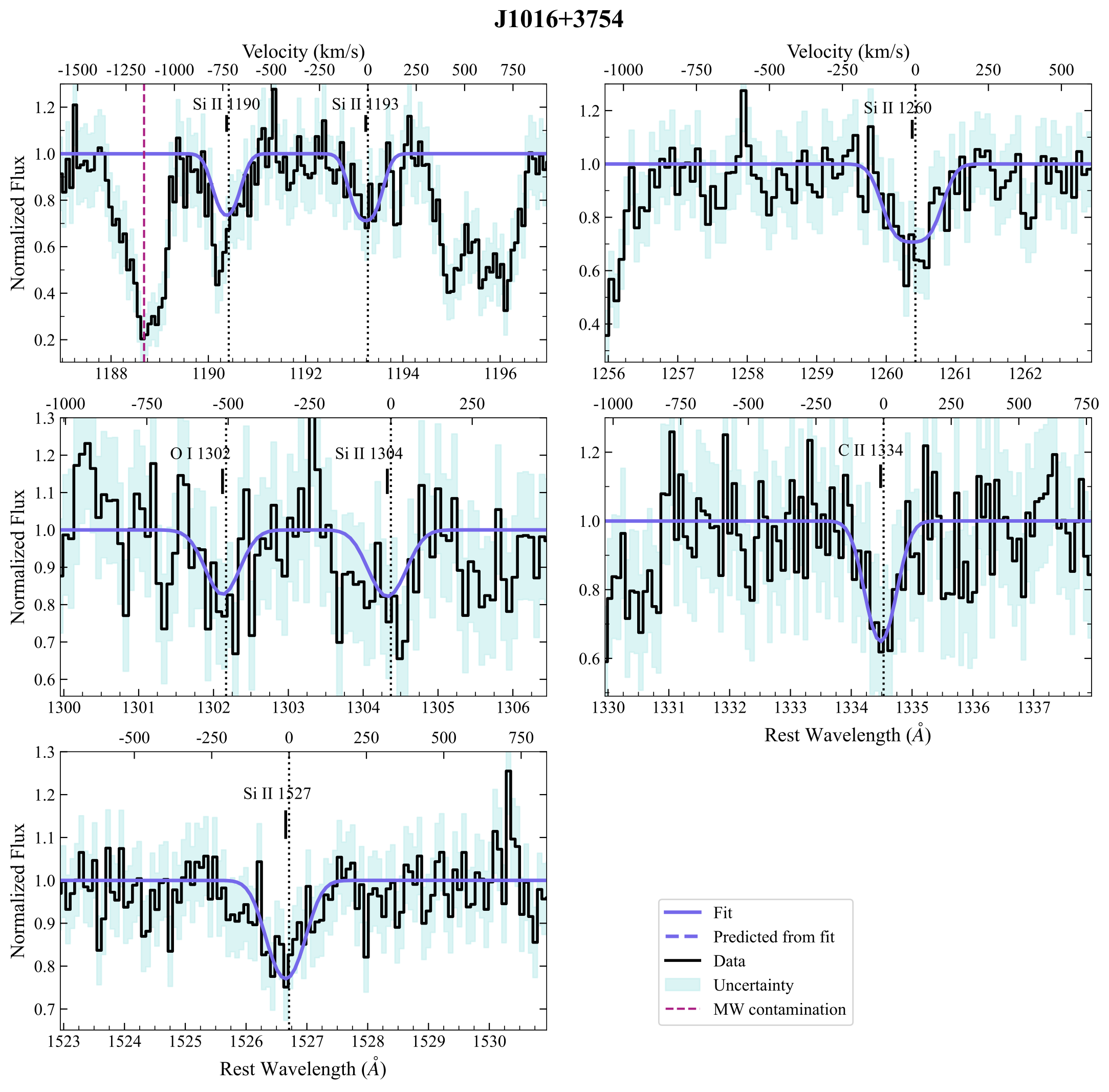}
    \figcaption{Simultaneous one-component fit to the LIS features in the CLASSY spectrum for J1016+3754. See description from Figure \ref{fig:firstfit} for details.}
\end{figure*}

\begin{figure*}
    \centering
    \includegraphics[width=\linewidth]{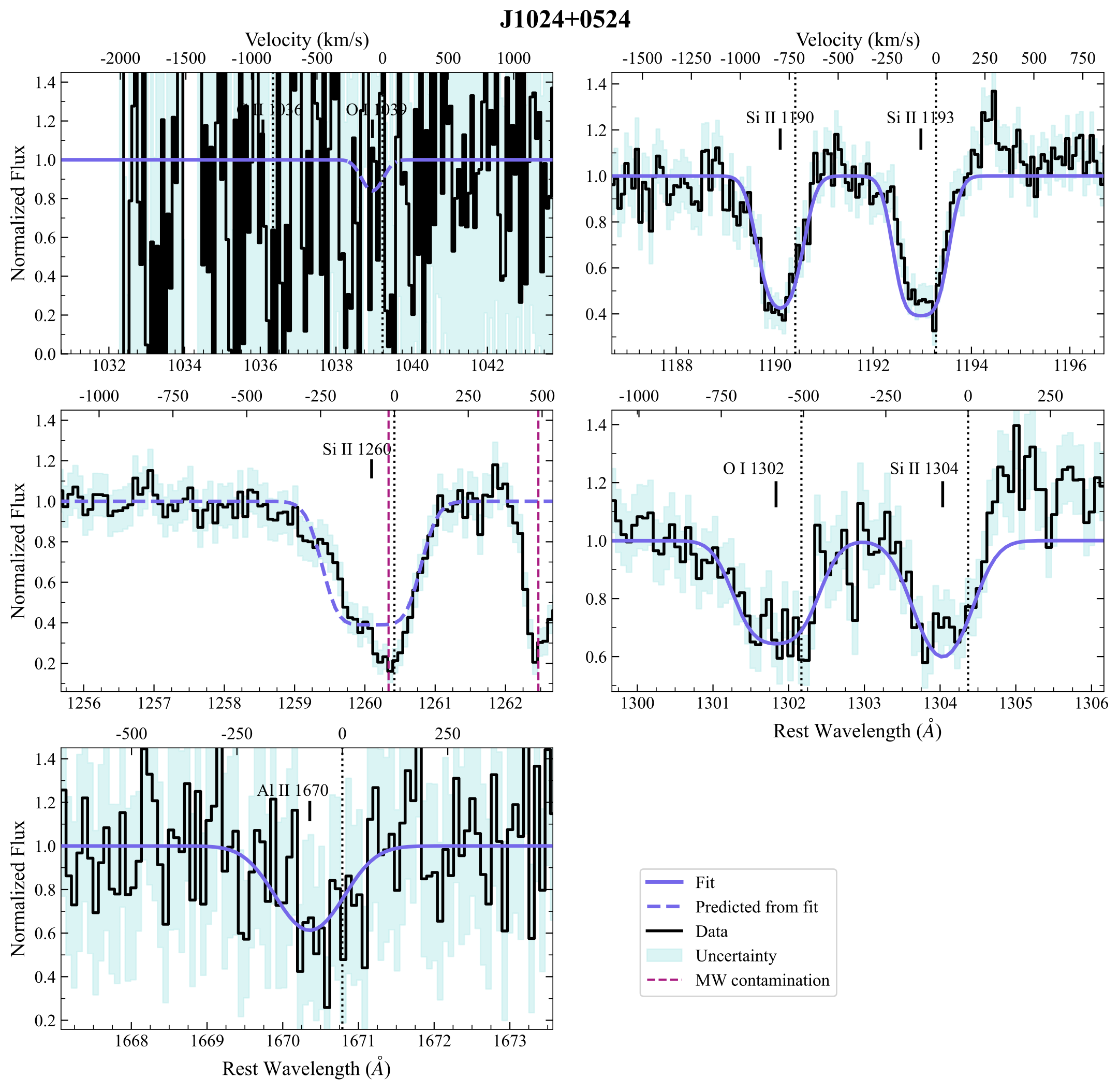}
    \figcaption{Simultaneous one-component fit to the LIS features in the CLASSY spectrum for J1024+0524. See description from Figure \ref{fig:firstfit} for details.}
\end{figure*}

\begin{figure*}
    \centering
    \includegraphics[width=\linewidth]{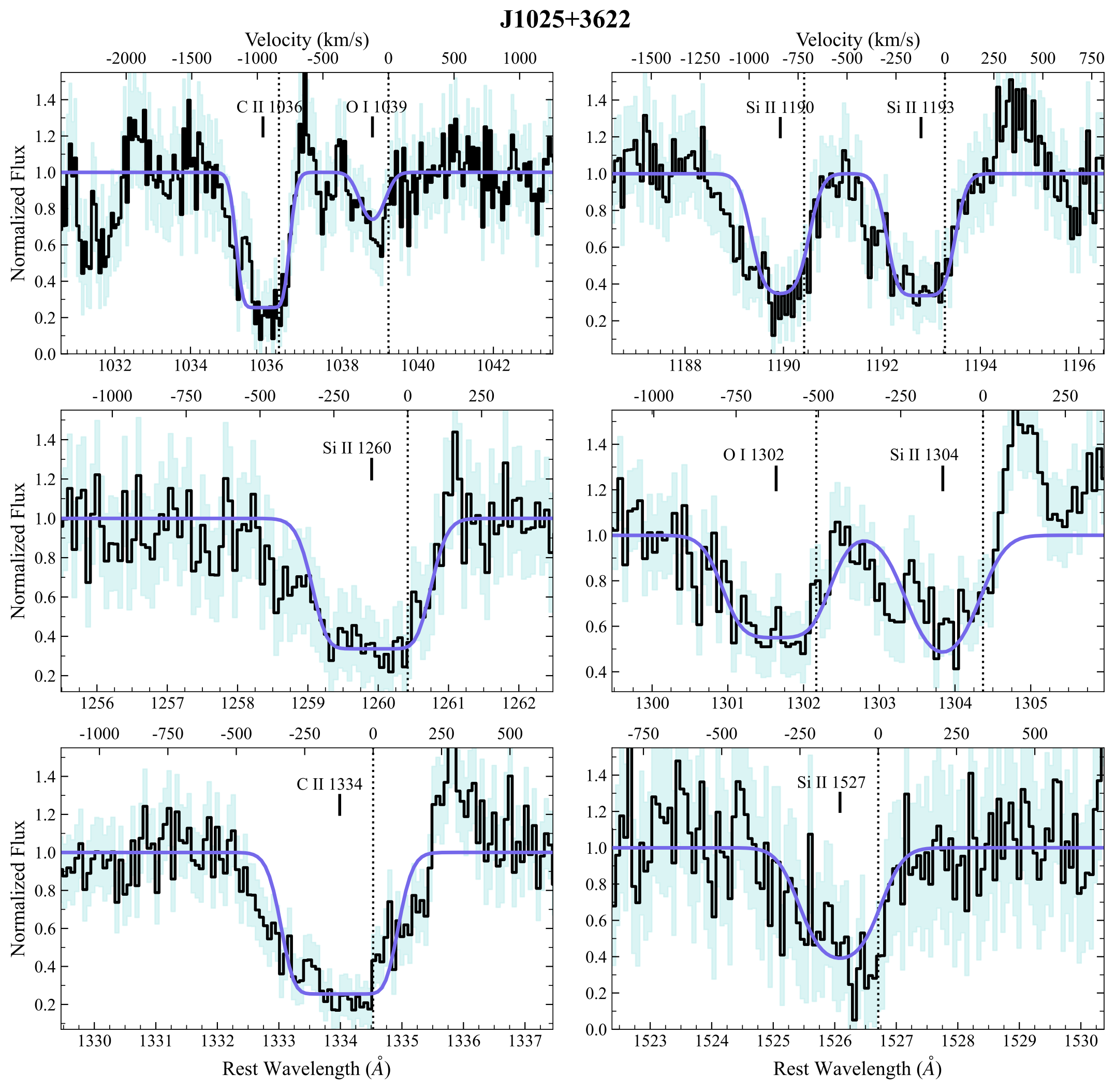}
    \figcaption{Simultaneous one-component fit to the LIS features in the CLASSY spectrum for J1025+3622. See description from Figure \ref{fig:firstfit} for details.}
\end{figure*}

\begin{figure*}
    \centering
    \includegraphics[width=\linewidth]{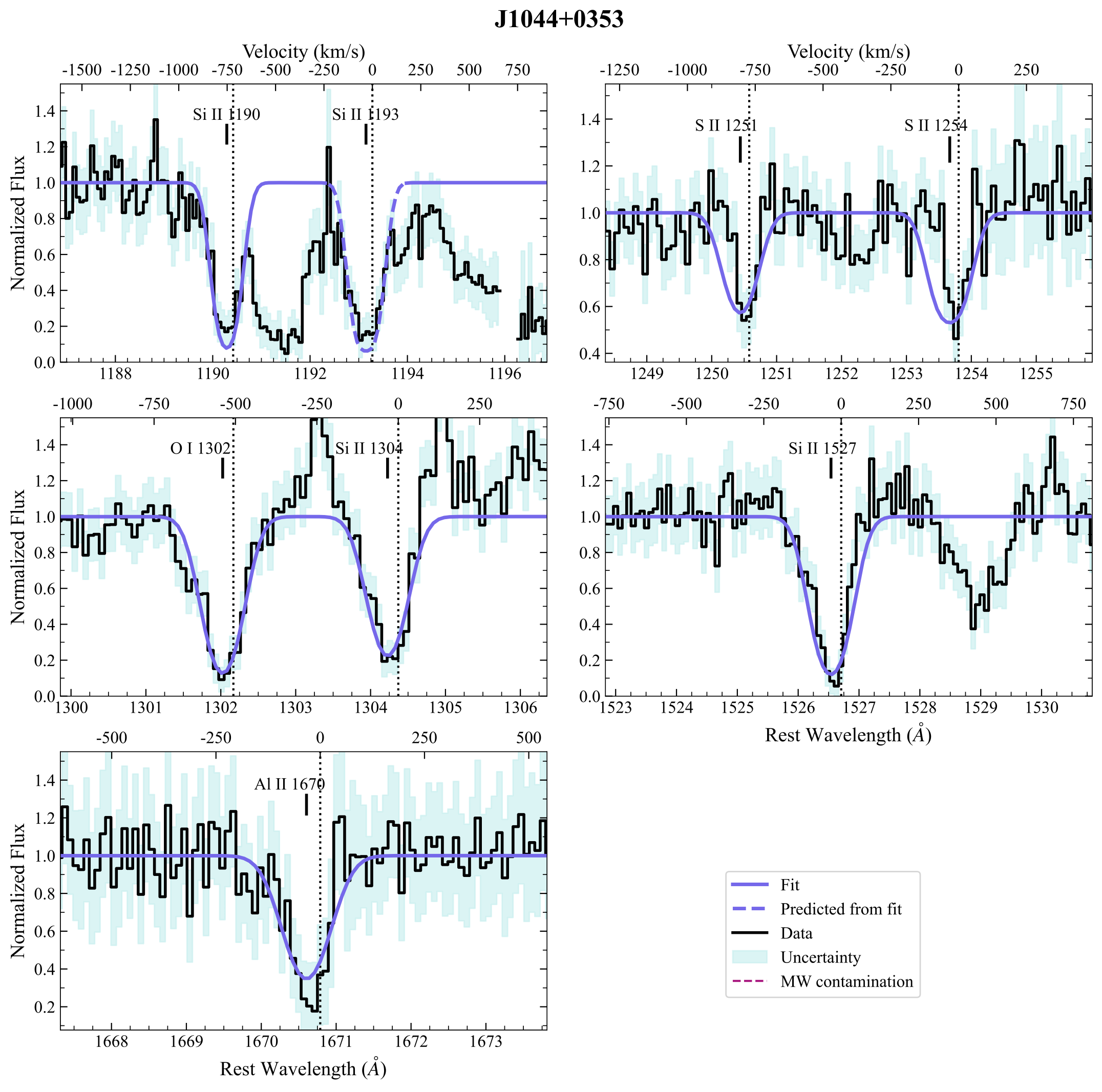}
    \figcaption{Simultaneous one-component fit to the LIS features in the CLASSY spectrum for J1044+0353. See description from Figure \ref{fig:firstfit} for details.}
\end{figure*}

\begin{figure*}
    \centering
    \includegraphics[width=\linewidth]{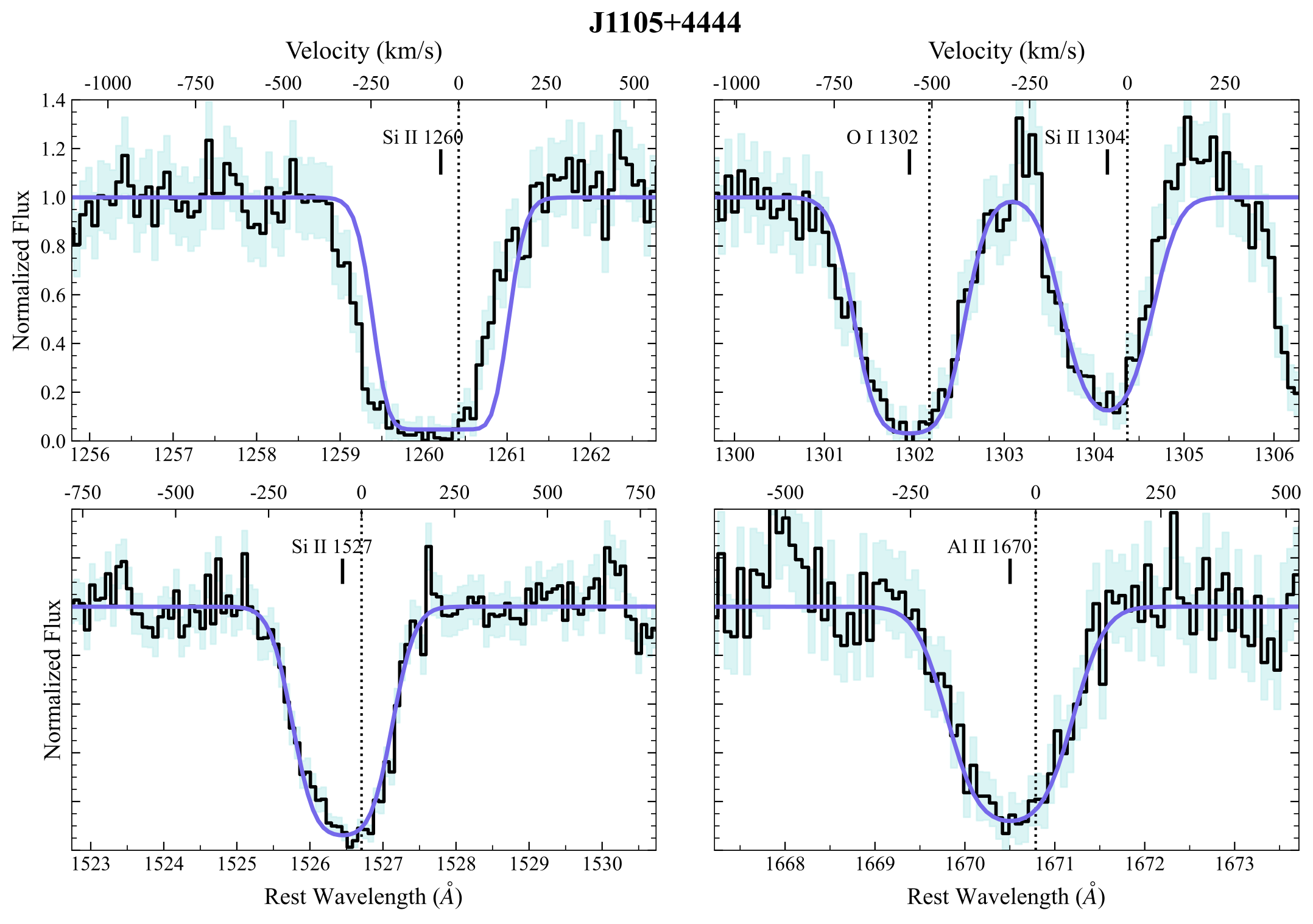}
    \figcaption{Simultaneous one-component fit to the LIS features in the CLASSY spectrum for J1105+4444. See description from Figure \ref{fig:firstfit} for details.}
\end{figure*}

\begin{figure*}
    \centering
    \includegraphics[width=\linewidth]{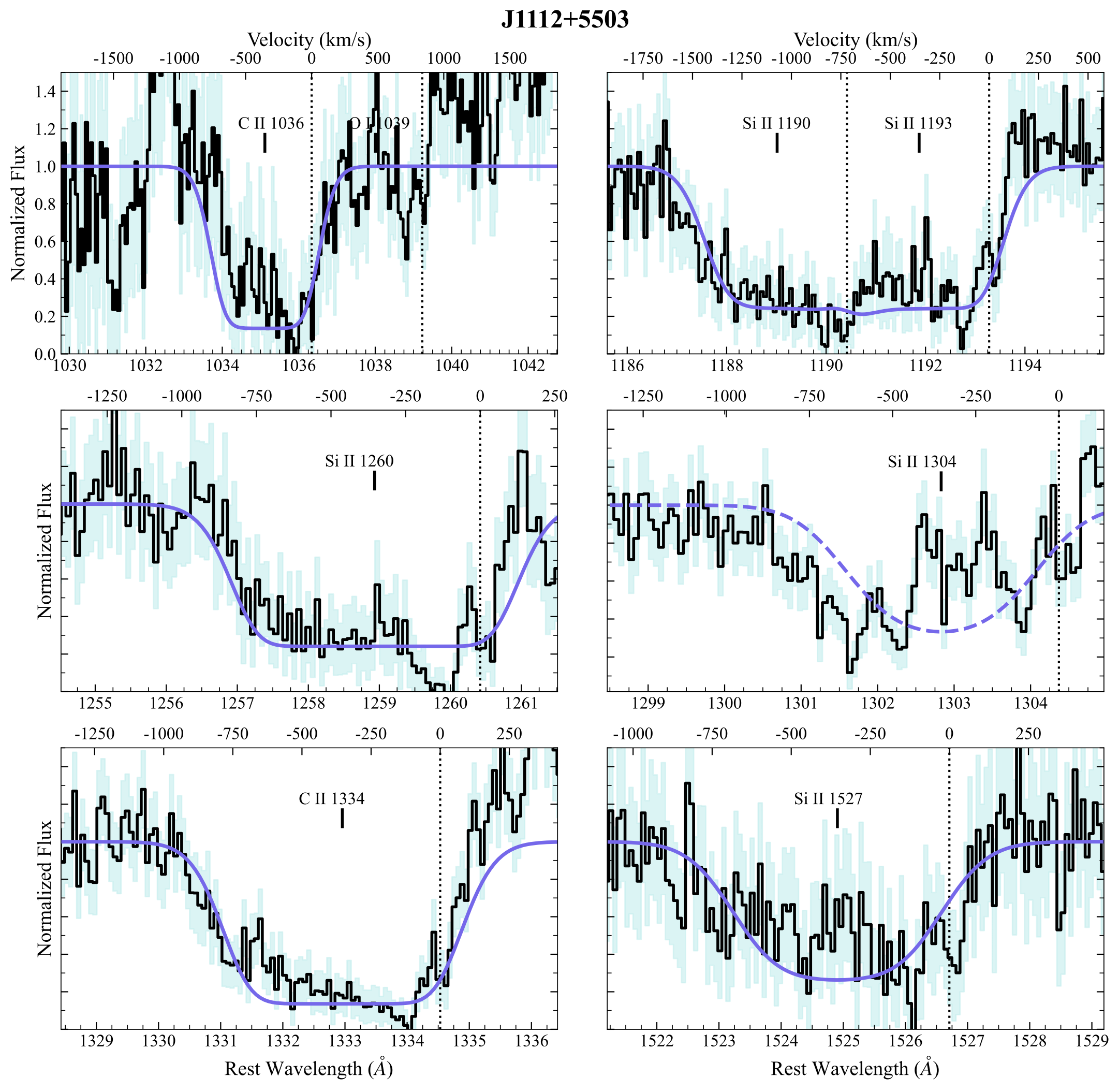}
    \figcaption{Simultaneous one-component fit to the LIS features in the CLASSY spectrum for J1112+5503. See description from Figure \ref{fig:firstfit} for details.}
\end{figure*}

\begin{figure*}
    \centering
    \includegraphics[width=\linewidth]{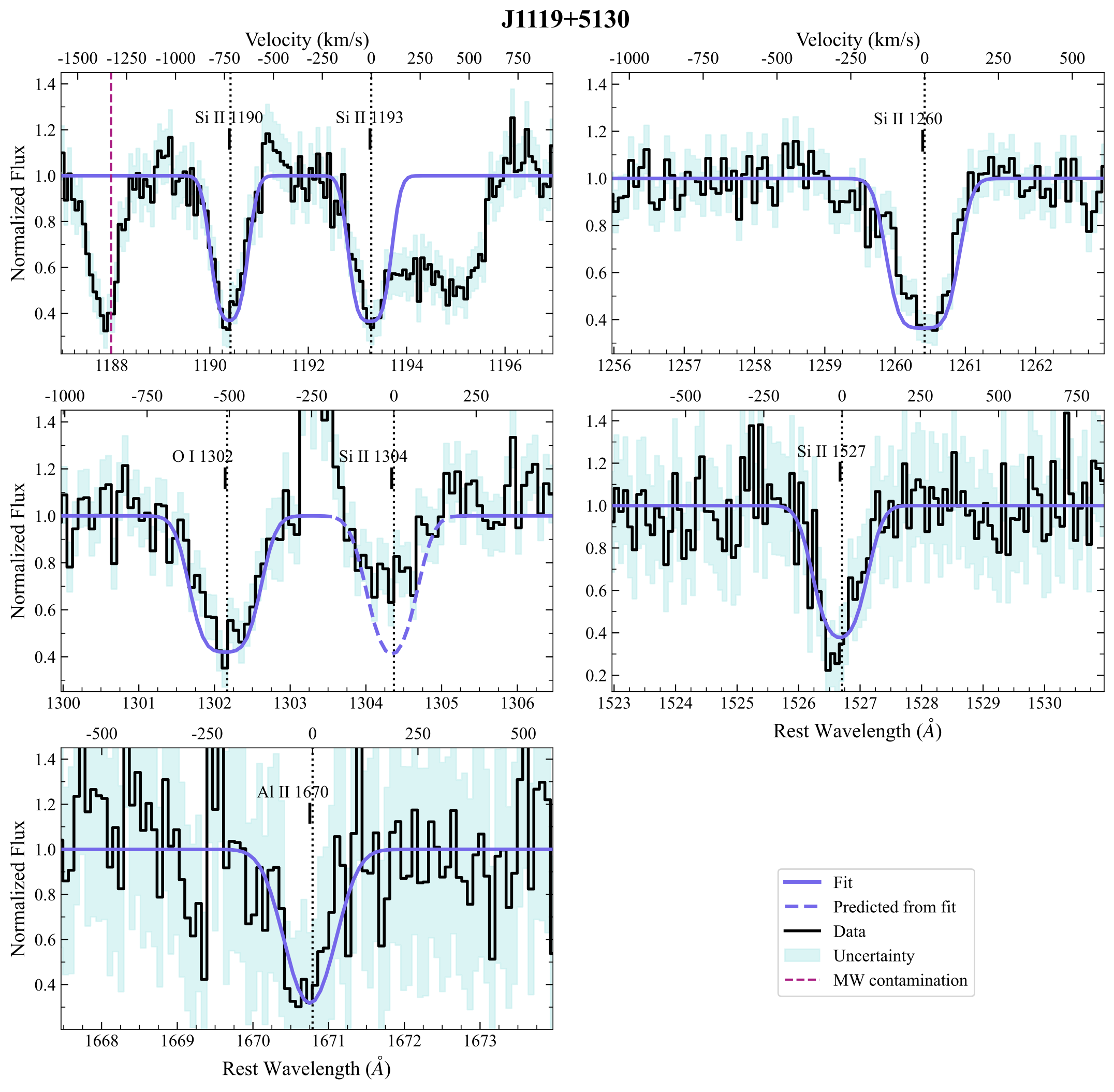}
    \figcaption{Simultaneous one-component fit to the LIS features in the CLASSY spectrum for J1119+5130. See description from Figure \ref{fig:firstfit} for details.}
\end{figure*}

\begin{figure*}
    \centering
    \includegraphics[width=\linewidth]{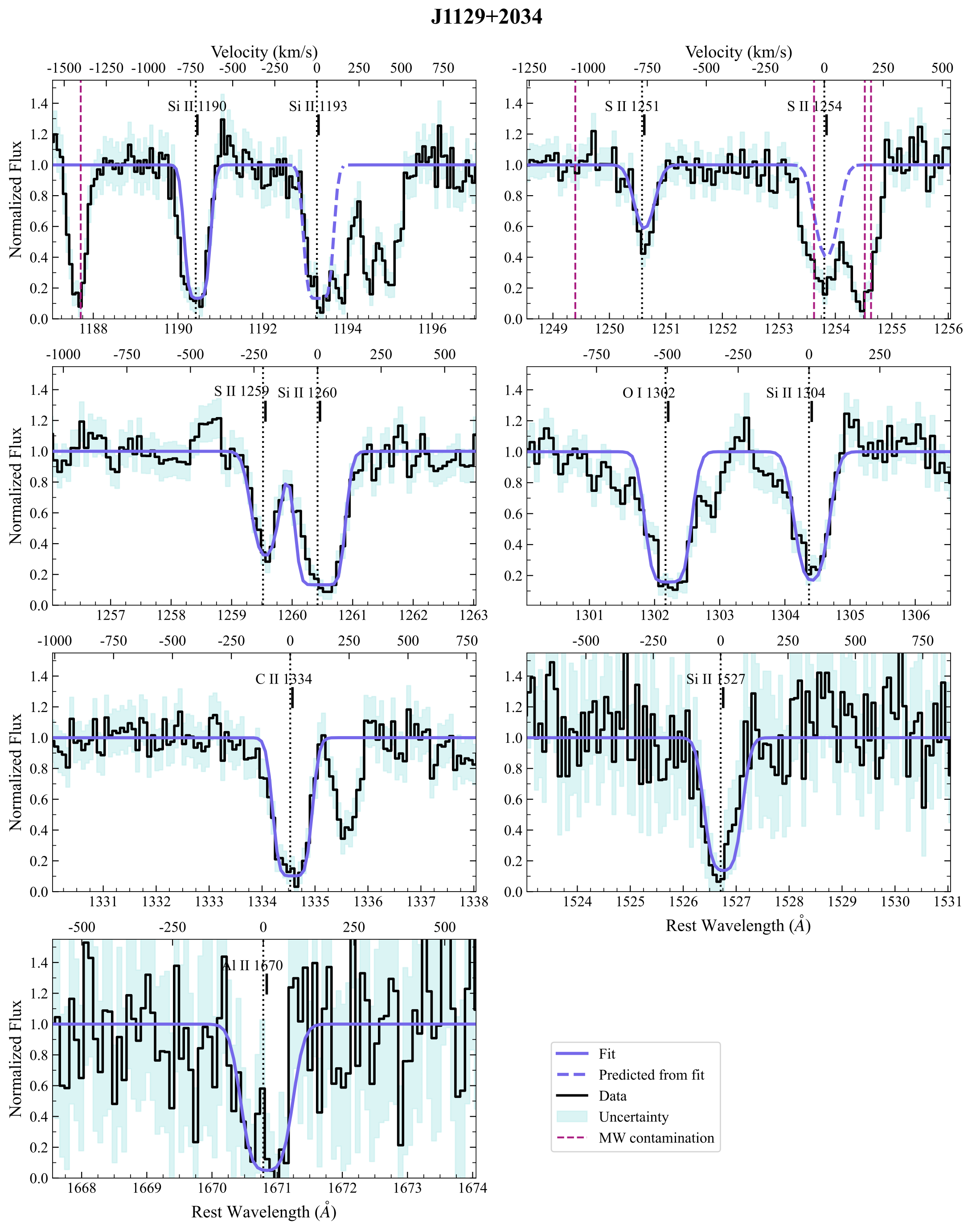}
    \figcaption{Simultaneous one-component fit to the LIS features in the CLASSY spectrum for J1129+2034. See description from Figure \ref{fig:firstfit} for details.}
\end{figure*}

\begin{figure*}
    \centering
    \includegraphics[width=\linewidth]{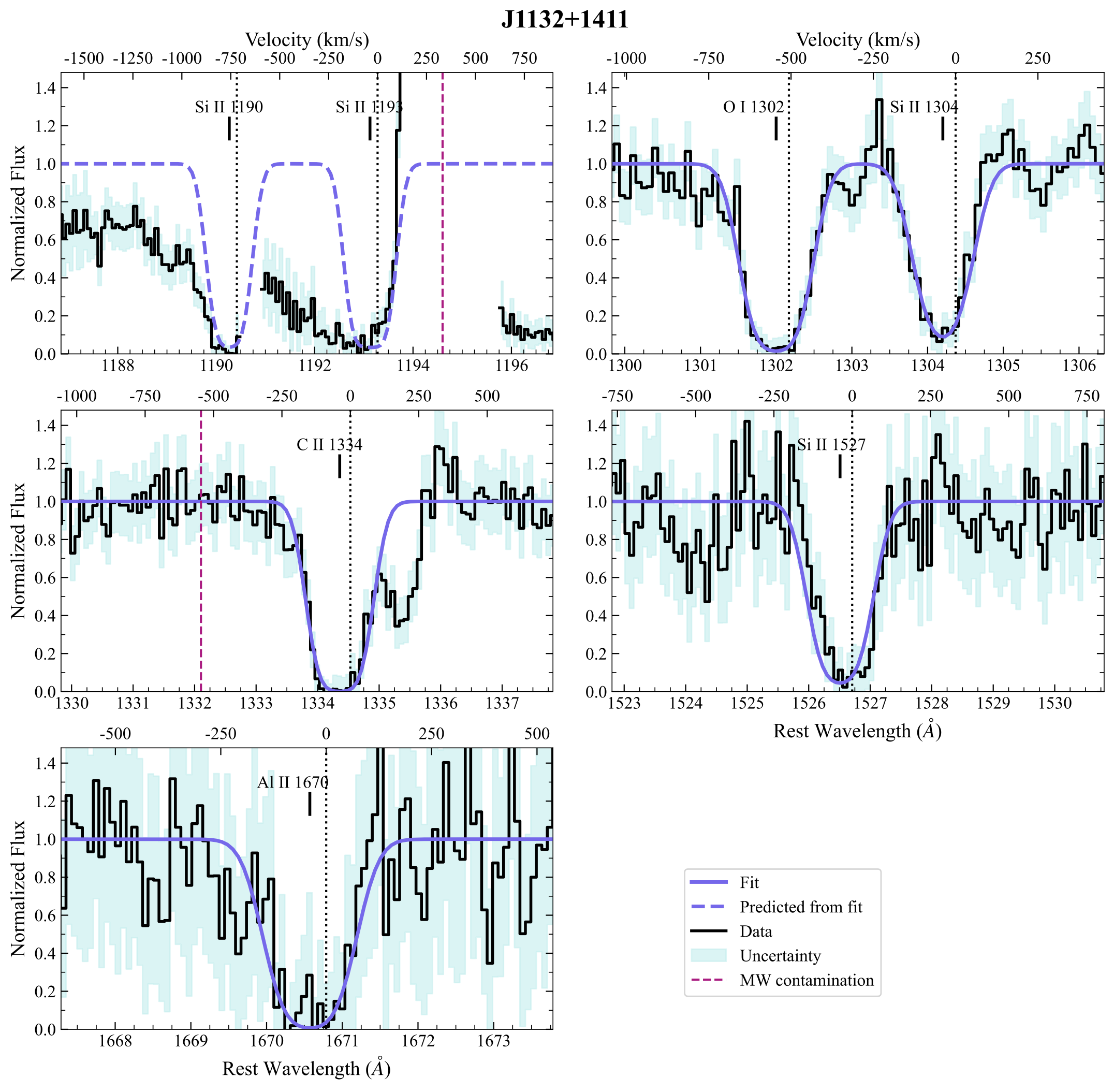}
    \figcaption{Simultaneous one-component fit to the LIS features in the CLASSY spectrum for J1132+1411. See description from Figure \ref{fig:firstfit} for details.}
\end{figure*}

\begin{figure*}
    \centering
    \includegraphics[width=\linewidth]{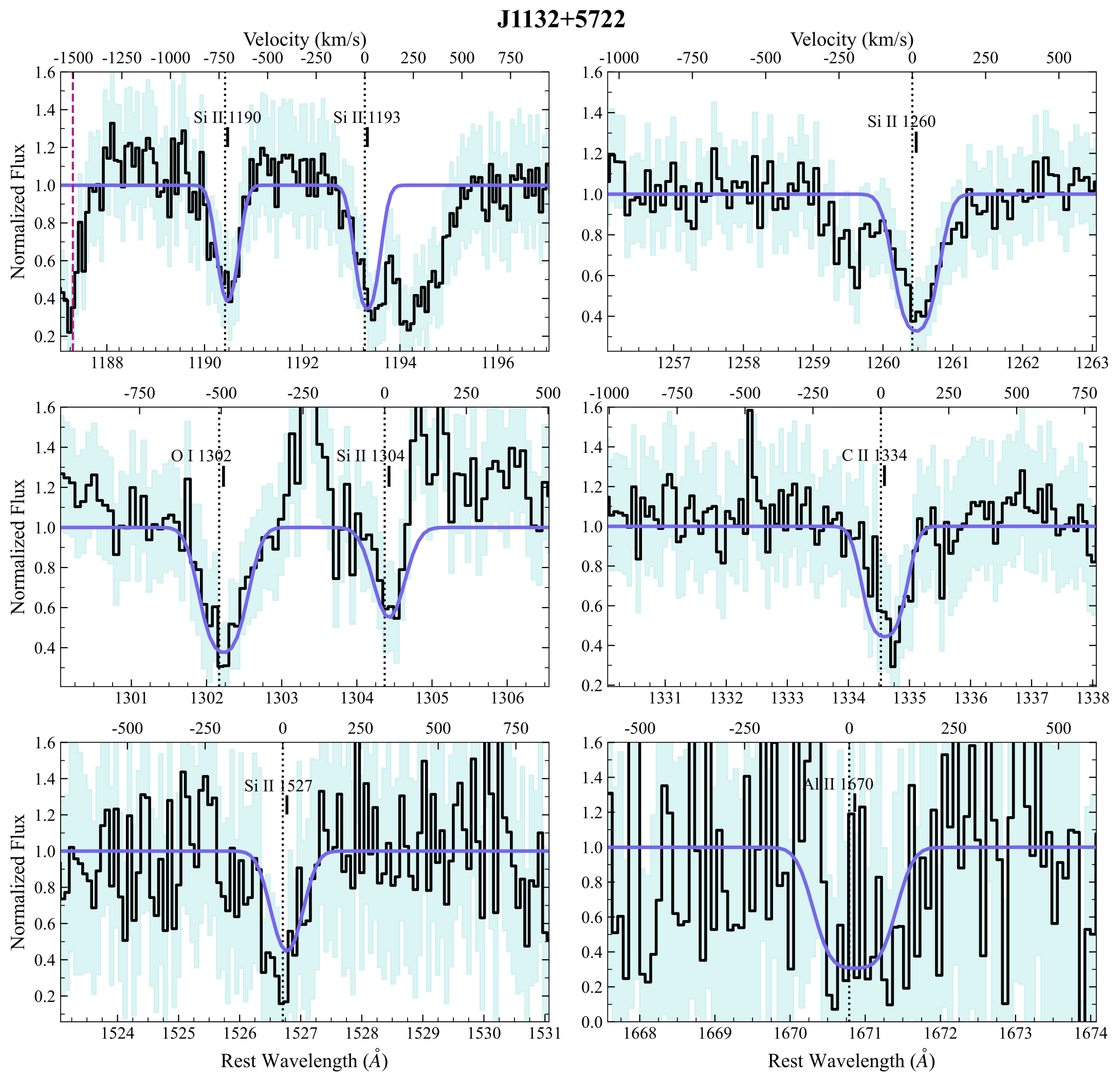}
    \figcaption{Simultaneous one-component fit to the LIS features in the CLASSY spectrum for J1132+5722. See description from Figure \ref{fig:firstfit} for details.}
\end{figure*}

\begin{figure*}
    \centering
    \includegraphics[width=\linewidth]{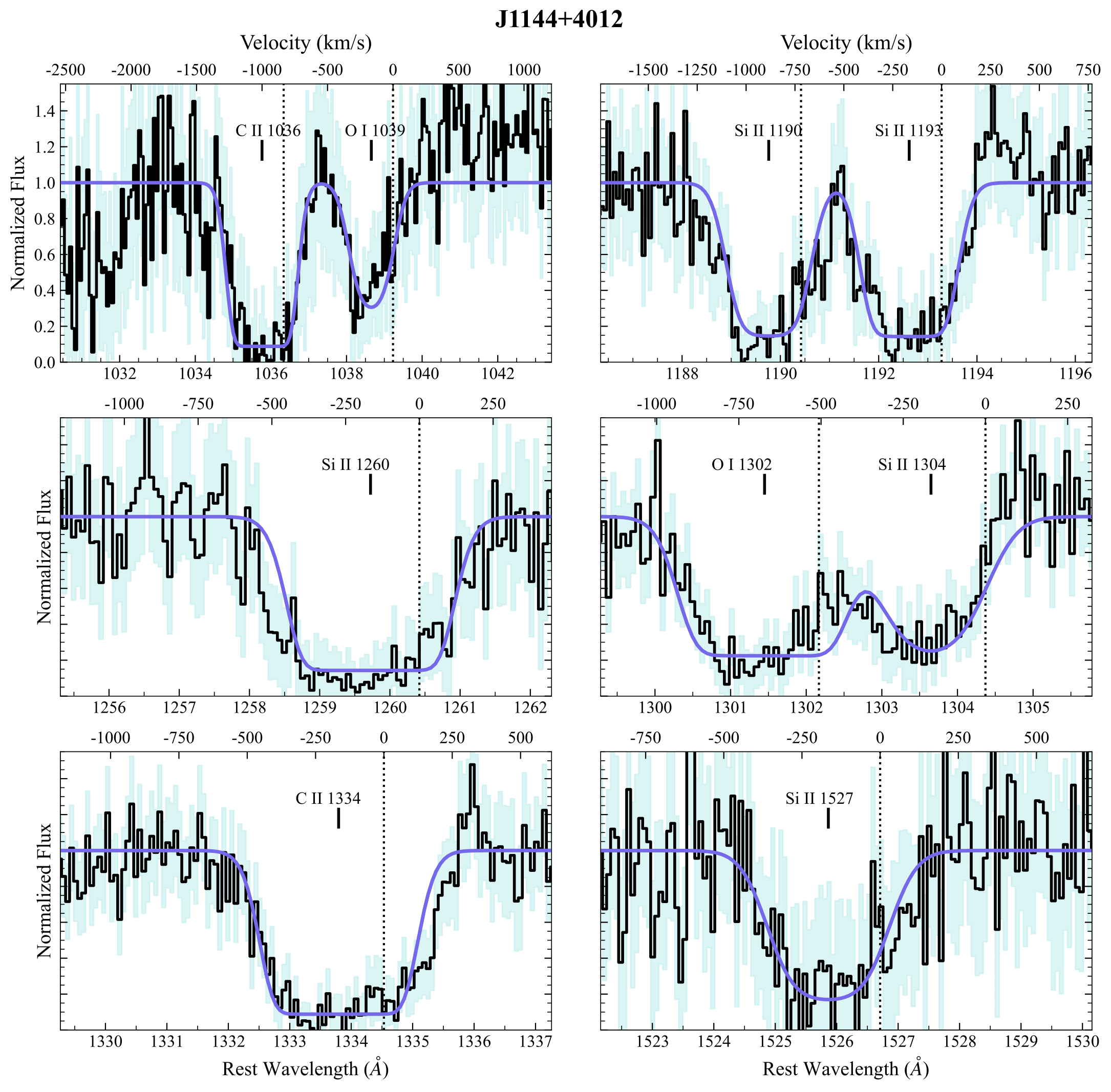}
    \figcaption{Simultaneous one-component fit to the LIS features in the CLASSY spectrum for J1144+4012. See description from Figure \ref{fig:firstfit} for details.}
\end{figure*}

\begin{figure*}
    \centering
    \includegraphics[width=\linewidth]{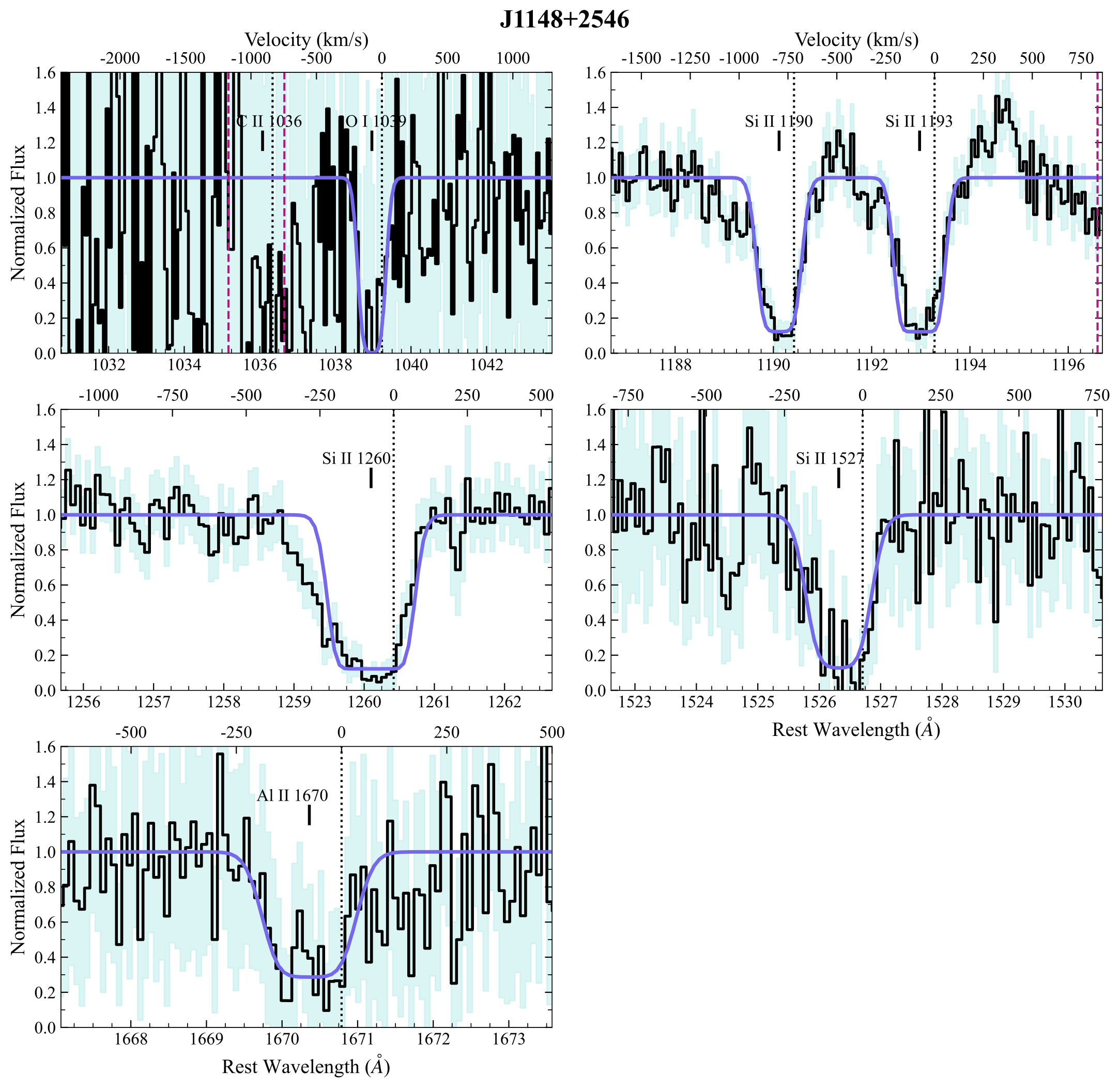}
    \figcaption{Simultaneous one-component fit to the LIS features in the CLASSY spectrum for J1148+2546. See description from Figure \ref{fig:firstfit} for details.}
\end{figure*}

\begin{figure*}
    \centering
    \includegraphics[width=\linewidth]{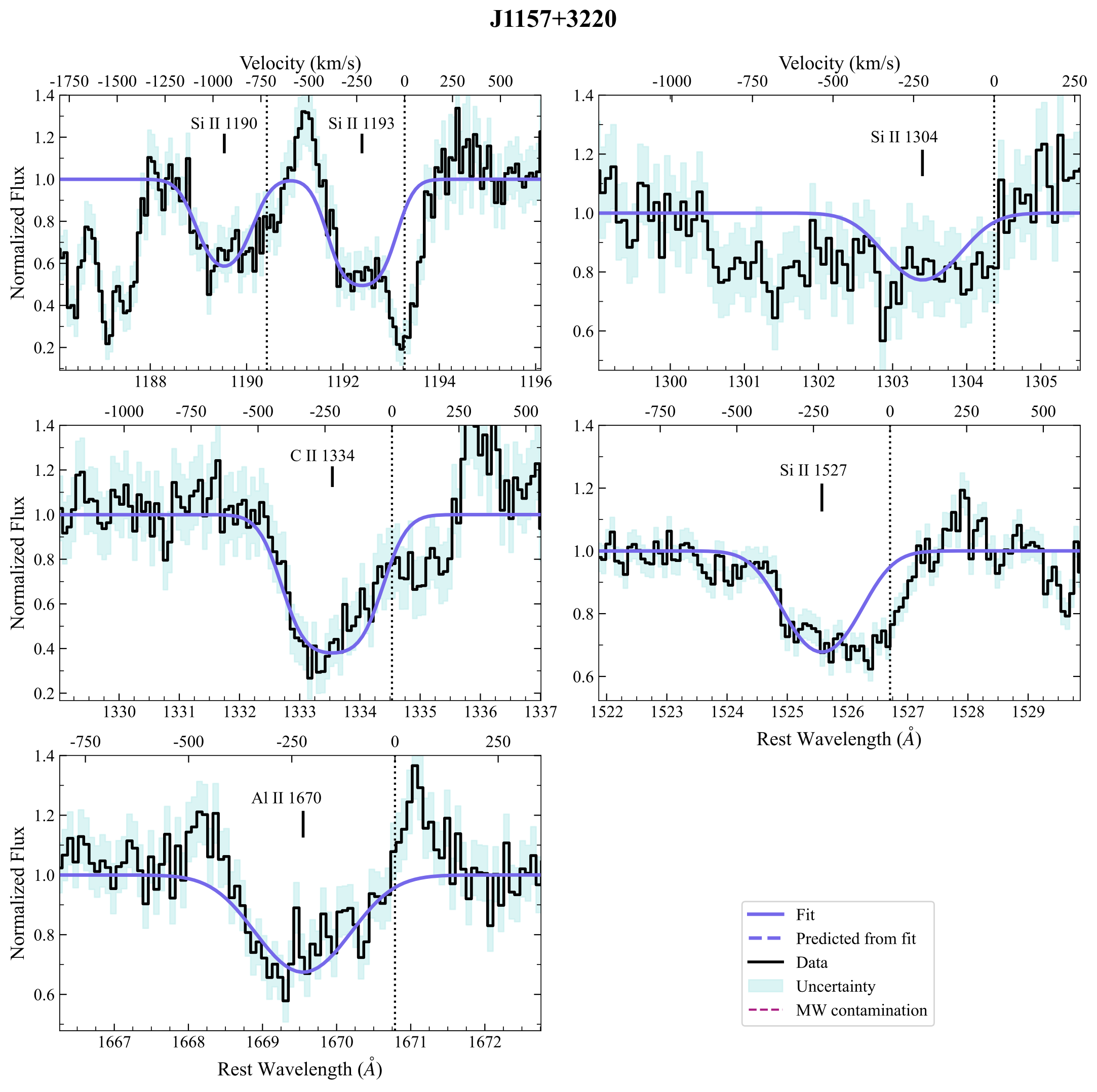}
    \figcaption{Simultaneous one-component fit to the LIS features in the CLASSY spectrum for J1157+3220. See description from Figure \ref{fig:firstfit} for details.}
\end{figure*}

\begin{figure*}
    \centering
    \includegraphics[width=\linewidth]{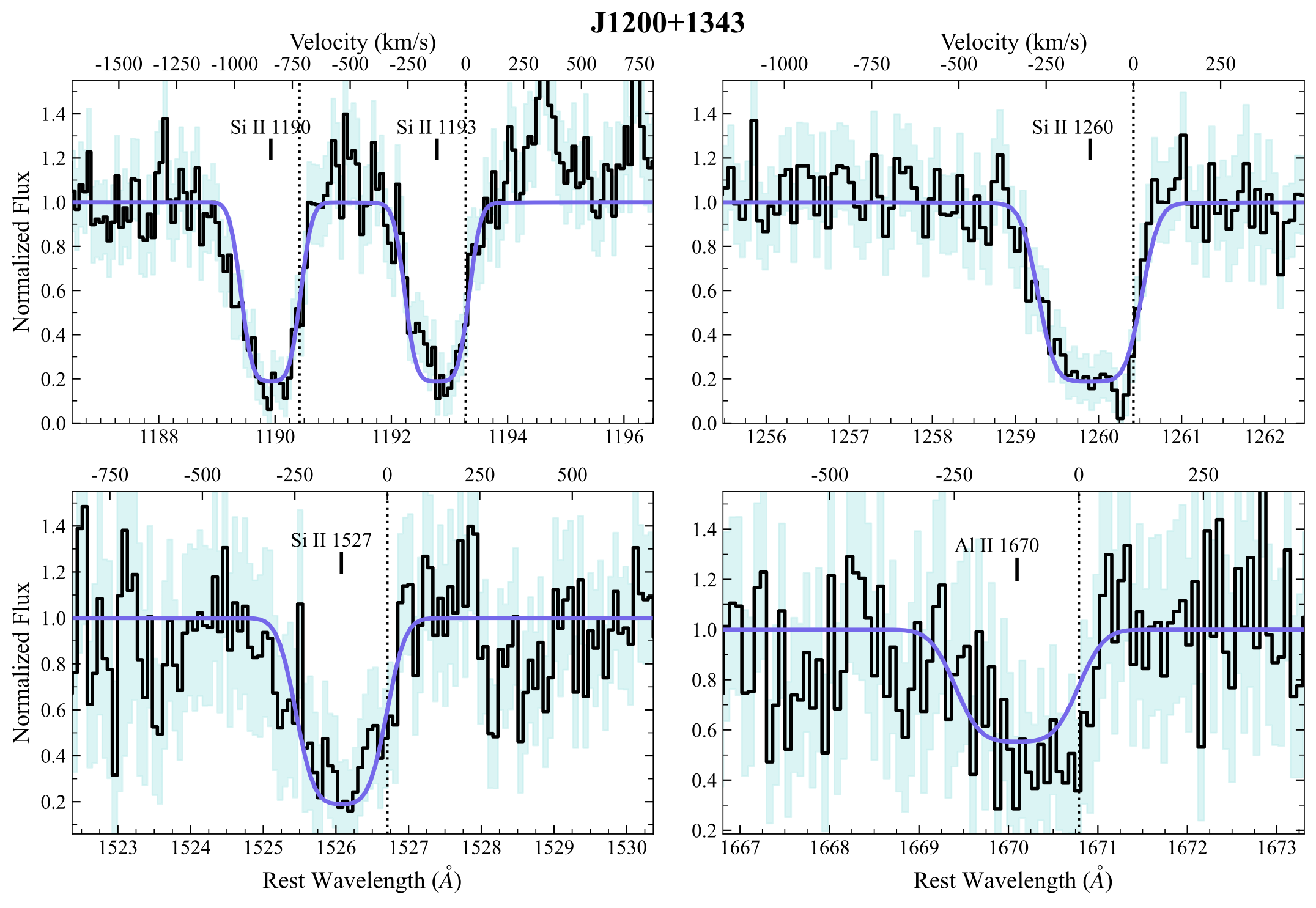}
    \figcaption{Simultaneous one-component fit to the LIS features in the CLASSY spectrum for J1200+1343. See description from Figure \ref{fig:firstfit} for details.}
\end{figure*}

\begin{figure*}
    \centering
    \includegraphics[width=\linewidth]{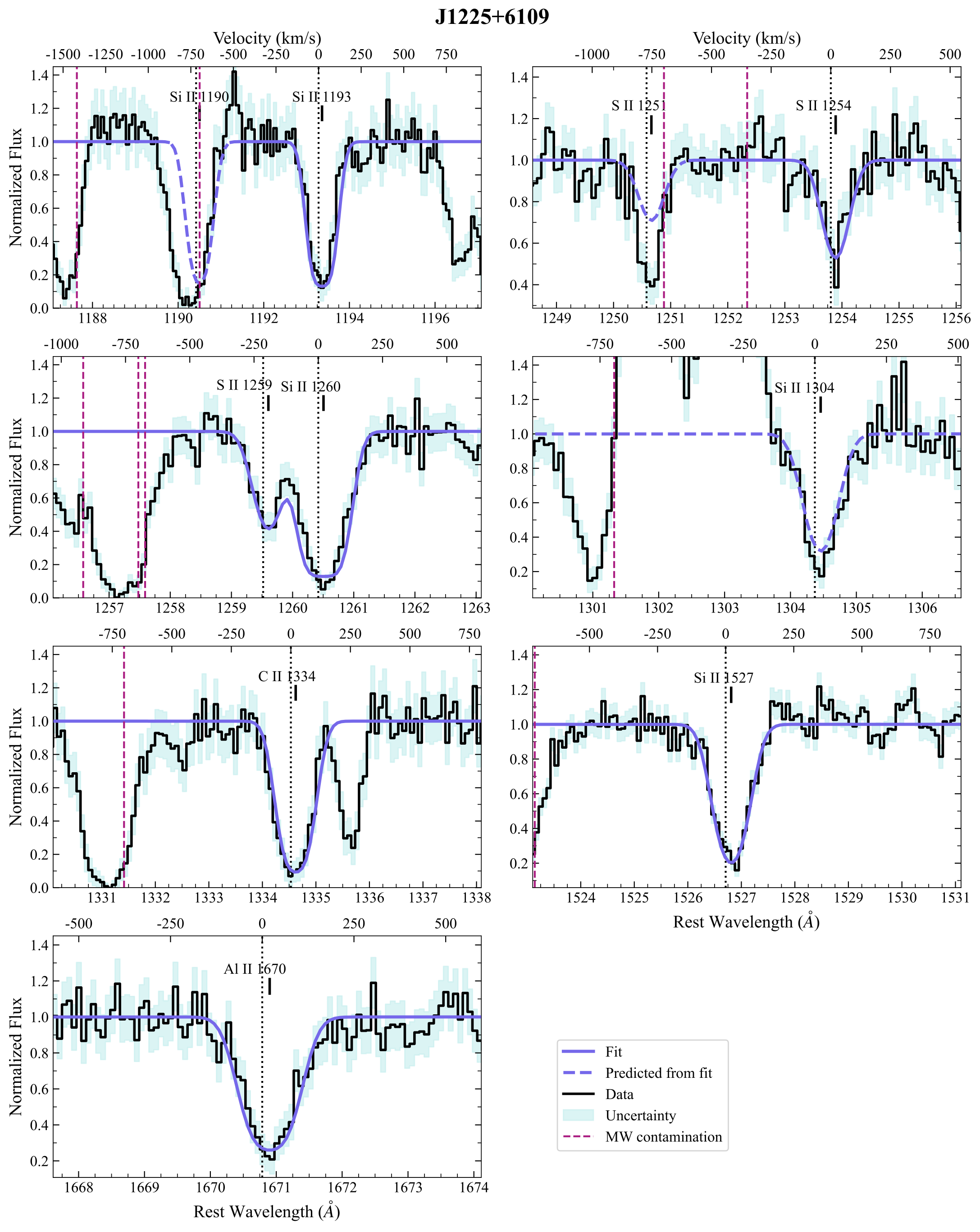}
    \figcaption{Simultaneous one-component fit to the LIS features in the CLASSY spectrum for J1225+6109. See description from Figure \ref{fig:firstfit} for details.}
\end{figure*}

\begin{figure*}
    \centering
    \includegraphics[width=\linewidth]{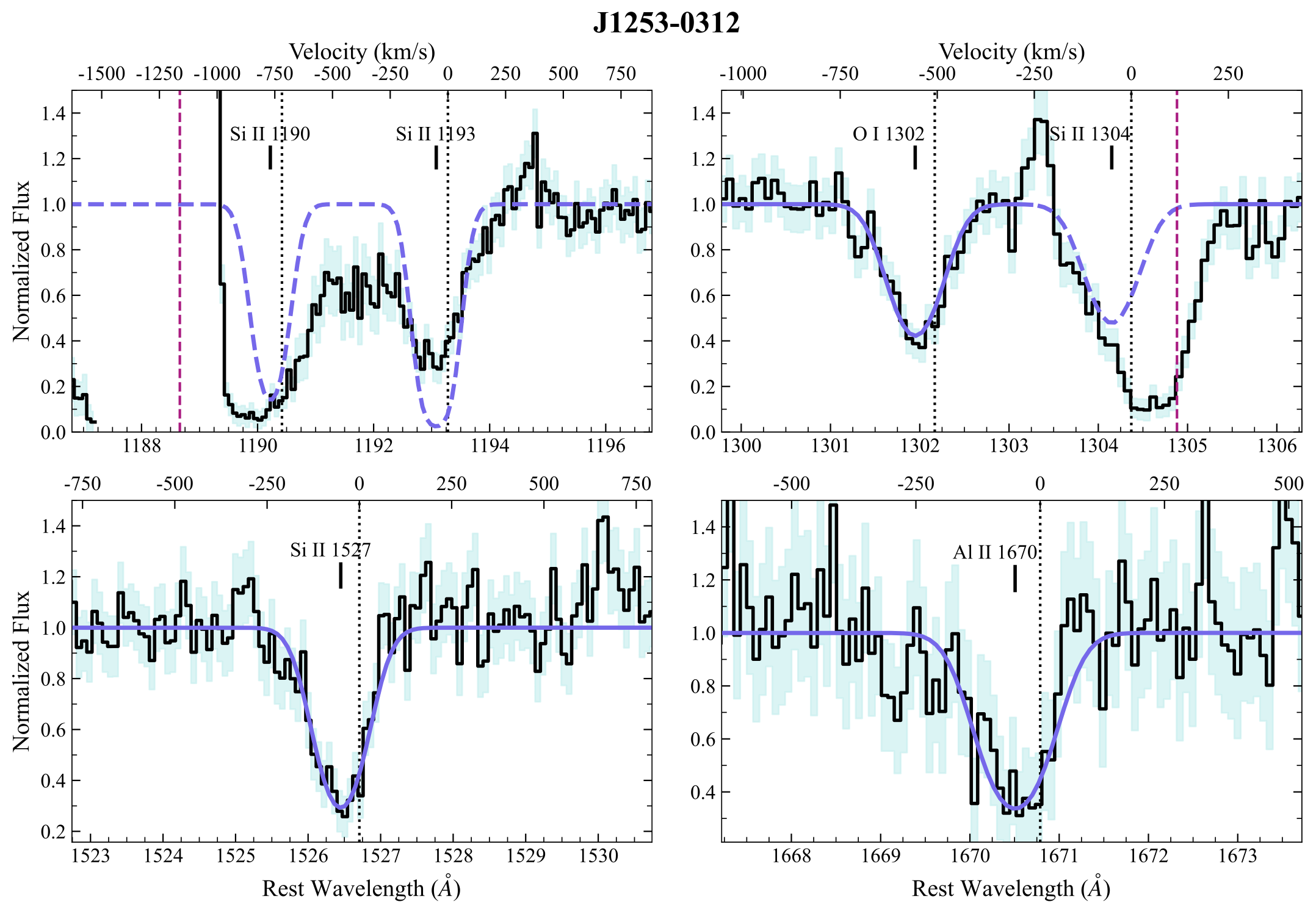}
    \figcaption{Simultaneous one-component fit to the LIS features in the CLASSY spectrum for J1253-0312. See description from Figure \ref{fig:firstfit} for details.}
\end{figure*}

\begin{figure*}
    \centering
    \includegraphics[width=\linewidth]{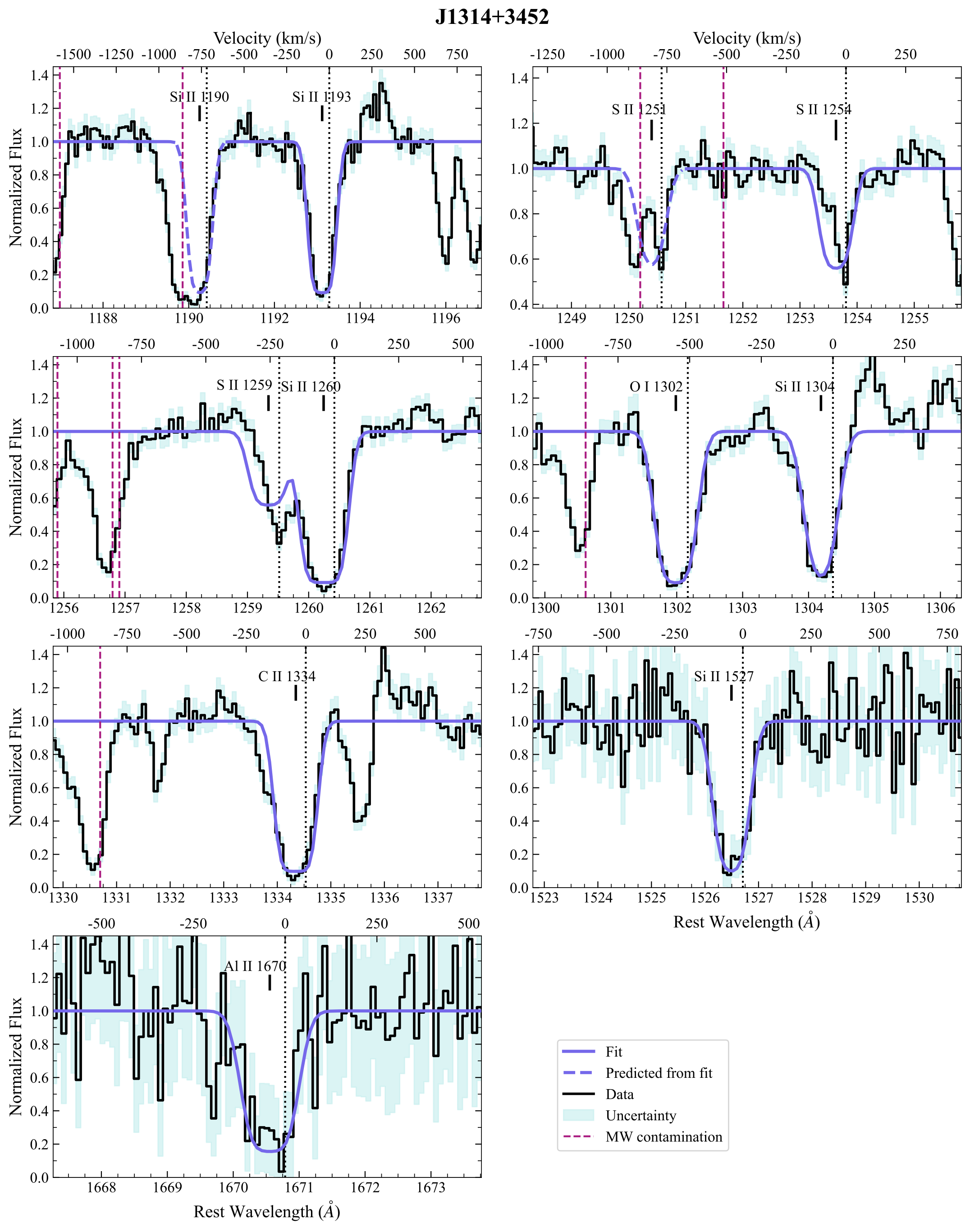}
    \figcaption{Simultaneous one-component fit to the LIS features in the CLASSY spectrum for J1314+3452. See description from Figure \ref{fig:firstfit} for details.}
\end{figure*}

\begin{figure*}
    \centering
    \includegraphics[width=\linewidth]{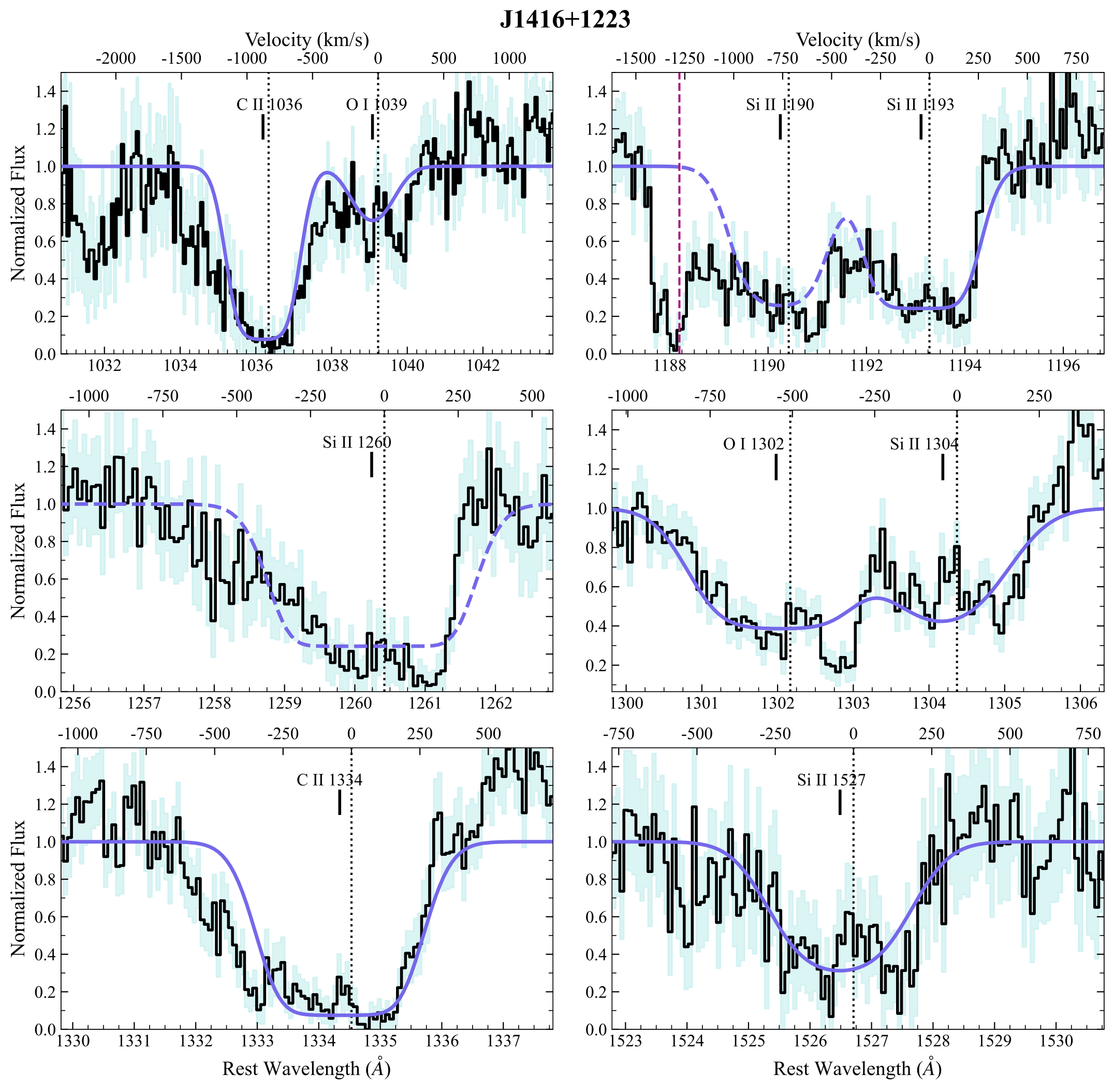}
    \figcaption{Simultaneous one-component fit to the LIS features in the CLASSY spectrum for J1416+1223. See description from Figure \ref{fig:firstfit} for details.}
\end{figure*}

\begin{figure*}
    \centering
    \includegraphics[width=\linewidth]{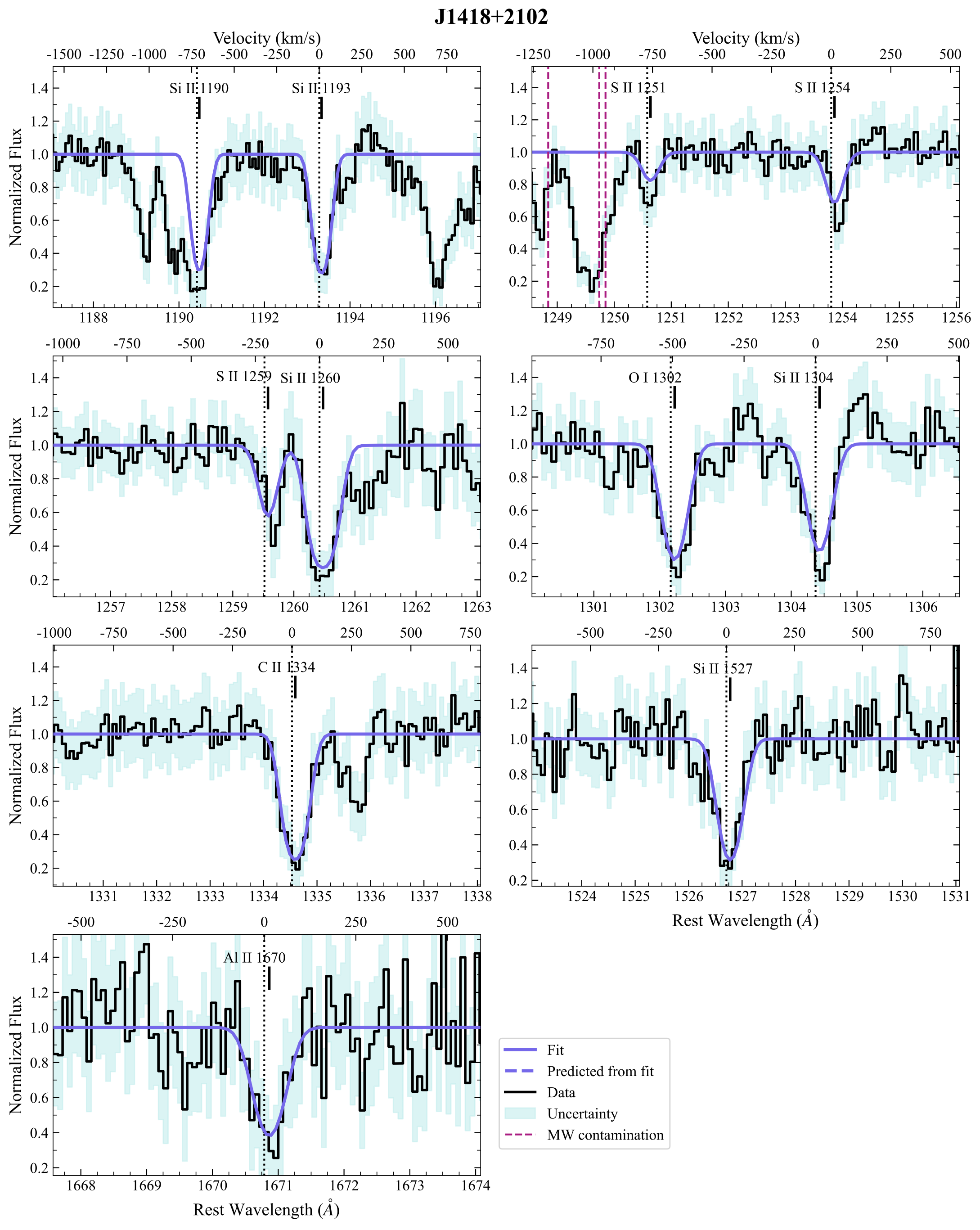}
    \figcaption{Simultaneous one-component fit to the LIS features in the CLASSY spectrum for J1418+2102. See description from Figure \ref{fig:firstfit} for details.}
\end{figure*}

\begin{figure*}
    \centering
    \includegraphics[width=\linewidth]{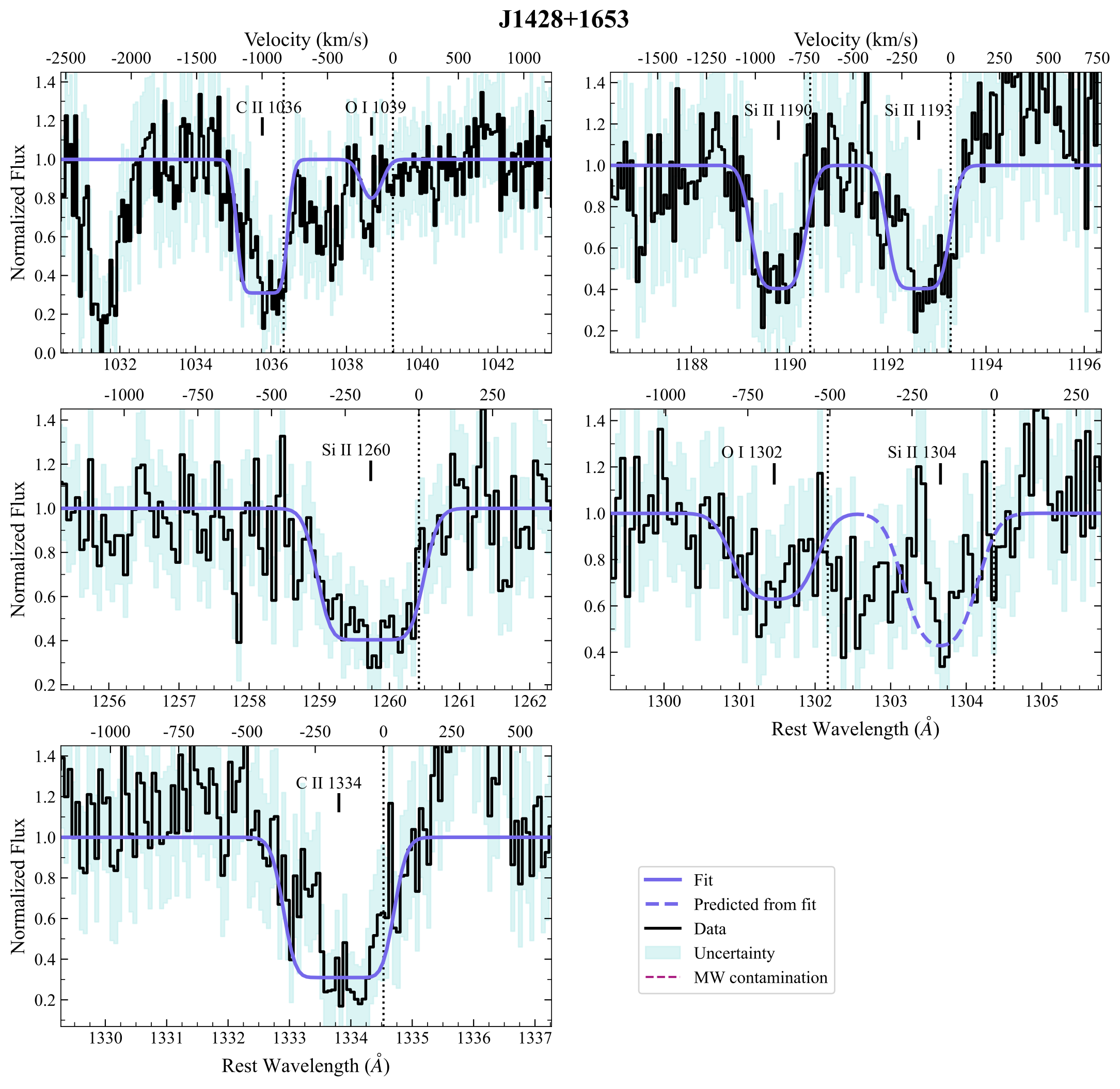}
    \figcaption{Simultaneous one-component fit to the LIS features in the CLASSY spectrum for J1428+1653. See description from Figure \ref{fig:firstfit} for details.}
\end{figure*}

\begin{figure*}
    \centering
    \includegraphics[width=\linewidth]{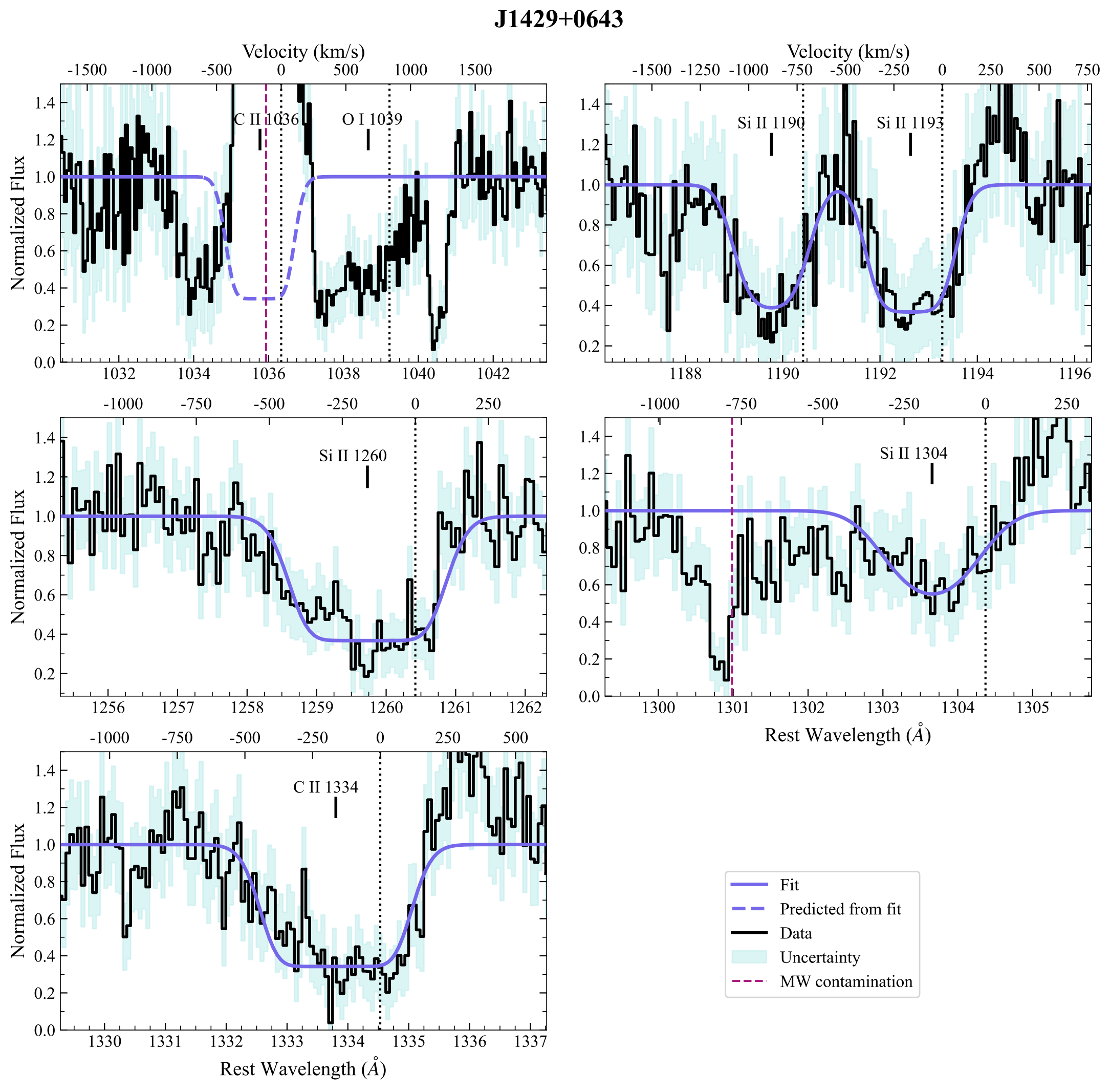}
    \figcaption{Simultaneous one-component fit to the LIS features in the CLASSY spectrum for J1429+0643. See description from Figure \ref{fig:firstfit} for details.}
\end{figure*}

\begin{figure*}
    \centering
    \includegraphics[width=\linewidth]{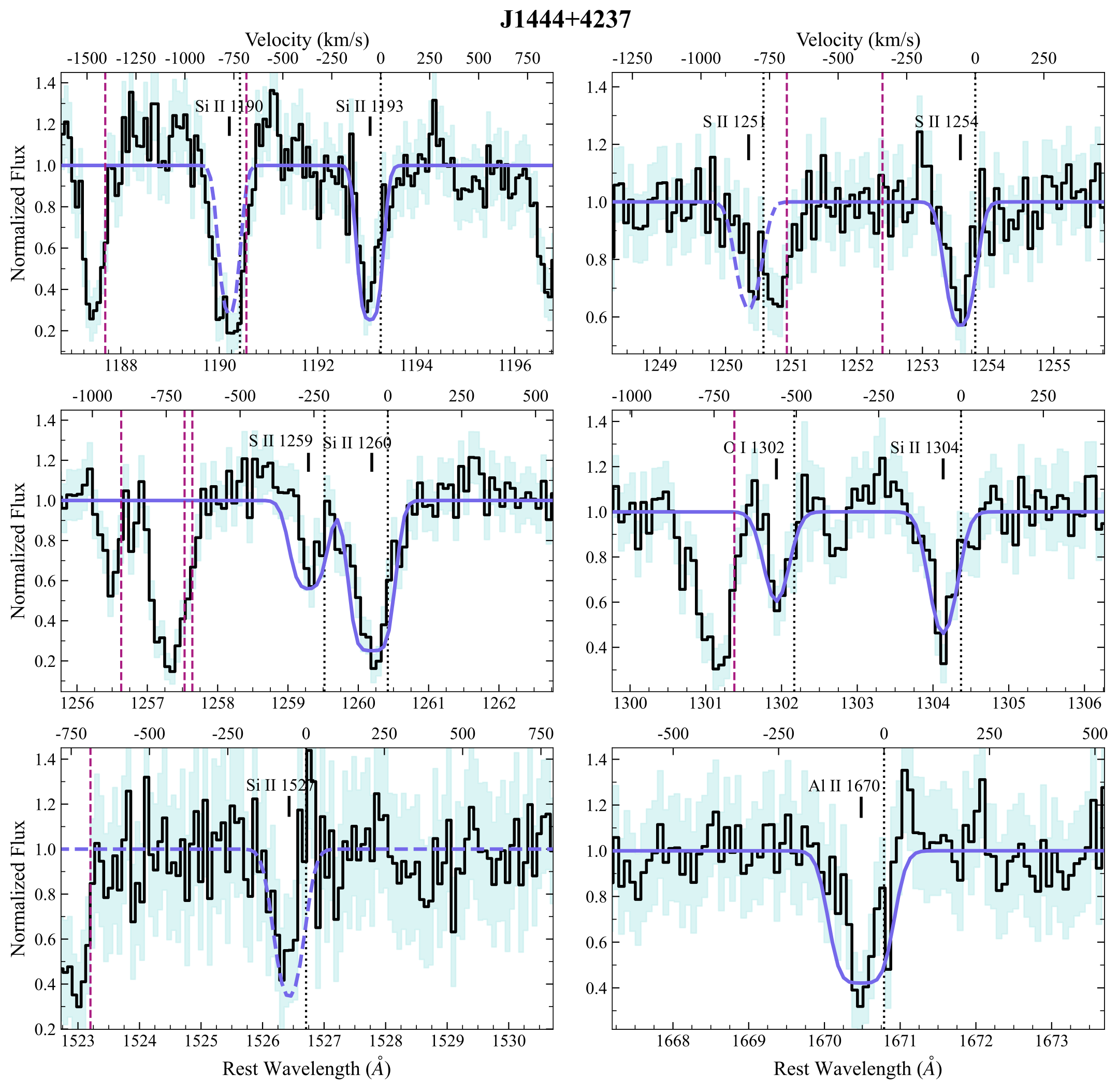}
    \figcaption{Simultaneous one-component fit to the LIS features in the CLASSY spectrum for J1444+4237. See description from Figure \ref{fig:firstfit} for details.}
\end{figure*}

\begin{figure*}
    \centering
    \includegraphics[width=\linewidth]{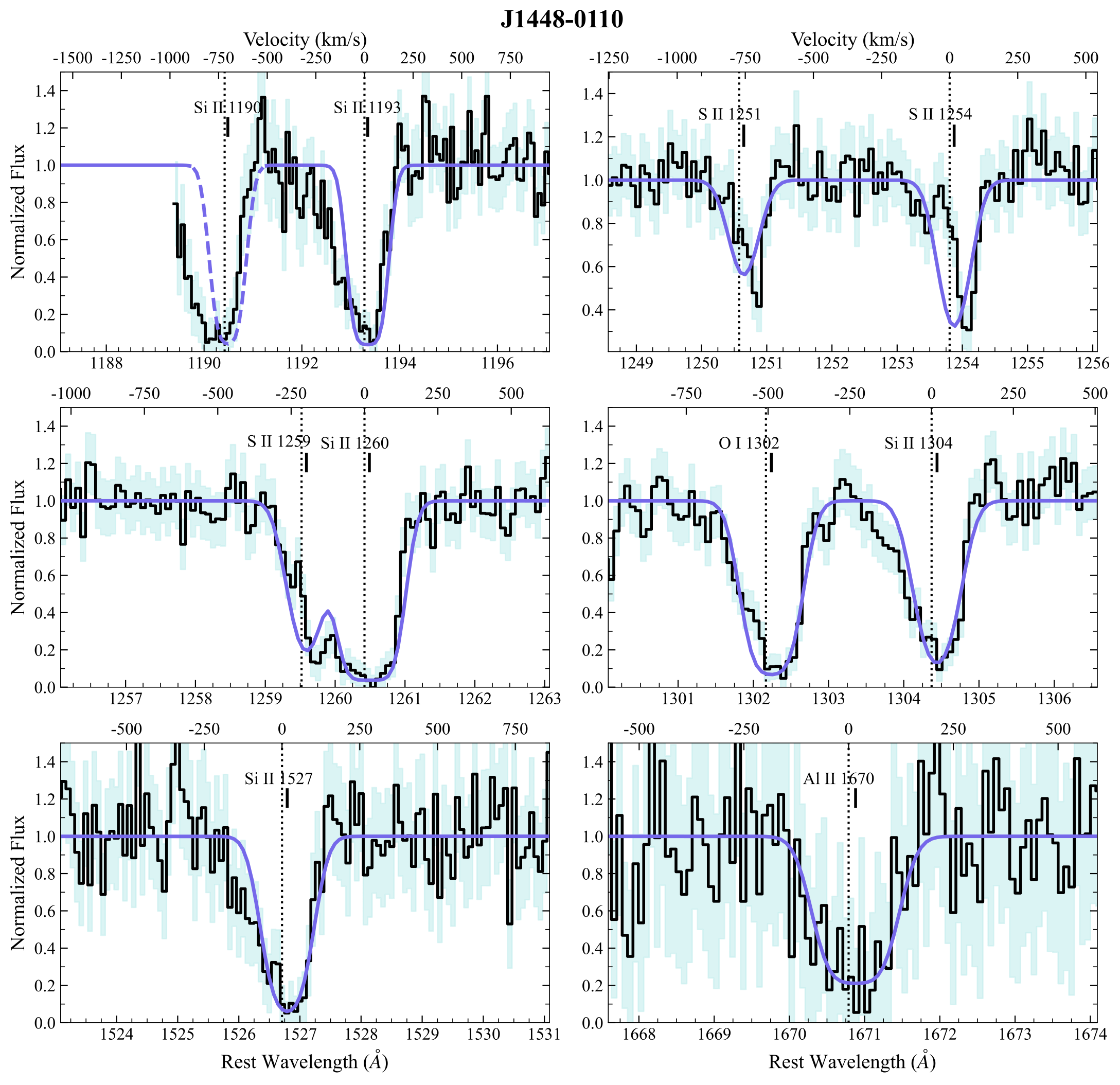}
    \figcaption{Simultaneous one-component fit to the LIS features in the CLASSY spectrum for J1448-0110. See description from Figure \ref{fig:firstfit} for details.}
\end{figure*}

\begin{figure*}
    \centering
    \includegraphics[width=\linewidth]{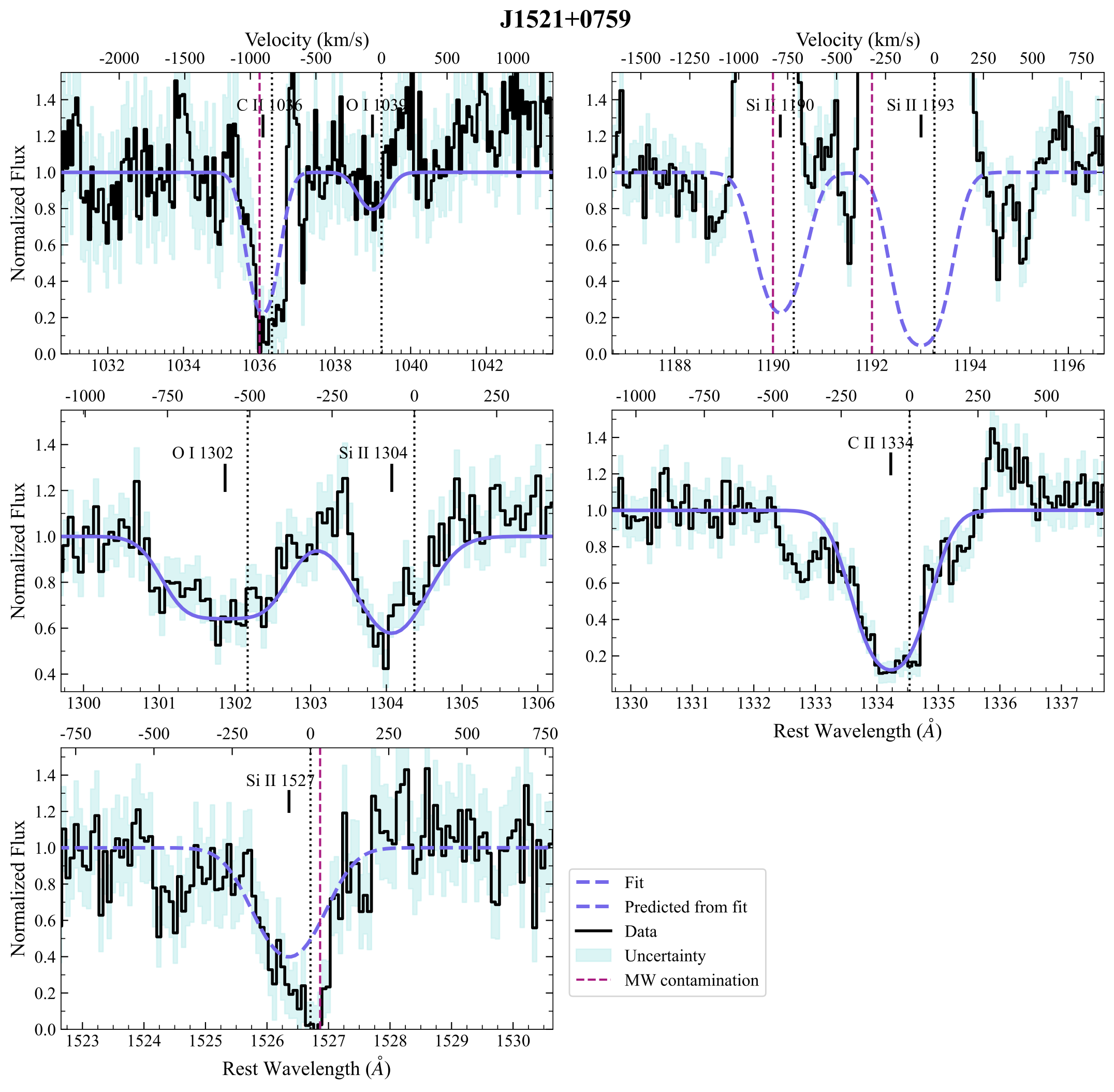}
    \figcaption{Simultaneous one-component fit to the LIS features in the CLASSY spectrum for J1521+0759. See description from Figure \ref{fig:firstfit} for details.}
\end{figure*}

\begin{figure*}
    \centering
    \includegraphics[width=\linewidth]{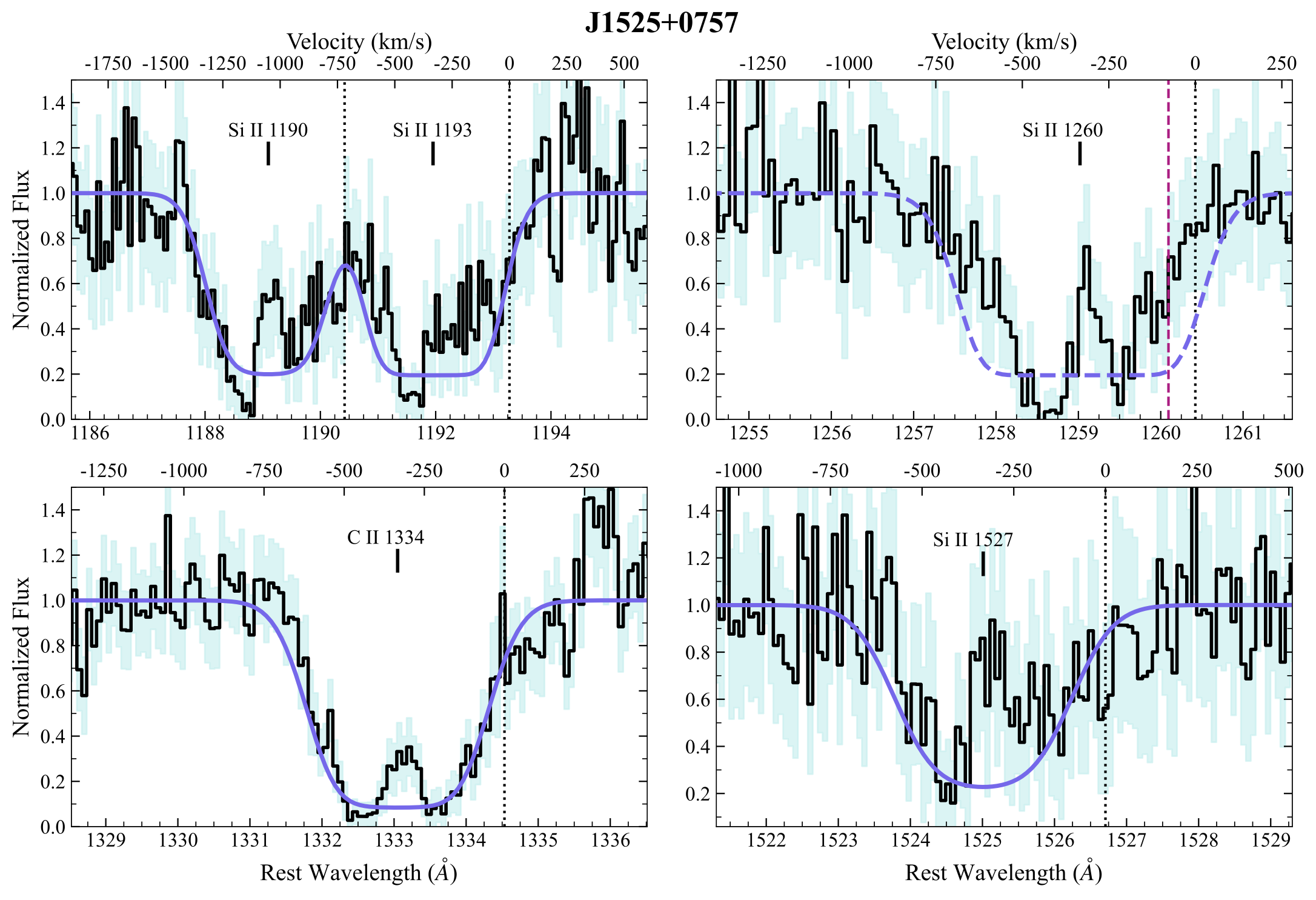}
    \figcaption{Simultaneous one-component fit to the LIS features in the CLASSY spectrum for J1525+0757. See description from Figure \ref{fig:firstfit} for details.}
\end{figure*}

\begin{figure*}
    \centering
    \includegraphics[width=\linewidth]{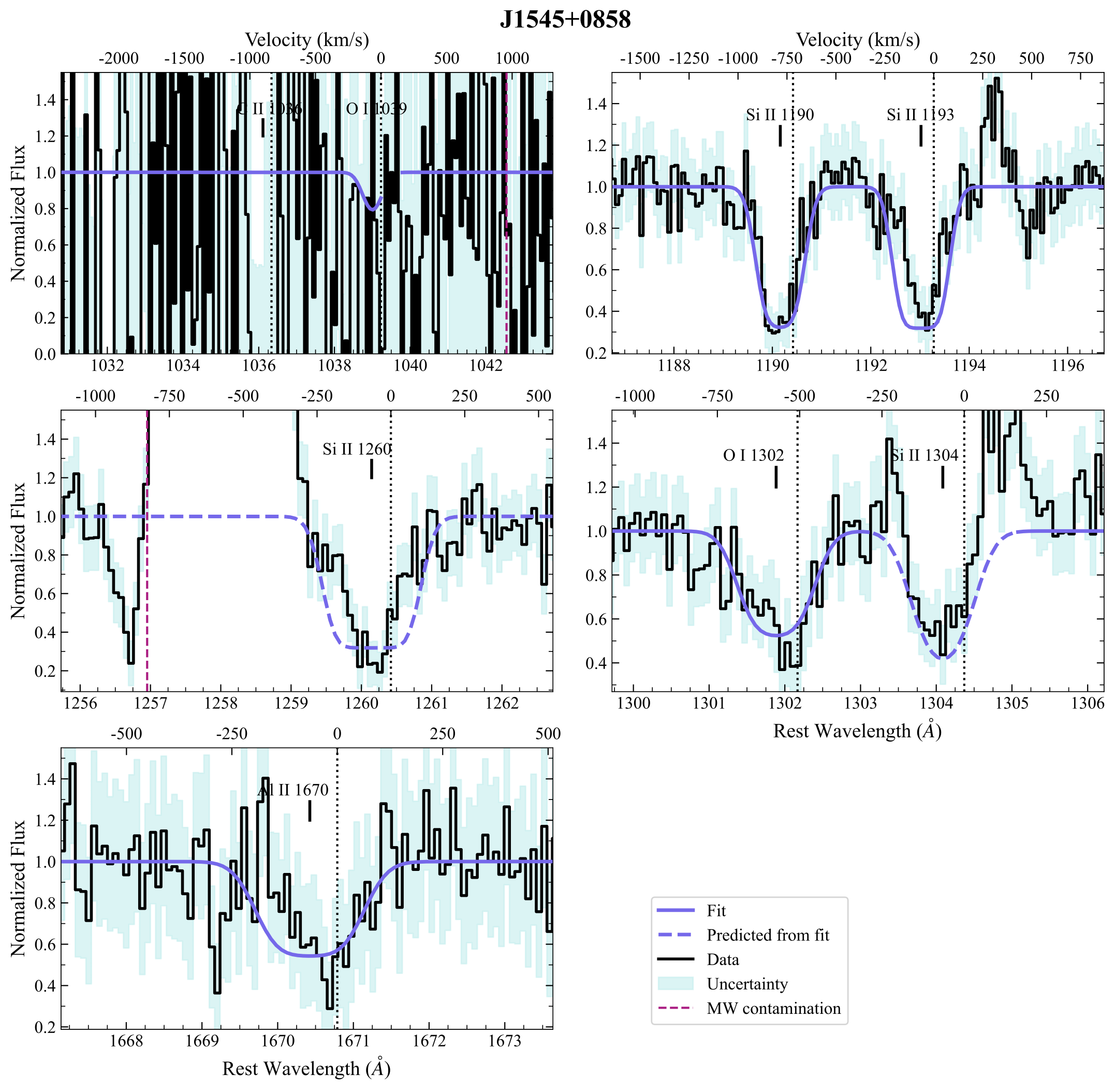}
    \figcaption{Simultaneous one-component fit to the LIS features in the CLASSY spectrum for J1545+0858. See description from Figure \ref{fig:firstfit} for details.}
\end{figure*}

\begin{figure*}
    \centering
    \includegraphics[width=\linewidth]{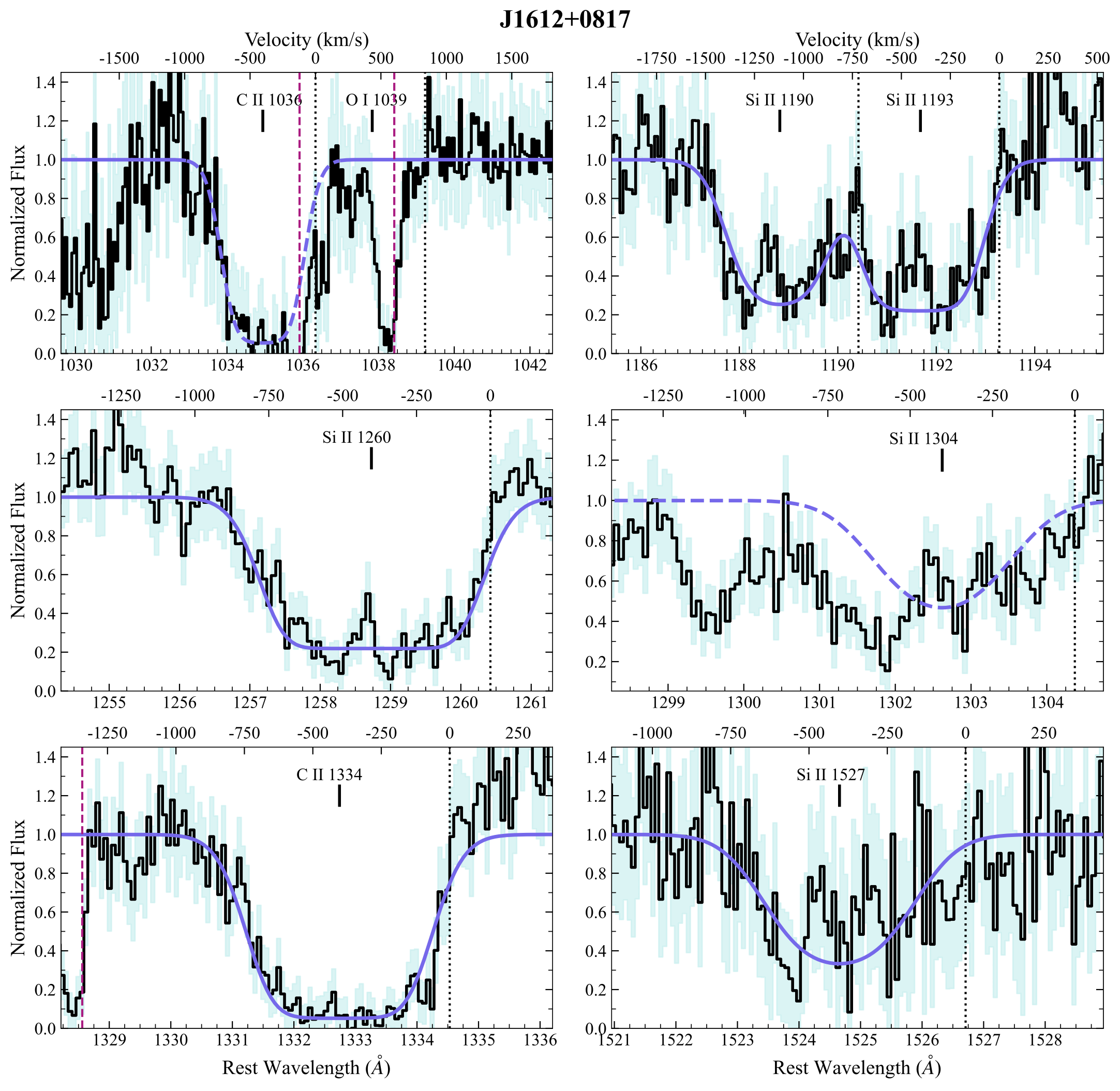}
    \figcaption{Simultaneous one-component fit to the LIS features in the CLASSY spectrum for J1612+0817. See description from Figure \ref{fig:firstfit} for details. \label{fig:lastfit}}
\end{figure*}

\end{appendix}
\end{document}